\documentclass[preprintnumbers,amsmath,amssymb]{revtex4}
\usepackage{color}
\usepackage{bm, graphicx, amsmath}
\usepackage{hyperref}
\usepackage{times}
\newenvironment{myfont}{\fontfamily{<cmtt>}\selectfont}{\par}
\DeclareTextFontCommand{\textmf}{\myfont}

\begin{document}

\title{Introduction to Coding Quantum Algorithms: \\ A Tutorial Series Using Pyquil}
\author{ Daniel Koch$^{\ast}$, Laura Wessing, Paul M. Alsing}
\affiliation{Air Force Research Lab, Information Directorate, Rome, New York}

\begin{abstract}

As the field of Quantum Computing continues to grow, so too has the general public's interest in testing some of the publicly available quantum computers. However, many might find learning all of the supplementary information that goes into quantum algorithms to be a daunting task, and become discouraged.  This tutorial is a series of lessons, aimed to teach the basics of quantum algorithms to those who may have little to no background in quantum physics and/or minimal knowledge of coding in python. Each lesson covers select physics/coding topics needed for writing quantum algorithms, eventually building up a toolset for tackling more and more challenging quantum algorithms. This tutorial series is designed to provide readers from any background with two services: 1) A concise and thorough understanding of some of the most popular/academically important quantum algorithms. 2) A fluent understanding of how to write code for quantum algorithms, using Rigetti's publicly available Pyquil.

\end{abstract}

\maketitle

\vspace*{1cm}

*corresponding author: daniel.koch.10.ctr@us.af.mil
\\

Code files available upon request.

\vspace*{\fill}

$\hspace{7.5cm}$ Approved for public release: 88ABW-2019-0567; distribution unlimited.

\pagebreak

\vspace*{3cm}

\section*{\Large{Table of Contents}}

Lesson 1 - Intro to Programs and Measurements.............................................................................................................................3
\\

Lesson 2 - Creating More Complex Programs...............................................................................................................................14
\\

Lesson 3 - Gates Provided by Pyquil..............................................................................................................................................21
\\

Lesson 4 - Our Custom Functions...................................................................................................................................................36
\\

Lesson 5.1 - Intro to Quantum Algorithms (Deutsch)....................................................................................................................47
\\

Lesson 5.2 - Deutsch-Jozsa \& Bernstein-Vazirani Algorithms.......................................................................................................58
\\

Lesson 5.3 - Simon's Algorithm.....................................................................................................................................................72
\\

Lesson 5.4 - The Grover Search.....................................................................................................................................................82
\\

Lesson 6 - Quantum Fourier Transformation.................................................................................................................................99
\\

Bibliography.................................................................................................................................................................................112
\\

Appendix (Our\_Pyquil\_Functions.py)..........................................................................................................................................113
\\

\pagebreak

\section*{\Large{Lesson 1 - Intro to Programs and Measurements}}
--------------------------------------------------------------------------------------------------------------------------------------------------------
\\

Welcome to lesson 1 in this tutorial series.  These lessons are designed to supplement existing literature in the field of Quantum Computing and Quantum Information \cite{NC,KLM,C,NO,YM,BH}, with an emphasis in coding quantum algorithms.
\\

This first lesson is designed to introduce you to Pyquil's formalism for running quantum circuits, specifically creating quantum systems using Programs and measurements. This lesson is recommended for first time users of Pyquil. If you do not have Pyquil ready for use on your computer, please check out the installation guide:
\\

http://docs.rigetti.com/en/stable/start.html
\\

https://github.com/rigetticomputing/pyquil
\\

For those who are already familiar with the basics of Pyquil, I recommend starting with lesson 4, which will cover some additional custom functions necessary before proceeding onto the quantum algorithms.
\\

--------------------------------------------------------------------------------------------------------------------------------------------------------
\\

In order to make sure that all cells of code run properly throughout this lesson, please run the following cell of code below:

\begin{figure}[h]
\centering
\includegraphics[scale=.65]{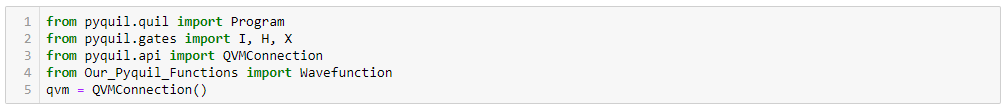}
\end{figure}

\section*{\large{ Creating Our First Quantum State }}
\centerline{---------------------------------------------------------------------------------------------------------------------------------}

Pyquil is a python language that allows us to create algorithms for a quantum computer. These algorithms tell the quantum computer what kinds of quantum systems to create, and then manipulate them with gates. Compared to classical algorithms, we will find that programming for a quantum computer is quite different, requiring us to face many new limitations posed on us by quantum systems. In turn however, these quantum algorithms allow us to solve problems much faster than any classical approach.
\\

Let's start with the simplest quantum system there is:

$$ |\Psi \rangle = | 0 \rangle $$

This is a quantum system of 1 qubit, in the state $|0\rangle$. Not terribly exciting, but we have to start somewhere! Consider this the "Hello World!" to programing with qubits.
\\

Let's see the code that generates this system, and then dissect its components:

\begin{figure}[h]
\centering
\includegraphics[scale=.65]{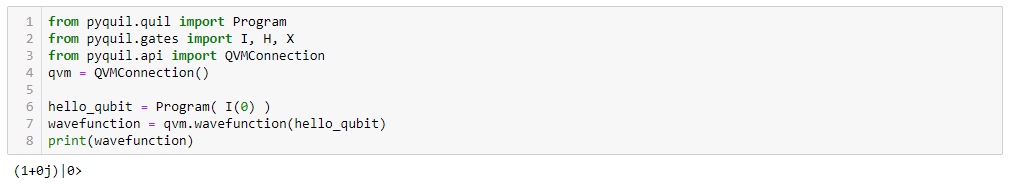}
\end{figure}

Congrats, you've just created your first quantum system using Pyquil!

$$*\textit{ crickets chirping}*$$

Okay, it's not a very exciting result, but there are already a lot of things going on in this code. Starting with our imports:
\\

--------------------------------------------------------------------------------------------

from pyquil.quil import $\textbf{Program}$

from pyquil.api import $\textbf{QVMConnection}$

--------------------------------------------------------------------------------------------
\\

These imports are what allow us to create and see the quantum system we are working with. The underlying quantum language behind Pyquil is Quil (Quantum Instruction Language), which Pyquil allows us to work with in a much more user-friendly syntax.
\\

$\textbf{Program}$ -- this is a class that can be thought of as our "instructions" for the quantum system. We will $\textit{store}$ operations into programs, sort of like a list, and then call upon other functions to run these operations later.
\\

$\textbf{QVMConnection}$ -- this is a cloud-based virtual machine that will $\textit{run}$ our quantum algorithms. It is a classical simulator, which means that it comes with a lot of handy features that are normally impossible on real quantum computers (QPUs - quantum processing units).
\\

The goal of this lesson is to become familiar with some of the basics of building and running $\textmf{programs}$, so don't worry if all of these new terms don't make sense just yet. Now, let's see the remaining three lines of code:
\\

--------------------------------------------------------------------------------------------

qvm = QVMConnection()

hello\_qubit = Program()

print(qvm.wavefunction(hello\_qubit))

--------------------------------------------------------------------------------------------
\\

The first line of code stores the $\textmf{QVMConnection}$ as '$\textmf{qvm}$', for later use. The second line of code creates a new $\textmf{program}$, with no instructions, and stores it to a variable we've called 'hello\_qubit'. And the last line of code uses a function called $\textbf{wavefunction}$, an extension of $\textmf{qvm}$, to create a string that shows our quantum state stored in hello\_qubit. This string is a user-friendly version of the state of our quantum system, meant for viewing.
\\

By default, when we create a $\textmf{program}$ with no instructions, Pyquil creates the state $|0\rangle$. Thus, when we use $\textmf{wavefunction}$() to look at the system, we get exactly the state $ |\Psi \rangle = | 0 \rangle $ we wanted.

\section*{\large{ Let's Bump Up the Qubits}}
\centerline{---------------------------------------------------------------------------------------------------------------------------------}

1 qubit is pretty exciting (I know), so let's try creating a state of multiple qubits.
\\

In the previous example we created a system of a single qubit, which by default is started in the state $01\rangle$. This was done by simply calling the function $\textmf{Program}$ with no arguments. While this is okay, it's better to be more precise when we initialize a quantum state. Consider the following two ways of creating the same system:

\begin{figure}[h]
\centering
\includegraphics[scale=.65]{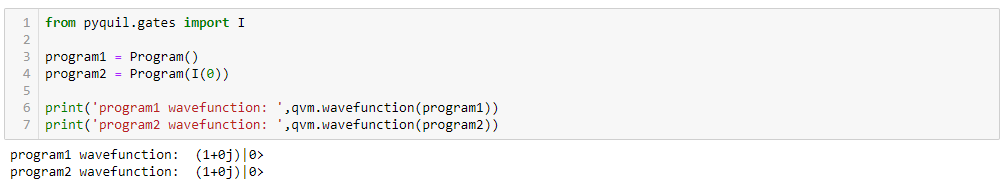}
\end{figure}

As we can see, both ways produce the same result, but the second $\textmf{program}$ has the argument $\textmf{I}$(0). This argument is actually a gate, the Identity operator, which we will discuss shortly, but essentially leaves the state of a qubit unchanged.
\\

Note: we are using python notation here, where all ordering starts from 0,1,2,3... So the first qubit will always be 'qubit $0$', the second qubit will be 'qubit $1$', and so on. Thus, $\textmf{I}$(0) is a gate operation on qubit 0.
\\

To verify that program2 is really different from program1, Pyquil allows us to print $\textmf{programs}$ to console. This is simply done by passing the $\textmf{program}$ variable in a print function:

\begin{figure}[h]
\centering
\includegraphics[scale=.65]{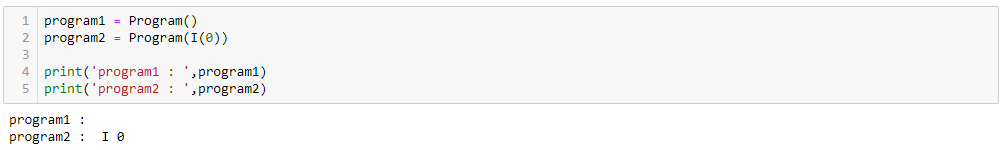}
\end{figure}

Both $\textmf{programs}$ produce the same wavefunction in the end, but program2 definitely has an extra step: apply the identity gate to qubit 0.
\\

So why would we prefer to include this extra operator if it doesn't do anything? In short, it is to make our code easier to read. $\textmf{Program}$( ) with no arguments is a little... sloppy, but $\textmf{Program}$( $\textmf{I}$(0) ) is much more clear - put qubit 0 into the system and leave it in the state $|0\rangle$.
\\

More importantly however, if we want to create systems with multiple qubits, we need a basic way of telling our program to initialize them. The identity gate allows us to call as many qubits as we want, in the form: $\textmf{I}$(0), $\textmf{I}$(1), $\textmf{I}$(2), ... $\textmf{I}$($n$), which in turn creates a basic quantum state of all the qubits in the state $|000...0\rangle$.

\begin{figure}[h]
\centering
\includegraphics[scale=.65]{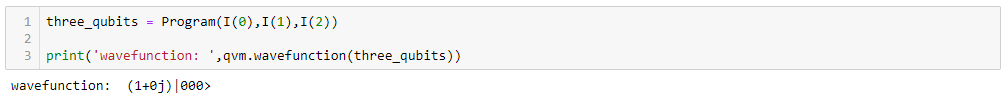}
\end{figure}

This code creates a quantum system of three qubits, all in the state $|0\rangle$. Note that there actually is a limit to how many qubits we can create, dictated by $\textmf{QVM}$, but in principle we can consider that this coding $\textit{syntax}$ is limitless.

\section*{\large{ Our First Superposition State }}
\centerline{---------------------------------------------------------------------------------------------------------------------------------}

Now that we can create multiple qubits, we want to $\textit{do something}$ with them. In quantum algorithms, that something is applying gates and making measurements. We've already seen one gate to far, $\textmf{I}$ -- the identity operator, so let's add two more to our toolbox.
\\

Pyquil comes with several pre-programed gates from the file $\textbf{pyquil.gates}$. For a complete list and explanation of all these gates, see lesson 3. Here, we are only going to need the following gates:

\begin{figure}[h]
\centering
\includegraphics[scale=.65]{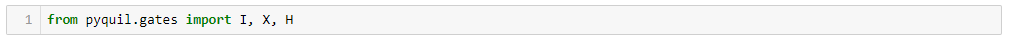}
\end{figure}

For this intro lesson we will only be using the gates I, X, and H, and we will briefly explain them here:

\section*{\large{ I (Identity Gate) }}

As we already saw, this gate acts on a single qubit, and leaves its state unchanged. The matrix for this gate is:

$$
\begin{bmatrix}
1 & 0\\
0 & 1
\end{bmatrix}
$$

While perhaps uninteresting in itself, the I gate is still an essential component for algorithms. We've already seen it used so far to initialize nice simple quantum states, and later we shall see it used in conjunction with other gates, for larger multi-qubit operations.

\section*{\large{ X (NOT Gate) }}

This gate is our quantum analog to the NOT gate, which flips a classical bit between 0 and 1. Here, it achieves the same effect with the states $|0 \rangle$ and $|1 \rangle$. The matrix for this operation is given by:

$$
\begin{bmatrix}
0 & 1\\
1 & 0
\end{bmatrix}
$$

Although it may appear that the $\textmf{X}$ gate is perfect analog to the classical NOT gate, quantum mechanics prevents it from being so. In particular, when we start to create superposition states, we will see that using this gate to flip qubits becomes a bit tricky.

\section*{\large{ H (Hadamard Gate) }}

This gate is going to allow us to create our first superposition state. In particular, the Hadamard gate takes a qubit and splits it into a 50-50 probability distribution between the states $|0\rangle$ and $|1\rangle$. Mathematically, it looks like this:

$$ H \hspace{.08cm} |\hspace{.05cm}0\rangle \hspace{.15cm} = \hspace{.15cm} \frac{1}{\sqrt{2}} \big{(} \hspace{.1cm}|\hspace{.05cm}0\rangle \hspace{.1cm}+\hspace{.1cm} |\hspace{.05cm}1\rangle \hspace{.06cm} \big{)} $$

$$ H \hspace{.08cm} |\hspace{.05cm}1\rangle \hspace{.15cm} = \hspace{.15cm} \frac{1}{\sqrt{2}} \big{(} \hspace{.1cm}|\hspace{.05cm}0\rangle \hspace{.1cm}-\hspace{.1cm} |\hspace{.05cm}1\rangle \hspace{.06cm} \big{)} $$

which is accomplished by the following matrix:

$$
\begin{bmatrix}
1 & 1\\
1 & -1
\end{bmatrix}
$$

Let's see an example:

\begin{figure}[h]
\centering
\includegraphics[scale=.65]{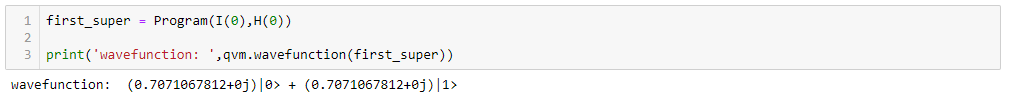}
\end{figure}

Sure enough, our qubit is in a superposition state! Our qubit has a 50\% chance of being in the state $|0\rangle$ or $|1\rangle$.
\\

Note: The numbers attached to the states here are the system's amplitudes, not probabilities. When working with quantum states, probabilities are always the $\textit{observables}$ that we see, but the amplitudes are the inner workings that really matter. Here, each state has an amplitude of $\frac{1}{\sqrt{2}}$, which when squared, tells us that each state has a probability of $\frac{1}{2}$.
\\

Now let's try making a superposition state of 2 qubits:

\begin{figure}[h]
\centering
\includegraphics[scale=.65]{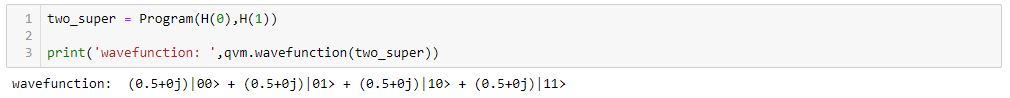}
\end{figure}

The wavefunction printed above shows an equal superposition of four states. These four states come from the following mathematical state:

$$
H \hspace{.08cm}|\hspace{.05cm}0\rangle \otimes H\hspace{.08cm}|\hspace{.05cm}0\rangle
$$

which is the tensor product of two seperate quantum states (one for each qubit in our system). A more common way of writing this is:

$$ (\hspace{.06cm} H \hspace{.06cm}|\hspace{.05cm}0\rangle \hspace{.06cm})\cdot( \hspace{.06cm}H \hspace{.06cm}|\hspace{.05cm}0\rangle \hspace{.06cm}) $$

$$\hspace{1.4cm}=\frac{1}{\sqrt{2}} \big{(} \hspace{.1cm}|\hspace{.05cm}0\rangle \hspace{.1cm}+\hspace{.1cm} |\hspace{.05cm}1\rangle \hspace{.06cm} \big{)} \hspace{.1cm} \cdot \hspace{.1cm} \frac{1}{\sqrt{2}} \big{(} \hspace{.1cm}|\hspace{.05cm}0\rangle \hspace{.1cm}+\hspace{.1cm} |\hspace{.05cm}1\rangle \hspace{.06cm} \big{)}$$

$$= \frac{1}{2} \big{(} \hspace{.1cm}|\hspace{.05cm}0\rangle \hspace{.1cm}+\hspace{.1cm} |\hspace{.05cm}1\rangle \hspace{.06cm} \big{)} \hspace{.02cm} \cdot \hspace{.02cm} \big{(} \hspace{.1cm}|\hspace{.05cm}0\rangle \hspace{.1cm}+\hspace{.1cm} |\hspace{.05cm}1\rangle \hspace{.06cm} \big{)}$$

$$\hspace{1.8cm}=\frac{1}{2} \big{(} \hspace{.1cm} |\hspace{.05cm}0\rangle \hspace{.04cm} |\hspace{.05cm}0\rangle \hspace{.06cm}+\hspace{.06cm} |\hspace{.05cm}0\rangle \hspace{.04cm} |\hspace{.05cm}1\rangle \hspace{.06cm}+\hspace{.06cm} |\hspace{.05cm}1\rangle \hspace{.04cm} |\hspace{.05cm}0\rangle \hspace{.06cm}+\hspace{.06cm} |\hspace{.05cm}1\rangle \hspace{.04cm} |\hspace{.05cm}1\rangle \hspace{.06cm} \big{)}$$

which is typically written using the standard shorthand:

$$\hspace{.3cm}=\frac{1}{2} \big{(} \hspace{.1cm} |\hspace{.05cm}00\rangle \hspace{.05cm}+\hspace{.05cm} |\hspace{.05cm}01\rangle \hspace{.05cm}+\hspace{.05cm} |\hspace{.05cm}10\rangle \hspace{.05cm}+\hspace{.05cm} |\hspace{.05cm}11\rangle \hspace{.06cm} \big{)}$$

And voila! We have our mixed state resulting from two qubits, each with a Hadamard gate applied to them. Recall that a single $\textmf{H}$ gate put our qubit into a 50/50 state between $|0\rangle$ and $|1\rangle$. Now, having two qubits undergo this gate, both of them in this 50/50 state, we get a combined system where any of the four individual combinations has a 25\% probability.
\\

As a side note, you may have noticed I did something a little different in the first line of code above:
\\

--------------------------------------------------------------------------------------------

two\_mixed = Program( H(0), H(1) )

--------------------------------------------------------------------------------------------
\\

Remember, when we assign a new qubit to a $\textmf{program}$, it starts off in the state $|0\rangle$ by default. But, our first operator on a new qubit does not need to be the Identity operator. We can just assume that the state of the qubit is $|0\rangle$, and skip right to the next operation. The only time it is necessary to initialize a qubit with $\textmf{iden}$ is when we want to specifically start it out in the state $|0\rangle$.
\\

Consider the following example where we would like only one of the qubits to start off in a superposition:

\begin{figure}[h]
\centering
\includegraphics[scale=.65]{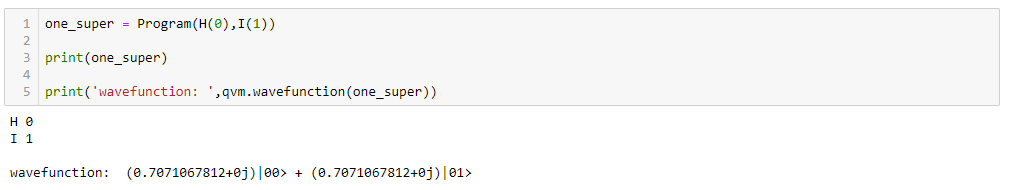}
\end{figure}

As shown above, qubit $0$ is initialized in a mixed state, while qubit $1$ remains in the state $|0\rangle$.
\\

If the ordering of the qubits looks weird to you, that's because Pyquil's $\textmf{wavefunction}$() writes the states of the qubits 'backwards' from normal convention. More specifically, Pyquil orders qubits in the same standard notation as classical bits, where the least significant bit (LSB) is the rightmost bit. For example, the number 5 in standard binary is 100, which can be understood as $ 4\cdot 1 + 2\cdot 0 + 1\cdot 0 $.
\\

In the example above, the left 0 in each state is actually qubit $1$, while the right values are qubit $0$. We can confirm this by printing the $\textmf{program}$, and seeing that qubit $0$ is indeed initialized with a Hadamard gate, and qubit $1$ with the Identity operator.
\\

While ordering qubits in this way is fine, there's a couple ways we can make our wavefunction reflect a more standard convention. The first method would be to actively switch around the values we assign to our qubits:

\begin{figure}[h]
\centering
\includegraphics[scale=.65]{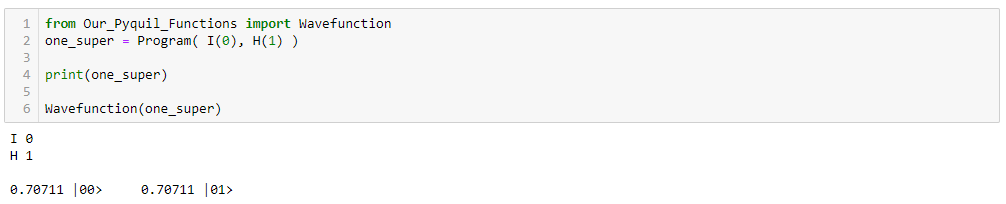}
\end{figure}

In this example, we are calling qubit $0$ 'q0', and assigning it the value 1, and vice versa for qubit $1$. This way, our code still reads pretty much the same way, and our output wavefunction matches standard convention, where qubit $0$ is the leftmost qubit. However, the major problem here is that we need to be $\textit{extra}$ careful with any and all steps involving our qubits, as to not accidentally operate on the wrong qubit (especially measurements). Also, now our printed $\textmf{program}$ looks weird, telling us that we applied our Hadamard gate to qubit $1$, when we know it's really on qubit $0$ (...what a headache).
\\

I am going to offer an alternative solution here. Using $\textmf{qvm.wavefunction}$ as its basis, let's import and use a function called $\textbf{Wavefunction}$, from the additional python file accompanying these tutorial lessons: $\textbf{Our\_Pyquil\_Functions}$.

\begin{figure}[h]
\centering
\includegraphics[scale=.65]{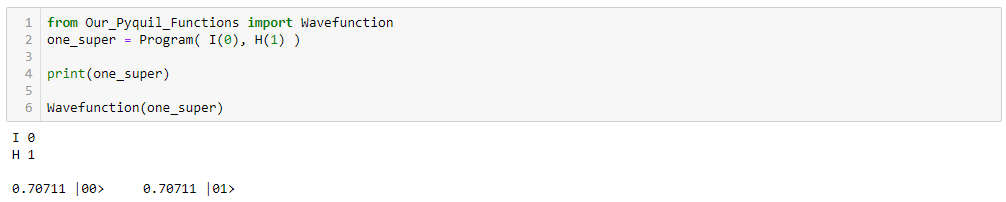}
\end{figure}

As we can see, our printed $\textmf{program}$ is in agreement with what we see as our final state.

Now, let's be clear about where this function came from. As part of this tutorial series, we will frequently be calling upon functions from the python file $\textmf{Our\_Pyquil\_Functions.py}$, which you should have stored in the same location as these .ipynb files and Pyquil. This python file is $\textit{not}$ a part of Pyquil. It is a python file filled with custom functions designed to coincide with these lessons, for learning purposes.

From this point on, we will be using $\textmf{Wavefunction}$ when we when to view the wavefunction of our quantum systems (sorry, no refunds). Keep in mind that we are still technically always using $\textmf{qvm.wavefunction}$(), just a customized version to suit our learning needs. In lesson 4 we will go through some of the additional features of $\textmf{Wavefunction}$, as well as other important custom functions.

\section*{\large{ Making a Measurement }}
\centerline{---------------------------------------------------------------------------------------------------------------------------------}

Now comes the final step for creating quantum algorithms -- measuring the quantum states that we create. To do this, Pyquil has a convenient way for us to measure and record the results of our quantum system, but the syntax is a little tricky at first.
\\

Let's see an example in action and then backtrack to understand each component:

\pagebreak

\begin{figure}[h]
\centering
\includegraphics[scale=.65]{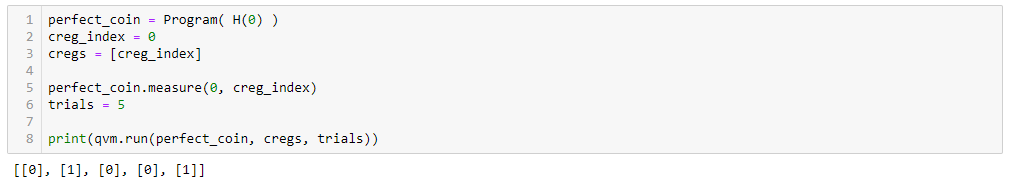}
\end{figure}

A lot to unpack here, so let's start with the first two lines of code, since they will be the easiest to understand:
\\

--------------------------------------------------------------------------------------------

perfect\_coin = Program( H(0) )

creg\_index = 0

--------------------------------------------------------------------------------------------
\\

In the first line, we are simply creating our $\textmf{program}$, just like we have done so far. Specifically, we are initializing a single qubit with a Hadamard gate, which we know results in an initial quantum system:

$$\frac{1}{\sqrt{2}} \big{(} \hspace{.1cm}|\hspace{.05cm}0\rangle \hspace{.1cm}+\hspace{.1cm} |\hspace{.05cm}1\rangle \hspace{.06cm} \big{)}$$

Next, we create a regular python variable called 'creg\_index' and store the integer 1. 'creg' is a shorthand version of 'classical register', and so creg\_index represents a particular index $\textit{in}$ the classical register. The classical register itself is just a list of bits, which will hold 0's and 1's, with the intention that each bit will hold a measurement results.
\\

In the next line we create our classical register using all of the indices we want (which in this case is just one):
\\

--------------------------------------------------------------------------------------------

creg = [ creg\_index ]

--------------------------------------------------------------------------------------------
\\

Now, in this example we've chosen our creg\_index to be the value 0, but it doesn't have to be. The classical register does not have to match the same indexing as our qubits. In general, we have the ability to measure any qubit, and store its result in any classical register index. For example, in the final code example for this section we will change creg\_index to 2, electing to store qubit $0$'s measurement result in the classical register index 2.
\\

Next, let's take a look at the line of code that is $\textit{actually}$ doing the measurement:
\\

--------------------------------------------------------------------------------------------

perfect\_coin.measure( 0, creg\_index )

--------------------------------------------------------------------------------------------
\\

$\textbf{measure}$ is a function that comes with the $\textmf{Program}$ class, so there was no need to import it directly. It takes two arguments -- a qubit index to measure, and a classical register index for storing. The second argument here is actually optional, as we are not forced to record a measurement result if we choose not to. Thus, the following code is perfectly acceptable:

\begin{figure}[h]
\centering
\includegraphics[scale=.65]{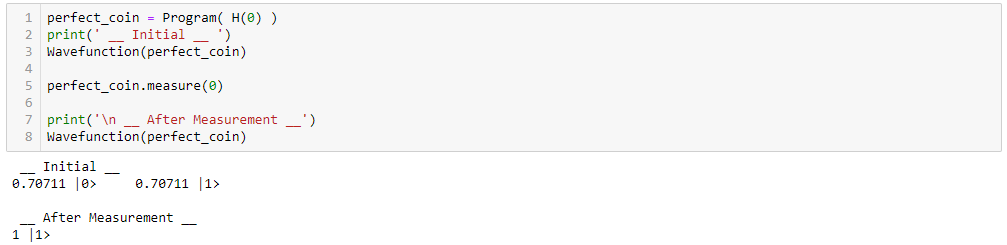}
\end{figure}

This code starts off with qubit 0 in an equal superposition of states $|0\rangle$ and $|1\rangle$, via the Hadamard gate, and then makes a measurement. We can peek at the wavefunction after the measurement (as we have done above), to see what state was measured, but this information isn't actually stored anywhere. That is to say, because we didn't pass a creg\_index integer argument in $\textmf{measure}$, the measurement result that we see displayed isn't physically saved anywhere except being printed to console.
\\

And remember, in a real quantum system we don't have $\textmf{Wavefunction}$ to cheat and look at the system for us. So if we make a measurement and don't record the results, we better be sure we didn't need that information! It is worth noting that $\textmf{measure}$ is actually changing our $\textmf{program}$ here, appending an additional instruction. To see this, let's try printing our $\textmf{program}$ to console, before and after $\textmf{measure}$:

\begin{figure}[h]
\centering
\includegraphics[scale=.65]{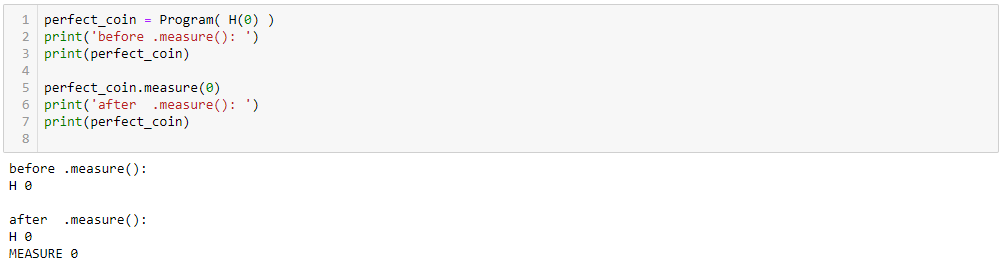}
\end{figure}

So technically, $\textmf{measure}$ is just adding another entry to our list of instructions stored in our $\textmf{program}$, not $\textit{actually}$ doing the measurement. The function responsible for that, which executes all of our $\textmf{program}$'s instructions, is $\textbf{qvm.run}$:
\\

--------------------------------------------------------------------------------------------

qvm.run(perfect\_coin, cregs, trials)

--------------------------------------------------------------------------------------------
\\

Let's take a closer look at $\textmf{run}$: The first argument is the $\textmf{program}$ whose instructions get executed, the second argument is the classical register locations to check afterwards, and the third argument is the number of times to repeat the process.
\\

Now, you may be wondering how earlier we were able to see the wavefunction after $\textmf{measure}$, without using $\textmf{run}$. This is because $\textmf{Wavefunction}$ uses the same classical simulator -- $\textmf{QVM}$, and are both independent ways of evaluating a $\textmf{program}$. Thus, whenever we use $\textmf{Wavefunction}$, we are essentially running a different simulation of all the $\textmf{program}$ instructions up to that point in the code.
\\

BUT, the real difference between these two functions is that the $\textmf{program}$ is not actually being executed when we use $\textmf{Wavefunction}$. We are essentially $\textit{simulating}$ the program and seeing what the state of the system $\textit{would}$ look like that at point. But we cannot extract any information out from $\textmf{Wavefunction}$ like we can from $\textmf{run}$.
\\

In addition, the two methods are independent of each other, so any measurement results between the two may not match. This is why $\textmf{Wavefunction}$ should be thought of as a simulation, and $\textmf{run}$ should be treated as the $\textit{real}$ running of our quantum code. Any measurement results that happen as a result $\textmf{run}$ are permanent to the quantum system.
\\

If you don't totally get what's going on with $\textmf{QVM}$ and the classical register, don't worry. In future lessons we will get plenty of examples. For now, let's wrap up this section by returning to our complete example (this time with an extra qubit):

\begin{figure}[h]
\centering
\includegraphics[scale=.65]{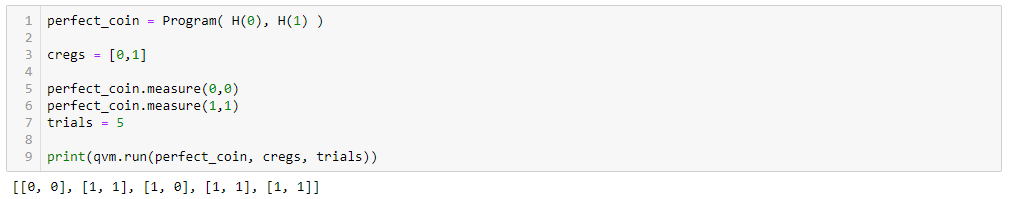}
\end{figure}

In this example, we've changed a few things to showcase how $\textmf{measure}$ and the classical register work with multiple qubits. Now, since we are dealing with two qubits, our classical register list $\textmf{cregs}$ contains the indices $0$ and $1$. Then, when we call upon the $\textmf{measure}$ function, we tell the code to measure qubits $0$ and $1$, and store them in the classical register indices respectively.
\\

In general, we will almost always store our measurement results in the same corresponding classical register indices. And if that is the case, then Pyquil provides us a way of measuring the entire system in a single line, via the $\textbf{measure\_all}$ function:

\begin{figure}[h]
\centering
\includegraphics[scale=.65]{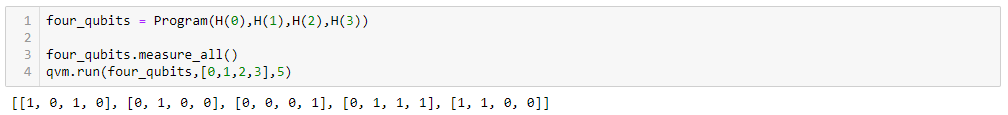}
\end{figure}

$\textmf{.measure\_all}$ works the same way as our single qubit measurements, but my default measures all of the qubits in the system, and stores them in their corresponding classical register indices. For example, in the code above, qubits $0$ - $3$ are all stored in the classical register locations 0 - 3 respectively.
\\

However, $\textmf{measure\_all}$ isn't limited to simply measuring all of the qubits in the system in order, it can also handle partial measurements. For example, suppose we would like to measure to measure all of our qubits, but store their measurement results in locations different from the qubit indices:

\begin{figure}[h]
\centering
\includegraphics[scale=.65]{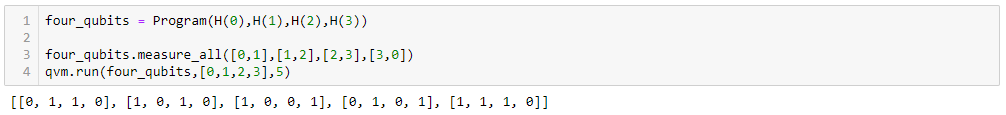}
\end{figure}

In the code above, we pass the arguments [0,1], [1,2], [2,3], and [3,0] into $\textmf{measure\_all}$, which can be understood as, "measure qubit $0$ and store it in the classical register location $1$ ... measure qubit $1$ and store it in the classical register location $2$ ... and so on".
\\

The advantage to using $\textmf{.measure\_all}$ then is to tidy up our code. All of the same classical register rules we covered earlier apply here as well. Thus, the $\textmf{measure\_all}$ function is just an alternative to using $\textmf{measure}$ when we want to condense our code.
\\

As another option to condensing our code, without using $\textmf{measure\_all}$, it's worth pointing out that we can string together multiple calls of the function $\textmf{measure}$, like so:

\begin{figure}[h]
\centering
\includegraphics[scale=.65]{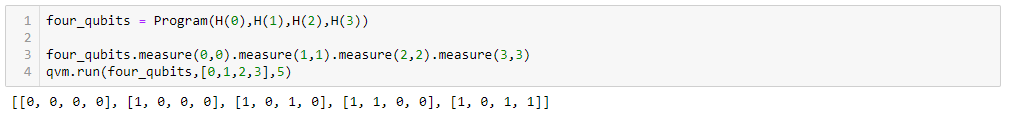}
\end{figure}

\section*{\large{ Perfect Coin Algorithm }}
\centerline{---------------------------------------------------------------------------------------------------------------------------------}

Now that we have a better understanding of how to make measurements and extract their results, let's put it to the test and write a simple function, using a quantum algorithm, to simulate a perfect coin:

\pagebreak

\begin{figure}[h]
\centering
\includegraphics[scale=.65]{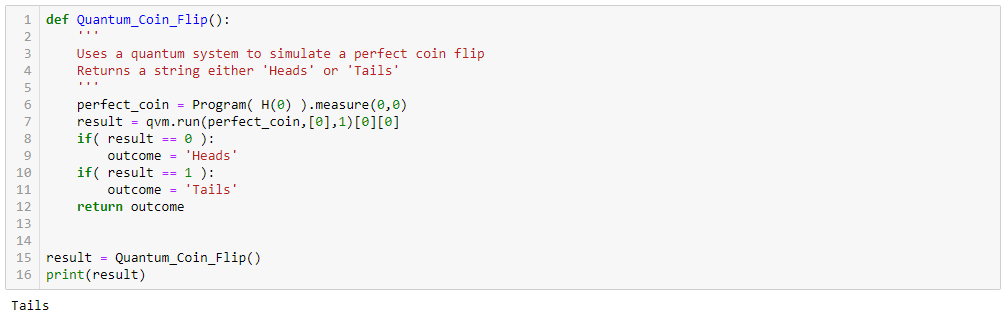}
\end{figure}

Run the cell of code above, and you should find that our function returns a string either 'Heads' or 'Tails'.
\\

Let's now take a look at the code, and see how we are able to turn our quantum system result and return to us a string. First off, we've combined multiple lines of code, essentially achieving all of the quantum operations in a single line of code:
\\

--------------------------------------------------------------------------------------------

perfect\_coin = Program( H(0) ).measure(0,0)

--------------------------------------------------------------------------------------------
\\

This line of code condensed the initialization and measurement entries of our program down to one line. The important point being that we do not need to write $\textmf{measure}$ as its own line of code, but rather, we can attach it to the end of our program. Technically speaking, this is allowed because $\textmf{measure}$ is a function of the $\textmf{Program}$ class, and all we were doing before was saving our program to a variable, and then calling $\textmf{measure}$ on that variable:
\\

--------------------------------------------------------------------------------------------

perfect\_coin = Program( )

perfect\_coin.measure( )

--------------------------------------------------------------------------------------------
\\

Note that this is a totally fine way of handling measurements, and for illustrative purposes it is very helpful in understanding the flow of the code: First we initialize our system... and then add a measurement to the program. But if we want to minimize our code, we can apply the $\textmf{measure}$ function directly after any line of code that is a program.
\\

Next, let's take a look at how we got our measurement result out of the classical register. To do this, let's first see the way in which the classical register's information comes back to us:

\begin{figure}[h]
\centering
\includegraphics[scale=.65]{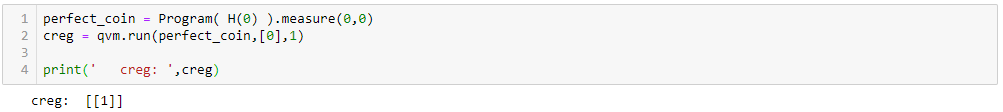}
\end{figure}

Our classical register comes out of the $\textmf{run}$ function as shown above. This is a list object, nested inside another list. Thus, our value out of these nested lists is as follows:

\begin{figure}[h]
\centering
\includegraphics[scale=.65]{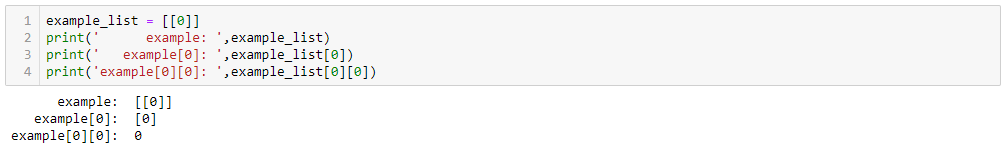}
\end{figure}

Using this method for our quantum code, we can extract the measurement result as follows:

\begin{figure}[h]
\centering
\includegraphics[scale=.65]{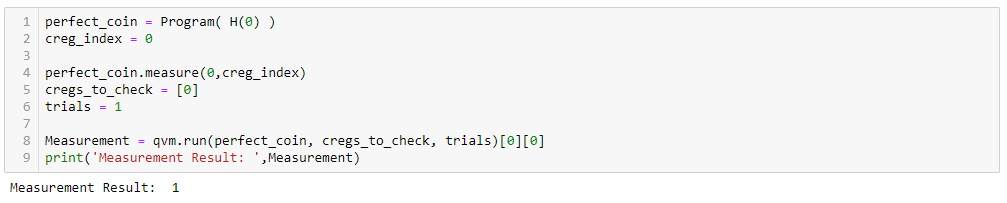}
\end{figure}

To summarize our Quantum\_Coin\_Flip function:
\\

1) We run the program 'perfect\_coin' exactly 1 time, storing the measurement result to the classical register location [0].
\\

2) The $\textmf{qvm.run}$ function then returns the classical register to us in the form of an array [[...]]
\\

3) We extract our measurement result using [0][0]. The first [0] grabs the inner list, while the second [0] grabs the 0$^{th}$ entry of that inner list, which is where we chose to store our measurement result.
\\

Now, let's see our perfect coin function in action! The code below is a silly example, whereby we use the result of the measurement to determine a gambling bet between Alice and Bob:

"Alice and Bob have recently gotten into an argument about the philosophy of picking the correct side of a coin flip. Bob was raised by the moto "Tails Never Fails", while Alice was taught "Tail Always Fails". Alice suggests that the two solve their disagreement with a series of coin flips, but Bob doesn't trust any coin that Alice owns, and vice versa for Alice. Thus, they agree to use a qubit as their coin. The loser of the bet must clean the other person's lab equipment for a month!"

\begin{figure}[h]
\centering
\includegraphics[scale=.65]{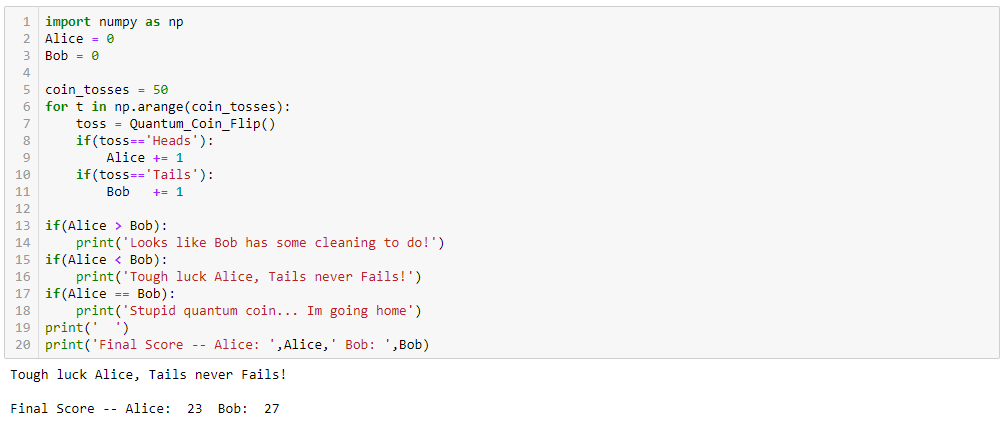}
\end{figure}

--------------------------------------------------------------------------------------------------------------------------------------------------------
\\

This concludes our introduction lesson to Pyquil! I hope that the examples in this lesson provide a good starting point for using Pyquil. Before moving on to the next tutorial, I strongly encourage you to write some simple code of your own, and test out all of the various functions and classes we studied here.
\\

--------------------------------------------------------------------------------------------------------------------------------------------------------


\pagebreak

\section*{\Large{Lesson 2 - Creating More Complex Programs}}
--------------------------------------------------------------------------------------------------------------------------------------------------------
\\

In this lesson, we will continue to cover some of the common tools provided by Pyquil for writing quantum algorithms. For a review on the basics of using Programs, QVM, etc. please check out lesson 1 in this tutorial series:
\\

Lesson 1 - Intro to Programs and Measurements
\\

--------------------------------------------------------------------------------------------------------------------------------------------------------
\\

In order to make sure that all cells of code run properly throughout this lesson, please run the following cell of code below:

\begin{figure}[h]
\centering
\includegraphics[scale=.65]{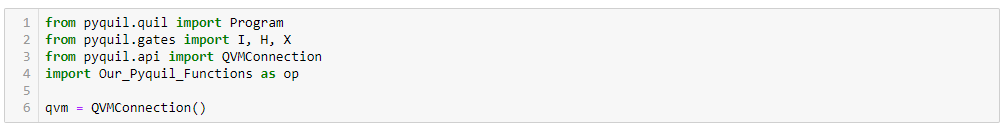}
\end{figure}

\section*{\large{ Observing / Editing Programs }}
\centerline{---------------------------------------------------------------------------------------------------------------------------------}

In lesson 1 we've already covered the basics of how to create quantum systems, view their wavefunctions, and make measurements. Now, we are going to cover some more advanced topics in Pyquil, as well as one custom function of our own, which will improve our coding abilities / provide us with some new tools. We will begin this tutorial with some functions that will be very handy for viewing / debugging our codes.

\section*{\large{ Measurements and Wavefunctions }}

To begin, let's quickly revist an example showcasing the two main ideas from lesson 1: viewing the wavefunction of our quantum system, and making measurements. In particular, we are going to look at an example of a partial measurement, one where we only measure some of the qubits in the system, but not all.
\\

By doing a partial measurement, we should find that our measured qubits collapse down to a single value, while the remaining qubits keep their superposition states. For example:

$$ \frac{1}{2} \big{(} \hspace{.1cm} |\hspace{.05cm}00\rangle \hspace{.08cm} + \hspace{.08cm} |\hspace{.05cm}01\rangle \hspace{.08cm} + \hspace{.08cm} |\hspace{.05cm}10\rangle \hspace{.08cm} + \hspace{.08cm} |\hspace{.05cm}11\rangle \hspace{.06cm} \big{)} $$

$$ \textmf{measure qubit }0 \hspace{.2cm} \rightarrow \hspace{.2cm} |\hspace{.05cm}1\rangle $$

$$ \frac{1}{\sqrt{2}} \big{(} \hspace{.1cm} |\hspace{.05cm}10\rangle \hspace{.08cm} + \hspace{.08cm} |\hspace{.05cm}11\rangle \hspace{.06cm} \big{)} $$

Now let's see it in code:

\pagebreak

\begin{figure}[h]
\centering
\includegraphics[scale=.65]{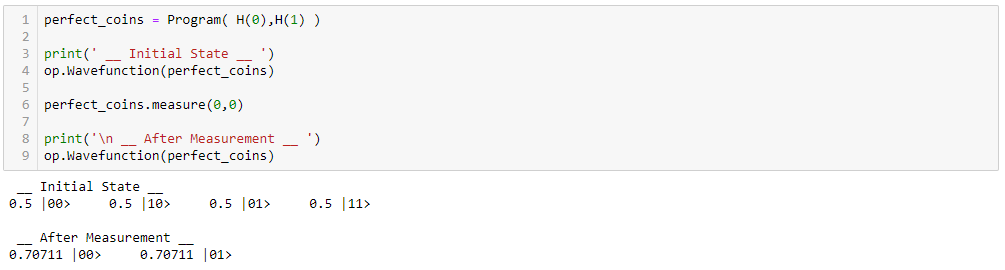}
\end{figure}

Run this cell of code a couple times until you see both outcomes. You should find that qubit $0$ is always collapsed down to a single state, while qubit $1$ remains in a superposition. Thus, the final state after the measurement will always be a superposition of two states.
\\

If in addition to observing the wavefunctions, we would also like to extract our measurement results, we can use the $\textmf{run}$ function:

\begin{figure}[h]
\centering
\includegraphics[scale=.65]{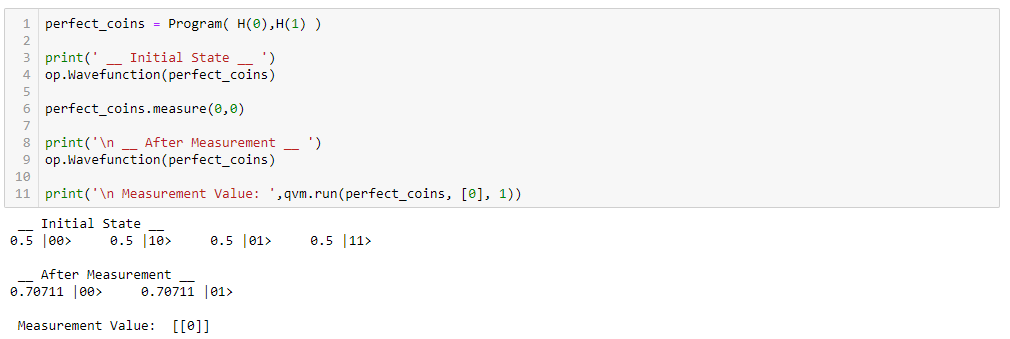}
\end{figure}

Note that the measurement result we extract matches with our wavefunction, which is nice. This is because we only ran 1 trial of our $\textmf{run}$ function. If we instead changed the code above to run multiple trials, what we would find is that our $\textmf{Wavefunctions}$ would match with the first trial, regardless of later trials. Similarly, if we were to look at the wavefunction after $\textmf{run}$, we would again observe a state in agreement with the first trial.

\section*{\large{ Viewing Many Measurements}}

So far we haven't covered anything new. In the example above, we simply create a quantum system, observe its wavefunction with $\textmf{Wavefunction}$, and then extract the measurement result. But suppose we didn't want to use $\textmf{Wavefunction}$ as a means of seeing which states are more probable, but rather measurements. In larger and more complex systems, often times it is a cumbersome task to calculate probabilities with every state's amplitude, especially when complex numbers are involved.
\\

Rather, it is easier to simulate many measurements, and simply see which states are the most probable. However, observing many measurement results with the $\textmf{run}$ function can be a bit tedious, and hard to grasp in a single glance:

\pagebreak

\begin{figure}[h]
\centering
\includegraphics[scale=.65]{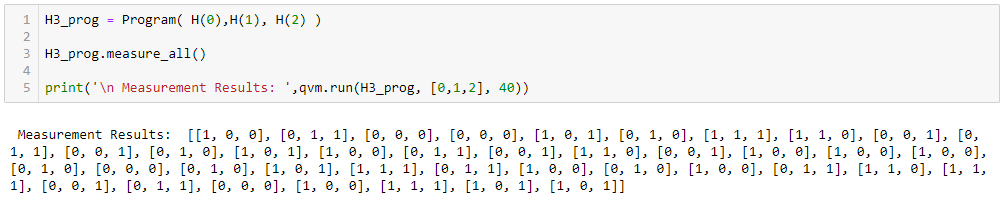}
\end{figure}

If what we're interested in is simply viewing the distribution of our measurement results, then I will propose an alternative way here, using a custom function called $\textbf{Measurement}$, from \textmf{Our\_Pyquil\_Functions}:

\begin{figure}[h]
\centering
\includegraphics[scale=.65]{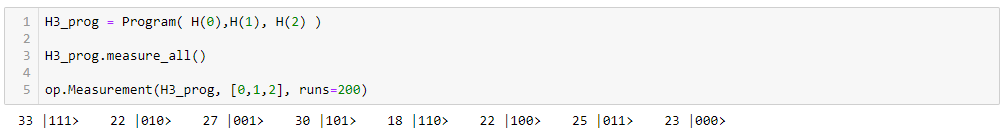}
\end{figure}

This $\textmf{Measurement}$ function still uses $\textmf{run}$ at its core, but simply contains some extra steps in order to display our measurement results back to us nicely (and a couple extra features which we will cover in lesson 4). Thus, if we are ever interested in visualizing the probabilities of our system through multiple measurements, we will be using $\textmf{Measurement}$ in these tutorials.

\section*{\large{ Amending Programs }}

As a quick reminder, Pyquil makes it very easy for us to view $\textmf{Programs}$ that we've written by simply printing them:

\begin{figure}[h]
\centering
\includegraphics[scale=.65]{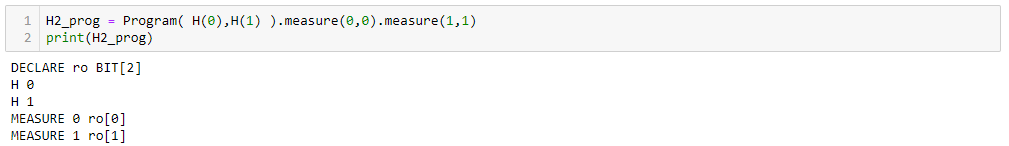}
\end{figure}

Note that in this program, we add all of our instructions in a single line of code. We assign all of our gates when we initialize the program using $\textmf{Program}$, and then chain together measurements using $\textmf{measure}$. Obviously it is unrealistic to always write entire quantum algorithms in one line, so now we are going to see some tools for writing more flexible quantum codes.

\section*{\large{ inst }}

If you take a look back at all of our examples up until now, including lesson 1, you will find that we only ever assign the gate operations of our programs in the first line. While there are certainly many instances where we will want to write many gates in our initialization step, Pyquil provides us with a tool for adding instructions at any time, via the $\textbf{inst}$ function.
\\

This function works in a similar way to $\textmf{measure}$, where we can chain it onto any $\textmf{Program}$ object in our code (this is because it is a function belonging to the $\textmf{Program}$ class). But the special thing about $\textmf{inst}$ is that we can pass it any instruction we want, including measurements:

\pagebreak

\begin{figure}[h]
\centering
\includegraphics[scale=.65]{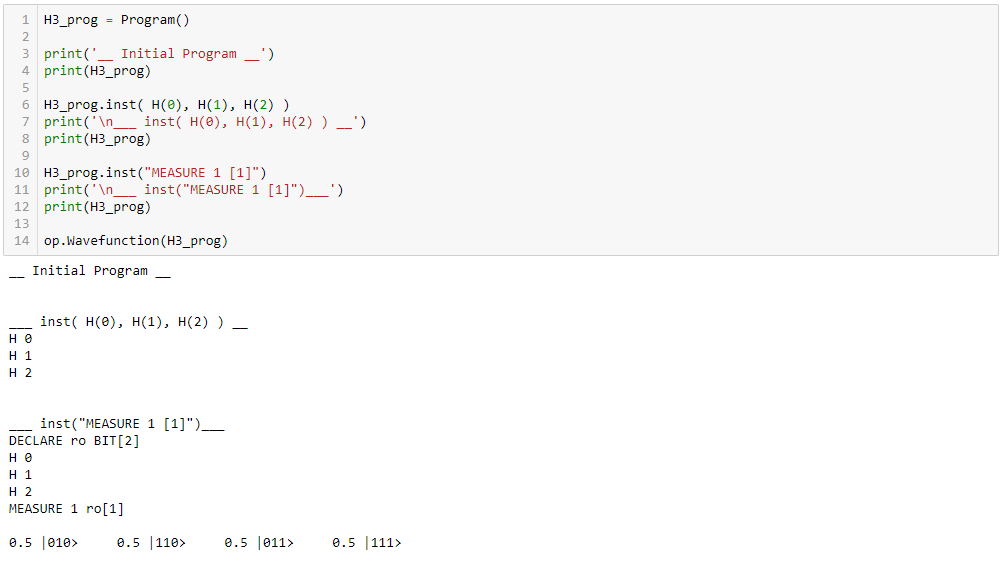}
\end{figure}

In the example above, we create an initial program that is empty. Then, we use the $\textmf{inst}$ function to initialize our three qubits with Hadamard gates, and later a measurement. Lastly, to confirm that our program produces the state we intended, we pass it to $\textmf{Wavefunction}$. And, sure enough, we find our system in a superposition of four states, where qubit $1$ is collapses to either $|0\rangle$ or $|1\rangle$ in every state.
\\

$\textmf{inst}$ has quite a bit of flexibility in taking arguments as well. Consider all of the following ways we can append two qubits:

\begin{figure}[h]
\centering
\includegraphics[scale=.65]{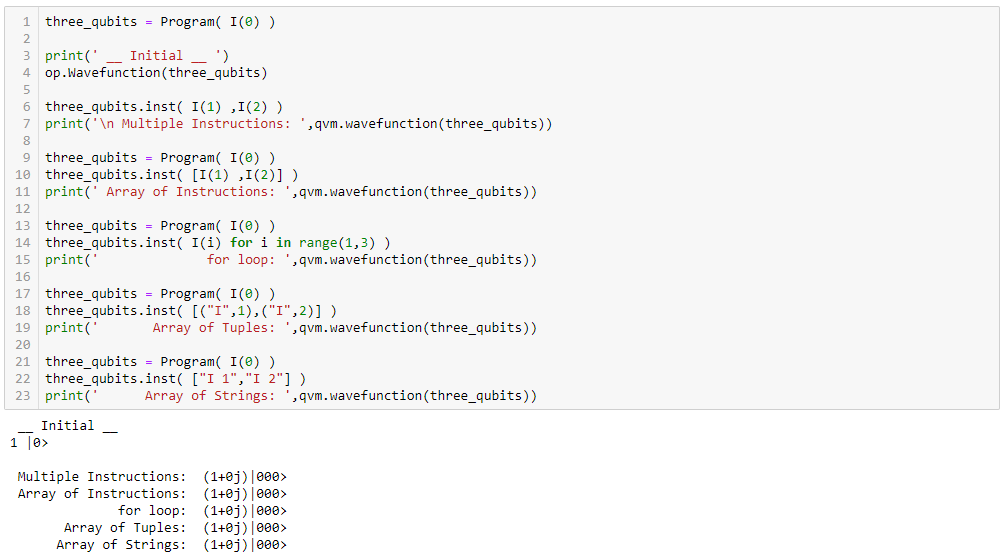}
\end{figure}

The $\textmf{inst}$ function gives us a great deal of flexibility in the style we want to append instructions to a program, and will really be the go-to function when we start writing longer algorithms. But at the end of the day, whether we feed $\textmf{inst}$ strings, tuples, arrays, etc., all of these methods achieve the same result, only differing in syntax.
\\

However, I will point out one of the methods from the code above that particularly helpful: $\textit{for loop}$. The ability to use a for-loop as a set of instructions is quite handy if we want to insert a long list of instructions, in a nice compact line of code.

\section*{\large{ pop }}

Typically, quantum algorithms are very rigid, where all of the steps in the algorithm are deliberate from start to finish. This is partly due to the nature of quantum systems and the way measurements collapse these systems, which in turn means they lack the ability to 'go back' and change based on measurement results (to a degree).
\\

Nevertheless, there are cases where we may want to amend our code fluidly, perhaps if we need to rerun an algorithm multiple times. We've just seen how to add instructions to our program as we go along, so now let's see how to remove them. In Pyquil, a $\textmf{Program}$ object is very similar to a list, and even supports some of the same functions we would normally be able to do with lists:

\begin{figure}[h]
\centering
\includegraphics[scale=.65]{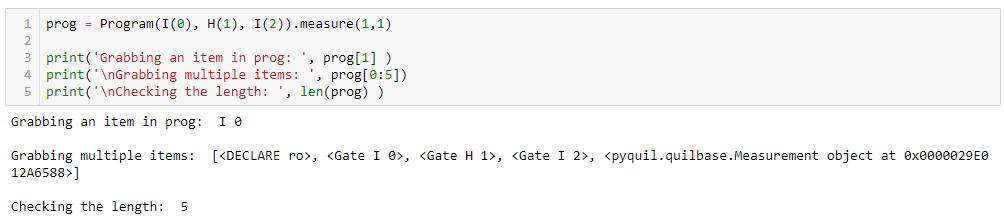}
\end{figure}

However, a program $\textit{is not}$ a list object, so we can't expect all of the normal python list functionalities. To see what $\textit{is}$ allowed, I encourage you to check out the source code for the $\textmf{Program}$ class for yourself.
\\

Here, our intention is removing instructions. Typically, if we wanted to remove objects from a list, the go-to functions would be $\textbf{remove}$, $\textbf{del}$, and $\textbf{pop}$. Of these three, the one that is supported by $\textmf{Program}$ objects, is $\textmf{pop}$:

\begin{figure}[h]
\centering
\includegraphics[scale=.65]{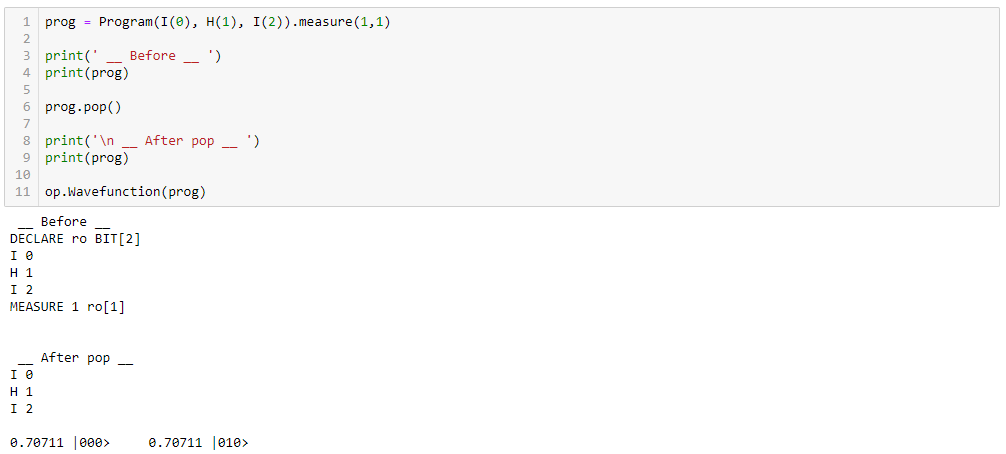}
\end{figure}

The $\textmf{pop}$ function works the same way for our programs as it does for lists, removing the last entry. In our example above, the last instruction in our program corresponds to our measurement. As shown in the printed program, as well as the final wavefunction, $\textmf{pop}$ did indeed remove this instruction from the program as intended.
\\

Now, the next logical question would be how to remove some intermediate instruction, instead of just the last one. To answer this, we are going to take a short detour. In particular, we need understand the interactions between multiple programs.

\section*{\large{Combining Programs}}

Up until now, all of our examples have been cases where we create and work with a single $\textmf{Program}$, from initialization to observation to measurements. In a sense, it's natural to equate "writing a quantum algorithm" to "designing a program." However, sometimes we will need the aid of extra programs to handle certain steps in our algorithms.
\\

Generally speaking, working with multiple programs is a very case-by-case problem. Certain algorithms may not need multiple programs, while other may require subroutines, where it is easier to define an entire secondary program to handle it. Other times we may only require a second program to handle some coding steps, which have no real impact on our physical system.
\\

In any case, we must always remember that the only thing that matters is whatever $\textmf{Program}$ we feed to the $\textmf{run}$ function. This is because the $\textmf{run}$ function should be thought of as $\textit{actually doing}$ the physical steps of a quantum algorithm. So regardless of how many edits it undergoes along the way, the only thing that will show up when we physically run a quantum algorithm, is the final $\textmf{Program}$.
\\

Knowing how to handle multiple programs will give us some more insight into the structure of the $\textmf{Programs}$ class. So to begin, let's start with how to combine two programs, via two different methods:

\begin{figure}[h]
\centering
\includegraphics[scale=.65]{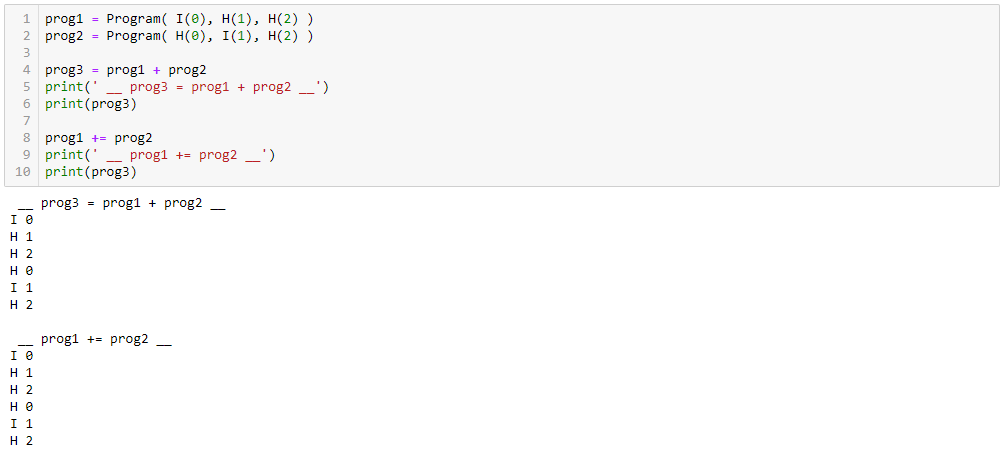}
\end{figure}

Both cases in this example produce the same final $\textmf{Program}$, combining $\textmf{prog1}$ and $\textmf{prog2}$. The difference between them, is that in the first case we store the combination of $\textmf{prog1}$ + $\textmf{prog2}$ as a new program $\textmf{prog3}$. In the second case, we actually append all of the instructions stored in $\textmf{prog2}$, into $\textmf{prog1}$.
\\

Thus, the '+' functionality combines two programs, to then be stored in whatever variable we choose, leaving both of the original programs unchanged. Conversely,'+=' appends all of the instructions from the second program onto the first, leaving the second program unaltered.
\\

Note that when using '+' and '+=', the ordering of the programs matters. For example, $\textmf{prog1}$ + $\textmf{prog2}$ will not be the same as $\textmf{prog2}$ + $\textmf{prog1}$. When combing programs with either method, all of the instructions of the second program get amended onto the end of the first.
\\

Now, suppose we only want to combine $\textit{some}$ of the instructions in two programs. To do this, we will use our ability to grab items from a $\textmf{Program}$, in a list-like manner, just as we did before. However, simply grabbing instructions from a $\textmf{Program}$ does not automatically make them into a $\textmf{Program}$. This is best shown in an example:

\begin{figure}[h]
\centering
\includegraphics[scale=.65]{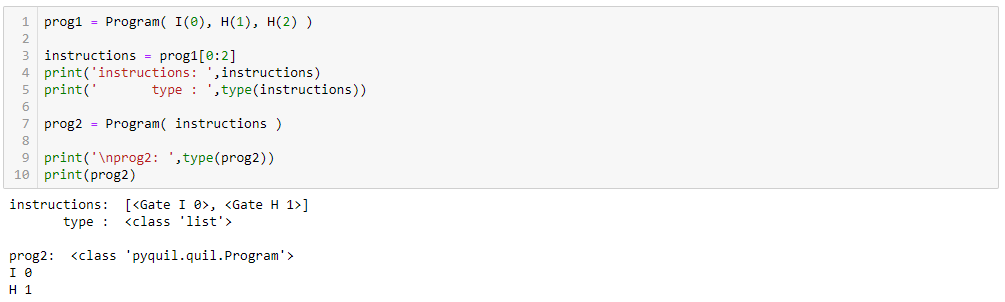}
\end{figure}

Take a moment to carefully examine the cell of code above, since there's quite a bit going on. To summarize, grabbing elements from a $\textmf{Program}$ returns a list of the instructions (an actual list object). Thus, grabbing $\textmf{Program}$ elements does not produce a $\textmf{Program}$.
\\

Then, in order to turn our $\textmf{Program}$ elements into a new $\textmf{Program}$ of their own, all we need to do is simply pass them through the $\textmf{Program}$ function, as shown above. The result will be a $\textmf{Program}$ object, which we can then assign to a new variable, or even our same program variable from before.

\section*{\large{ Removing a Program Element }}

Having completed our detour into combing / creating new programs, maybe you have a hunch about how we're going to solve our problem from earlier: removing a desired element from a $\textmf{Program}$ (hint hint, the last sentence before this section). While there are certainly many ways of handling this, let's see one way which uses the technique we just learned:

\begin{figure}[h]
\centering
\includegraphics[scale=.65]{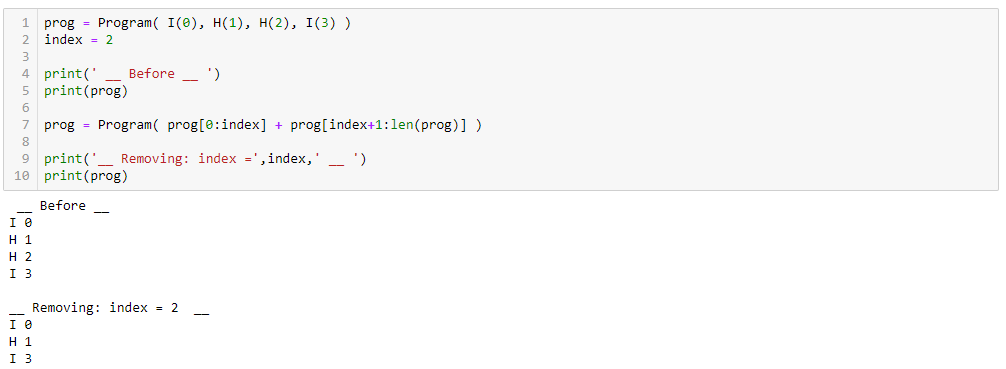}
\end{figure}

Rather than using $\textmf{pop}$ at all, we can simply remove a program's element by creating two new $\textmf{Programs}$ before and after the index value, combine them, and define a new $\textmf{Program}$. This method could be used to remove entire sections of instructions as well if needed.
\\

--------------------------------------------------------------------------------------------------------------------------------------------------------
\\

This concludes our second lesson of learning Pyquil! We now have most of the tools we need to start studying some famous quantum algorithms, which begin in lesson 5. The last major component missing from our toolbox is all of the standard gates provided by Pyquil, which we will cover next lesson!
\\

--------------------------------------------------------------------------------------------------------------------------------------------------------


\pagebreak

\section*{\Large{ Lesson 3 - Gates Provided by Pyquil}}
--------------------------------------------------------------------------------------------------------------------------------------------------------
\\

The goal of this lesson is to introduce some of the predefined gates that come standard with Pyquil. We will go through each gate, accompanied with a short explanation and working example.
\\

Before proceeding, please consider reading the previous lessons in this series, which covers all of the basics for working with Pyquil:
\\

Lesson 1 - Intro to Programs and Measurements
\\

Lesson 2 - Creating More Complex Programs
\\

--------------------------------------------------------------------------------------------------------------------------------------------------------
\\

In order to make sure that all cells of code run properly throughout this lesson, please run the following cell of code below:

\begin{figure}[h]
\centering
\includegraphics[scale=.65]{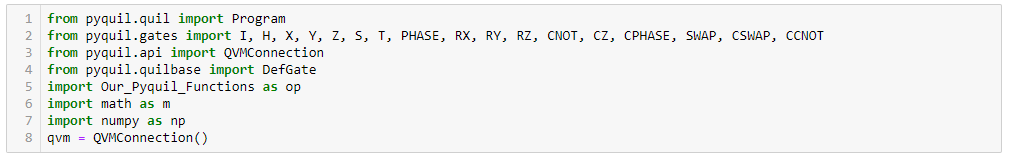}
\end{figure}

\section*{\large{ Single Qubit Gates }}
\centerline{---------------------------------------------------------------------------------------------------------------------------------}

\section*{\large{ I}}

The Identity Operator

$$
\begin{bmatrix}
1 & 0\\
0 & 1
\end{bmatrix}
$$

The effect of this gate renders the qubit's state unchanged.

\begin{figure}[h]
\centering
\includegraphics[scale=.65]{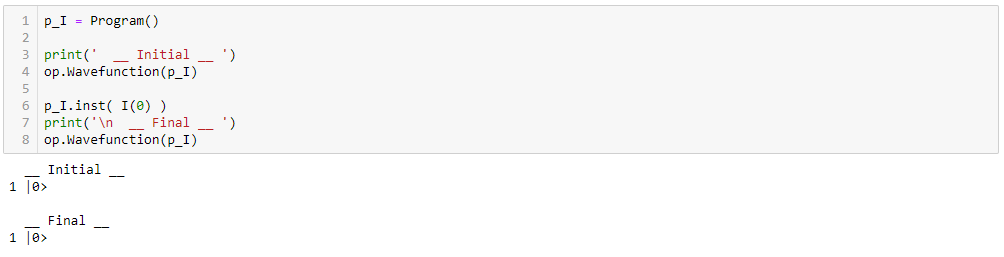}
\end{figure}

\section*{\large{ Hadamard (H)}}

$$
\begin{bmatrix}
1 & 1\\
1 & -1
\end{bmatrix}
$$

The effect of this gate is as follows:

$$ H \hspace{.08cm} |\hspace{.05cm} 0\rangle \hspace{.1cm} = \hspace{.1cm} \frac{1}{\sqrt{2}}\big{(} \hspace{.1cm}|\hspace{.05cm}0\rangle \hspace{.06cm}+\hspace{.06cm} |\hspace{.05cm}1\rangle \hspace{.06cm} \big{)}$$

$$ H \hspace{.08cm} |\hspace{.05cm} 0\rangle \hspace{.1cm} = \hspace{.1cm} \frac{1}{\sqrt{2}}\big{(} \hspace{.1cm}|\hspace{.05cm}0\rangle \hspace{.06cm}-\hspace{.06cm} |\hspace{.05cm}1\rangle \hspace{.06cm} \big{)}$$

This gate results in a qubit being in a 50 / 50 superposition of states $|0\rangle$ and $|1\rangle$. While this may seem simple enough, the importance of the Hadamard gate cannot be understated. In the coming lessons, we shall see that the Hadamard gate is largely responsible for the success of many quantum algorithms.

\begin{figure}[h]
\centering
\includegraphics[scale=.65]{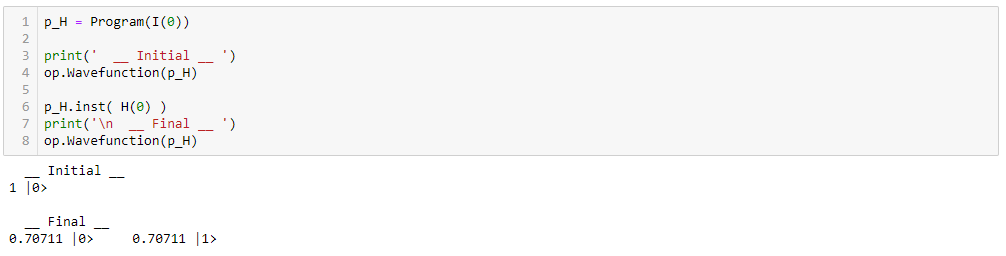}
\end{figure}

\section*{\large{ Pauli Operators }}
\centerline{---------------------------------------------------------------------------------------------------------------------------------}

\section*{\large{ X }}

$$
\begin{bmatrix}
0 & 1\\
1 & 0
\end{bmatrix}
$$

The effect of this gate is to flip a qubit's state between $|0\rangle$ and $|1\rangle$. This gate can be thought of the quantum analog to flipping a classical bit (the NOT gate). In systems with many superposition states, this gate will be very useful in isolating particular states for future operations.

\pagebreak

\begin{figure}[h]
\centering
\includegraphics[scale=.65]{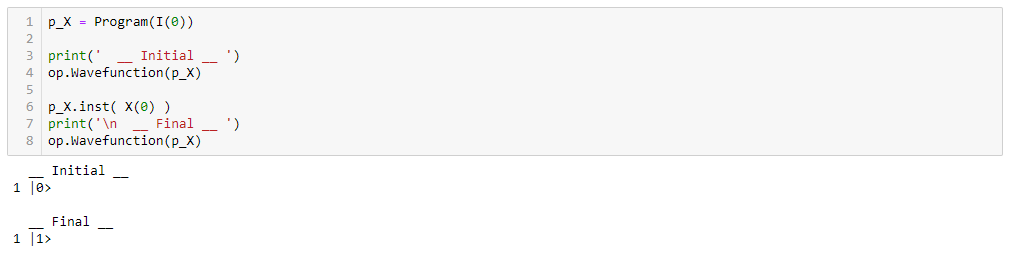}
\end{figure}

\section*{\large{ Y }}

$$
\begin{bmatrix}
0 & -i\\
i & 0
\end{bmatrix}
$$

The effect of this gate is to flip a qubit's $|0\rangle$ and $|1\rangle$ amplitudes and multiplies by an imaginary number (phase). From a probabilities perspective, this gate has the same effect as the X gate. However, the additional phase makes this gate very useful in creating certain constructive / deconstructive interferences.

\begin{figure}[h]
\centering
\includegraphics[scale=.65]{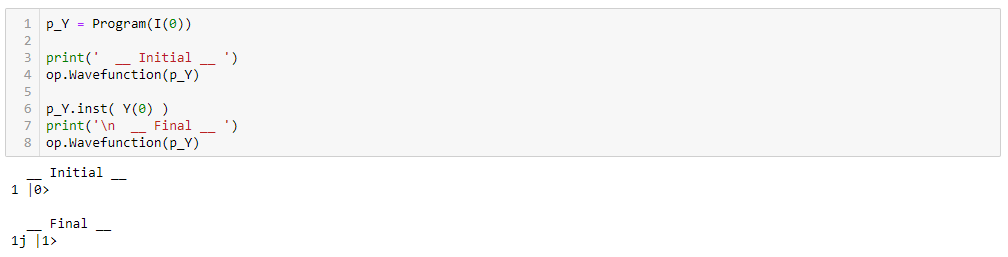}
\end{figure}

\section*{\large{ Z }}

$$
\begin{bmatrix}
1 & 0\\
0 & -1
\end{bmatrix}
$$

The effect of this gate leaves a qubit's $|0\rangle$ amplitude unchanged, while multiplying by -1 (phase) to a qubit's $|1\rangle$ amplitude. The power of this gate comes from the fact that it only affects the $|1\rangle$ component, which will be frequently used for picking out certain states in the system while leaving others unaltered.

\pagebreak

\begin{figure}[h]
\centering
\includegraphics[scale=.65]{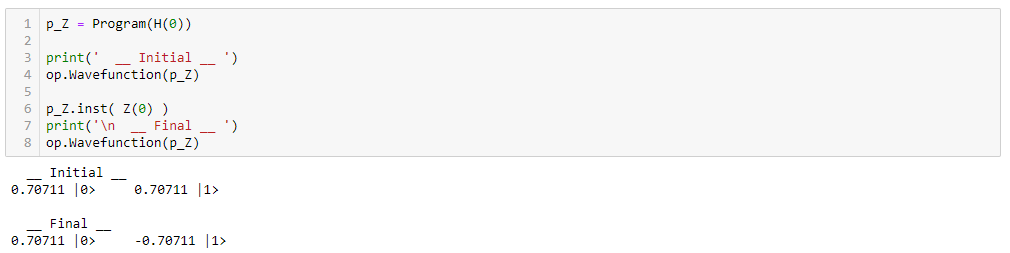}
\end{figure}

\section*{\large{ Phase Gates }}
\centerline{---------------------------------------------------------------------------------------------------------------------------------}

The following series of gates are all single qubit operations, which multiply a qubit's $|1\rangle$ state component by a phase. Doing so does not change the probability of the system, but is an essential component for algorithms that rely on particular kinds of interference.

\section*{\large{ PHASE (R$_{\phi}$) }}

$$
\begin{bmatrix}
1 & 0\\
0 & e^{i\phi}
\end{bmatrix}
$$

A gate similar to the Z gate. It leaves a qubit's $|0\rangle$ amplitude unchanged, while multiplying by a phase $e^{i\phi}$ to a qubit's $|1\rangle$ amplitude. In Pyquil, this gate goes by the name 'PHASE'. This gate will find many of the same uses as the Z gate, picking out certain states while leaving others unchanged. However, the extra degree of phase is a powerful tool for creating certain interference effects.

\begin{figure}[h]
\centering
\includegraphics[scale=.65]{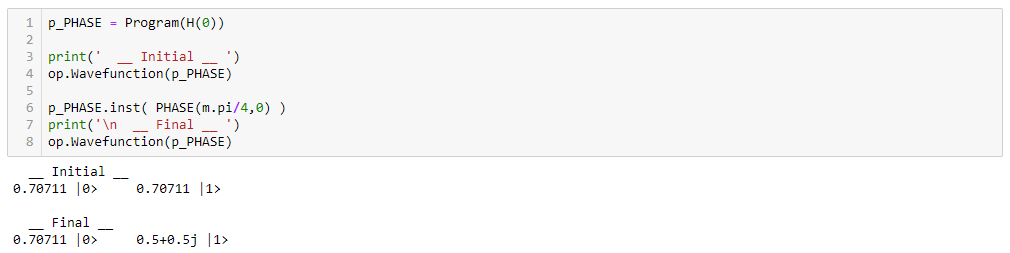}
\end{figure}

\section*{\large{ S }}

$$
\begin{bmatrix}
1 & 0\\
0 & i
\end{bmatrix}
$$

A pre-defined gate for R$_{\phi}$, $\phi$=$\frac{\pi}{2}$. It leaves a qubit's $|0\rangle$ amplitude unchanged, while multiplying by i (phase) to a qubit's $|1\rangle$ amplitude.

\begin{figure}[h]
\centering
\includegraphics[scale=.65]{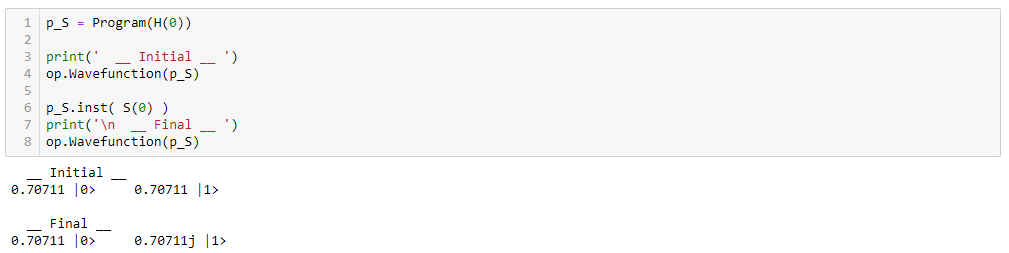}
\end{figure}

\section*{\large{ T }}

A pre-defined gate for R$_{\phi}$, $\phi$=$\frac{\pi}{4}$

$$
\begin{bmatrix}
1 & 0\\
0 & e^{i\frac{\pi}{4}}
\end{bmatrix}
$$

A pre-defined gate for R$_{\phi}$, $\phi$=$\frac{\pi}{2}$. It leaves a qubit's $|0\rangle$ amplitude unchanged, while multiplying by i (phase) to a qubit's $|1\rangle$ amplitude.

\begin{figure}[h]
\centering
\includegraphics[scale=.65]{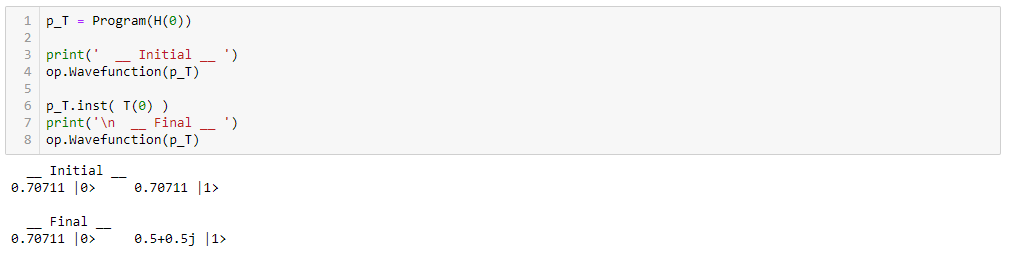}
\end{figure}

\section*{\large{ Rotation Gates }}
\centerline{---------------------------------------------------------------------------------------------------------------------------------}

The follow gates all represent rotations of a state on a Bloch Sphere. A Bloch sphere is a visual representation that maps the state of a qubit to a location on the surface of a sphere, radius = 1. An image of a Bloch sphere and it's axes is given below:

\pagebreak

\begin{figure}[h]
\centering
\includegraphics[scale=1]{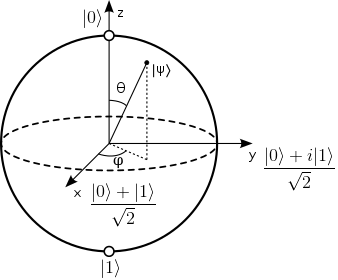}
\end{figure}

Note that the opposite ends of the x and y axis are:
\\

-x = $\frac{1}{\sqrt{2}} \big{(}$ $|\hspace{.06cm} 0\rangle$ - $|\hspace{.06cm} 1\rangle$ $\big{)}$ $\hspace{.3cm}$
-y = $\frac{1}{\sqrt{2}} \big{(}$ $|\hspace{.06cm} 0\rangle$ - $i\hspace{.05cm}$$|\hspace{.06cm} 1 \rangle$ $\big{)}$
\\

Thus, opposite axes on a Bloch Sphere represent orthogonal states.

\section*{\large{ R$_x$($\theta$) }}

$$
\begin{bmatrix}
cos(\frac{\theta}{2}) & -i \cdot sin(\frac{\theta}{2})\\
-i \cdot sin(\frac{\theta}{2}) & cos(\frac{\theta}{2})
\end{bmatrix}
$$

A rotation gate where the initial and final states can be represented as $\theta$ rotation around the x-axis on a Bloch Sphere.

\begin{figure}[h]
\centering
\includegraphics[scale=.65]{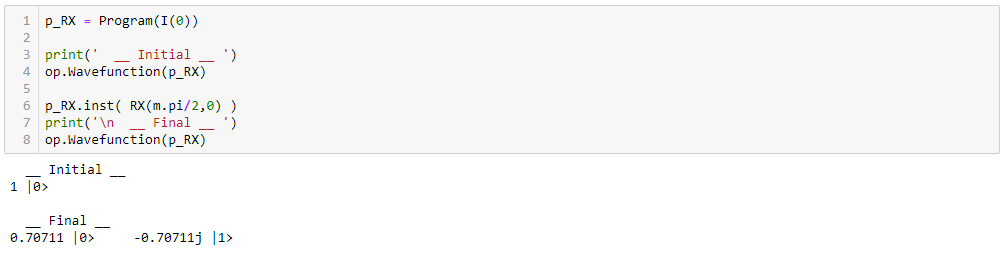}
\end{figure}

\section*{\large{ R$_y$($\theta$) }}

$$
\begin{bmatrix}
cos(\frac{\theta}{2}) & -sin(\frac{\theta}{2})\\
sin(\frac{\theta}{2}) & cos(\frac{\theta}{2})
\end{bmatrix}
$$

A rotation gate where the initial and final states can be represented as $\theta$ rotation around the y-axis on a Bloch Sphere.

\begin{figure}[h]
\centering
\includegraphics[scale=.65]{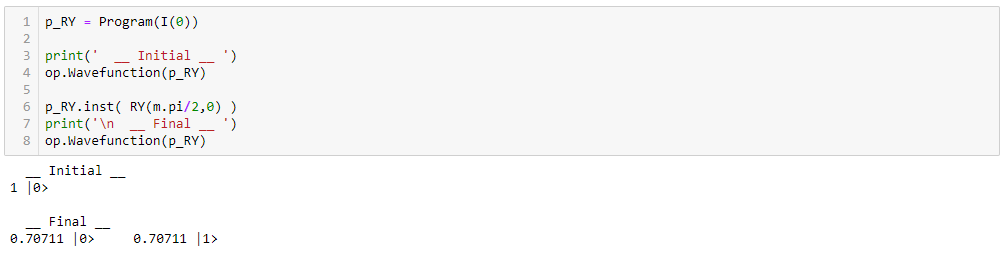}
\end{figure}

\section*{\large{ R$_z$($\theta$) }}

$$
\begin{bmatrix}
e^{\frac{-i\theta}{2}} & 0\\
0 & e^{\frac{i\theta}{2}}
\end{bmatrix}
$$

A rotation gate where the initial and final states can be represented as $\theta$ rotation around the z-axis on a Bloch Sphere.

\begin{figure}[h]
\centering
\includegraphics[scale=.65]{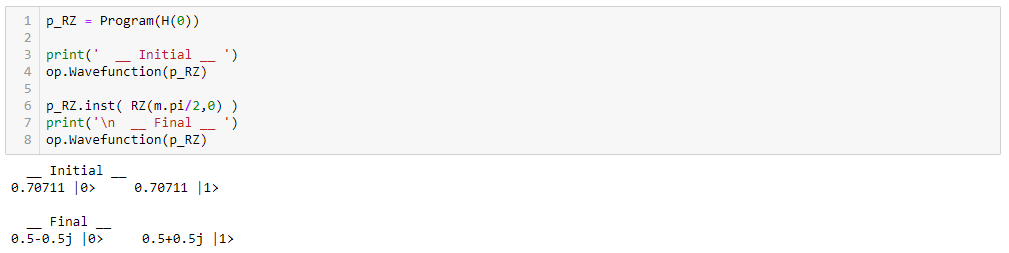}
\end{figure}

\section*{\large{ Two Qubit Control Gates }}
\centerline{---------------------------------------------------------------------------------------------------------------------------------}

All of the following gates act on 2 qubits. In particular, each gate uses a 'target qubit' and a 'control qubit'. The role of the control qubit is to determine whether or not a particular operation is applied to the target qubit. If the control qubit is in the state $|1\rangle$, then the operation is carried out on the target qubit. Conversely, if the control qubit is in the state $|0\rangle$, then the target qubit remains unchanged.

\section*{\large{ CNOT }}

The effect of the CNOT gate can be described as follows:

$$\textbf{CNOT} \hspace{.2cm} |\hspace{.05cm}00\rangle \hspace{.25cm} \rightarrow \hspace{.25cm} |\hspace{.05cm}00\rangle $$
$$\textbf{CNOT} \hspace{.2cm} |\hspace{.05cm}01\rangle \hspace{.25cm} \rightarrow \hspace{.25cm} |\hspace{.05cm}01\rangle $$
$$\textbf{CNOT} \hspace{.2cm} |\hspace{.05cm}10\rangle \hspace{.25cm} \rightarrow \hspace{.25cm} |\hspace{.05cm}11\rangle $$
$$\textbf{CNOT} \hspace{.2cm} |\hspace{.05cm}11\rangle \hspace{.25cm} \rightarrow \hspace{.25cm} |\hspace{.05cm}10\rangle $$

$$
\begin{bmatrix}
1 & 0 & 0 & 0\\
0 & 1 & 0 & 0\\
0 & 0 & 0 & 1\\
0 & 0 & 1 & 0\\
\end{bmatrix}
$$

In this notation, the first qubit is the control and the second qubit is the target. Another way to think of this gate is as a 'control-X' gate, where the state of the control qubit determines whether or not an X gate is applied to the target qubit. This gate sometimes goes by the name 'CX'.
\\

The CNOT gate is perhaps one of the most important tools in our quantum computing arsenal. Since we cannot have a purely 'NOT', the CNOT gate is our closest match. In combination with other gates, it will allow us to construct all manners of multi-qubit operations.

\begin{figure}[h]
\centering
\includegraphics[scale=.65]{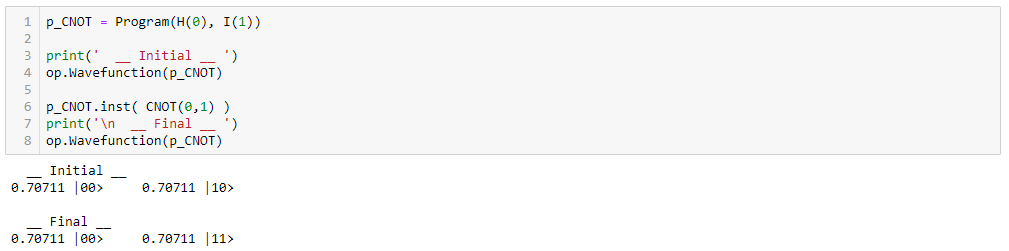}
\end{figure}

\section*{\large{ CZ (Control-Z) }}

The control-Z gate works similarly to the CNOT gate, only instead of flipping the target qubit (applying an X gate), we apply a Z gate:

$$\textbf{CZ} \hspace{.2cm} |\hspace{.05cm}00\rangle \hspace{.45cm} \rightarrow \hspace{.45cm} |\hspace{.05cm}00\rangle $$
$$\textbf{CZ} \hspace{.2cm} |\hspace{.05cm}01\rangle \hspace{.45cm} \rightarrow \hspace{.45cm} |\hspace{.05cm}01\rangle $$
$$\textbf{CZ} \hspace{.2cm} |\hspace{.05cm}10\rangle \hspace{.45cm} \rightarrow \hspace{.45cm} |\hspace{.05cm}10\rangle $$
$$\textbf{CZ} \hspace{.2cm} |\hspace{.05cm}11\rangle \hspace{.45cm} \rightarrow \hspace{.15cm} -|\hspace{.05cm}11\rangle $$

$$
\begin{bmatrix}
1 & 0 & 0 & 0\\
0 & 1 & 0 & 0\\
0 & 0 & 1 & 0\\
0 & 0 & 0 & -1\\
\end{bmatrix}
$$

Recall that a Z gates leaves a qubit in the state $|0\rangle$ untouched, while flipping the sign on a qubit in the state $|1\rangle$. Thus, the CZ gate performs a similar operation, only affecting the state $|11\rangle$, as shown above.

\pagebreak

\begin{figure}[h]
\centering
\includegraphics[scale=.65]{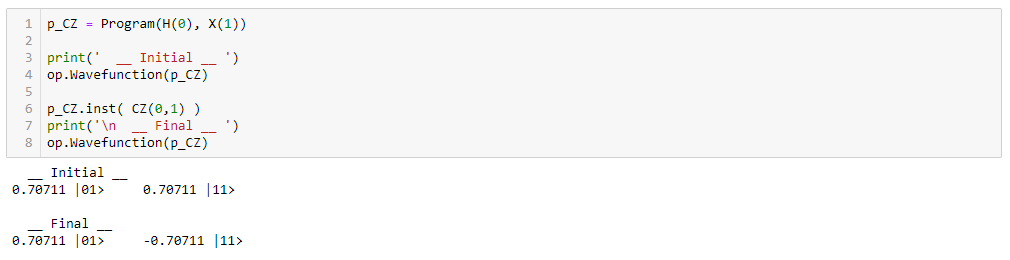}
\end{figure}

\section*{\large{ Control Phase Gates }}
\centerline{---------------------------------------------------------------------------------------------------------------------------------}

The following four gates will be bundled together, since they all represent the same general effect.

\section*{\large{ CPHASE($\phi$) }}

The control-phase gate, also referred to as a CPHASE gate, uses a control qubit to apply a $R_{\phi}$ gate to a target qubit. The net effect is similar to that of the control-Z gate, only differing by the phase that gets multiplied to the state $|111\rangle$:

$$
\begin{bmatrix}
1 & 0 & 0 & 0\\
0 & 1 & 0 & 0\\
0 & 0 & 1 & 0\\
0 & 0 & 0 & e^{i\phi}\\
\end{bmatrix}
$$

Pyquil comes with addition CPHASE operators, one for each of the four states in a 2-qubit systems.

\section*{\large{ CPHASE10($\phi$) }}

$$
\begin{bmatrix}
1 & 0 & 0 & 0\\
0 & 1 & 0 & 0\\
0 & 0 & e^{i\phi} & 0\\
0 & 0 & 0 & 1\\
\end{bmatrix}
$$

\section*{\large{ CPHASE01($\phi$) }}

$$
\begin{bmatrix}
1 & 0 & 0 & 0\\
0 & e^{i\phi} & 0 & 0\\
0 & 0 & 1 & 0\\
0 & 0 & 0 & 1\\
\end{bmatrix}
$$

\section*{\large{ CPHASE00($\phi$) }}

$$
\begin{bmatrix}
e^{i\phi} & 0 & 0 & 0\\
0 & 1 & 0 & 0\\
0 & 0 & 1 & 0\\
0 & 0 & 0 & 1\\
\end{bmatrix}
$$

Since all of these operators achieve the same result, we will only show the case for CPHASE here:

\begin{figure}[h]
\centering
\includegraphics[scale=.65]{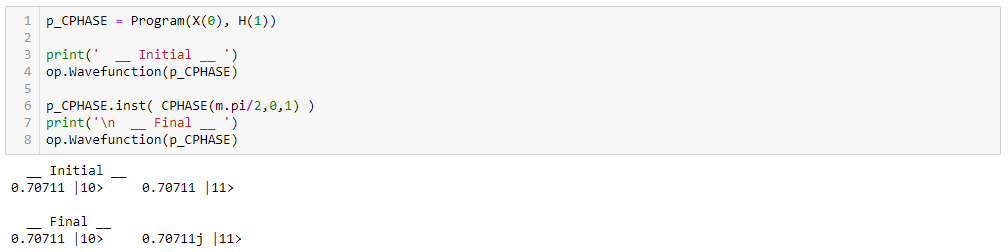}
\end{figure}

\section*{\large{ SWAP }}

The SWAP gate causes two qubits to trade states.

$$\textbf{SWAP} \hspace{.2cm} |\hspace{.05cm}00\rangle \hspace{.25cm} \rightarrow \hspace{.25cm} |\hspace{.05cm}00\rangle $$
$$\textbf{SWAP} \hspace{.2cm} |\hspace{.05cm}01\rangle \hspace{.25cm} \rightarrow \hspace{.25cm} |\hspace{.05cm}10\rangle $$
$$\textbf{SWAP} \hspace{.2cm} |\hspace{.05cm}10\rangle \hspace{.25cm} \rightarrow \hspace{.25cm} |\hspace{.05cm}01\rangle $$
$$\textbf{SWAP} \hspace{.2cm} |\hspace{.05cm}11\rangle \hspace{.25cm} \rightarrow \hspace{.25cm} |\hspace{.05cm}11\rangle $$

$$
\begin{bmatrix}
1 & 0 & 0 & 0\\
0 & 0 & 1 & 0\\
0 & 1 & 0 & 0\\
0 & 0 & 0 & 1\\
\end{bmatrix}
$$

A simple way of viewing the effect of this gate is that all of the $0$'s and $1$'s in each state switch places. As a result, we can see that the SWAP gate has no effect on the states $|00\rangle$ and $|11\rangle$.

\pagebreak

\begin{figure}[h]
\centering
\includegraphics[scale=.65]{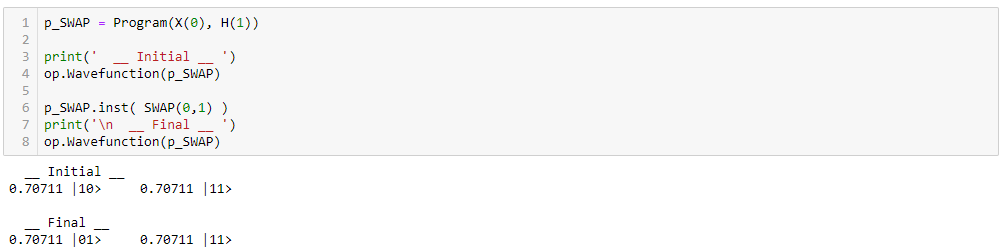}
\end{figure}

\section*{\large{ 3 Qubit Control Gates }}
\centerline{---------------------------------------------------------------------------------------------------------------------------------}

The following two gates take 3 qubits as inputs. They are essentially higher order versions of the CNOT and SWAP gates, adding one extra control qubit to each.

\section*{\large{ CSWAP }}

The control-swap gate uses a control qubit to determine whether or not to apply a SWAP gate to two target qubits. If the control qubit is in the state $|1\rangle$, then a SWAP gate is performed. Examples:

$$\textbf{CSWAP} \hspace{.2cm} |\hspace{.05cm}010\rangle \hspace{.25cm} \rightarrow \hspace{.25cm} |\hspace{.05cm}010\rangle $$
$$\textbf{CSWAP} \hspace{.2cm} |\hspace{.05cm}101\rangle \hspace{.25cm} \rightarrow \hspace{.25cm} |\hspace{.05cm}110\rangle $$
$$\textbf{CSWAP} \hspace{.2cm} |\hspace{.05cm}110\rangle \hspace{.25cm} \rightarrow \hspace{.25cm} |\hspace{.05cm}101\rangle $$
$$\textbf{CSWAP} \hspace{.2cm} |\hspace{.05cm}111\rangle \hspace{.25cm} \rightarrow \hspace{.25cm} |\hspace{.05cm}111\rangle $$

$$
\begin{bmatrix}
1 & 0 & 0 & 0 & 0 & 0 & 0 & 0\\
0 & 1 & 0 & 0 & 0 & 0 & 0 & 0\\
0 & 0 & 1 & 0 & 0 & 0 & 0 & 0\\
0 & 0 & 0 & 1 & 0 & 0 & 0 & 0\\
0 & 0 & 0 & 0 & 1 & 0 & 0 & 0\\
0 & 0 & 0 & 0 & 0 & 0 & 1 & 0\\
0 & 0 & 0 & 0 & 0 & 1 & 0 & 0\\
0 & 0 & 0 & 0 & 0 & 0 & 0 & 1\\
\end{bmatrix}
$$

This gate is also sometimes referred to as a Fredkin Gate.

\pagebreak

\begin{figure}[h]
\centering
\includegraphics[scale=.65]{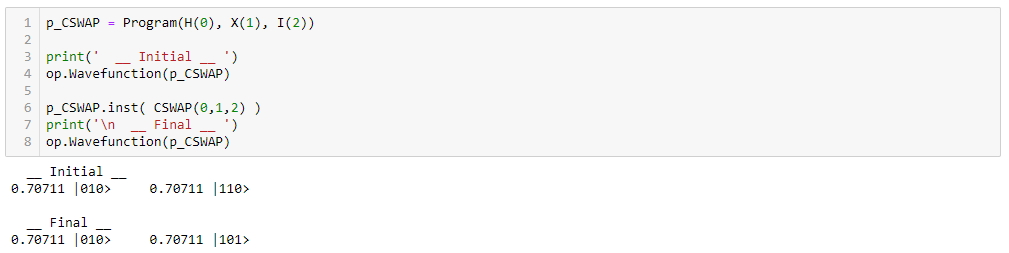}
\end{figure}

\section*{\large{ CCNOT }}

The control-control not gate uses two control qubits to determine if an X gate is applied to a single target qubit. Examples:

$$\textbf{CCNOT} \hspace{.2cm} |010\rangle \hspace{.25cm} \rightarrow \hspace{.25cm} |010\rangle $$
$$\textbf{CCNOT} \hspace{.2cm} |101\rangle \hspace{.25cm} \rightarrow \hspace{.25cm} |101\rangle $$
$$\textbf{CCNOT} \hspace{.2cm} |110\rangle \hspace{.25cm} \rightarrow \hspace{.25cm} |111\rangle $$
$$\textbf{CCNOT} \hspace{.2cm} |111\rangle \hspace{.25cm} \rightarrow \hspace{.25cm} |110\rangle $$

$$
\begin{bmatrix}
1 & 0 & 0 & 0 & 0 & 0 & 0 & 0\\
0 & 1 & 0 & 0 & 0 & 0 & 0 & 0\\
0 & 0 & 1 & 0 & 0 & 0 & 0 & 0\\
0 & 0 & 0 & 1 & 0 & 0 & 0 & 0\\
0 & 0 & 0 & 0 & 1 & 0 & 0 & 0\\
0 & 0 & 0 & 0 & 0 & 1 & 0 & 0\\
0 & 0 & 0 & 0 & 0 & 0 & 0 & 1\\
0 & 0 & 0 & 0 & 0 & 0 & 1 & 0\\
\end{bmatrix}
$$

This gate is also sometimes referred to as a Toffoli Gate. Much like the CNOT gate, the effect of this gate is equivalent to an X gate on the states $|110\rangle$ and $|111\rangle$.

\begin{figure}[h]
\centering
\includegraphics[scale=.65]{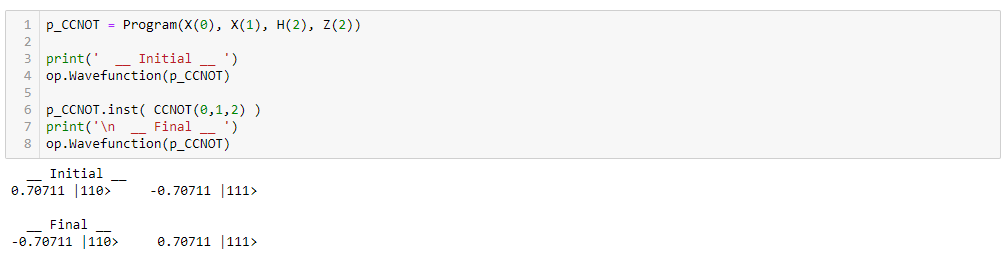}
\end{figure}

\section*{\large{ Defining Gates }}
\centerline{---------------------------------------------------------------------------------------------------------------------------------}

Pyquil allows the user to create more gates beyond the basic set provided above, so long as one can specify the gate's matrix representation. In order for a defined gate to be valid, it must be a unitary matrix of size $2^{n} \times 2^{n}$, where n is the number of qubits the gate operators on. Unitary matrices are defined as follows:

$$ UU^{\dagger} = I $$

where I is the identity matrix, and $\dagger$ is the complex conjugate of a matrix.
\\

As a simple example, let's try defining I$^2$, the 2-qubit identity matrix:

\begin{figure}[h]
\centering
\includegraphics[scale=.65]{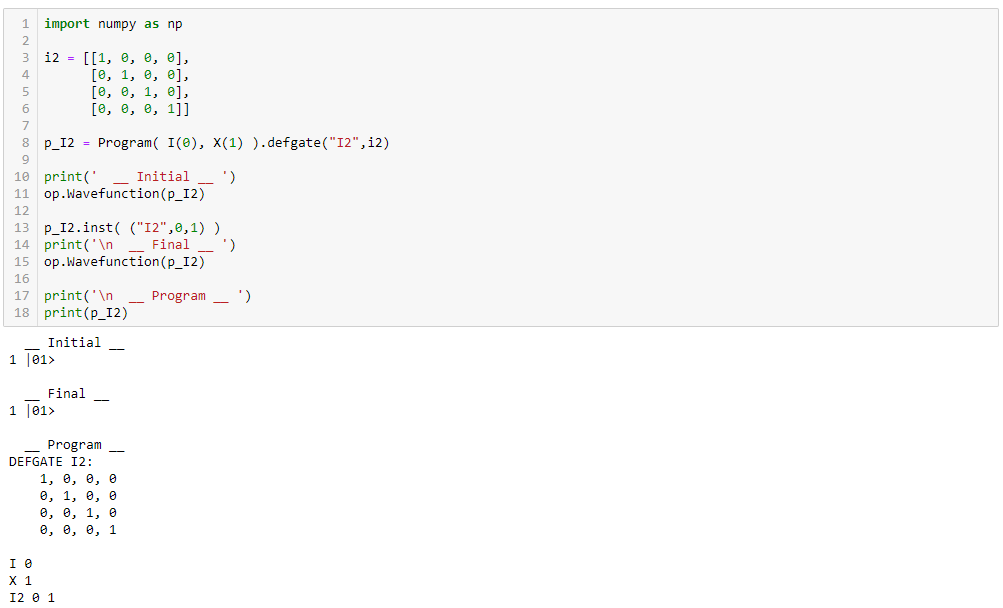}
\end{figure}

The I$^2$ operator works as intended, leaving both states unchanged.
\\

The function that allows us to define our own gates is $\textbf{DefGate}$. This function takes two arguments: $\hspace{.1cm}$ 1) a string to store the name of the operator $\hspace{.1cm}$ 2) a unitary matrix, which can be a list, array, numpy array, etc. $\textmf{DefGate}$ is a part of the $\textmf{Program}$ class, so we can attach it after any $\textmf{Program}$ object:
\\

--------------------------------------------------------------------------------------------

.DefGate("NAME",matrix)

--------------------------------------------------------------------------------------------
\\

Once defined within a program, we can apply our custom operator using a tuple:
\\

--------------------------------------------------------------------------------------------

("NAME",qubit0,qubit1,...)

--------------------------------------------------------------------------------------------
\\

The first entry in the tuple will be the name we've given to the operator, followed by all of the qubits that we intend to operate on. Once defined and a part of the program, we can view it by printing the program, as shown above.
\\

Let's try one more example, creating a 4$\times$4 matrix for $X_0 \otimes Z_1$, an X gate on qubit $0$ and a Z gate on qubit $1$. The matrix for this gate is given by:

$$ X_0 \otimes Z_1 $$

$$
\begin{bmatrix}
0 & 1 \\
1 & 0 \\
\end{bmatrix} \hspace{.2cm} \otimes \hspace{.2cm}
\begin{bmatrix}
1 & 0 \\
0 & -1 \\
\end{bmatrix}
$$

$$
\begin{bmatrix}
0 \cdot \begin{bmatrix}
1 & 0 \\
0 & -1 \\
\end{bmatrix} & 1 \cdot \begin{bmatrix}
1 & 0 \\
0 & -1 \\
\end{bmatrix} & \\
1 \cdot \begin{bmatrix}
1 & 0 \\
0 & -1 \\
\end{bmatrix} & 0 \cdot \begin{bmatrix}
1 & 0 \\
0 & -1 \\
\end{bmatrix} & \\
\end{bmatrix}
$$

$$
\begin{bmatrix}
0 & 0 & 1 & 0 \\
0 & 0 & 0 & -1 \\
1 & 0 & 0 & 0 \\
0 & -1 & 0 & 0 \\
\end{bmatrix}
$$

where the steps shown here are a tensor product of matrices.
\\

Let's use this matrix and compare it to applying the X and Z gates separately:

\begin{figure}[h]
\centering
\includegraphics[scale=.65]{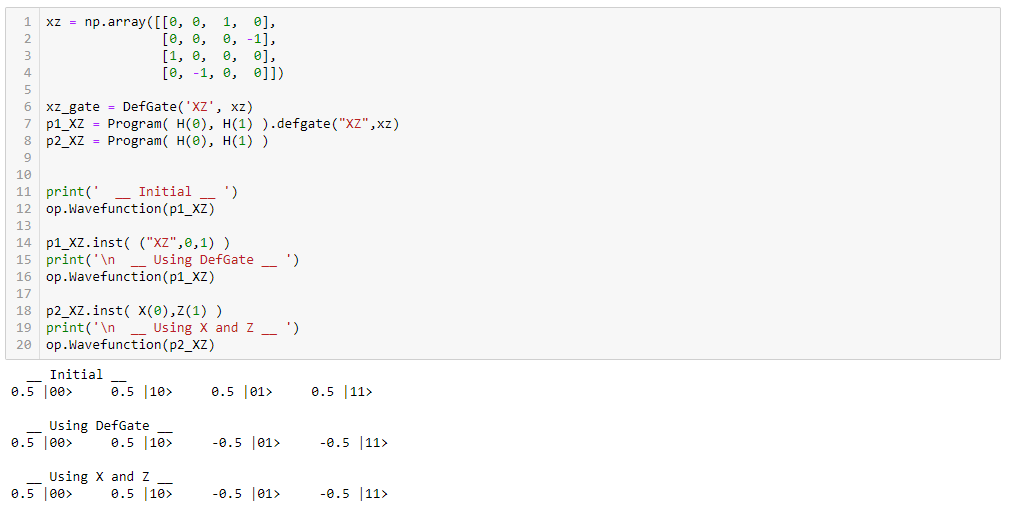}
\end{figure}

Both states are the same, which means our XZ gate did it's job correctly. Using a tensor product to combine any two individually unitary gates into a single higher order gate will always be unitary as well. Thus, we can make arbitrarily large unitary operators, so long as we construct them from smaller unitary matrices.

--------------------------------------------------------------------------------------------------------------------------------------------------------
\\

This concludes all of the quantum gates provided by Pyquil that we will cover here. For the complete list of standard gates, check out:
\\

http://docs.rigetti.com/en/stable/apidocs/gates.html
\\

Using the gates covered in this lesson, along with all of Pyquil knowledge from lessons 1 \& 2, we are now ready to begin coding up some of the most academically important quantum algorithms. But first, I encourage you to take a look at lesson 4, which covers some additional tools not provided by Pyquil, but will be immensely helpful in our education purposes.
\\

--------------------------------------------------------------------------------------------------------------------------------------------------------


\pagebreak

\section*{\Large{ Lesson 4 - Our Custom Functions }}
--------------------------------------------------------------------------------------------------------------------------------------------------------
\\

In this lesson, we will be covering some functions and operators that are not a part of Pyquil, but will be used frequently in the coming lessons. In particular, we will go through some important functions from the python file $\textmf{Our\_Pyquil\_Functions}$. The motivation for using these customs functions will be in the hopes that they make the learning endeavor of future lessons easier.
\\

Before proceeding, please consider reading the previous lessons in this series, which covers all of the Pyquil basics of programs and measurements needed for this lesson:
\\

Lesson 1 - Intro to Programs and Measurements
\\

Lesson 2 - Creating More Complex Programs
\\

Lesson 3 - Gates Provided by Pyquil
\\

--------------------------------------------------------------------------------------------------------------------------------------------------------
\\
In order to make sure that all cells of code run properly throughout this lesson, please run the following cell of code below:

\begin{figure}[h]
\centering
\includegraphics[scale=.65]{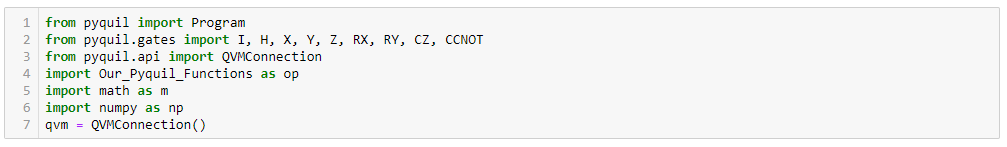}
\end{figure}

Throughout all of the coming lessons, the two custom functions that we will be using constantly are $\textbf{Wavefunction}$ and $\textbf{Measurement}$. These functions will allow us to view our quantum systems with ease. In particular, these functions will handle many of the tedious steps needed to display wavefunctions and measurement results. In addition, both of these functions come with some optional arguments, which give us more control of how we would like to view our results.
\\

Both of these functions do not amend our code in any way. In essence, they are designed for learning purposes only, providing the user tools for viewing quantum systems in a more understandable way.

\section*{\large{Wavefunction}}
\centerline{---------------------------------------------------------------------------------------------------------------------------------}

$\textmf{Wavefunction}$ will hands down be the most common function we call upon from $\textmf{Our\_Pyquil\_Functions}$. In fact, we've already seen its default use numerous times in lessons 2 and 3. This is because viewing the quantum systems we create is very important if we want to understand what is going on! In the previous lessons, we have only seen the default use of this function, which takes a $\textmf{Program}$ and prints the amplitudes associated with each state:

\begin{figure}[h]
\centering
\includegraphics[scale=.65]{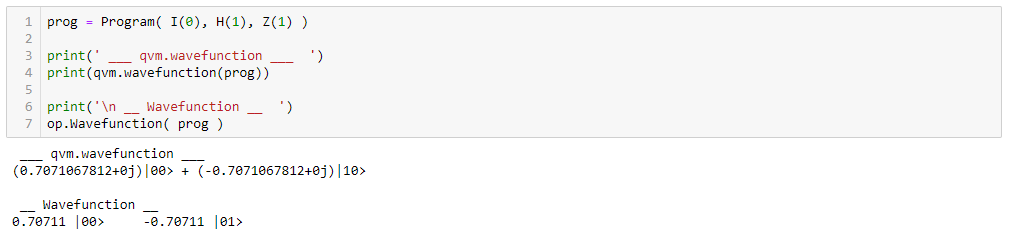}
\end{figure}

Take a look at the cell of code above and notice the differences between $\textbf{qvm.wavefunction}$ and $\textmf{Wavefunction}$. Ignoring the difference in decimal precisions, which can be changed for both, the biggest difference is display simplicity. At its core, $\textmf{Wavefunction}$ is still calling upon QVM in the normal way, but with some extra lines of code to remove some of the clutter.
\\

But this alone isn't enough to warrant using $\textmf{Wavefunction}$. The true intention for using this custom function is in the additional options it provides us.

\section*{\large{ Precision }}

This first optional argument that we will take a look at will control the decimal precision for our amplitudes. This is done using the argument $\textbf{precision}$, which takes an integer:

\begin{figure}[h]
\centering
\includegraphics[scale=.65]{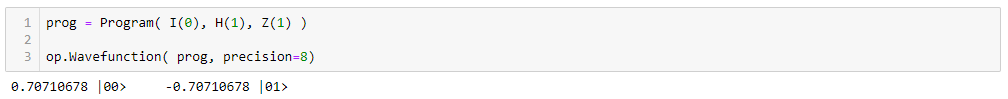}
\end{figure}

Please note that this is also possible with QVM, using the function $\textbf{pretty\_print}$:

\begin{figure}[h]
\centering
\includegraphics[scale=.65]{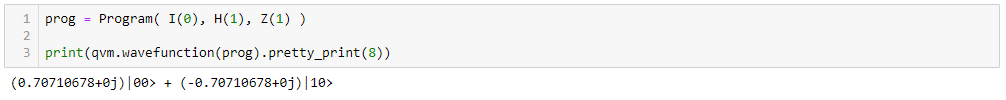}
\end{figure}

Thus, this feature isn't anything new, but it is still important nevertheless. If the precision argument isn't given, $\textmf{Wavefunction}$ will go to 5 decimals of precision by default.

\section*{\large{ Column }}

The next argument is a change in the layout that the states are displayed, but one that can be quite necessary as our quantum systems grow larger. By passing the $\textbf{column}$ argument and setting it to True, each state in the wavefunction will be display as its own line. This will be quite handy when we want to focus on a single state amongst the clutter of many.

\begin{figure}[h]
\centering
\includegraphics[scale=.65]{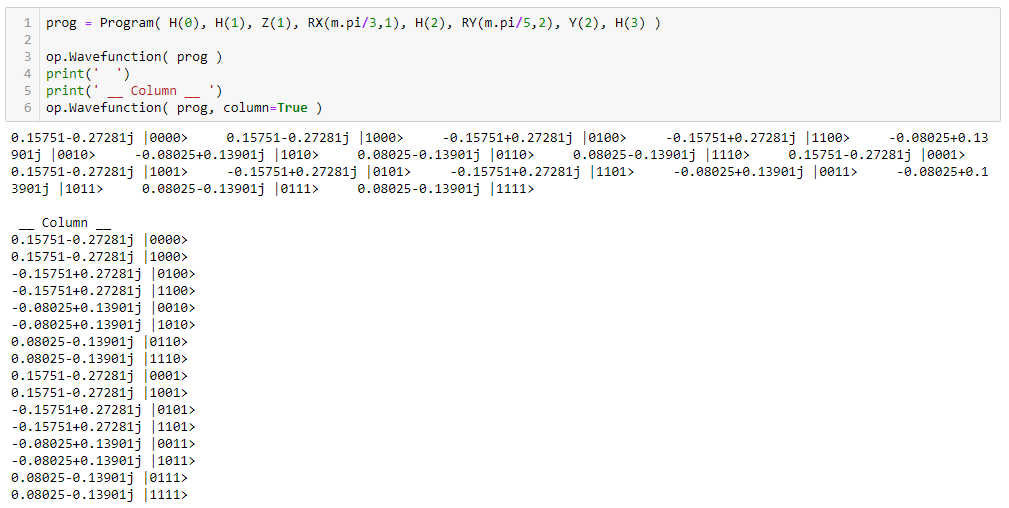}
\end{figure}

\section*{\large{ Systems and Show\_Systems }}

In many of the coming algorithms, we will be dealing with ancilla qubits. These qubits make up secondary systems, which serve specific purposes, but ultimately are unimportant in the final measurement. When dealing with these ancilla systems, we may wish to sometimes view their qubits for learning purposes, and other times choose to simply ignore them. As a tool for separating ancilla systems, and choosing whether or not to display them, we will use the arguments $\textbf{systems}$ and $\textbf{show\_systems}$.
\\

When we pass the argument $\textmf{systems}$, we must set it equal to a list containing the groupings of qubits. The sum of this list must equal the total number of qubits in the system. For example:

\begin{figure}[h]
\centering
\includegraphics[scale=.65]{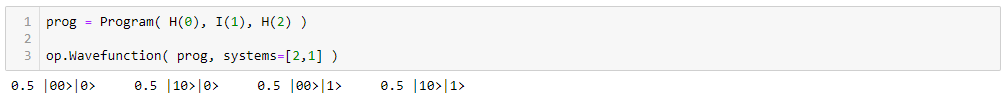}
\end{figure}

In this example, our quantum system is a state of 3 qubits. But perhaps the third qubit is an ancilla, and we would like to monitor is separately from qubits $0$ and $1$. By passing the argument $\textmf{systems}$ = [2,1], we get the displayed wavefunction above, which puts the ancilla qubit as its own state. Specifically, this argument groups qubits together by their numerical order, according to the length and values of the list. In this example, [2,1] tells the function to display qubits $0$ and $1$ as a group, followed by qubit $2$ as its own group.
\\

Mathematically, we must remember that separating qubits off like this is still the same physical state:

$$ |\hspace{.06cm} 10 \rangle \hspace{.05cm} | \hspace{.06cm} 1 \rangle \hspace{.25cm} = \hspace{.25cm} |\hspace{.06cm} 101 \rangle $$

Thus, this is once again just a cosmetic change, but a very useful one in lessons to come.
\\

Now, suppose we have a group of ancilla qubits, or even several groups, which we do not want to display when we call upon $\textmf{Wavefunction}$. For example, we often times deal with ancilla qubits that always remain in the state of all $0$'s before and after an operation. Thus, viewing all these qubits in the $|0\rangle$ state becomes repetitive, and adds a lot of unnecessary clutter. To avoid this, we can use the $\textmf{show\_systems}$ argument to choose which systems we want to view:

\begin{figure}[h]
\centering
\includegraphics[scale=.65]{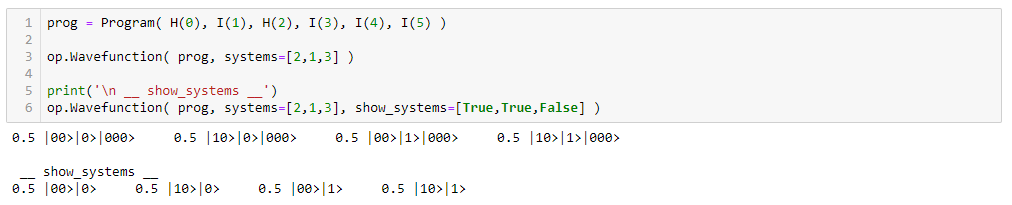}
\end{figure}

As shown here, every state in the system carries the same $|000\rangle$ system of ancilla qubits. By passing the argument $\textmf{show\_systems}$, and setting it equal to a list of equal length to $\textmf{systems}$, containing truth values, we can opt to remove certain systems from our display. Specifically, the index locations of each True or False in the $\textmf{show\_systems}$ argument correspond to the same groups in the $\textmf{systems}$ argument. We will use this argument quite frequently in later lessons to avoid any extra clutter on our systems.
\\

Some important points about using $\textmf{systems}$ and $\textmf{show\_systems}$:
\\

1) Both arguments must be lists of equal length, containing only integers and truth values respectively.
\\

2) $\textmf{show\_systems}$ will only work if $\textmf{systems}$ is also an argument. Passing $\textmf{show\_systems}$ by itself will result in no change to the display of $\textmf{Wavefunction}$.
\\

3) Using $\textmf{show\_systems}$ to remove a system from the display of a wavefunction should only be used when all states in the system have the same ancilla state. Otherwise, the printed wavefunction will show duplicates of the same state:

\pagebreak

\begin{figure}[h]
\centering
\includegraphics[scale=.65]{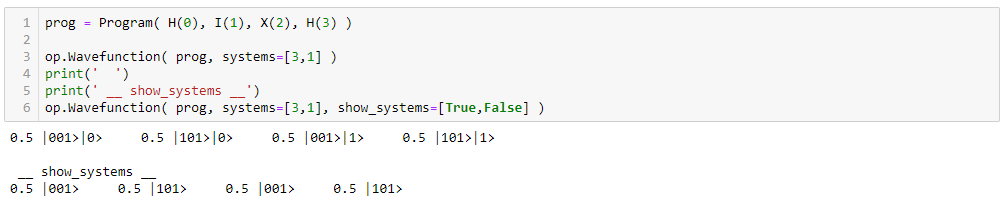}
\end{figure}

Compare the two lines above, and note that choosing to not display the ancilla qubit is problematic. Since the ancilla qubit is in a superposition, choosing to not display its results in the wavefunction looks very odd. Mathematically, this odd-looking state is still technically correct in some sense, showing the associated amplitudes with each possible state of the main system (so long as we don't add states together). However, this could lead to some potentially very confusing results, and should be avoided.

\section*{\large{Measurement}}
\centerline{---------------------------------------------------------------------------------------------------------------------------------}

For instances where we will need to make a measurement on our quantum system, or many, we will call upon the $\textmf{Measurement}$ function to handle several aspects of the measurement. In particular, this function contains within it the $\textmf{run}$ function, and all of the necessary lines of code to run a simulated measurement.
\\

However, the one thing that it won't do is add a measurement instruction to the program. This is a step that we must do manually before calling upon $\textmf{Measurement}$. The $\textmf{Measurement}$ function requires two arguments: $\hspace{.1cm}$ 1) the program $\hspace{.1cm}$2) a list containing indices. This list acts as the classical register, thus it should contain the indices of the qubits being measured:

\begin{figure}[h]
\centering
\includegraphics[scale=.65]{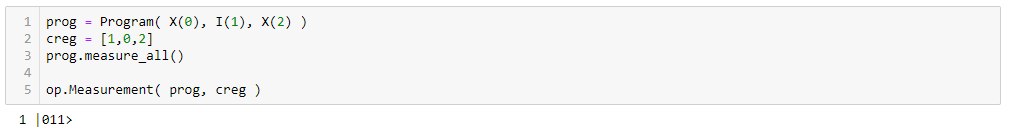}
\end{figure}

As we can see in this example, the measurement results are correct. Just like before, if we change the indices in the list $\textmf{creg}$ then our results will reflect the order in which we've chosen to view the qubits (as shown in the example above). Thus, if we are interested in viewing accurate measurement results, we must be sure to pass a list that corresponds with the ordering of our qubits.

\section*{\large{ runs }}

The most common argument we will pass to the $\textmf{Measurement}$ function will be $\textbf{runs}$, which is the argument we referred to as 'trials' in previous lessons when working with the normal $\textmf{run}$ function. Here, passing $\textmf{runs}$ and setting it equal to an integer will tell the function how many times to measure the system:

\begin{figure}[h]
\centering
\includegraphics[scale=.65]{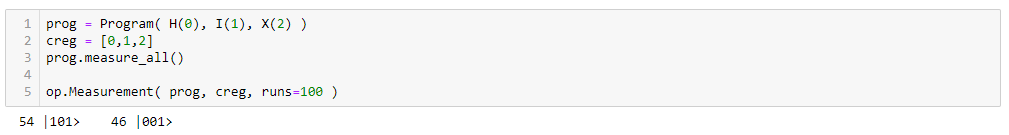}
\end{figure}

Passing the $\textmf{runs}$ argument is useful when we would like to simulate measurements on our system, and gain some insight into the different probabilities of measuring certain states.

\section*{\large{ print\_M }}

Most of the time we will want to display the measurement results of our qubits, but not always. Perhaps we would like to perform a measurement on our system, store the results, but not necessarily view them. Although the intention behind $\textmf{Measurement}$ is for displaying results, we can opt to display nothing by passing the $\textbf{print\_M}$ argument, and setting it to False:

\begin{figure}[h]
\centering
\includegraphics[scale=.65]{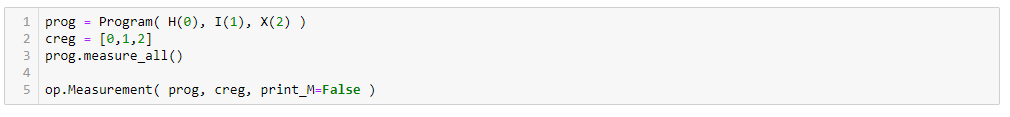}
\end{figure}

Running this code should result in nothing displayed. By itself, setting $\textmf{print\_M}$ to False is sort of a mute step. Nothing is displayed, and nothing is returned to us from the function.
\\

However, often times the motivation behind making a measurement and not viewing the results is to extract these results for some other purpose. For example, in lesson $1$ we used the results of a coin flipping algorithm to determine a winner. If this is our intent, then in order to get back measurement results, we need only pass the argument $\textbf{return\_M}$ as True, which will return a dictionary object to us:

\begin{figure}[h]
\centering
\includegraphics[scale=.65]{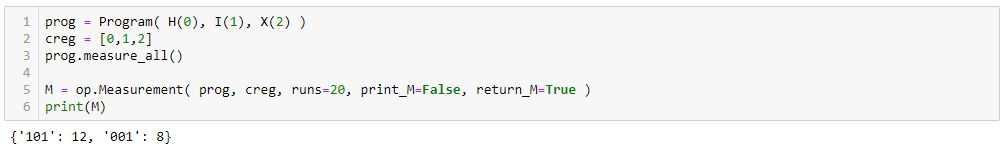}
\end{figure}

This dictionary object contains all of the simulated measurement results, for whatever use an algorithm may require. Please note that when using the $\textmf{return\_M}$ argument, we must create a variable to store the returned dictionary object. And also, $\textmf{return\_M}$ and $\textmf{print\_M}$ do not need to be used together. Both arguments are independent of each other, which means we can choose to display results / extract results in any manner we like.
\\

In addition to the arguments already covered, $\textmf{Measurement}$ also takes the $\textmf{column}$ argument as well. Passing the $\textmf{column}$ argument and setting it to True results in the same style of display as shown above.

\section*{\large{ Higher Order Control Gates }}
\centerline{---------------------------------------------------------------------------------------------------------------------------------}

The next two functions that we are about to study will be used very frequently in the coming lessons, but not necessarily always on display. In particular, many of the algorithms we will study in lesson 5 require higher order control gates as a smaller component, typically written into other functions that we will call upon in $\textmf{Our\_Pyquil\_Functions}$. As we saw in lesson 3, Pyquil's standard gates only comes with a handful of control gates. In particular, the largest of which use two control qubits (CCNOT). While the gates provided by Pyquil are all that we need for a universal set, it will be helpful to define our own function for constructing these higher order control gates.
\\

For any control gate, we want to perform some operation $\textit{only}$ if the control qubit(s) is in the $|1\rangle$ state. For higher order control gates, any state in the system where a qubit is in the $|0\rangle$ state will not receive the operation. The problem then becomes, how do we construct arbitrarily large control gates using only the tools provided to us by Pyquil. One option would be to use $\textmf{DefGate}$, if we knew how to represent our operator as a matrix.
\\

While $\textmf{DefGate}$ is a great tool, relying on it too much is problematic for our learning purposes. Specifically, a major component to learning how to program quantum algorithms is overcoming both conceptual and technological challenges. $\textmf{DefGate}$ side steps technological hurdles of quantum algorithms by allowing us to perform operators without knowing how to construct them.

\section*{\large{ The Strategy }}

The way we are going to construct our higher order control gates is a straightforward strategy, using only CCNOT and CNOT gates. This approach is by no means optimal for many cases, but it is a good foundation for how a higher order control gate $\textit{should}$ function.
\\

The major hurdle to overcome is that we must use the conditions on $N$ qubits in order to invoke a single operation. A single CCNOT gate can only take two control qubits at a time, thus leaving us well short of our goal. However, the trick to this strategy will be to use the aid of ancilla qubits. In essence, these ancilla qubits will allow us to temporarily store information about our control qubits, and ultimately determine whether or not to apply the operation.
\\

First, let's see the general strategy written in terms of classical code:

\begin{figure}[h]
\centering
\includegraphics[scale=.65]{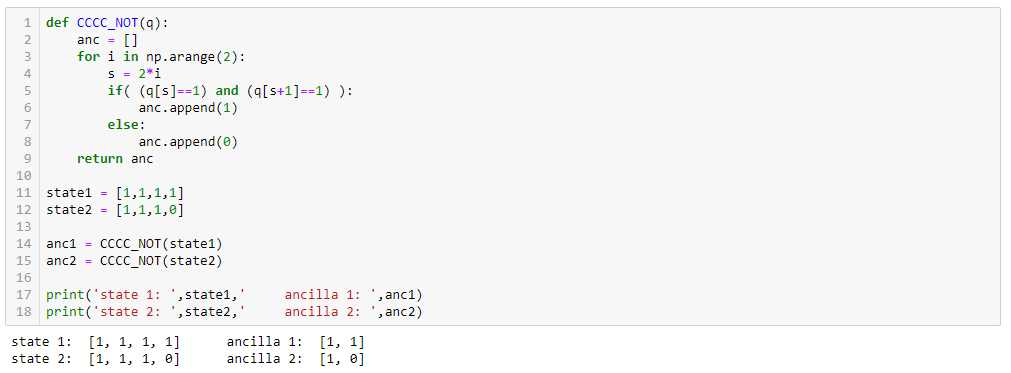}
\end{figure}

In this example, we can see that 'state1' results in the state of all $1$'s for the ancilla system, while 'state2' results in a $0$ for the second ancilla qubit. If we were to use these two ancilla qubits as the control qubits for a CCNOT, the first system would receive the operation, while the second wouldn't. If we compare this result to our initial states, we would have achieved exactly a 4-qubit control gate: the state $|1111\rangle$ receives the operation while the state $|1110\rangle$ does not.
\\

The key point to this example is the way in which we reduced our 4-qubit problem down to 2. Specifically, we work through our control qubits in groups of 2, putting an ancilla qubit in the state $|1\rangle$ or $|0\rangle$ depending on if the two control qubits are in the state $|11\rangle$ or not. For our quantum code, we are going to invoke this strategy using CCNOT gates to reduce the dimension of our problem by 1 per CCNOT gate, until we eventually arrive at only 2 ancilla qubits:

\begin{figure}[h]
\centering
\includegraphics[scale=.6]{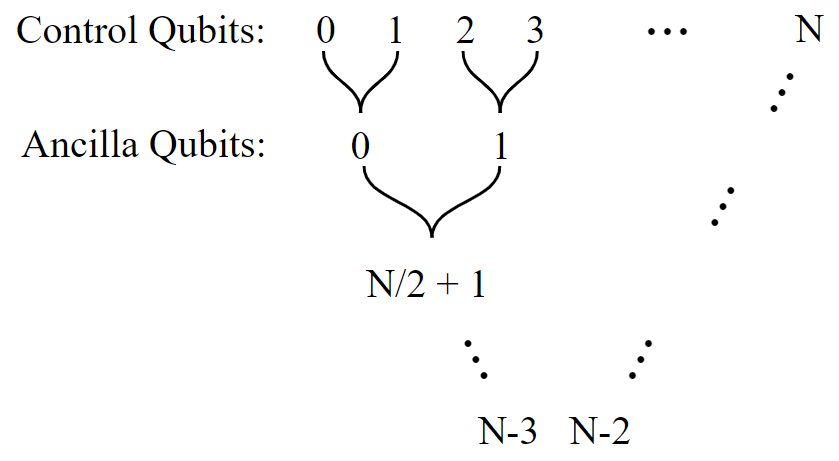}
\end{figure}

In this diagram, each layer moving downward holds the information about the preceding layer, where the control qubits are the highest layer (and ultimately the source of the operation). The junctions connecting two qubits in the diagram represent a CCNOT gate, where the resulting state of the ancilla qubit is $|1\rangle$ if both control qubits are in the $|1\rangle$ state, and a $|0\rangle$ if either of the controlling qubits are in the state $|0\rangle$. Using this recursive process, if a single qubit from the highest layer is in the $|0\rangle$ state, it will trickle down all the way to the final layer, which in turn will mean the control-operation does not happen.
\\

Now, in order for this strategy to work, we need all of the ancilla qubits to be initialized in the state $|0\rangle$. This is because each operation is a CCNOT gate, which is effectively a 2-qubit control X gate. Thus, in order for this CCNOT gate to leave the target ancilla qubits in the state $|1\rangle$, only when control qubits are in the $|1\rangle$ state, the target qubit must initially be in the $|0\rangle$ state.
\\

In total, in order to condense the information of $N$ control qubits down to two ancilla qubits, we require $N-2$ ancilla qubits. From there, if we wish to combine these two final ancilla qubits down to one, for a single qubit control gate, we will require 1 additional ancilla qubit, bringing our total to $N-1$.
\\

Now, requiring up to $N-1$ ancilla qubits just to perform an operation on $N$ qubits may seem like a steep price to pay. Truthfully, it is. Qubits are not exactly plentiful on current quantum computers, which makes this general strategy a little too resource intensive to be practical in most cases. However, this is a problem to keep in mind, but ignore for the time being. Our goal in these lessons is to learn the basics of quantum algorithms, not solve current research efforts (that's your job afterwards). In the coming lessons we will be using this higher order control strategy frequently, because we have the luxury of using simulated qubits. And, thanks to the argument $\textmf{show\_systems}$ from earlier, we can effectively ignore as many ancilla qubits as we want, and focus on the important results.
\\

Now then, let's take a look at what the diagram above looks like in terms of a quantum circuit diagram, using $N=4$ as our example:

\begin{figure}[h]
\centering
\includegraphics[scale=.8]{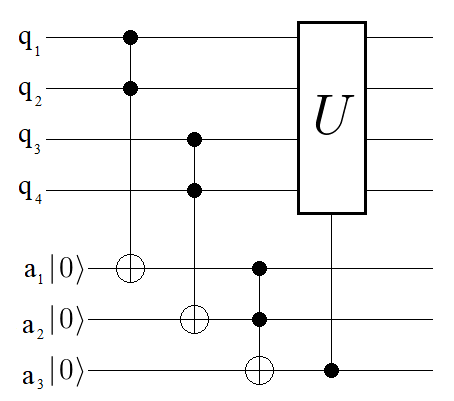}
\end{figure}

In this diagram, we are performing the general $N$-control gate strategy. By using three ancilla qubits, we are able to condense the information of all the states stored on our 4-qubit system, down to a single control ancilla. The operator $U$ in the diagram represents any single control gate we've already seen thus far in lesson 3. As an example, let's implement this diagram in code, replacing $U$ with a control-Z gate:

\pagebreak

\begin{figure}[h]
\centering
\includegraphics[scale=.65]{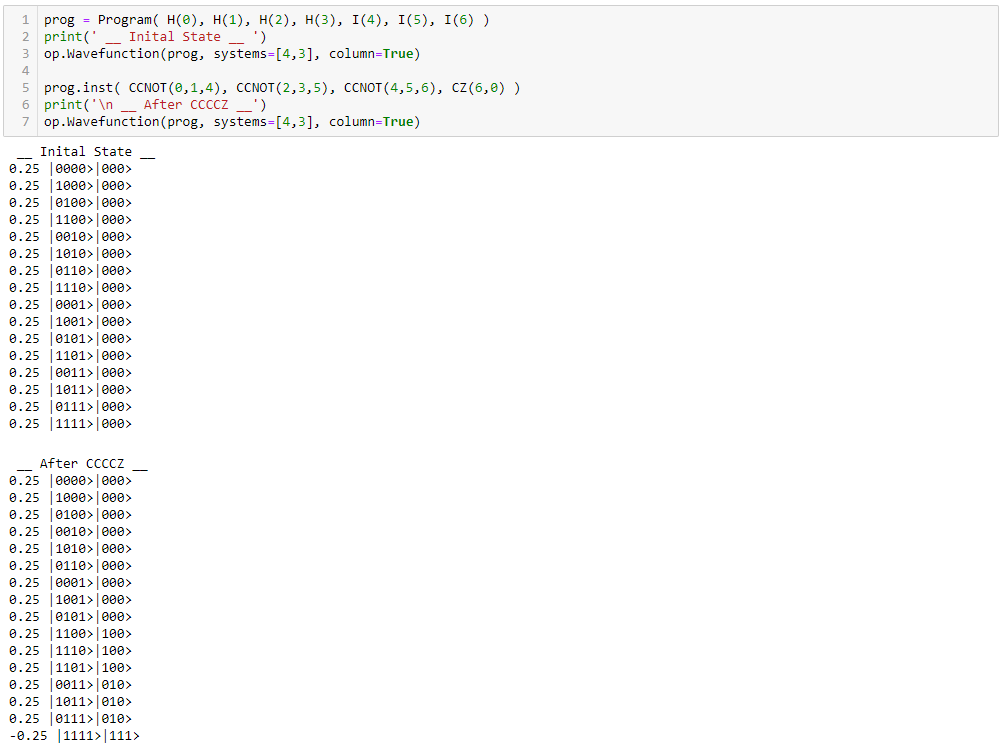}
\end{figure}

Success, the code above successfully picks out the $|1111\rangle$ state and applies a control-Z gate. In this code, we've arbitrarily chosen to apply the CZ gate to qubit $0$, but any of them will result in the same effect. This is because only the state $|1111\rangle$ will pick up the effect, which correspondingly means any of the qubits in this state are candidates to be the target qubit.
\\

While the coding example above works as intended, there is one detail we've overlooked. Namely, the final state of all our ancilla qubits. Take a look at the results above, and notice which ancilla qubits are in the $|1\rangle$ and $|0\rangle$ states. If there were no further steps in our algorithm, we could in principle leave them as they are currently, since they do not affect a measurement on the main system. But if we wanted to apply any further steps, the fact that all of the states in the system have varying ancilla states is problematic. Specifically, states in our main system will no longer undergo superpositions as we may intend.
\\

The remedy for this problem is that we need to return all of the ancilla qubits back to their original state of all $0$'s. To do this, we need only apply all of the CCNOT gates in reverse:

\pagebreak

\begin{figure}[h]
\centering
\includegraphics[scale=.65]{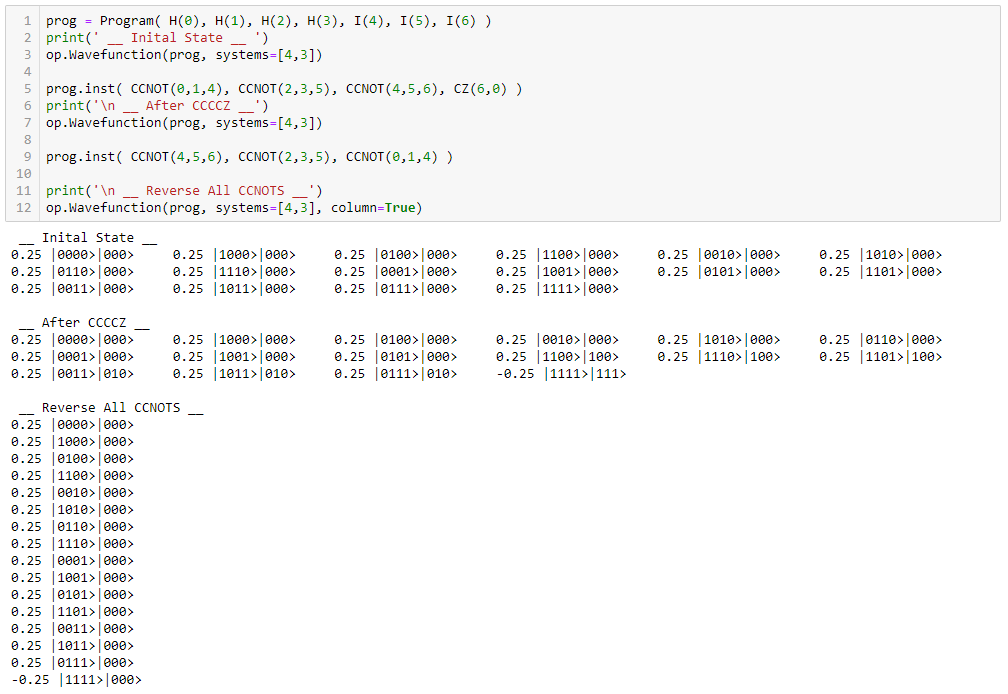}
\end{figure}

The example above is the complete 4-qubit control gate template. In general, this same strategy can be expanded to construct any $N$-control gate.
\\

For our coming tutorials, we will avoid overcrowding our code with all of these steps every time we wish to use such an $N$-control gate (which will be a lot). Thus, we will instead call upon the $\textbf{n\_NOT}$ and $\textbf{n\_Control\_U}$ functions from $\textmf{Our\_Pyquil\_Functions}$ to condense our code. We will use $\textmf{n\_NOT}$ specifically when we want to implement a $N$-control NOT gate, and $\textmf{n\_Control\_U}$ for everything else.

\section*{\large{ n\_NOT }}

The $\textmf{n\_NOT}$ function takes the following arguments:
\\

$\hspace{.1cm} \big{(}$ program, [control qubits], target qubit, [ancilla qubits] $\big{)}$
\\

where the brackets indicate that the argument is a list of the integers corresponding to those particular qubits. For example:

\pagebreak

\begin{figure}[h]
\centering
\includegraphics[scale=.65]{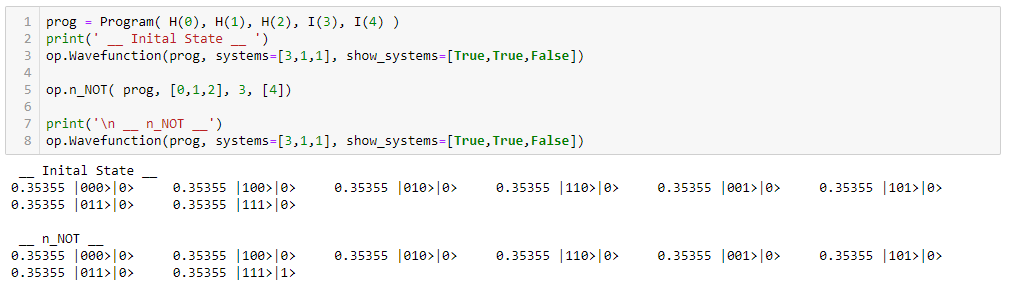}
\end{figure}

In this example we've applied a 3-control NOT gate, where the control qubits are [0, 1, 2], and the target is qubit 3. As shown above, all states initially start with the target qubit in the $|0\rangle$ state. But after we apply our $\textmf{n\_NOT}$ function, the state $|111\rangle$ receives an X gate to its target qubit, flipping it to $|1\rangle$. In addition, all of the ancilla qubit are returned to the $|0\rangle$ state (change the last False in each Wavefunction to verify this for yourself).

\section*{\large{ n\_Control\_U }}

The $\textmf{n\_Control\_U}$ function takes the following arguments:
\\

$\hspace{.1cm} \big{(}$ program, [control qubits], [ancilla qubits], [gates] $\big{)}$
\\

where $\textbf{gates}$ refers to the single-qubit control operations you would like to invoke (can be more than one). Specifically, there are four control-operations supported by this function, with the following formats:
\\

$\textbf{CNOT}: \hspace{.6cm} \big{(}$ 'X', target $\big{)}$
\\

$\textbf{CZ}: \hspace{1.1cm} \big{(}$ 'Z', target $\big{)}$
\\

$\textbf{CPHASE}: \hspace{.1cm} \big{(}$ 'PHASE', target, angle $\big{)}$
\\

$\textbf{CSWAP}: \hspace{.3cm} \big{(}$ 'SWAP', target1, target2 $\big{)}$
\\

Thus, the $\textmf{gates}$ argument for this function is a list of tuples in the forms shown above. Let's see an example of using a control-Z gate, followed by a control-X:

\begin{figure}[h]
\centering
\includegraphics[scale=.65]{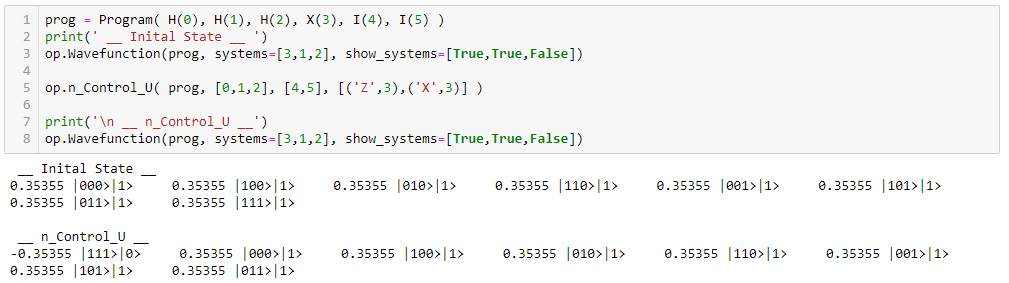}
\end{figure}

Just as intended,this operation first picks out the state $|111\rangle$ and applies a Z gate to the target qubit, followed by an X gate. The result is that the target qubit first picks up a negative phase, and is then flipped to the $|0\rangle$ state.
\\

Since we only need to use all of the CCNOT gates once in order to obtain the control ancilla qubits, we can perform as many control operations as we want before applying all of the CCNOT gates in reverse.

--------------------------------------------------------------------------------------------------------------------------------------------------------
\\

This concludes lesson 4, and all of the most relevant functions that we will be using from $\textmf{Our\_Pyquil\_Functions}$. There are plenty more functions in this python file, and I encourage you to check them out for yourself. From this point forward, everytime we need access to new custom functions, we will discuss them in their respective lessons, when we first encounter them.
\\

--------------------------------------------------------------------------------------------------------------------------------------------------------


\pagebreak

\section*{\Large{ Lesson 5.1 - Into to Quantum Algorithms (Deutsch) }}
--------------------------------------------------------------------------------------------------------------------------------------------------------
\\

This tutorial is the first of four, all labeled 'Lesson 5'. The theme of these lessons is to introduce and explain several 'easier' quantum algorithms. These algorithms are all of historical / academic importance, although perhaps not terribly relevant for application purposes. We shall see several threads of commonality between all four algorithms, which make them a good set of algorithms to learn together.
\\

In this tutorial, we will begin by briefly discussing the context for when we want to use quantum algorithms, and what makes a quantum algorithm 'faster'. Then, we will proceed to the main topic: the Deutsch Algorithm.
\\

Before proceeding, please consider reading the previous lessons in this series, which covers all of the Pyquil basics of programs and measurements needed for this lesson:
\\

Lesson 1 - Intro to Programs and Measurements
\\

Lesson 2 - Creating More Complex Programs
\\

Lesson 3 - Gates Provided by Pyquil
\\

Original publication of the algorithm: \cite{D}
\\

--------------------------------------------------------------------------------------------------------------------------------------------------------
\\
In order to make sure that all cells of code run properly throughout this lesson, please run the following cell of code below:

\begin{figure}[h]
\centering
\includegraphics[scale=.65]{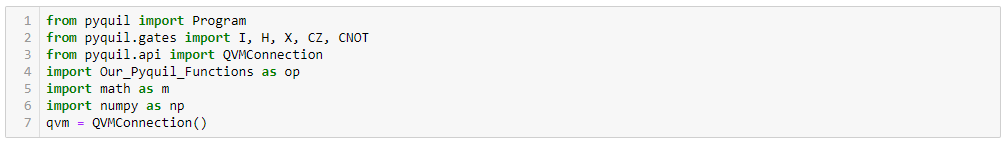}
\end{figure}

\section*{\large{ " The Quantum Advantage " }}
\centerline{---------------------------------------------------------------------------------------------------------------------------------}

Perhaps you've come across this phrase before - 'The Quantum Advantage' - and you weren't sure what it meant, but it sure sounded cool! The Quantum Advantage is referring to the goal that a quantum computer will be able to outperform a classical computer, for some certain process. Hence, the reason we're here, studying quantum algorithms! Currently, there are already a handful of known $\textit{mathematical}$ cases where a quantum computer $\textit{should}$ provide us with a speedup.
\\

As you may suspect based on my use of italics, the realization of these speedups has yet to happen. We're all pretty sure it $\textit{will}$ happen, but not so sure $\textit{when}$, or $\textit{what}$ the algorithm will be. Many believe we are certainly very close, especially with bigger and better quantum computers just on the horizon. The Quantum Advantage is going to take a great deal of collaborative effort from physics / engineering / mathematics / computer science / etc. Thus, the goal of these tutorial lessons is to bring us up to speed on everything we need to know in order to start contributing to this effort.
\\

One disclaimer before we get started however, is that the algorithms we are about to work through do not directly translate to real quantum computers. Or in other words, the algorithms we will be writing assume 'perfect' quantum computers. Much later in this tutorial series we will discuss what it means to design algorithms around current quantum computing hardware, and the new challenges that arise. Thus, take these algorithms with a grain of salt, keeping in mind that we are studying them for their academic value, rather than practical purposes. In particular, the math behind some of these algorithms is quite challenging, and simply understanding how each algorithm works is a major milestone. Then, once you've seen all the 'perfect scenario' quantum algorithms, you will be in a much better position to start designing algorithms on real quantum chips.

\section*{\large{ The Deutsch Algorithm }}
\centerline{---------------------------------------------------------------------------------------------------------------------------------}

In terms of simplicity and elegance, there is no better starting point than the Deutsch Algorithm. It's simple to understand, simple to implement, and gets the point across of what it means to outperform a classical algorithm. So let's begin by framing our problem:

Suppose we are given a 'black box function' $f$. By this we mean that we are given some function $f$, which we can use, but we don't know its effect. Specifically, $f$ acts on a bit of information, either $0$ or $1$, and returns an output, also either $0$ or $1$. Thus, when we feed $f$ the inputs $0$ and $1$, the function will be describable by two out of the four following possibilities:

$$ f(0) \rightarrow 0 \hspace{.5cm} f(0) \rightarrow 1 $$

$$ f(1) \rightarrow 0 \hspace{.5cm} f(1) \rightarrow 1 $$

Based on these possibilities, we can say that $f$ is guaranteed to be either a 'balanced' or 'constant' function. A balanced function means that $f$'s outputs will be half $0$'s and half $1$'s, ex: $\hspace{.15cm} f(0) \rightarrow 1 \hspace{.35cm} f(1) \rightarrow 0 $. A constant function means that the output will be either all $0$'s or all $1$'s, ex: $\hspace{.15cm} f(0) \rightarrow 1 \hspace{.35cm} f(1) \rightarrow 1 $. So then, given this mysterious $f$, what is the minimum number of uses by which we can determine whether it is a balanced or constant function?
\\

Well, let's take a look at the classical approach. Since we can only work with classical bits, let's say we feed the function a $0$, and we get back a $1$. We now have one piece of information: $ f(0) \rightarrow 1 $. But based on this one result, can we conclude what will happen for $f(1)$?
\\

The answer is no. The information we got from one call of the function $f$ is insufficient to determine whether $f$ is a balanced or constant function. If we get $\hspace{.1cm} f(1) \rightarrow 1 $, we will conclude that $f$ is constant, while if we get $\hspace{.1cm} f(1) \rightarrow 0 $, we will conclude that it is balanced. Thus, $\textit{classically}$, we use the black box function $f$ twice in order to determine its nature. If you are still a little unsure, the diagram below represents a flow chart of all the possibilities:

\begin{figure}[h]
\centering
\includegraphics[scale=.6]{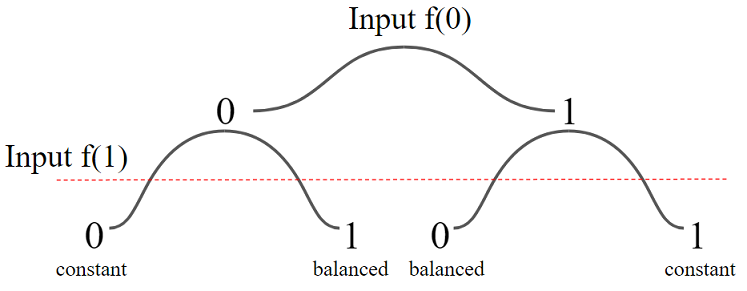}
\end{figure}

Let's write up a simple code to simulate this problem:

\pagebreak

\begin{figure}[h]
\centering
\includegraphics[scale=.65]{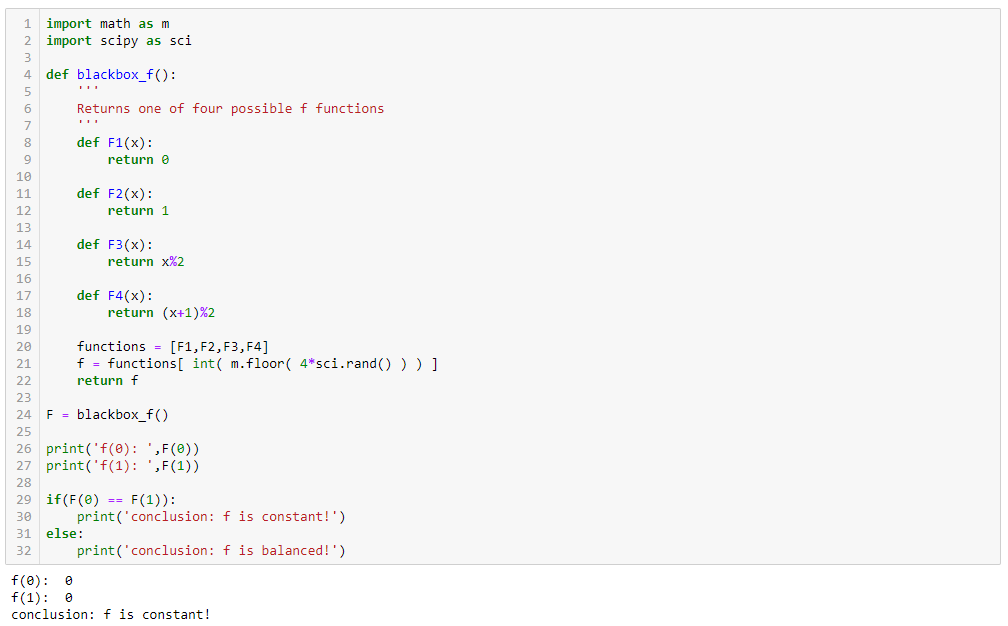}
\end{figure}

The cell of code above randomly generates one of the four possible black box functions, tests it with the inputs $f(0)$ and $f(1)$, and concludes whether the function is balanced or constant based on the results. It's kind of a silly example, but it gets the point across.
\\

Now, let's see if we can do any better with our powerful quantum computers! As you might imagine, there's really only one way to be $\textit{faster}$ than the classical approach here. We need to be able to determine if $f$ is constant or balanced in only one function call.
\\

When we move out of the realm of classical computing, and into quantum computing, what we gain are qubits over bits. Thus, we are going to give the function $f$ a qubit as an input. However, we assume that $f$ is a classical function, meaning that we can't actually feed it a qubit in the same way we can feed it a regular bit. There's a couple valid reasons why $f$ shouldn't be able to handle a qubit, but the best is perhaps a simple mathematical argument. Consider what would happen if we sent in a qubit in a superposition state between $|0\rangle$ and $|1\rangle$, for a constant $f$:

$$ f \Big{(} \hspace{.15cm} \frac{1}{\sqrt{2}} \big{(} \hspace{.08cm} |\hspace{.06cm}0\rangle + |\hspace{.06cm}1\rangle \hspace{.08cm} \big{)} \hspace{.15cm} \Big{)} \rightarrow \frac{1}{\sqrt{2}} \big{(} \hspace{.08cm} |\hspace{.06cm}1\rangle + |\hspace{.06cm}1\rangle \hspace{.08cm} \big{)} = \frac{2}{\sqrt{2}} |\hspace{.06cm}1\rangle \hspace{.4cm} \textmf{Not Unitary!} $$

Remember that quantum systems always must be unitary (it's not our rule, blame physics!). Thus, $f$ is a strictly classical function that only operates on classical bits.
\\

So then, in order to use a quantum computer, we must side-step the problem of using $f$. To do this, we will define a quantum operation $g$, which will incorporate our classical function $f$ in such a way that we still have a unitary operator. For the same reason shown above, there's just no way of creating a unitary operator that incorporates $f$ $\textit{and}$ acts on 1 qubit. Thus, the best we can do is an operator that acts on two qubits:

$$ g \hspace{.08cm}| \hspace{.08cm} q_1 \rangle \hspace{.06cm} | \hspace{.08cm} q_2\rangle \hspace{.24cm} \longrightarrow \hspace{.24cm} | \hspace{.08cm} q_1 \rangle \hspace{.06cm} | \hspace{.08cm} q_2 \oplus f( \hspace{.05cm} q_1 ) \hspace{.05cm} \rangle $$

where the symbol $\oplus$ means addition modulo 2: $\hspace{.18cm} 0 \oplus 1 = 1 \hspace{.6cm} 1 \oplus 1 = 0$ (basically if a number adds up to 2, it becomes a 0). Let's see a quick example:

\pagebreak

$$ f(\hspace{.06cm}0,1) \rightarrow (\hspace{.06cm}0,1) $$

$$ g \hspace{.15cm} \frac{1}{\sqrt{2}} \big{(} \hspace{.1cm} | \hspace{.06cm} 10\rangle + |\hspace{.06cm} 01\rangle \big{)} \hspace{.2cm} \longrightarrow \hspace{.2cm} \frac{1}{\sqrt{2}} \big{(} \hspace{.1cm} | \hspace{.06cm} 11\rangle + |\hspace{.06cm} 01\rangle \big{)} $$

More specifically:

$$ g \hspace{.15cm}| \hspace{.06cm} 10\rangle \hspace{.5cm} = \hspace{.5cm} g \hspace{.15cm} | \hspace{.06cm} 1\rangle \hspace{.08cm} | \hspace{.06cm} 0 \rangle \hspace{.5cm} \longrightarrow \hspace{.5cm} | \hspace{.06cm} 1\rangle \hspace{.08cm} | \hspace{.06cm} 0 \oplus f(1) \hspace{.06cm} \rangle \hspace{.3cm} = \hspace{.3cm} |\hspace{.06cm} 11 \rangle $$

Now, you may be wondering where the heck this addition modulo 2 came from. This is part of the trick that comes with solving classical problems via quantum algorithms. Sometimes we need to introduce new ways of approaching the problem. Here, we are able to incorporate $f$ into our quantum operation by using $\oplus$, which will guarantee everything stays unitary.
\\

Most importantly however, by using this $g$, we can see that based on what kind of function $f$ is, we get different final states:

$$ f(\hspace{.06cm}0,1) \rightarrow (\hspace{.06cm}0,1) \hspace{5cm} f(\hspace{.06cm}0,1) \rightarrow (\hspace{.06cm}1,0) $$

$$ g(\hspace{.08cm} |\hspace{.06cm}00\rangle \hspace{.08cm}) \rightarrow |\hspace{.06cm}00\rangle \hspace{5cm} g(\hspace{.08cm} |\hspace{.06cm}00\rangle \hspace{.08cm}) \rightarrow |\hspace{.06cm}01\rangle $$
$$ g(\hspace{.08cm} |\hspace{.06cm}01\rangle \hspace{.08cm}) \rightarrow |\hspace{.06cm}01\rangle \hspace{5cm} g(\hspace{.08cm} |\hspace{.06cm}01\rangle \hspace{.08cm}) \rightarrow |\hspace{.06cm}00\rangle $$
$$ g(\hspace{.08cm} |\hspace{.06cm}10\rangle \hspace{.08cm}) \rightarrow |\hspace{.06cm}11\rangle \hspace{5cm} g(\hspace{.08cm} |\hspace{.06cm}10\rangle \hspace{.08cm}) \rightarrow |\hspace{.06cm}10\rangle $$
$$ g(\hspace{.08cm} |\hspace{.06cm}11\rangle \hspace{.08cm}) \rightarrow |\hspace{.06cm}10\rangle \hspace{5cm} g(\hspace{.08cm} |\hspace{.06cm}11\rangle \hspace{.08cm}) \rightarrow |\hspace{.06cm}11\rangle $$

$$ f(\hspace{.06cm}0,1) \rightarrow 0 \hspace{5.8cm} f(\hspace{.06cm}0,1) \rightarrow 1 $$

$$ g(\hspace{.08cm} |\hspace{.06cm}00\rangle \hspace{.08cm}) \rightarrow |\hspace{.06cm}00\rangle \hspace{5cm} g(\hspace{.08cm} |\hspace{.06cm}00\rangle \hspace{.08cm}) \rightarrow |\hspace{.06cm}01\rangle $$
$$ g(\hspace{.08cm} |\hspace{.06cm}01\rangle \hspace{.08cm}) \rightarrow |\hspace{.06cm}01\rangle \hspace{5cm} g(\hspace{.08cm} |\hspace{.06cm}01\rangle \hspace{.08cm}) \rightarrow |\hspace{.06cm}00\rangle $$
$$ g(\hspace{.08cm} |\hspace{.06cm}10\rangle \hspace{.08cm}) \rightarrow |\hspace{.06cm}10\rangle \hspace{5cm} g(\hspace{.08cm} |\hspace{.06cm}10\rangle \hspace{.08cm}) \rightarrow |\hspace{.06cm}11\rangle $$
$$ g(\hspace{.08cm} |\hspace{.06cm}11\rangle \hspace{.08cm}) \rightarrow |\hspace{.06cm}11\rangle \hspace{5cm} g(\hspace{.08cm} |\hspace{.06cm}11\rangle \hspace{.08cm}) \rightarrow |\hspace{.06cm}10\rangle $$

Just like how our classical $f$ can map the bits (0,1) to one of four possibilities, our $g$ operator can map our two qubit states to one of four final states. Also like the classical case, if we use only one state as an input, we cannot determine whether $f$ is a balanced or constant function. For example, using the state $|00\rangle$ as an input will give us one of two results: $|00\rangle$ or $|01\rangle$. Based on which result we get, we've eliminated two out of the four categories above, but still are left with with two possibilities, one balanced and one constant.
\\

Now that we have our $g$ function mathematically defined, it's time to create it in our code. In the $\textmf{Our\_Pyquil\_Functions}$ file, our blackbox $g$ has already been created for us (you're welcome). We will go into the specifics of this $g$ later in this tutorial, but for now let's just see it in action. Run the cell of code below a few times, and verify the effect that $g$ is having on our initial state:

\pagebreak

\begin{figure}[h]
\centering
\includegraphics[scale=.65]{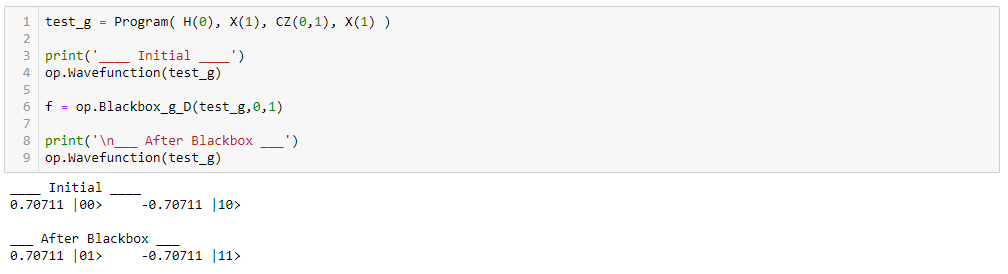}
\end{figure}

As mentioned before, if we want to to do better than the classical case, we need to be able to determine whether $f$ is constant or balanced, with only one call of $g$. As shown above, using only one state as an input does not get the job done. So then, we will send in a superposition of states, since that is the big advantage to using qubits over bits!
\\

Now, let's use the state in our code example to demonstrate what we can do with our final state, say for the case $f(0,1) \rightarrow (1,0)$:

$$ g\Big{(} \hspace{.12cm} \frac{1}{\sqrt{2}}\big{(} \hspace{.08cm} |\hspace{.06cm}00\rangle - |\hspace{.06cm}10\rangle \hspace{.08cm} \big{)} \hspace{.12cm} \Big{)} \hspace{.45cm}\rightarrow \hspace{.45cm} \frac{1}{\sqrt{2}}\big{(} \hspace{.12cm} |\hspace{.06cm}01\rangle - |\hspace{.06cm}10\rangle \hspace{.12cm} \big{)} $$

Compare the result we have here, with the four possibilities above. By sending in this superposition state, our output states corresponds to exactly one of the possible $f$'s (we did it!). Thus, if we could read out the entire final state, we would be done. But alas, we can't see wavefunctions, only measurements:

$$ \textmf{possibility 1} \hspace{7cm} \textmf{possibility 2} \hspace{.8cm} $$

$$ g\Big{(} \hspace{.25cm} \frac{1}{\sqrt{2}}\big{(} \hspace{.08cm} |\hspace{.06cm}00\rangle + |\hspace{.06cm}10\rangle \hspace{.08cm} \big{)} \hspace{.1cm} \Big{)} \hspace{.3cm} \rightarrow \hspace{.3cm} |\hspace{.06cm}01\rangle \hspace{3cm} g\Big{(} \hspace{.25cm} \frac{1}{\sqrt{2}} \big{(} \hspace{.08cm} |\hspace{.06cm}00\rangle + |10\rangle \hspace{.08cm} \big{)} \hspace{.1cm} \Big{)} \hspace{.3cm} \rightarrow \hspace{.3cm} |\hspace{.06cm}10\rangle $$

Two things are problematic here: 1) Individually, neither measurement result is conclusive as to what kind of function $f$ is. 2) Even if one of the measurements $\textit{could}$ tell us about $f$, there's only a 50\% chance we get that measurement result.
\\

Never fear, for there is a correct input state still to come. This example was just meant to demonstrate the potential that qubits and superposition states have to offer. The trick is that we need to be thinking of what kind of final wavefunction we will get $\textit{and}$ the information we can extract from a measurement on that state. So, without further ado, let's take a look at the input state that is going to solve our problem:

$$ |\hspace{.08cm} \psi \rangle_{in} = \frac{1}{2} \big{(}\hspace{.14cm} |\hspace{.06cm}00\rangle - |\hspace{.06cm}01\rangle + |\hspace{.06cm}10\rangle - |\hspace{.06cm}11\rangle \hspace{.1cm} \big{)} $$

Which is obtainable by the following sequence of gates:

\begin{figure}[h]
\centering
\includegraphics[scale=.65]{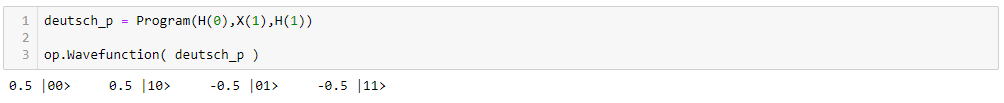}
\end{figure}

Alright, let's see what happens when we apply our gate $g$ to this input state:

$$ f(\hspace{.05cm}0,1) \rightarrow (\hspace{.05cm}0,1) \hspace{8cm} f(\hspace{.05cm}0,1) \rightarrow (\hspace{.05cm}1,0\hspace{.05cm}) $$
$$ g\hspace{.08cm} |\hspace{.08cm} \psi\rangle_{in} \hspace{.08cm} \hspace{.2cm} \rightarrow \hspace{.2cm} \frac{1}{2} \big{(}\hspace{.08cm} |\hspace{.06cm}00\rangle - |\hspace{.06cm}01\rangle - |\hspace{.06cm}10\rangle + |\hspace{.06cm}11\rangle \hspace{.04cm}\big{)}\hspace{4cm} g\hspace{.08cm} |\hspace{.08cm}\psi\rangle_{in} \hspace{.08cm} \hspace{.2cm} \rightarrow \hspace{.2cm} \frac{1}{2} \big{(}\hspace{.08cm} -|\hspace{.06cm}00\rangle + |\hspace{.06cm}01\rangle + |\hspace{.06cm}01\rangle - |\hspace{.06cm}11\rangle \hspace{.08cm}\big{)} $$

$$ f(\hspace{.05cm}0,1) \rightarrow 0 \hspace{8.7cm} f(\hspace{.05cm}0,1) \rightarrow 1 $$
$$ g\hspace{.08cm} |\hspace{.08cm} \psi\rangle_{in} \hspace{.08cm} \hspace{.2cm} \rightarrow \hspace{.2cm} \frac{1}{2} \big{(}\hspace{.08cm} |\hspace{.06cm}00\rangle - |\hspace{.06cm}01\rangle + |\hspace{.06cm}10\rangle - |\hspace{.06cm}11\rangle \hspace{.04cm}\big{)}\hspace{4cm} g\hspace{.08cm} |\hspace{.08cm}\psi\rangle_{in} \hspace{.08cm} \hspace{.2cm} \rightarrow \hspace{.2cm} \frac{1}{2} \big{(}\hspace{.08cm} -|\hspace{.06cm}00\rangle + |\hspace{.06cm}01\rangle - |\hspace{.06cm}01\rangle + |\hspace{.06cm}11\rangle \hspace{.08cm}\big{)} $$

Now, based on the four results above, we can start to see an interesting result emerging: the output states for both cases where $f$ is balanced are equal, up to a phase difference. And the same holds true for both output states when $f$ is constant. However, as we noted before, we can only see this when looking at the wavefunctions, but a measurement result will not give us this same information. In fact, all four states will produce the same measurement probabilities.
\\

Sooo, let's do one more thing: apply Hadamard gates to both qubits. Now, I will setup the algebra below, but skip most of the steps. I encourage you to go through one of the calculations for yourself. If you plan to follow along through the rest of the lesson 5 tutorials, I $\textit{strongly}$ recommend working through the algebra steps, as we will be using Hadamard transformations like this one $\textit{a lot}$:

$$ H \hspace{.06cm} \frac{1}{2} \big{(}\hspace{.08cm} |\hspace{.05cm}00\rangle - |\hspace{.05cm}01\rangle + |\hspace{.05cm}10\rangle - |\hspace{.05cm}11\rangle \hspace{.08cm}\big{)} \hspace{.3cm} = \hspace{.3cm} \frac{1}{2} \Big{(}\hspace{.16cm} \frac{1}{\sqrt{2}}\big{(} \hspace{.08cm} |\hspace{.05cm}0\rangle + |\hspace{.05cm}1\rangle \hspace{.08cm} \big{)} \cdot \frac{1}{\sqrt{2}}\big{(} \hspace{.08cm} |\hspace{.05cm}0\rangle + |\hspace{.05cm}1\rangle \hspace{.08cm} \big{)} \hspace{.12cm} - \hspace{.28cm} \cdot \cdot \cdot \Big{)} $$

$$ = |11\rangle \hspace{1.5cm} $$

Doing all the algebra for the four possible $f$ functions, we get the following final states:

$$ f(0,1) \rightarrow (0,1) \hspace{6.2cm} f(0,1) \rightarrow (1,0) $$
$$ |11\rangle \hspace{7.6cm} -|11\rangle $$

$$ f(0,1) \rightarrow 0 \hspace{6.7cm} f(0,1) \rightarrow 1 $$
$$ |01\rangle \hspace{7.6cm} -|01\rangle $$

Let's confirm this with our code:

\begin{figure}[h]
\centering
\includegraphics[scale=.65]{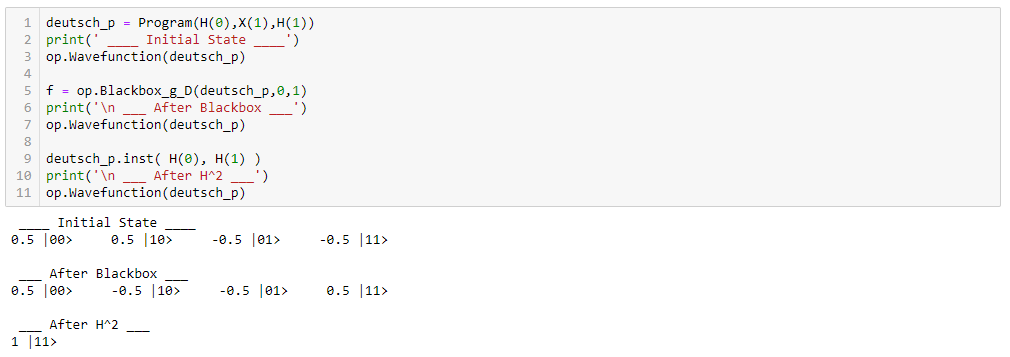}
\end{figure}

So then, how do we extract our information from these finals states is the final question. If we take a look at all four possibilities, we can see that qubit $1$ is always in the $|1\rangle$ state, so that's no good to us. However, when we make a measurement on qubit $0$, we will get one of two possibilities. If we measure a $|1\rangle$, we can conclude that $f$ is a balanced function, and if we measure a $|0\rangle$, we can conclude that $f$ is constant. Thus, we have successfully identified what kind of function $f$ is, with only one function call!
\\

We have now completed the Deutsch Algorithm, in all its glory. In order to determine if a black box function $f$ is constant or balanced, we do the following:

$$ \textmf{prepare} \hspace{.1cm} |\hspace{.06cm}01\rangle \hspace{.3cm} \rightarrow \hspace{.3cm} H^2 \hspace{.06cm} |\hspace{.06cm}01\rangle \hspace{.3cm} \rightarrow \hspace{.3cm} g \hspace{.08cm} |\hspace{.06cm} \psi_{in} \rangle \hspace{.3cm} \rightarrow \hspace{.3cm} H^2 \hspace{.06cm} | \hspace{.06cm}\psi_{out} \rangle \hspace{.3cm} \rightarrow \hspace{.3cm} \textmf{measure qubit} \hspace{.1cm}0 $$

And as shown above, the measurement result on qubit $0$ will perfectly determine $f$'s nature for us. In fact, since we never bother to check qubit $1$, we can actually get the same results by only applying a single Hadamard gate on qubit $0$, after $g$. This will result in qubit $0$ becoming either $|0\rangle$ or $|1\rangle$, while leaving qubit $1$ still in a superposition. This is just a slight optimization.
\\

Since the steps to solving the Deutsch Algorithm are always the same, we can create a function that will always apply the steps for us. And, there is already one waiting for us in $\textmf{Our\_Pyquil\_Functions}$, called $\textbf{Deutsch}$:

\begin{figure}[h]
\centering
\includegraphics[scale=.65]{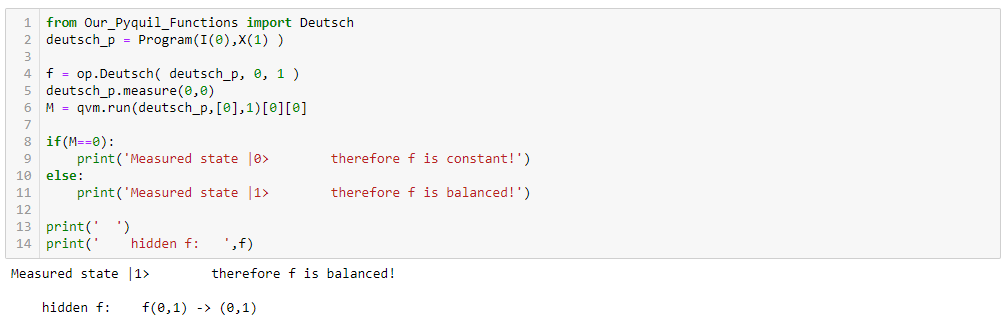}
\end{figure}

Running this cell of code a couple times should convince you that we have indeed solved our blackbox $f$ problem using the Deutsch Algorithm. Our conclusion about $f$ is always 100\% correct, and we can even check the $f$ function to prove it!

\section*{\large{ Further Analysis of the Deutsch Algorithm }}
\centerline{---------------------------------------------------------------------------------------------------------------------------------}

The last code example is the full Deutsch Algorithm, but our work isn't finished yet. In the next three algorithms, we are going to be encountering similar tricks over and over, and they're all related to the Hadamard gate. Specifically, the 'Hadamard Transformation', which just means we apply $H$ gates to all of the qubits in the system. In this final section, we are going to cover why this transformation works, and also briefly on how we constructed our blackbox $g$ operator.

\section*{\large{ The Hadamard Transformation }}

Now, if you followed through with the algebra steps we left out previously, then you may have noticed the underlying pattern. To begin, let's imagine for a second that we exclude the $g$ step of our algorithm. Following our state's wavefunction through each step, we get:

$$ |\hspace{.06cm}01\rangle \hspace{.24cm} - H^2 \rightarrow \hspace{.24cm} \frac{1}{2} \big{(} |\hspace{.06cm}00\rangle - |\hspace{.06cm}01\rangle + |\hspace{.06cm}10\rangle - |\hspace{.06cm}11\rangle \big{)} \hspace{.24cm} - H^2 \rightarrow \hspace{.24cm} |\hspace{.06cm}01\rangle $$

Now, if $f$ happens to be constant, this is essentially the whole process. The effect of $g$ will either leave the state completely unchanged ($f \rightarrow 0$), or apply an overall phase -1 ($f \rightarrow 1$). Thus, in both cases we can see that the second $H^2$ operation maps us back to either $|01\rangle$ or $-$$|01\rangle$. Note how two applications of $H^2$ takes us back to our original state.
\\

By contrast, if $f$ is balanced, the net effect of $g$ appears in the form of moving the two negative signs around. Or more specifically, moving the negative signs onto a different pair of states:

$$ f(\hspace{.04cm}0,1) \rightarrow (\hspace{.04cm}0,1) $$
$$ \frac{1}{2}\big{(}\hspace{.12cm} |\hspace{.06cm}00\rangle - |\hspace{.06cm}01\rangle + |\hspace{.06cm}10\rangle - |\hspace{.06cm}11\rangle \hspace{.08cm} \big{)} \hspace{.34cm} - g \rightarrow \hspace{.34cm} \frac{1}{2}\big{(}\hspace{.12cm} |\hspace{.06cm}00\rangle - |\hspace{.06cm}01\rangle - |\hspace{.06cm}10\rangle + |\hspace{.06cm}11\rangle \hspace{.08cm} \big{)} $$

$$ f(\hspace{.04cm}0,1) \rightarrow (1,0\hspace{.04cm}) $$
$$\hspace{.3cm} \frac{1}{2}\big{(} \hspace{.12cm} |\hspace{.06cm}00\rangle - |\hspace{.06cm}01\rangle + |\hspace{.06cm}10\rangle + |\hspace{.06cm}11\rangle \hspace{.08cm} \big{)} \hspace{.34cm} - g \rightarrow \hspace{.34cm} \frac{1}{2}\big{(}\hspace{.04cm} -|\hspace{.06cm}00\rangle + |\hspace{.06cm}01\rangle + |\hspace{.06cm}10\rangle - |\hspace{.06cm}11\rangle \hspace{.08cm} \big{)} $$

For the top example, we can see that the states $|10\rangle$ and $|11\rangle$ switch amplitudes. And for the second case, we have states $|00\rangle$ and $|01\rangle$ switch. The net effect is that in both balanced cases the states $|00\rangle$ \& $|11\rangle$ always have the same sign, as well as $|01\rangle$ \& $|10\rangle$. By contrast, for both constant cases, our 'paired' states that always have the same sign are $|00\rangle$ \& $|10\rangle$ , and $|01\rangle$ \& $|11\rangle$. Keep this in mind, as next we are going to show the full Hadamard transformation map on two qubits:

\begin{figure}[h]
\centering
\includegraphics[scale=.65]{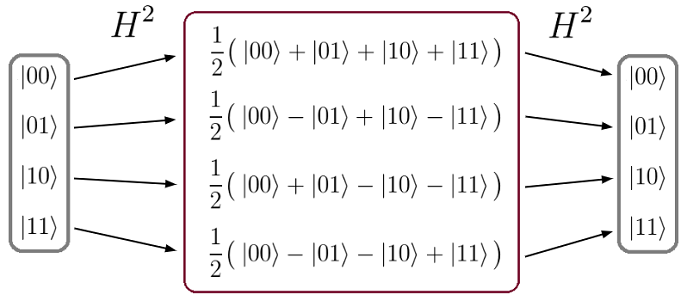}
\end{figure}

As shown above, the Hadamard Transformation maps each of the four possible two qubit states to a unique superposition state, AND, all of these states are orthogonal (easy to check for yourself). So then, take a look at the mapping above, and compare it to the effect of $g$ we pointed out earlier. You should find that the final states we obtain after $g$ correspond to the ones that map back to $|01\rangle$ for a constant $f$, and $|11\rangle$ for a balanced $f$.
\\

Now that we've seen how the $H^2$ transformation works, let's take a look at the steps of our algorithm again. Run the cell of code below a few times, and confirm for yourself that $g$ always puts our state in one of the superposition states that will get mapped to either $|01\rangle$ or $|11\rangle$ (and don't forget about an overall phase of -1):

\pagebreak

\begin{figure}[h]
\centering
\includegraphics[scale=.65]{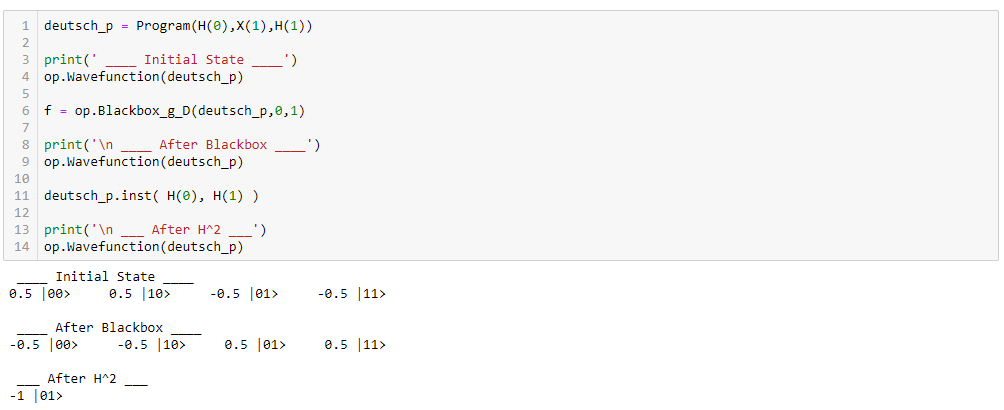}
\end{figure}

\section*{\large{ Constructing $g$ }}

We are going to be seeing the Hadamard transformation $\textit{a lot}$ in the coming tutorials, so the section above is a sufficient first exposure. If it didn't full sink in, don't worry. Each lesson we will see the Hadamard transformation used for a slightly different trick, and in time you will come to appreciate why it's such a powerful tool.
\\

In this final section, we will briefly cover how we constructed the blackbox function $g$, as it is also insightful into the thought process behind constructing quantum operations from a gate level perspective. To begin, let's see the four possible $g$ operators in matrix form, followed by their gate instructions:

$$ f(\hspace{.04cm}0,1) \longrightarrow (\hspace{.04cm}0,1) \hspace{2.9cm} f(1,0\hspace{.04cm}) \longrightarrow (\hspace{.04cm}0,1) $$

$$
\begin{bmatrix} 1 & 0 & 0 & 0 \\ 0 & 1 & 0 & 0 \\ 0 & 0 & 0 & 1 \\ 0 & 0 & 1 & 0 \\ \end{bmatrix} \hspace{3.8cm} \begin{bmatrix} 0 & 1 & 0 & 0 \\ 1 & 0 & 0 & 0 \\ 0 & 0 & 1 & 0 \\ 0 & 0 & 0 & 1 \\ \end{bmatrix}
$$

$$ f(\hspace{.04cm}0,1) \longrightarrow 0 \hspace{3.7cm} f(1,0\hspace{.04cm}) \longrightarrow 1 $$

$$
\begin{bmatrix} 1 & 0 & 0 & 0 \\ 0 & 1 & 0 & 0 \\ 0 & 0 & 1 & 0 \\ 0 & 0 & 0 & 1 \\ \end{bmatrix} \hspace{3.8cm} \begin{bmatrix} 0 & 1 & 0 & 0 \\ 1 & 0 & 0 & 0 \\ 0 & 0 & 0 & 1 \\ 0 & 0 & 1 & 0 \\ \end{bmatrix}
$$

where these matrices are operating on the basis:

$$ \begin{bmatrix} | \hspace{.06cm} 00\rangle \\ |\hspace{.06cm} 01\rangle \\ |\hspace{.06cm} 10\rangle \\ |\hspace{.06cm} 11\rangle \\ \end{bmatrix} $$

Take a look at these four matrices and see if you recognize any of them (hint: two of them are gates we studied in lesson 3). For starters, the matrix corresponding to $f(\hspace{.02cm}0,1) \rightarrow 0$ is just the Identity matrix on two qubits: $I^2$. And also, the matrix corresponding to $f(\hspace{.02cm}0,1) \rightarrow (\hspace{.02cm}0,1)$ is just a CNOT gate! Scroll $\textit{all}$ the way back up to where we first outlined the full effect of $g$ on each possible input state, and confirm for yourself that the operations that describe $\hspace{.1cm} f(\hspace{.02cm}0,1) \rightarrow 0 \hspace{.1cm}$ and $\hspace{.1cm} f(\hspace{.02cm}0,1) \rightarrow (\hspace{.02cm}0,1) \hspace{.1cm}$ are indeed just $I^2$ and CNOT.
\\

Now, the case for $\hspace{.1cm} f(\hspace{.02cm}0,1) \rightarrow 0 \hspace{.1cm}$ is the easiest to understand. Since everything maps to $0$, and our $g$ operator is addition modulo 2: $\hspace{.1cm} |\hspace{.06cm}q_1 \oplus 0 \rangle \hspace{.1cm}$, every state remains unchanged. For the remaining other three matrices, they are all categorizable by a single pattern: which inputs get mapped to $1$. Specifically, if either $0$ or $1$ gets mapped to $1$, we find a corresponding 2$\times$2 matrix located along the diagonal: $ \begin{bmatrix} 0 & 1 \\ 1 & 0 \\\end{bmatrix} $. And conversely, if a particular input gets mapped to $0$, we find a: $ \begin{bmatrix} 1 & 0 \\ 0 & 1 \\\end{bmatrix} $. Both of these matrices should be recognizable, as they are just the $X$ and $I$ single qubit gates.
\\

So what is this telling us about $g$? Well, we are now seeing why we chose to incorporate $f$ into our $g$ operator via $\oplus$ (addition modulo 2). Loosely speaking, adding $\oplus \hspace{.04cm} 1$ to a qubit state is equivalent to applying an $X$, and adding $\oplus \hspace{.07cm} 0$ is equivalent to doing nothing (which is what an $I$ gate does). But remember that $g$ only affects our second qubit, which means none of these matrices should ever change the state of qubit $0$. Or another way of saying that is, the only transformations that are allowed are: $\hspace{.1cm}$ $|\hspace{.06cm}00\rangle \leftrightarrow$ $|01\rangle \hspace{.08cm}$ and $\hspace{.08cm}$ $|10\rangle \leftrightarrow $ $|11\rangle$.
\\

So then, how can we deduce what $g$ will look like, based on which inputs map to $0$ and $1$? Well, for the balanced cases where the input get mapped to one of each output, ask yourself what kind of gate operation flips the states on one qubit, contingent on on the state of the other qubit. That's a CNOT! And for the constant case $\hspace{.1cm} f(\hspace{.02cm}0,1) \rightarrow 1 \hspace{.1cm}$, what kind of gate operation flips the states on a particular qubit, regardless of all other qubit states. An $X$ gate!
\\

Starting with the $\hspace{.1cm} f(\hspace{.02cm}0,1) \rightarrow 1 \hspace{.1cm}$ case, you may be looking at the matrix representation above and thinking, "that doesn't look like a regular $X$ gate". True, because it's a single qubit $X$ gate applied to a 2-qubit system. Remember, $g$ needs to operate on the $\textit{whole}$ system, so all matrix representations must be 4$\times$4. Thus, to see what a single qubit operation looks like on a 2-qubit system, we need to take the tensor product: $I \otimes X$. This operation leaves qubit $0$ unchanged thanks to the Identity gate, and applies our $X$ gate to qubit $1$. If you're new to 'outer product' matrix multiplication, here's a step-by-step walk through:

$$ I \otimes X $$

$$\begin{bmatrix} 1 & 0 \\ 0 & 1 \\\end{bmatrix} \hspace{.2cm} \otimes \hspace{.2cm}\begin{bmatrix} 0 & 1 \\ 1 & 0 \\\end{bmatrix}$$

$$\begin{bmatrix} 1 \cdot \begin{bmatrix} 0 & 1 \\ 1 & 0 \\\end{bmatrix} & 0 \cdot \begin{bmatrix} 0 & 1 \\ 1 & 0 \\\end{bmatrix} & \\ 0 \cdot \begin{bmatrix} 0 & 1 \\ 1 & 0 \\\end{bmatrix} & 1 \cdot \begin{bmatrix} 0 & 1 \\ 1 & 0 \\\end{bmatrix} & \\\end{bmatrix}$$

$$\begin{bmatrix} 0 & 1 & 0 & 0 \\ 1 & 0 & 0 & 0 \\ 0 & 0 & 0 & 1 \\ 0 & 0 & 1 & 0 \\ \end{bmatrix}$$

For a reference: https://en.wikipedia.org/wiki/Tensor\_product
\\

Luckily, matrix representations are not a necessary ingredient for creating quantum algorithms, only gates (although sometimes matrix representations are very insightful). Thus, when we go to write our code, we need only use a single $X$ gate on qubit $1$, and not $I \otimes X$.
\\

Lastly, let's talk about how to construct the case for $\hspace{.1cm} f(\hspace{.02cm}0,1) \rightarrow (\hspace{.02cm}0,1) \hspace{.1cm}$. If we look at its matrix representation, it kind of looks like CNOT gate, only backwards. In fact, its effect is exactly like a 'backwards' CNOT gate:
\\

$\hspace{.1cm} $ $|00\rangle \rightarrow$ $|01\rangle \hspace{1.3cm}$ $|01\rangle \rightarrow$ $|00\rangle \hspace{1.3cm}$ $|10\rangle \rightarrow$ $|10\rangle \hspace{1.3cm}$ $|11\rangle \rightarrow$ $|11\rangle \hspace{.1cm}$.
\\

In essence, it functions like a CNOT gate, where if the control qubit is in the state $|0\rangle$, an $X$ gate is applied to the target.
\\

Since we don't have a gate operation that has this exact effect, we'll have to build one! And to do it, we will essentially borrow a CNOT gate, and 'trick' it into applying an $X$ gate when qubit $0$ is in the $|0\rangle$ state. To do this, we will use an $X$ gate on qubit $0$ first, then apply a CNOT, and lastly flip qubit $0$ back with another $X$ gate:

\begin{figure}[h]
\centering
\includegraphics[scale=.65]{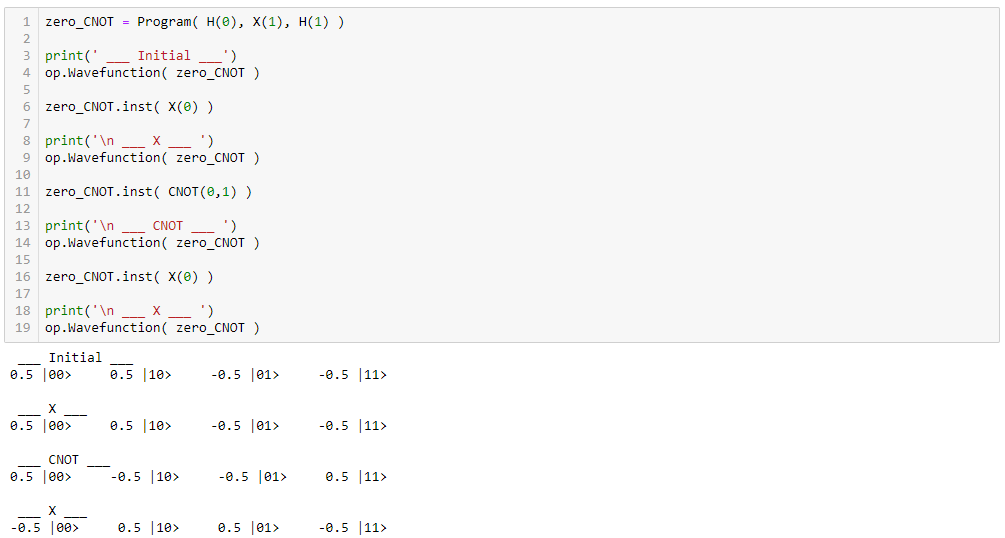}
\end{figure}

Compare the first and last wavefunctions, and confirm for yourself that we have indeed achieved the desired operation. By applying $X$ gates before and after the CNOT gate, we are able to effectively use the state $|0\rangle$ as the control. We will see this trick used in future lessons, as it is a very common way for getting a 'control operation' on any specific state.
\\

We are now officially done with Deutsch Algorithm! Although it's just a one-step process, the Deutsch Algorithm is an important first hurdle in terms of understanding how quantum algorithms can outperform classical counterparts, and the subtle math tricks involved. As you've probably already guessed, there's a lot that goes into even the simplest of quantum algorithms! But have no fear, many of the topics covered in this section will be seen again in future algorithms.
\\

--------------------------------------------------------------------------------------------------------------------------------------------------------
\\

This concludes lesson 5.1!$\hspace{.03cm}$ As we move through several algorithms over the next couple lessons, you may find that what these quantum algorithm achieves at face value isn't terribly complicated. For example, "determine if $f$ is constant or balanced in one step". But understanding $\textit{why}$ and $\textit{how}$ these quantum algorithms work is much more challenging. This is why our primary focus for these lesson 5 tutorials will be on explaining their inner workings, rather than racing straight to a final code that works.
\\

--------------------------------------------------------------------------------------------------------------------------------------------------------


\pagebreak

\section*{\Large{ Lesson 5.2 - Deutsch-Jozsa \& Bernstein-Vazirani Algorithms }}
--------------------------------------------------------------------------------------------------------------------------------------------------------
\\

This tutorial continues the series of in-depth guides to some of the most popular quantum algorithms, all labeled 'Lesson 5'. In this tutorial, we will cover the Deutsch-Jozsa and Bernstein-Vazirani Algorithms, which are problems very closely related to the Deutsch Algorithm. Both algorithms use a Hadamard Transformation as the core to their success, and are in fact solved with the same circuit.
\\

For any reminders / refreshers on Pyquil notation and basics, check out lessons 1 - 4. Also, please consider reading Lesson 5.1 - Intro to Quantum Algorithms (Deutsch), which covers many topics that we will be skipping over for this lesson.
\\

Original publications of the algorithms: \cite{DJ} \& \cite{BV}
\\

--------------------------------------------------------------------------------------------------------------------------------------------------------
\\
In order to make sure that all cells of code run properly throughout this lesson, please run the following cell of code below:

\begin{figure}[h]
\centering
\includegraphics[scale=.65]{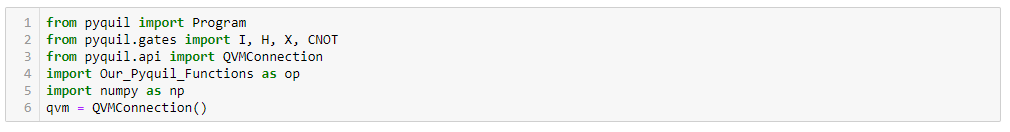}
\end{figure}

\section*{\large{ The Deutsch-Jozsa Algorithm }}
\centerline{---------------------------------------------------------------------------------------------------------------------------------}

In the first part to this tutorial, we will be studying the Deutsch-Jozsa Algorithm, which is very closely related to the Deutsch Algorithm. The difference in this new problem is that instead of a function $f$ which maps a single bit of $0$ or $1$ as either a constant or balanced function, we now have an $f$ which maps an $\textit{entire}$ string of bits to either $0$ or $1$. More specifically, let's say we have a string of bits labeled { x$_0$, x$_1$, ... , x$_n$ }, where each x$_i$ is either a $0$ or a $1$. When we pass our string of bits through the function $f$, it returns a single value of $0$ or $1$:

$$ f(\hspace{.08cm}\{x_0, x_1, x_2,...\} \hspace{.08cm}) \hspace{.2cm} \rightarrow \hspace{.2cm} 0 \hspace{.18cm} \textmf{or} \hspace{.18cm} 1$$

For a sting of $n$ bits, there are a total of 2$^n$ possible combinations. But, just like our previous problem, we are promised that $f$ is either a constant or balanced function. Here, a constant $f$ returns all $0$'s or $1$'s for any input, while a balanced $f$ returns $0$'s and $1$'s for exactly half of all inputs. For example:

$$ f_{constant} \hspace{6.4cm} f_{balanced} $$

$$ \{0,0,0\} \rightarrow 0 \hspace{1cm} \{1,0,0\} \rightarrow 0 \hspace{3cm} \{0,0,0\} \rightarrow 0 \hspace{1cm} \{1,0,0\} \rightarrow 1 $$
$$ \{0,0,1\} \rightarrow 0 \hspace{1cm} \{1,0,1\} \rightarrow 0 \hspace{3cm} \{0,0,1\} \rightarrow 1 \hspace{1cm} \{1,0,1\} \rightarrow 1 $$
$$ \{0,1,0\} \rightarrow 0 \hspace{1cm} \{1,1,0\} \rightarrow 0 \hspace{3cm} \{0,1,0\} \rightarrow 0 \hspace{1cm} \{1,1,0\} \rightarrow 0 $$
$$ \{0,1,1\} \rightarrow 0 \hspace{1cm} \{1,1,1\} \rightarrow 0 \hspace{3cm} \{0,1,1\} \rightarrow 1 \hspace{1cm} \{1,1,1\} \rightarrow 0 $$

As shown above, our balanced function returns $0$'s and $1$'s for exactly half of the inputs (4 for the case of $n=3$). And, the ruleset governing which inputs return $0$'s and $1$'s is completely independent of the individual bits. That is to say, if our $f$ is truly a randomized balanced function, that takes $n$ bit strings as input, then there are $\frac{2^n !}{2 \cdot 2^{n-1}!}$ possible variations! Thus, there is no way to conclude any information about $f$ by studying trends with individual bits. In fact, if we instead rewrite our strings of bits as base 10 numbers, then our $f$ function looks like the following:

$$ f_{constant} \hspace{5cm} f_{balanced} $$

$$ \{1\} \rightarrow 0 \hspace{1cm} \{5\} \rightarrow 0 \hspace{3cm} \{1\} \rightarrow 0 \hspace{1cm} \{5\} \rightarrow 1 $$
$$ \{2\} \rightarrow 0 \hspace{1cm} \{6\} \rightarrow 0 \hspace{3cm} \{2\} \rightarrow 1 \hspace{1cm} \{6\} \rightarrow 1 $$
$$ \{3\} \rightarrow 0 \hspace{1cm} \{7\} \rightarrow 0 \hspace{3cm} \{3\} \rightarrow 0 \hspace{1cm} \{7\} \rightarrow 0 $$
$$ \{4\} \rightarrow 0 \hspace{1cm} \{8\} \rightarrow 0 \hspace{3cm} \{4\} \rightarrow 1 \hspace{1cm} \{8\} \rightarrow 0 $$

Written this way, it should be more clear that individual bits are meaningless to $f$. Only the string as a whole determines how $f$ operates on the input. Hopefully the setup for our new problem is clear, and how it varies from our previous one in lesson 5.1. Now then, the question we want to pose to both a classical and quantum computer is: 'how fast can you determine if $f$ is constant or balanced?'

\section*{\large{ Classical }}

From the classical perspective, we must check each possible string of bits one by one until we can make a conclusion. In the best case scenario, we get different outputs on our first two inputs, example: {0,0,0,..} $\rightarrow$ 0 and {1,0,0,...} $\rightarrow$ 1. In fact, if we ever get two different outputs, then we can 100\% conclude our $f$ is balanced. Conversely, if we continue to see the same output for each input we try, for example:

$$\{0,0,0,0\} \rightarrow 1$$
$$\{0,0,0,1\} \rightarrow 1$$
$$\{0,0,1,0\} \rightarrow 1$$
$$\{0,0,1,1\} \rightarrow 1$$
$$.$$
$$.$$

then we must check exactly 2$^{n-1}$ + 1 combinations in order to conclude that $f$ is constant, which is one more than half the total. To see why we need so many, consider an $f$ where we checked 8 out of 16 total combinations, getting all 0's, only to then find that the 9$^{th}$ input returns a 1. Probabilistically, this is a very unlikely event. In fact, if we get the same result continually in succession, we can express the probability that our $f$ is constant as a function of $k$ inputs as:

$$ \textmf{P}_{constant}(k) \hspace{.12cm} = \hspace{.12cm} 1 - \frac{1}{2^{k-1}} \hspace{2cm} \textmf{for} \hspace{.15cm} k \leq 2^{n-1} $$

Perhaps more realistically, we can opt to truncate our classical algorithm early, say if we are over X\% confident. But if we want the full 100\%, then we are stuck checking 2$^{n-1}$ + 1 entries.

\section*{\large{Quantum}}

For our quantum computer, we will solve this problem with 100\% confidence after only one function call of $f$. As we shall see, we achieve this result nearly the exact same way as the Deutsch Algorithm, with only a slight twist.
\\

To show how similar the flow of this algorithm is to the previous one, let's write the complete Deutsch and Deutsch-Jozsa Algorithms side by side:

$$ \textmf{Deutsch:} \hspace{1cm} \textmf{prepare} \hspace{.1cm} |\hspace{.06cm} 01\rangle \hspace{.24cm} \rightarrow \hspace{.24cm} H^2 \hspace{.1cm} |\hspace{.06cm} 01\rangle \hspace{.3cm} \rightarrow \hspace{.3cm} g \hspace{.08cm} | \hspace{.06cm}\psi \rangle \hspace{.24cm} \rightarrow \hspace{.24cm} H^2 \hspace{.06cm} | \hspace{.06cm}\psi \rangle \hspace{.3cm} \rightarrow \hspace{.3cm} \textmf{measure qubit 0} \hspace{1.54cm}$$

$$ \textmf{Deutsch-Jozsa:} \hspace{1cm} \textmf{prepare} \hspace{.1cm} |\hspace{.06cm}0\rangle ^{\otimes n}|1\rangle \hspace{.3cm} \rightarrow \hspace{.3cm} H^{n+1} |\hspace{.06cm}0\rangle ^{\otimes n}|\hspace{.06cm}1\rangle \hspace{.3cm} \rightarrow \hspace{.3cm} g \hspace{.08cm} | \hspace{.06cm}\psi \rangle \hspace{.3cm} \rightarrow \hspace{.3cm} H^{n+1} | \hspace{.06cm}\psi \rangle \hspace{.3cm} \rightarrow \hspace{.3cm} \textmf{measure qubits} \hspace{.06cm} ^{\otimes n} $$

where once again $g$ is the unitary function that contains our mystery function $f$.
\\

Comparing these two algorithms, they're nearly identical. The only difference here is that instead of using a single qubit in the state $|0\rangle$, we go through all of the steps with $|0\rangle^{\otimes n}$. Recall that in the Deutsch Algorithm we used a single qubit paired with an ancilla qubit to solve our problem ('ancilla' refers to a qubit(s) that is used but doesn't matter in the final measurement). Here, for our new $f$ that takes an $n$ bit string, we use $n$ qubits paired again with just a single ancilla.
\\

Also like before, we will embed our function $f$ into a unitary operator $g$, via addition modulo 2 $\oplus$. Let X$_i$ represent some string {x$_0$, x$_1$, ...}, then our $g$ operator acts as follows:

$$ \textmf{Classical} \hspace{5cm} \textmf{Quantum} $$

$$ \hspace{1.4cm} f\hspace{.05cm}( \hspace{.06cm} X_i \hspace{.06cm} ) \rightarrow 0 \hspace{.1cm} \textmf{or} \hspace{.1cm} 1 \hspace{1cm} \iff \hspace{1cm} g \hspace{.06cm}| \hspace{.06cm}X_i \rangle \hspace{.04cm} | \hspace{.06cm}\alpha \rangle \rightarrow \hspace{.06cm}| \hspace{.06cm} X_i \rangle \hspace{.04cm} | \hspace{.06cm} \alpha \oplus f\hspace{.05cm}( \hspace{.06cm} X_i \hspace{.06cm} ) \hspace{.04cm} \rangle $$

where the state $|X_i \rangle$ refers to the state of the $n$ individual qubits:

$$ | \hspace{.06cm} X_i \rangle \hspace{.2cm} = \hspace{.2cm} | \hspace{.06cm} x_1 x_2 x_3 \cdot \cdot \rangle \hspace{.2cm} = \hspace{.2cm} | \hspace{.06cm} x_1\rangle \otimes | \hspace{.06cm} x_2\rangle \otimes | \hspace{.06cm} x_3 \rangle \cdot \cdot $$

For example, suppose we had a particular $f$ with the following: $\hspace{1cm} f(\hspace{.05cm} 010 \hspace{.05cm}) \hspace{.1cm} \rightarrow \hspace{.1cm}$ $1$ $\hspace{.55cm} f(\hspace{.05cm} 011 \hspace{.05cm}) \hspace{.1cm} \rightarrow \hspace{.1cm}$ $0$.
\\

The corresponding $g$ operator would then:

$$ g \hspace{.08cm} | \hspace{.06cm} 010 \rangle \hspace{.04cm} | \hspace{.06cm} \alpha \rangle \hspace{.4cm} \longrightarrow \hspace{.4cm} | \hspace{.06cm} 010 \rangle \hspace{.04cm} | \hspace{.06cm} \alpha \oplus 0 \rangle \hspace{.4cm} = \hspace{.4cm} | \hspace{.06cm} 010 \rangle \hspace{.04cm} | \hspace{.06cm} \alpha \rangle $$

$$ \hspace{.36cm} g \hspace{.08cm} | \hspace{.06cm} 011 \rangle \hspace{.04cm} | \hspace{.06cm} \alpha \rangle \hspace{.4cm} \longrightarrow \hspace{.4cm} | \hspace{.06cm} 011 \rangle \hspace{.04cm} | \hspace{.06cm} \alpha \oplus 1 \rangle \hspace{.4cm} = \hspace{.4cm} | \hspace{.06cm} 011 \rangle \hspace{.08cm} X \hspace{.04cm} | \hspace{.06cm} \alpha \rangle $$

Note that we are using the fact that addition modulo 2 is equivalent to an X gate in this example, a result we showed in lesson 5.1. Thus, we can view the net effect of our $g$ operator as picking out states at random and applying $X$ gates to their ancilla state. For the constant cases, we will have either all or none of the states in the system recieve $X$ gates to their ancilla, while for the balanced cases, exactly half of the states will receive the operation.
\\

Just like the Deutsch Algorithm, a key component is the state $|\alpha \rangle$. As we've laid out our algorithm above, we initialize our ancilla qubit in the $|1\rangle$ state, and then apply a $H$ to it, causing it to be in the state $|-x \rangle$ before our $g$ operator. This means that the effect from the $g$ operation will be as follows:

\pagebreak

$$ f\hspace{.04cm}( \hspace{.1cm} X_i \hspace{.1cm} ) \hspace{.2cm} \rightarrow \hspace{.2cm} 0 \hspace{7cm} f\hspace{.04cm}( \hspace{.1cm} X_j \hspace{.1cm} ) \hspace{.2cm} \rightarrow \hspace{.2cm} 1$$

$$ g \hspace{.08cm} | \hspace{.08cm} X_i \rangle \hspace{.04cm} | -x \rangle \hspace{.5cm} \longrightarrow \hspace{.5cm} | \hspace{.08cm} X_i \rangle \hspace{.04cm} | -x \rangle \hspace{4cm} g \hspace{.08cm} | \hspace{.08cm} X_j \rangle \hspace{.04cm} | -x \rangle \hspace{.5cm} \longrightarrow \hspace{.5cm} -| \hspace{.08cm} X_j \rangle \hspace{.04cm} | -x \rangle $$

where this result comes from what happens when we apply an $X$ gate to the state $|-x\rangle$:

$$ X \hspace{.06cm} |-x\rangle = -|-x\rangle $$

Proving this result is a nice exercise that I recommend doing at least once. Write out $|-x\rangle$ in the $\big{\{}$ $|0\rangle$, $|1\rangle$ $\big{\}}$ basis, and it's only a couple line proof.

If you followed along with the Deutsch Algorithm in 5.1, we used this exact trick. Here, we are using this technique again, only on larger qubit states. Let's see this effect in an example, by importing the function $\textbf{Blackbox\_g\_DJ}$ from $\textmf{Our\_Pyquil\_Functions}$:

\begin{figure}[h]
\centering
\includegraphics[scale=.65]{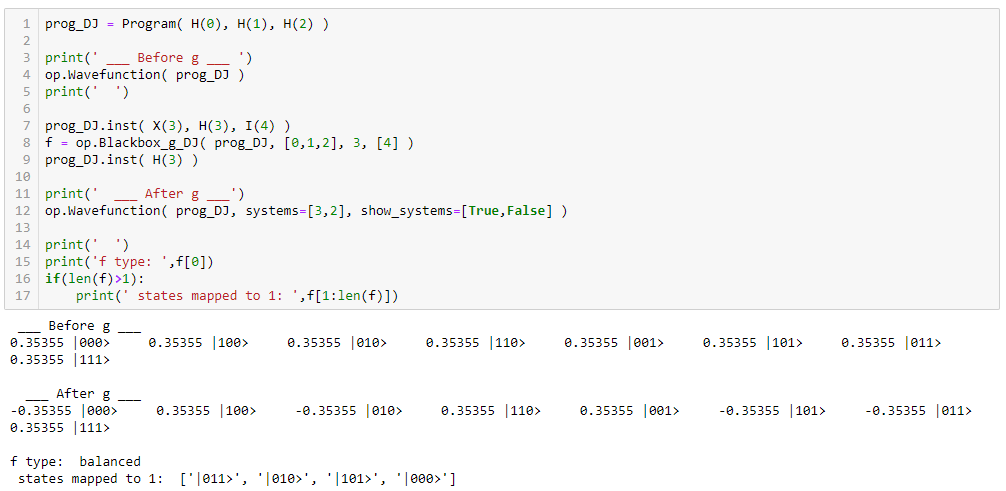}
\end{figure}

We're purposely skipping some of the details regarding the function $\textmf{Blackbox\_g\_DJ}$ here, in favor of just seeing the important result from $g$ (for example, there's an extra ancilla qubit here that we will explain later).
\\

The code above performs the Deutsch-Jozsa Algorithm up to the step where we apply our blackbox $g$. Run the cell of code a couple times until you come across a case where $f$ is balanced. When you do, you should notice that exactly half of the states in the system pick up a negative sign. And, these states exactly match up with the states that get mapped to $1$ by the embedded $f$, printed at the bottom.
\\

Next then, our algorithm calls for another Hadamard Transformation of our system, followed by a measurement:

\pagebreak

\begin{figure}[h]
\centering
\includegraphics[scale=.65]{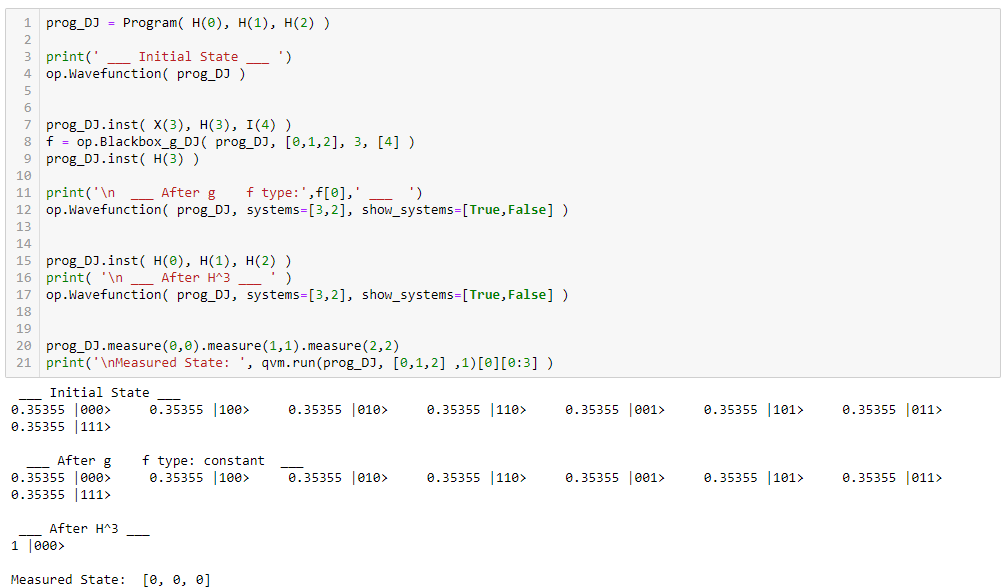}
\end{figure}

Take a moment to carefully check each step in our 'DJ\_qc' printed above, and you should find that all of our steps are in agreement with the algorithm steps we outlined earlier:

$$ \textmf{prepare} \hspace{.1cm} |\hspace{.06cm}0\rangle ^{\otimes n}|1\rangle \hspace{.3cm} \rightarrow \hspace{.3cm} H^{n+1} |\hspace{.06cm}0\rangle ^{\otimes n}|\hspace{.06cm}1\rangle \hspace{.3cm} \rightarrow \hspace{.3cm} g \hspace{.08cm} | \hspace{.06cm}\psi \rangle \hspace{.3cm} \rightarrow \hspace{.3cm} H^{n+1} | \hspace{.06cm}\psi \rangle \hspace{.3cm} \rightarrow \hspace{.3cm} \textmf{measure qubits} \hspace{.06cm} ^{\otimes n} $$

The only thing slightly out of order is that we apply the $H$ gate to the ancilla qubit a little early. But, you can verify for yourself that it is applied after $g$, so it is in agreement with the outlined steps.
\\

Now for the final piece, what to do with our measured state. Recall that the solution to the Deutsch Algorithm problem was based on whether we found qubit $0$ in the state $|0\rangle$ or $|1\rangle$:

$$ |0\rangle \leftrightarrow f_{constant} \hspace{2.6cm} |1\rangle \leftrightarrow f_{balanced} $$

Here, we will make the same conclusion about $f$ based on the measured state of our $n$ qubits:

$$ |000...\rangle \leftrightarrow f_{constant} \hspace{2cm} \textmf{any qubit in state} \hspace{.22cm} |1\rangle \leftrightarrow f_{balanced} $$

If we measure our $n$ qubit system and we find that any of our qubits are in the $|1\rangle$ state, we can conclude that $f$ is balanced. Conversely, if we find that all of the qubits are in the state $|0\rangle$, we can conclude that $f$ is constant. We can make both of these conclusions with 100\% certainty. With this explanation now in hand, I encourage you to return to the cell of code above, and confirm these results.
\\

And that's the full Deutsch-Jozsa Algorithm! In the next section, we will go back over multiple parts that we skipped over, for a deeper understanding as to why it works.

\section*{\large{ Deeper Look at the Deutsch-Jozsa }}

At a quick glance, the three most important keys to the success of the Deutsch-Jozsa algorithm are as follows:

$$ 1) \hspace{.25cm} \textmf{the ancilla qubit: } \hspace{4.3cm} |\hspace{.06cm}1 \rangle \hspace{.2cm} -H \rightarrow \hspace{.2cm} |-x\rangle $$

$$ \hspace{.25cm} 2) \hspace{.15cm} \textmf{ Hadamard transformation: } \hspace{2.6cm} H^{n+1} \hspace{.3cm} (stuff) \hspace{.3cm} H^{n+1} $$

$$ \hspace{1cm} 3) \hspace{.15cm} \textmf{ effect of }\oplus \textmf{ in }g\textmf{: } \hspace{4.2cm} g\hspace{.1cm} | \hspace{.1cm} X_i \rangle \hspace{.06cm} | \hspace{.08cm} \alpha \rangle \hspace{.2cm} \equiv \hspace{.2cm} | \hspace{.1cm} X_i \rangle \hspace{.1cm} X \hspace{.04cm} | \hspace{.08cm} \alpha \rangle $$

It's the combination of these three components that really makes the algorithm tick. We've already seen in the previous section the interplay between keys (1) and (3), and how they result in the flipping of signs on certain states. Now, we're going to focus in particular on the role of the Hadamard Transformation, and why the negative signs make all the difference.
\\

Let's start off with the constant cases, where the claim is that we will always measure the state $|00...0\rangle$. These cases correspond to all states in the system receiving the same action from $g$, and can be represented as the following matrices:

$$ \begin{bmatrix}
1 & 0 & 0 & & & & & \\
0 & 1 & 0 & & & & &\\
0 & 0 & 1 & & & & &\\
& & & . & & & &\\
& & & & . & & &\\
& & & & & . & & \\
& & & & & & . & \\
& & & & & & & . \\
\end{bmatrix} \hspace{.2cm} \equiv \hspace{.2cm} I^{n+1} \hspace{1.5cm} and \hspace{2cm} \begin{bmatrix}
0 & 1 & & & & & & \\
1 & 0 & & & & & &\\
& & 0 & 1 & & & &\\
& & 1 & 0 & & & &\\
& & & & . & & &\\
& & & & & . & & \\
& & & & & & . & \\
& & & & & & & . \\
\end{bmatrix} \hspace{.2cm} \equiv \hspace{.2cm} I^{n} \otimes X_{n+1}
$$

The first case is just the identity matrix, which makes sense considering that the final state we measure is the exact same as the one we prepare:

$$ |0\rangle ^{\otimes n}|1\rangle \hspace{.3cm} \rightarrow \hspace{.3cm} H^{n+1} \hspace{.3cm} \rightarrow \hspace{.3cm} I^{\otimes n+1} \hspace{.3cm} \rightarrow \hspace{.3cm} H^{n+1} \hspace{.3cm} \rightarrow \hspace{.3cm} |0\rangle ^{\otimes n}|1\rangle $$

The second case will also result in the same final state, but with a negative sign. To see this, let's work with the case of $n=2$, plus one ancilla qubit, and consider the effect of the matrix on the states $|000\rangle$ and $|001\rangle$ (where the last qubit is the ancilla):

$$ \begin{bmatrix}
0 & 1 & & & & & & & \\
1 & 0 & & & & & &\\
& & . & & & & &\\
& & & . & & & &\\
& & & & . & & &\\
& & & & & & & \\
& & & & & & & \\
& & & & & & & \\
\end{bmatrix} \hspace{.2cm} \begin{bmatrix}
\alpha_{000} \\ \alpha_{001} \\ . \\ . \\ . \\ \\ \\ \\
\end{bmatrix} \hspace{.6cm} = \hspace{.6cm} \begin{bmatrix}
\alpha_{001} \\ \alpha_{000} \\ . \\ . \\ . \\ \\ \\ \\
\end{bmatrix} $$

The effect of $g$ swaps the amplitudes for the states $|000\rangle$ and $|001\rangle$ (equivalent to an $X$ gate). Now, consider that this $g$ operator is happening between our two Hadamard Transformations. Initially, we only prepare the state $|00\rangle$$|1\rangle$, but the first $H^{n+1}$ transformation puts us in an equal superposition of $\textit{all}$ possible states, including $|000\rangle$ and $|001\rangle$. As a result, we observe the following effect from $g$:

$$ \frac{1}{\sqrt{2}} \big{(} |000\rangle - |001\rangle \big{)} \hspace{.3cm} -g \rightarrow \hspace{.3cm} \frac{1}{\sqrt{2}} \big{(} -|000\rangle + |001\rangle \big{)}$$

This example only follows the states $|000\rangle$ and $|001$, but if we consider the pattern of the second $f_{constant}$ matrix above, it should be clear that this effect will happen to all of states in our main qubit system:

\pagebreak

$$ |\hspace{.06cm} 00\rangle \hspace{.1cm} |-x\rangle \hspace{1.6cm} |\hspace{.08cm} 000\rangle \hspace{.1cm} - \hspace{.1cm} |\hspace{.08cm} 001\rangle \hspace{3.5cm} -|\hspace{.08cm} 000\rangle \hspace{.1cm} + \hspace{.1cm} |\hspace{.08cm} 001\rangle \hspace{1.8cm} -|\hspace{.06cm} 00\rangle \hspace{.1cm} |-x\rangle $$
$$ |\hspace{.06cm} 01\rangle \hspace{.1cm} |-x\rangle \hspace{1.6cm} |\hspace{.08cm} 010\rangle \hspace{.1cm} - \hspace{.1cm} |\hspace{.08cm} 011\rangle \hspace{3.5cm} -|\hspace{.08cm} 010\rangle \hspace{.1cm} + \hspace{.1cm} |\hspace{.08cm} 011\rangle \hspace{1.8cm} -|\hspace{.06cm} 01\rangle \hspace{.1cm} |-x\rangle $$
$$ |\hspace{.06cm} 10\rangle \hspace{.1cm} |-x\rangle \hspace{.65cm} = \hspace{.5cm} |\hspace{.08cm} 100\rangle \hspace{.1cm} - \hspace{.1cm} |\hspace{.08cm} 101\rangle \hspace{1.055cm} -g\rightarrow \hspace{1.25cm} - \hspace{.1cm}|\hspace{.08cm} 100\rangle \hspace{.1cm} + \hspace{.1cm} |\hspace{.08cm} 101\rangle \hspace{.69cm} = \hspace{.8cm} -\hspace{.1cm}|\hspace{.06cm} 10\rangle \hspace{.1cm} |-x\rangle $$
$$ |\hspace{.06cm} 11\rangle \hspace{.1cm} |-x\rangle \hspace{1.6cm} |\hspace{.08cm} 110\rangle \hspace{.1cm} - \hspace{.1cm} |\hspace{.08cm} 101\rangle \hspace{3.5cm} -|\hspace{.08cm} 110\rangle \hspace{.1cm} + \hspace{.1cm} |\hspace{.08cm} 111\rangle \hspace{1.8cm} -|\hspace{.06cm} 11\rangle \hspace{.1cm} |-x\rangle $$

Another way of visualizing what is happening here, is that we are essentially picking up a negative global phase in between our two $H^{n+1}$ transformation. And since a global phase does nothing to our overall system, the state that comes out from the second Hadamard Transformation will be the negative of the state coming into the first:

$$ |\hspace{.1cm} \psi \rangle \hspace{.6cm} - H^{n+1} \rightarrow \hspace{.6cm} |\hspace{.1cm} \phi \rangle \hspace{.2cm} \rightarrow \hspace{.2cm} -|\hspace{.1cm} \phi \rangle \hspace{.6cm} - H^{n+1} \rightarrow \hspace{.6cm} -|\hspace{.1cm} \psi \rangle $$

And since our initial state for the main qubit system is $|00...0\rangle$, we will get back $-$$|00...0\rangle$. Thus, we will measure all of the qubits in the $|0\rangle$ state, and conclude that $f$ is constant. In the cell of code below, we follow our quantum system through all the steps of the two constant $g$ cases:

\begin{figure}[h]
\centering
\includegraphics[scale=.65]{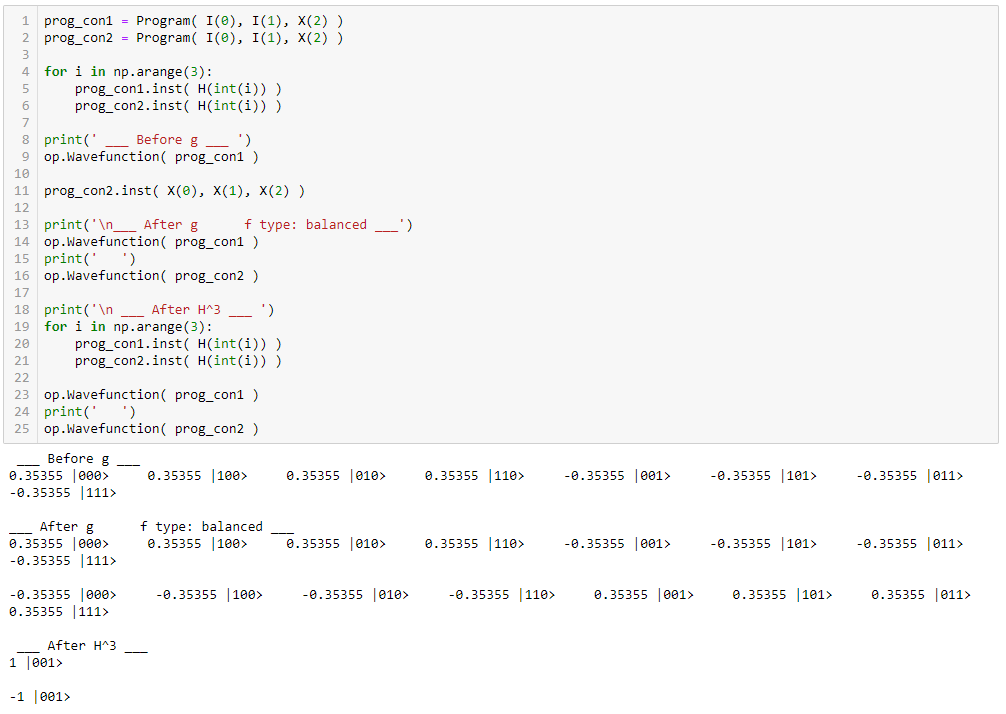}
\end{figure}

As for the balanced cases, their $g$ matrices also have the same $\begin{bmatrix} 0 & 1\\ 1 & 0\end{bmatrix}$ structure appearing along the diagonal. However, instead of the full diagonal, only half of the entries will have this $X$ gate structure. That is to say, if we break up the diagonal of a balanced $g$ into 2$^{n}$ 2$\times$2 blocks, we will find that exactly half of them are $\begin{bmatrix} 0 & 1\\ 1 & 0\end{bmatrix}$, and the other half are $\begin{bmatrix} 1 & 0\\ 0 & 1\end{bmatrix}$ (an Identity matrix).
\\

What this means for our system, is that if we revisit the the exercise we just did for the states $|000\rangle$ and $|001$ above, we can apply to same logic to half our system. For example:

$$ |\hspace{.06cm} 00\rangle \hspace{.1cm} |-x\rangle \hspace{1.6cm} |\hspace{.08cm} 000\rangle \hspace{.1cm} - \hspace{.1cm} |\hspace{.08cm} 001\rangle \hspace{3.2cm} -|\hspace{.08cm} 000\rangle \hspace{.1cm} + \hspace{.1cm} |\hspace{.08cm} 001\rangle \hspace{1.4cm} -|\hspace{.06cm} 00\rangle \hspace{.1cm} |-x\rangle $$
$$ |\hspace{.06cm} 01\rangle \hspace{.1cm} |-x\rangle \hspace{1.6cm} |\hspace{.08cm} 010\rangle \hspace{.1cm} - \hspace{.1cm} |\hspace{.08cm} 011\rangle \hspace{3.7cm} |\hspace{.08cm} 010\rangle \hspace{.1cm} - \hspace{.1cm} |\hspace{.08cm} 011\rangle \hspace{1.9cm} |\hspace{.06cm} 01\rangle \hspace{.1cm} |-x\rangle $$
$$ |\hspace{.06cm} 10\rangle \hspace{.1cm} |-x\rangle \hspace{.6cm} = \hspace{.5cm} |\hspace{.08cm} 100\rangle \hspace{.1cm} - \hspace{.1cm} |\hspace{.08cm} 101\rangle \hspace{1.155cm} -g\rightarrow \hspace{1.35cm} |\hspace{.08cm} 100\rangle \hspace{.1cm} - \hspace{.1cm} |\hspace{.08cm} 101\rangle \hspace{.6cm} = \hspace{.8cm} |\hspace{.06cm} 10\rangle \hspace{.1cm} |-x\rangle $$
$$ |\hspace{.06cm} 11\rangle \hspace{.1cm} |-x\rangle \hspace{1.6cm} |\hspace{.08cm} 110\rangle \hspace{.1cm} - \hspace{.1cm} |\hspace{.08cm} 101\rangle \hspace{3.2cm} -|\hspace{.08cm} 110\rangle \hspace{.1cm} + \hspace{.1cm} |\hspace{.08cm} 111\rangle \hspace{1.4cm} -|\hspace{.06cm} 11\rangle \hspace{.1cm} |-x\rangle $$

which would be the result of the $g$ matrix:

$$ \begin{bmatrix}
0 & 1 & & & & & & \\
1 & 0 & & & & & &\\
& & 1 & & & & &\\
& & & 1 & & & &\\
& & & & 1 & & &\\
& & & & & 1 & & \\
& & & & & & 0 & 1\\
& & & & & & 1 & 0 \\
\end{bmatrix} \hspace{.2cm} \begin{bmatrix}
\alpha_{000} \\ \alpha_{001} \\ \alpha_{010} \\ \alpha_{011} \\ \alpha_{100} \\ \alpha_{101} \\ \alpha_{110} \\ \alpha_{111}
\end{bmatrix}$$

Now, if you take the final state above: $\hspace{.25cm} \frac{1}{2} \big{(} \hspace{.02cm} -|\hspace{.06cm} 00\rangle + |\hspace{.06cm} 01\rangle + |\hspace{.06cm} 10\rangle - |\hspace{.06cm} 11\rangle \hspace{.08cm} \big{)}\hspace{.1cm}|-x\rangle$, and apply the second $H^{n+1}$, you should get -$|11\rangle$ $|1\rangle$ as your final state. Which, if we were to make a measurement on, would definitely find at least one qubit in the state $|1\rangle$. Thus, we would conclude that our $f$ is balanced.
\\

But, rather than working through every possible $f_{balanced}$ and verifying that we never get the state $|00\rangle$, all we need to do is notice the trend $\textit{why}$ we will never get it. If we recall why a balanced $f$ in the Deutsch Algorithm would always lead to the state $|11\rangle$, we can apply the same logic here. Specifically, in the Deutsch case, a balanced $f$ would always lead to qubit $0$ being in the $|1\rangle$ state because of the second Hadamard Transformation always being applied to either:
\\

$\hspace{.25cm} \frac{1}{2} \big{(} \hspace{.02cm} -|\hspace{.06cm} 00\rangle + |\hspace{.06cm} 01\rangle + |\hspace{.06cm} 10\rangle - |\hspace{.06cm} 11\rangle \hspace{.08cm} \big{)}\hspace{.1cm}$ or $\hspace{.25cm} \frac{1}{2} \big{(} \hspace{.02cm} |\hspace{.06cm} 00\rangle - |\hspace{.06cm} 01\rangle - |\hspace{.06cm} 10\rangle + |\hspace{.06cm} 11\rangle \hspace{.08cm} \big{)}\hspace{.1cm}$, which in turn gives us either $\hspace{.1cm}$ $|11\rangle \hspace{.08cm}$ or $\hspace{.08cm} -$$|11\rangle$.
\\

So why does that matter. Remember, our condition for concluding if $f$ is constant or balanced is based entirely around measuring the state $|00...0\rangle$. If $f$ is constant, it will be the $\textit{only}$ final state, and if $f$ is balanced, it will $\textit{never}$ be a part of the final system. So then, the question becomes: under what conditions will the second $H^{n}$ transformation yield only $|00...0\rangle$, or not at all. If you've worked through some of the algebra examples thus far (good for you, gold star!), you may have a hunch as to how the state $|00...0\rangle$ comes out of a Hadamard Transformation. If not, consider all of the individual contributions to the state $|00...0\rangle$ in the example below:

$$ H^{n} \hspace{.06cm}|\hspace{.06cm} 01101...\rangle \hspace{.3cm} \rightarrow \hspace{.3cm} \Big{(} \frac{1}{\sqrt{2}} \Big{)}^n \big{(}\hspace{.1cm} |\hspace{.06cm}0\rangle + |\hspace{.06cm}1\rangle \hspace{.08cm} \big{)} \big{(}\hspace{.1cm} |\hspace{.06cm}0\rangle - |\hspace{.06cm}1\rangle \hspace{.08cm} \big{)} \big{(}\hspace{.1cm} |\hspace{.06cm}0\rangle - |\hspace{.06cm}1\rangle \hspace{.08cm} \big{)} \cdot \cdot \cdot \cdot $$

No matter how many qubits there are, and no matter which ones are in the state $|0\rangle$ or $|1\rangle$, the state $|00...0\rangle$ will $\textit{always}$ be positive. More specifically, $H$$|0\rangle$ and $H$$|1\rangle$ both contribute a positive $|0\rangle$ to the final superposition state. So then, the only way in which we can get a negative sign on the state $|00...0\rangle$ is if the entire state is negative before the Hadamard gate! If there is an overall negative sign before the H$^n$, then it will carry over to the final state, turning $|00...0\rangle$ negative.
\\

Now then, combine this result with the one just prior: that a balanced $f$ will leave exactly half of the states negative before the second $H^{n+1}$, and we have our answer. Exactly half of the $|00...0\rangle$ states will come out positive, and the other half will be negative. SO, they cancel out to zero, and our final system will not contain the state $|00...0\rangle$.
\\

No matter to what higher order we go, if $f$ is balanced, this process will $\textit{always}$ cause the state $|00...0\rangle$ to perfectly deconstructively interfere. Thus, we will always measure a state in the system where at least one qubit is in the state $|1\rangle$! And conversely, as we've already shown, the two constant $f$ cases lead to final systems that $\textit{only}$ contain the state $|00...0\rangle$. This is how we can be 100\% certain of $f$ based on our measurement result.

\pagebreak

\begin{figure}[h]
\centering
\includegraphics[scale=.65]{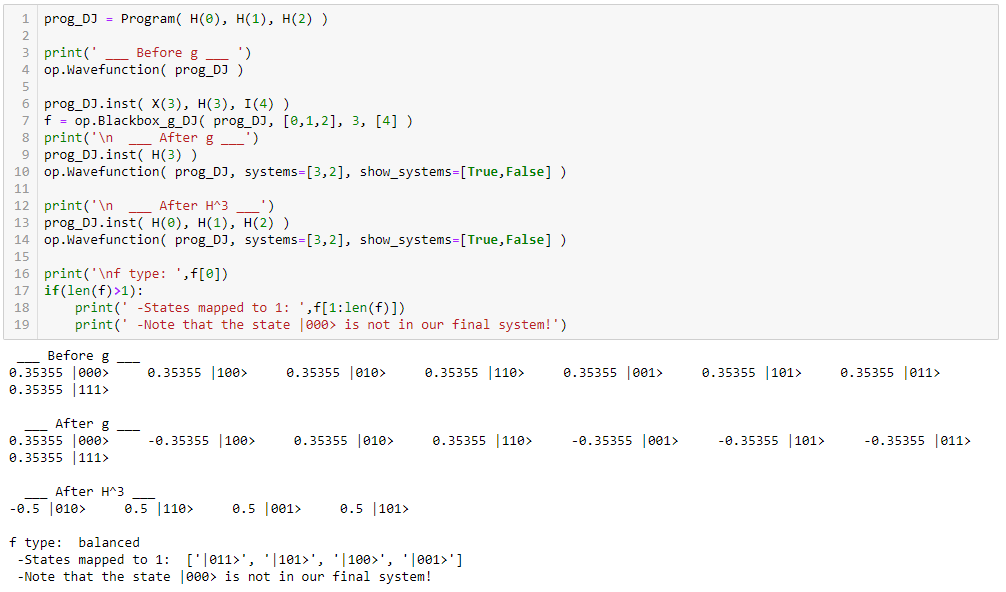}
\end{figure}

\section*{\large{ A closer look at the Blackbox\_g\_DJ }}

To implement the $g$ operator we've been using thus far, we will follow a trick very similar to the one we used in the Deutsch operator. We've already covered both of the constant cases in the previous section, so here we will focus solely on how to create a balanced $g$. Specifically, we need to be able to encode the following operation on exactly half of the states in the system:

$$ \begin{bmatrix}
0 & 1 & & & & & & & \\
1 & 0 & & & & & &\\
& & . & & & & &\\
& & & . & & & &\\
& & & & . & & &\\
& & & & & & & \\
& & & & & & & \\
& & & & & & & \\
\end{bmatrix} \hspace{.2cm} \begin{bmatrix}
\alpha_{000} \\ \alpha_{001} \\ . \\ . \\ . \\ \\ \\ \\
\end{bmatrix} \hspace{.6cm} = \hspace{.6cm} \begin{bmatrix}
\alpha_{001} \\ \alpha_{000} \\ . \\ . \\ . \\ \\ \\ \\
\end{bmatrix} $$

which for systems larger than $n=2$, this will require a higher order CNOT operation. For example: $ \hspace{.2cm} CCCNOT \hspace{.16cm}|\hspace{.08cm}1110\rangle \hspace{.18cm} \rightarrow \hspace{.18cm} |\hspace{.08cm}1111\rangle$.
\\

We've already covered how to implement these higher order CNOT gates in lesson 4, so please refer to that lesson for a more detailed explanation. For our balanced $g$ operators here, all we need to do is pick out half of the states in the system and apply our $\textbf{n\_NOT}$ gate, using the $n$ qubits as the control, and our ancilla qubit in the $|-x\rangle$ state as the target.
\\

But, since the $\textmf{n\_NOT}$ gate only operates on the state of all $|1\rangle$'s, we must use some $X$ gates before and after. Let's see an example of this, where we will use $|010\rangle$ as our control state:

\pagebreak

\begin{figure}[h]
\centering
\includegraphics[scale=.65]{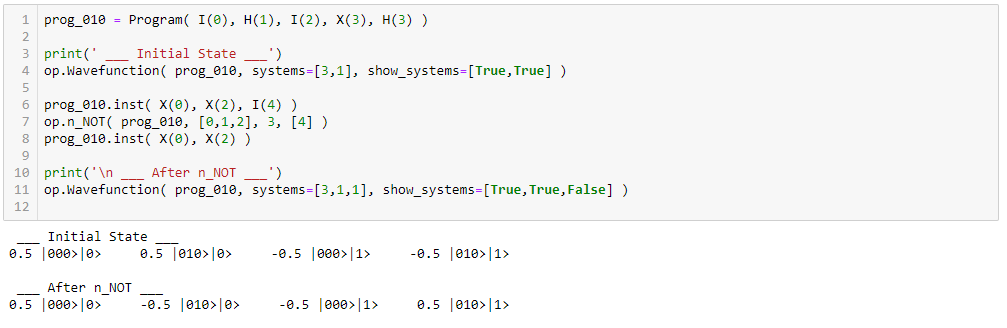}
\end{figure}

As we can see in this code example, we've achieved the desired effect: $\hspace{.18cm}$ $|0100\rangle \hspace{.1cm} \rightarrow \hspace{.1cm} $ $|0101\rangle \hspace{.12cm}$ and $\hspace{.12cm}$ $|0101\rangle \hspace{.1cm} \rightarrow \hspace{.1cm} $ $|0100\rangle \hspace{.12cm}$ by using $X$ gates on qubits $0$ and $2$ before and after the $\textmf{n\_NOT}$ operation. These $X$ gates transform our desired control state into the state of all $1$'s, such that the $\textmf{n\_NOT}$ operation will work, and then back to the original state:

$$ |\hspace{.08cm}010\rangle \hspace{.06cm}|\hspace{.08cm}0\rangle \hspace{.4cm} - X_0 \otimes X_2 \rightarrow \hspace{.4cm} |\hspace{.08cm}111\rangle \hspace{.06cm}|\hspace{.08cm}0\rangle \hspace{.4cm} - \textmf{n\_NOT} \rightarrow \hspace{.4cm} |\hspace{.08cm}111\rangle \hspace{.06cm}|\hspace{.08cm}1\rangle \hspace{.4cm} - X_0 \otimes X_2 \rightarrow \hspace{.4cm} |\hspace{.08cm}010\rangle \hspace{.06cm}|\hspace{.08cm}1\rangle $$

And, we must not forget that the $\textmf{n\_NOT}$ operation uses an additional $n-2$ ancilla qubits. Thus, in the example above we add an extra qubit to our system in the $|0\rangle$ state before the $\textmf{n\_NOT}$ step.
\\

By using this technique, we can effectively pick out any state in the system to be our control, which translates to the operation required of our $g$ matrix:

$$\hspace{.2cm} f \hspace{.04cm} (\hspace{.08cm} X_i \hspace{.08cm}) \hspace{.2cm} \rightarrow \hspace{.2cm} 1 \hspace{1.2cm} \longleftrightarrow \hspace{1.2cm} g\hspace{.08cm} | \hspace{.1cm} X_i \rangle \hspace{.06cm} |-x\rangle \hspace{.2cm} \rightarrow \hspace{.2cm} -| \hspace{.1cm} X_i \rangle \hspace{.06cm} |-x\rangle $$

With this trick in hand, and some classical code for picking out half the states at random, we have our Deutsch-Jozsa $g$ operator!

\section*{\large{ Bernstein-Vazirani Algorithm }}
\centerline{---------------------------------------------------------------------------------------------------------------------------------}

To quickly recap, we just solved the problem of an unknown $f$, in which we were told it is either constant or balanced, using only one application of the blackbox followed by a measurement:

$$ |000...\rangle \leftrightarrow f_{constant} \hspace{2cm} \textmf{any qubit in state} \hspace{.22cm} |1\rangle \leftrightarrow f_{balanced} $$

Now, in the second part to this tutorial, we will look at a different problem that can be solved using the same quantum circuit. Specifically, we are given a $\textit{new}$ blackbox function $f$:

$$ f(\hspace{.06cm}X_i \hspace{.06cm}) = a\cdot X_i \oplus b $$

where $X_i$ is the same string of bits as before: {x$_0$, x$_1$, ...}.
\\

Inside this $f$, we have two unknown quantities: $a$ and $b$, where $a$ is a string of bits (same length as X), and $b$ is just a single bit. $X_i$ and $a$ are multiplied together via a standard dot product of vectors, and $\oplus$ still refers to addition modulo 2 here. Let's look at an example:

$$ X = \{ 1, 0, 1, 1 \} \hspace{.8cm} a = \{1, 0, 0, 1\} \hspace{.8cm} b = 1 $$

$$ f\hspace{.05cm}(\hspace{.06cm} X \hspace{.06cm}) = \big{(}\hspace{.12cm} 1 \cdot 1 \hspace{.1cm} + \hspace{.1cm} 0\cdot 0 \hspace{.1cm} + \hspace{.1cm} 1 \cdot 0 \hspace{.1cm} + \hspace{.1cm} 1 \cdot 1 \hspace{.12cm} \big{)} \oplus 1 \hspace{.25cm} = \hspace{.25cm} 2 \oplus 1 \hspace{.25cm} = \hspace{.25cm} 1 $$

So then, the problem which we are going to solve is: how quickly can we determine $a$ with a quantum computer? Determining the constant $b$ can be achieved in one step by passing an $X_i$ of all $0$'s, by both a classical and quantum computer. Thus, the real challenge is in determining $a$. Classically, we would have to evaluate $f$, $n$ times, where $n$ is the length of the bit string $a$. As we shall see, by using our quantum circuit, we will be able to fully determine $a$ in just one step!
\\

Just like the Deutsch-Jozsa problem, we will solve this new $f$ using the exact same steps:

$$ \textmf{prepare} \hspace{.1cm} |\hspace{.06cm}0\rangle ^{\otimes n}|1\rangle \hspace{.3cm} \rightarrow \hspace{.3cm} H^{n+1} |\hspace{.06cm}0\rangle ^{\otimes n}|\hspace{.06cm}1\rangle \hspace{.3cm} \rightarrow \hspace{.3cm} g \hspace{.08cm} | \hspace{.06cm}\psi \rangle \hspace{.3cm} \rightarrow \hspace{.3cm} H^{n+1} | \hspace{.06cm}\psi \rangle \hspace{.3cm} \rightarrow \hspace{.3cm} \textmf{measure qubits} \hspace{.06cm} ^{\otimes n} $$

We also embed our new $f$ into a unitary operator $g$ in the same manner as before:

$$f \hspace{.05cm}| \hspace{.08cm} X_i\rangle \hspace{.06cm}|\hspace{.08cm} \alpha \rangle \hspace{.4cm} \rightarrow \hspace{.4cm}|\hspace{.08cm} X\rangle \hspace{.06cm}| \hspace{.08cm} \alpha \oplus f\hspace{.04cm}(\hspace{.06cm} X_i \hspace{.06cm}) \hspace{.05cm}\rangle$$

Let's see the full example first, and then discuss why it works:

\begin{figure}[h]
\centering
\includegraphics[scale=.65]{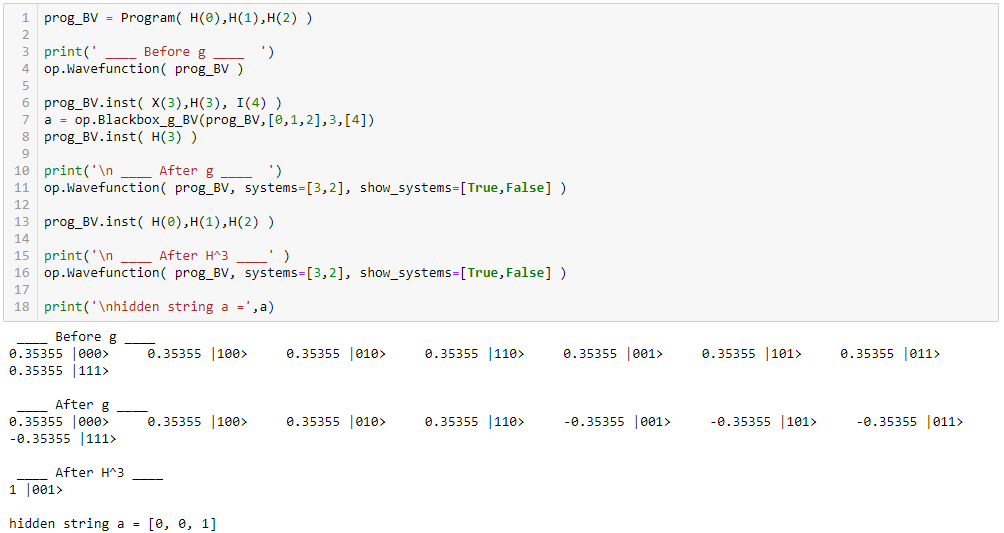}
\end{figure}

Run the code above a couple times, and you should find that the final state of our system is always $|a\rangle$. Or in other words, our final system is always guaranteed to be in the state exactly matching the string of bits $a$. Thus, a measurement on the system will reveal $a$ with 100\% accuracy, solving our problem in just one step!
\\

Alrighty, now to explain the magic.
\\

First off, let's show an example of an $f$(X) on two qubits, and the outputs it produces from each of the four states side by side with the corresponding $g$ operator:

\pagebreak

$$ a = \{1, 0 \} \hspace{.8cm} b = 1 $$

$$f( \hspace{.08cm} \{0,0\} \hspace{.08cm} ) \hspace{.16cm} \rightarrow \hspace{.16cm} \big{(} 0\cdot 1 + 0\cdot 0 \big{)} \oplus 1 \hspace{.16cm} = \hspace{.16cm} 1 \hspace{1.6cm} \leftrightarrow \hspace{1.6cm} g \hspace{.08cm}| \hspace{.06cm}00\rangle \hspace{.06cm} |-x\rangle \hspace{.16cm} \rightarrow \hspace{.12cm} - | \hspace{.06cm}00\rangle \hspace{.06cm} |-x\rangle $$
$$f( \hspace{.08cm} \{0,1\} \hspace{.08cm} ) \hspace{.16cm} \rightarrow \hspace{.16cm} \big{(} 0\cdot 1 + 1\cdot 0 \big{)} \oplus 1 \hspace{.16cm} = \hspace{.16cm} 1 \hspace{1.6cm} \leftrightarrow \hspace{1.6cm} g \hspace{.08cm}| \hspace{.06cm}01\rangle \hspace{.06cm} |-x\rangle \hspace{.16cm} \rightarrow \hspace{.12cm} - | \hspace{.06cm}01\rangle \hspace{.06cm} |-x\rangle $$
$$f( \hspace{.08cm} \{1,0\} \hspace{.08cm} ) \hspace{.16cm} \rightarrow \hspace{.16cm} \big{(} 1\cdot 1 + 0\cdot 0 \big{)} \oplus 1 \hspace{.16cm} = \hspace{.16cm} 0 \hspace{1.6cm} \leftrightarrow \hspace{1.6cm} g \hspace{.08cm}| \hspace{.06cm}10\rangle \hspace{.06cm} |-x\rangle \hspace{.16cm} \rightarrow \hspace{.48cm} | \hspace{.06cm}10\rangle \hspace{.06cm} |-x\rangle $$
$$f( \hspace{.08cm} \{1,1\} \hspace{.08cm} ) \hspace{.16cm} \rightarrow \hspace{.16cm} \big{(} 1\cdot 1 + 1\cdot 0 \big{)} \oplus 1 \hspace{.16cm} = \hspace{.16cm} 0 \hspace{1.6cm} \leftrightarrow \hspace{1.6cm} g \hspace{.08cm}| \hspace{.06cm}11\rangle \hspace{.06cm} |-x\rangle \hspace{.16cm} \rightarrow \hspace{.48cm} | \hspace{.06cm}11\rangle \hspace{.06cm} |-x\rangle $$

The thing to note about this pattern is that the effect of $f$ has produced a $1$ for exactly half of the states, which will in turn cause a sign flip for half of the states in the system. This result will hold true for all $a$'s, with only one exception: $a$={ $0$, $0$, $0$, ...}. If $a$ is all $0$'s, then the effect of $f$ is completely determined by $b$, which will either flip all of the states, or none.
\\

Causing exactly half of the states to pick up a negative sign ties in with the previous algorithm. Recall that when half of the states in the system undergo a sign flip, the amplitude on the state $|00...0\rangle$ always cancels to zero (deconstructively sums to zero):

$$ \hspace{.3cm} H^3\hspace{.08cm} \frac{1}{2\sqrt{2}} \big{(} \hspace{.12cm} |\hspace{.05cm}000\rangle - |\hspace{.05cm}001\rangle + |\hspace{.05cm}010\rangle - |\hspace{.05cm}011\rangle + |\hspace{.05cm}100\rangle - |\hspace{.05cm}101\rangle + |\hspace{.05cm}110\rangle - |\hspace{.05cm}111\rangle \hspace{.1cm} \big{)} \hspace{.26cm} \rightarrow \hspace{.26cm} |\hspace{.05cm}001\rangle \hspace{5.4cm} $$

$$ H^3\hspace{.08cm} \frac{1}{2\sqrt{2}} \big{(} |\hspace{.05cm}000\rangle + |\hspace{.05cm}001\rangle - |\hspace{.05cm}010\rangle - |\hspace{.05cm}011\rangle + |\hspace{.05cm}100\rangle - |\hspace{.05cm}101\rangle + |\hspace{.05cm}110\rangle - |\hspace{.05cm}111\rangle \big{)} \hspace{.26cm} \rightarrow \hspace{.26cm} \frac{1}{2} \big{(} |\hspace{.05cm}001\rangle + |\hspace{.05cm}010\rangle - |\hspace{.05cm}101\rangle + |\hspace{.05cm}110\rangle \big{)} $$

As we can see in these two examples, the state $|000\rangle$ is in neither of the final systems. More importantly however, notice that if a certain combination of states are negative before the $H^3$, all of the states in the system will deconstructively sum to zero, except for one. But if the combination of states isn't a 'special order', then our final system will be in a superposition state.
\\

So then, what dictates one of the 'special combinations', which will result in a single final state. To answer that question, we can work backwards and apply H$^{3}$ to any state and get our answer:

$$ | \hspace{.06cm} 001 \rangle \hspace{.5cm} \longleftarrow H^3 \longrightarrow \hspace{.5cm} \frac{1}{2\sqrt{2}} \big{(} \hspace{.12cm} |\hspace{.05cm}000\rangle - |\hspace{.05cm}001\rangle + |\hspace{.05cm}010\rangle - |\hspace{.05cm}011\rangle + |\hspace{.05cm}100\rangle - |\hspace{.05cm}101\rangle + |\hspace{.05cm}110\rangle - |\hspace{.05cm}111\rangle \hspace{.1cm} \big{)} $$

By applying a Hadamard gate to each qubit, and working through the algebra steps, we can find out which combinations of negatives corresponds to any state. Or, we can let our code do it for us:

\begin{figure}[h]
\centering
\includegraphics[scale=.65]{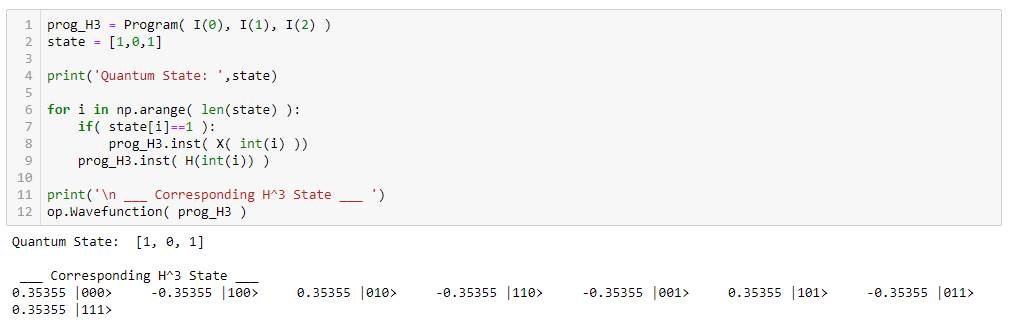}
\end{figure}

By changing the array 'state' in the example above, we can see what the corresponding state gets mapped to via the Hadamard Transformation.
\\

So then, why does our new blackbox $f$ always guarantee we get the correct negative sign flips that will lead to a single final state? The short answer: because it is consistent. That is to say, this $f$ applies the same rules to every $X_i$. Remember that for the $f$ in the Deutsch-Jozsa Algorithm, we were only guaranteed it was balanced or constant, but for the cases where it was balanced, we gained no information about the inner workings of which combinations got mapped to $0$ or $1$.
\\

Here, because the $f$ in this problem is a consistent set of rules: $\hspace{.15cm} X_i \cdot a \oplus b$, its effect will apply negative signs in a correspondingly consistent manner. If we think about the effect of the dot product $\hspace{.1cm} X_i \cdot a$ in particular, this function essentially picks out how many matches of $1$'s there are between $a$ and the state. For example, $ \hspace{.15cm} X_i \cdot a \hspace{.2cm} \rightarrow \hspace{.2cm} \{1,0,1,1\} \cdot \{1,0,0,1\} = 2$, where 2 is exactly the number of cases where $X_i$ and $a$ have $1$'s in the same position. Then for our $g$ operator: 2 mod 2 = 0, so ignoring $b$, this inner product would not result in a sign flip on the state $|1011\rangle$, because $f$( $X_i$ ) = 0.
\\

If that was a little hard to follow, consider the following two matrices, which show an exact 1-to-1 correlation between states where $ \hspace{.1cm} X_i \cdot a$ = odd $ \hspace{.3cm} \longleftrightarrow \hspace{.3cm}$ negative signs on states resulting from $H^3$:

$$\hspace{2.9cm}X_i = 000 \hspace{.09cm} 001 \hspace{.09cm} 010 \hspace{.09cm} 011 \hspace{.09cm} 100 \hspace{.09cm} 101 \hspace{.09cm} 110 \hspace{.09cm} 111 \hspace{8.2cm} $$

$$a = \begin{pmatrix}
000 \\
001 \\
010 \\
011 \\
100 \\
101 \\
110 \\
111 \\
\end{pmatrix} \hspace{.4cm} \begin{bmatrix}
& & & & & & & \\
& \hspace{.3cm} o \hspace{.3cm} & & \hspace{.3cm} o \hspace{.3cm} & & \hspace{.3cm} o \hspace{.3cm} & & \hspace{.3cm} o \hspace{.3cm} \\
& & o & o & & & o & o \\
& o & o & & & o & o & \\
& & & & o & o & o & o \\
& o & & o & o & & o & \\
& & o & o & o & o & & \\
& o & o & & o & & & o \\
\end{bmatrix} \hspace{1cm} \longleftrightarrow \hspace{1cm} \begin{bmatrix}
& & & & & & & \\
& - & & - & & - & & - \\
& & - & - & & & - & - \\
& - & - & & & - & - & \\
& & & & - & - & - & - \\
& - & & - & - & & - & \\
& & - & - & - & - & & \\
& - & - & & - & & & - \\
\end{bmatrix}$$

The matrix on left shows all of the combinations of $\hspace{.08cm} X_i \cdot a$ that result in an odd number, while the matrix on the right shows which states come out negative as a result of H$^3$ $|X_i\rangle$. What this is showing then, is that for all the states where $\hspace{.08cm} X_i \cdot a$ results in an odd number, these are the exact states that need to be negative in order for H$^3$ to map back to a single final state.
\\

Because our $f$ function contains addition modulo 2, it has the following effect based on whether we add an even or odd number:

$$ \textmf{odd} \oplus \{0,1\} = \{1,0\} \hspace{2cm} \textmf{even} \oplus \{0,1\} = \{0,1\}$$

This in turn determines which states pick up a negative signs after $g$, when used in combination with our $|-x\rangle$ ancilla:

$$| -x \hspace{.08cm} \oplus \hspace{.12cm} \textmf{odd} \hspace{.08cm} \rangle \hspace{.2cm} = \hspace{.2cm} -|-x\rangle \hspace{3cm} | -x \hspace{.08cm} \oplus \hspace{.12cm} \textmf{even} \hspace{.08cm} \rangle \hspace{.2cm} = \hspace{.2cm} |-x\rangle$$

Thus, we get our 1-to-1 correlation between the states that share an odd number of $1$'s with $a$, and the states that will pick up a negative sign before the second $H^N$ mapping. By using the clever trick of setting up our ancilla qubit in the state $|-x\rangle$, the net effect of our blackbox operator $g$ results in negative sign flips on just the right states, as shown above. And the only role that the constant $b$ has then is an overall negative sign:

$$ | \hspace{.1cm} \psi \hspace{.05cm} \rangle_{final} = (b)^{-1} \hspace{.08cm} | \hspace{.08cm} a\rangle$$

which is undetectable by a measurement. And that's the full algorithm!
\\

$\ast$ $\textit{The crowd leaps out of their seats in applause}$ $\ast$
\\

We won't go into any further analysis of the $g$ operator here, since it is essentially the same as the Deutsch-Jozsa one. The only difference is on which states receive the $\textmf{n\_NOT}$ gate. In the Deutsch-Jozsa case, the states were picked at random, while for the Bernstein-Vazirani $g$, the states are determined by a randomly picked $a$.
\\

To conclude this tutorial, two cells of code are presented for you to try out, each of which incorporates all of the steps outlined above for the respective two algorithms, written into the functions $\textbf{Deutsch\_Jozsa}$ and $\textbf{Bernstein\_Vazirani}$. In the examples below, change $Q$ to be the number of qubits you would like as the main system:

\pagebreak

\begin{figure}[h]
\centering
\includegraphics[scale=.65]{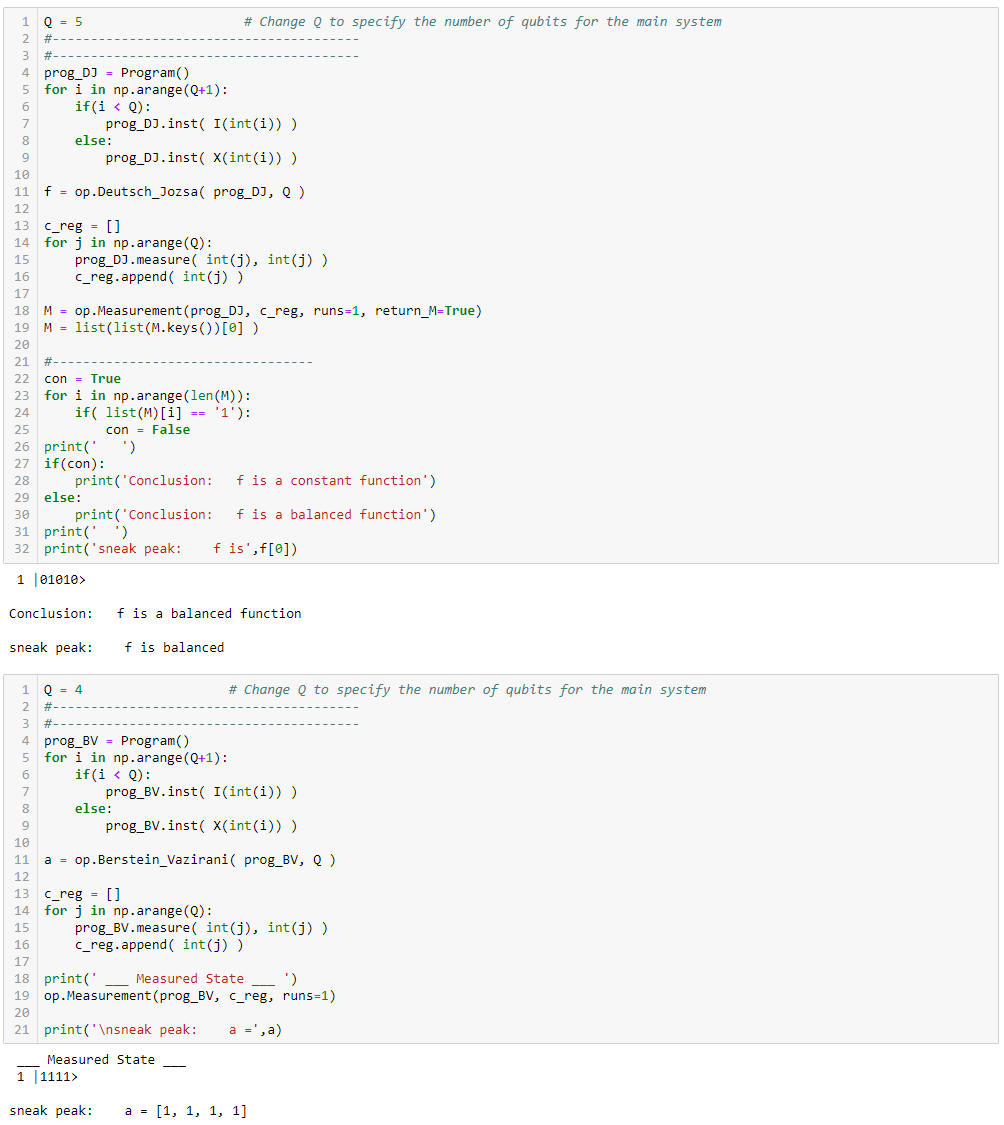}
\end{figure}

--------------------------------------------------------------------------------------------------------------------------------------------------------
\\

This concludes lesson 5.2! The two algorithms studied in this tutorial are important hurdles towards understanding some of the more complex ones to come. In particular, I encourage you to play around with the code examples, and make sure you fully understand why the Hadamard Transformation allowed us to solve these problems in just one step.
\\

--------------------------------------------------------------------------------------------------------------------------------------------------------


\pagebreak

\section*{\Large{ Lesson 5.3 - Simon's Algorithm }}
--------------------------------------------------------------------------------------------------------------------------------------------------------
\\

In this tutorial, we will cover Simon's Algorithm, which is another 'blackbox' style problem, similar to those we've seen in lessons 5.1 and 5.2. The key difference in solving this new problem, is that it will require multiple measurements as well as a classical computing component.
\\

For any reminders / refreshers on Pyquil notation and basics, check out lessons 1 - 4. Also, please consider reading Lesson 5.1 and 5.2, which cover many of the underlying mathematics that we will see in this lesson.
\\

Original publication of the algorithm: \cite{S}
\\

--------------------------------------------------------------------------------------------------------------------------------------------------------
\\
In order to make sure that all cells of code run properly throughout this lesson, please run the following cell of code below:

\begin{figure}[h]
\centering
\includegraphics[scale=.65]{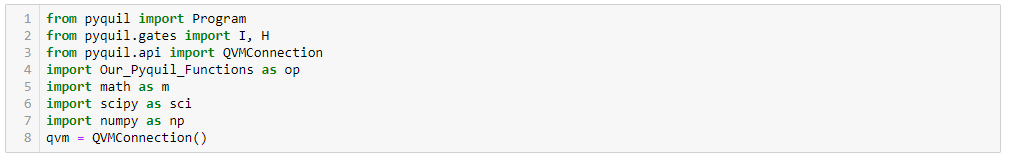}
\end{figure}

\section*{\large{ Simon's Algorithm }}
\centerline{---------------------------------------------------------------------------------------------------------------------------------}

Simon's Algorithm will be our final 'blackbox' style algorithm in these lesson 5 tutorials. It will share many of the same tricks as before, but with a unique final answer. Just like lessons 5.1 and 5.2, solving Simon's Algorithm revolves around using Hadamard gates before and after we call upon our blackbox function. The Hadamard Transformation will allow us to apply the effect of the blackbox function to all possible combinations at once, which we use to our advantage.

\section*{\large{ Quantum Component }}

Now, let's present Simon's problem: we are given an unknown blackbox function $f$, which is $\textit{guaranteed}$ to be either one-to-one or two-to-one, where one-to-one and two-to-one functions have the following properties:

$$ one-to-one \hspace{6.7cm} two-to-one $$

$$ f(1) \rightarrow 1 \hspace{8cm} f(1) \rightarrow 1$$
$$ f(2) \rightarrow 3 \hspace{8cm} f(2) \rightarrow 2$$
$$ f(3) \rightarrow 2 \hspace{8cm} f(3) \rightarrow 1$$
$$ f(4) \rightarrow 4 \hspace{8cm} f(4) \rightarrow 2$$

One-to-one functions have exactly one unique output for every input, while two-to-one functions map exactly two inputs to every unique output. In addition, if our $f$ turns out to be two-to-one, we are also guaranteed that there is a 'key bit-string' $s$ which correlates which inputs map to the same output:

$$ \textmf{given} \hspace{.2cm} x_1, x_2: \hspace{.4cm} f(x_1) = f(x_2) $$

$$ \textmf{guaranteed :} \hspace{.4cm} x_1 \oplus x_2 = s $$

So then, given this blackbox $f$, how quickly can we determine if $f$ is one-to-one or two-to-one? Then, if $f$ turns out to be two-to-one, how quickly can we determine $s$? As it turns out, both cases boil down to the same problem of finding $s$, where a key bit-string of $s=\{ \hspace{.08cm}0, 0, 0 \hspace{.04cm} ... \hspace{.08cm} \}$ $\hspace{.06cm}$ represents the one-to-one $f$.
\\

Let's see a quick example of this kind of $f$:

\begin{figure}[h]
\centering
\includegraphics[scale=.65]{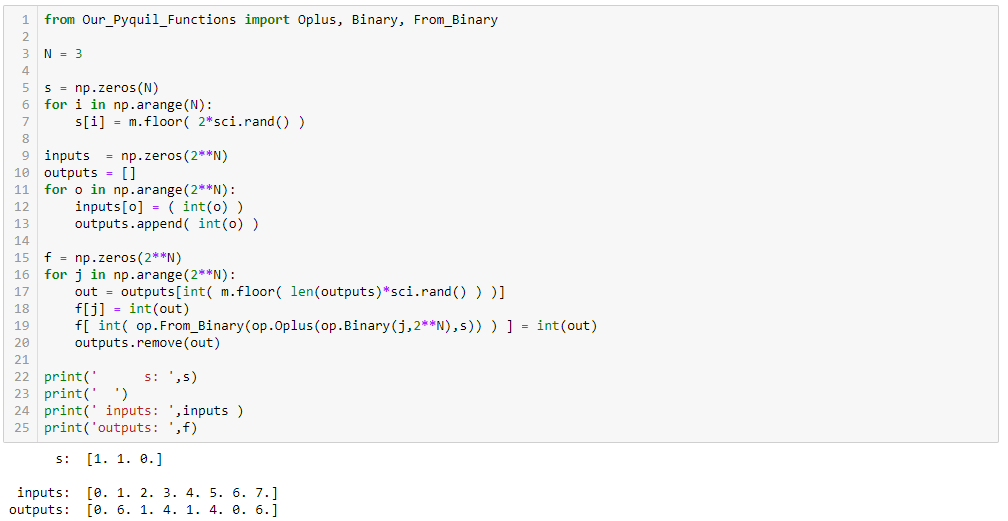}
\end{figure}

The cell of code above simulates a blackbox $f$ for a random key bit-string. Run the code a few times, and verify for yourself that all the correlated output states obey: $\hspace{.2cm} x_1 \oplus x_2 = s \hspace{.2cm}$ (hint, the correlation must be done in binary). Don't worry too much about understanding the lines of code, as the final result is really what we're after: an example of an $f$ function that maps inputs to outputs based on the string $s$.
\\

Classically, if we want to know what $s$ is for a given $f$, with 100\% certainty, we have to check up to 2$^{N-1}$+1 (just over half the total) inputs until we find two cases of the same output. Although, probabilistically the average number of inputs will be closer to the order of $O \big{(} \sqrt{2^N} \big{)}$. Much like the Deutsch-Jozsa problem, if we get lucky, we could solve the problem with our first two tries. But if we happen to get an $f$ that is one-to-one, or get $\textit{really}$ unlucky with an $f$ that's two-to-one, then we're stuck with the full 2$^{N-1}$+1.
\\

For our quantum computer, we shall see that our quantum circuit can solve the problem in one step, $\textit{some}$ of the time. Specifically, when we go to measure all the qubits in our second system, there is one measurement result for which we can conclude that $f$ is one-to-one. But if we get any other measurement result, then our work is not quite finished, and we need to do some additional measurements and calculations in order to determine $s$.
\\

Like our previous quantum algorithms, we will embed our blackbox $f$ into a unitary operator $g$:

$$ g \hspace{.08cm} | \hspace{.08cm} X_i\rangle \hspace{.06cm} | \hspace{.08cm} \alpha \rangle \hspace{.2cm} \longrightarrow \hspace{.2cm} |\hspace{.08cm} X_i\rangle \hspace{.06cm} | \hspace{.08cm} \alpha \oplus f \hspace{.05cm} (\hspace{.04cm} X_i\hspace{.04cm}) \hspace{.06cm} \rangle$$

where $\alpha$ will be the state $|00...0\rangle$.
\\

Let's take a look at an example of a particular $f$, and its corresponding $g$ operation:

\pagebreak

$$ s = \{ 1,0 \} \hspace{1.2cm} $$

$$ f( \hspace{.1cm} 00 \hspace{.1cm} ) \hspace{.3cm} \rightarrow \hspace{.3cm} 11 \hspace{5cm} g \hspace{.08cm} | \hspace{.06cm} 00 \rangle \hspace{.05cm} | \hspace{.06cm} 00\rangle \hspace{.3cm} \rightarrow \hspace{.3cm} | \hspace{.06cm} 00 \rangle \hspace{.05cm} | \hspace{.06cm} 11\rangle $$
$$ f( \hspace{.1cm} 01 \hspace{.1cm} ) \hspace{.3cm} \rightarrow \hspace{.3cm} 10 \hspace{5cm} g \hspace{.08cm} | \hspace{.06cm} 01 \rangle \hspace{.05cm} | \hspace{.06cm} 00\rangle \hspace{.3cm} \rightarrow \hspace{.3cm} | \hspace{.06cm} 01 \rangle \hspace{.05cm} | \hspace{.06cm} 10\rangle $$
$$ f( \hspace{.1cm} 10 \hspace{.1cm} ) \hspace{.3cm} \rightarrow \hspace{.3cm} 11 \hspace{2cm} \longleftrightarrow \hspace{2cm} g \hspace{.08cm} | \hspace{.06cm} 10 \rangle \hspace{.05cm} | \hspace{.06cm} 00\rangle \hspace{.3cm} \rightarrow \hspace{.3cm} | \hspace{.06cm} 10 \rangle \hspace{.05cm} | \hspace{.06cm} 11\rangle $$
$$ f( \hspace{.1cm} 11 \hspace{.1cm} ) \hspace{.3cm} \rightarrow \hspace{.3cm} 10 \hspace{5cm} g \hspace{.08cm} | \hspace{.06cm} 11 \rangle \hspace{.05cm} | \hspace{.06cm} 00\rangle \hspace{.3cm} \rightarrow \hspace{.3cm} | \hspace{.06cm} 11 \rangle \hspace{.05cm} | \hspace{.06cm} 10\rangle $$

Compare the function $f$ on the left, to the effect of $g$ on the right. The classical version of our $f$ function takes in a string of bits, and outputs a string of bits of equal length. Note that this is different from the Deutsch-Jozsa and Bernstein-Vazirani algorithms we saw in lesson 5.2. The consequence of having an $f$ that outputs a string of bits is that we need to increase the size of our second system. Thus, if our $f$ is a function of an N-bit input, then we need N qubits for our second system.
\\

Just to illustrate this point, let's focus on one particular state:

$$ g \hspace{.12cm} | \hspace{.1cm} 01 \rangle \hspace{.08cm} | \hspace{.1cm} 00\rangle \hspace{.4cm} \longrightarrow \hspace{.4cm} | \hspace{.1cm} 01 \rangle \hspace{.08cm} | \hspace{.1cm}f(\hspace{.06cm} 01 \hspace{.06cm}) \hspace{.05cm} \rangle = | \hspace{.1cm} 01 \rangle \hspace{.08cm} | \hspace{.1cm} 10\rangle $$

Note that the effect of $f$ is not consistent among individual qubits. Only the string as a whole determines which states get mapped to where.
\\

Now, let's use this 2-qubit example to showcase the role of $s$ in this problem. We've already said that $s$ is a string of bits that correlates inputs, such that $\hspace{.1cm} x_1 \oplus x_2 = s$. Looking at which input states share the same outputs, we have: $\hspace{.2cm}$ $|00\rangle \leftrightarrow$ $|10\rangle \hspace{.1cm}$ and $\hspace{.1cm}$ $|01\rangle \leftrightarrow$ $|11\rangle \hspace{.06cm}$ as our correlated inputs. If we add these states together (modulo 2), we get:

$$ \{ 0,0 \} \hspace{.15cm} \oplus \hspace{.15cm} \{ 1,0 \} \hspace{.3cm} = \hspace{.3cm} \{ 1,0 \}$$
$$ \{ 0,1 \} \hspace{.15cm} \oplus \hspace{.15cm} \{ 1,1 \} \hspace{.3cm} = \hspace{.3cm} \{ 1,0 \}$$

which is indeed the $s$ for this particular $f$. This string of bits $s$ doesn't provide us any information about the outputs we will get, only the inputs that will share the same output. Which outputs result from correlated input pairs is still completely hidden within $f$, which means that if we want a complete picture for a given $f$, more work needs to be done.
\\

Turning now to some code, let's see an example of applying this $g$ operator to a 2-qubit system:

\pagebreak

\begin{figure}[h]
\centering
\includegraphics[scale=.65]{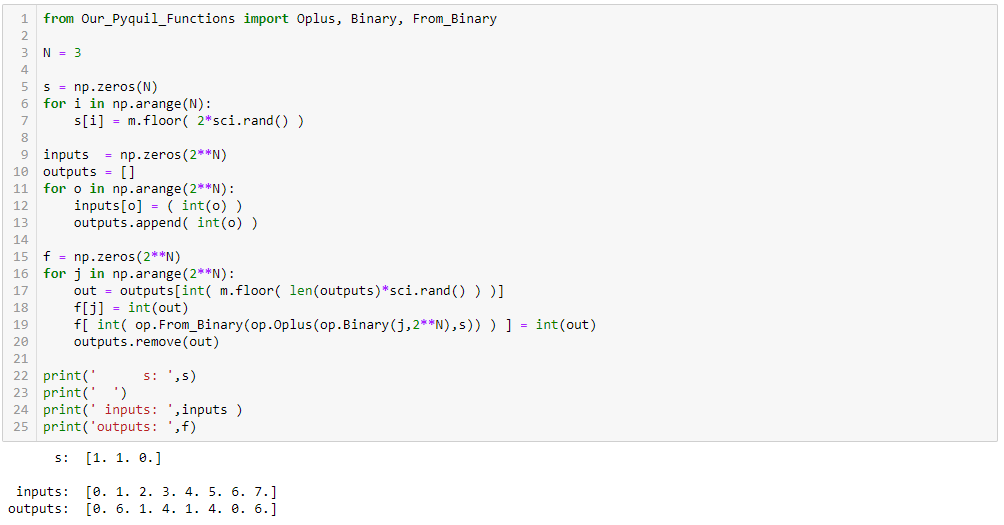}
\end{figure}

Running the cell of code above should produce the following initial state:

$$ \frac{1}{2} \big{(}\hspace{.1cm} |\hspace{.06cm} 00\rangle \hspace{.05cm}|\hspace{.06cm}00\rangle + |\hspace{.06cm} 01\rangle \hspace{.05cm}|\hspace{.06cm}00\rangle + |\hspace{.06cm} 10\rangle \hspace{.05cm}|\hspace{.06cm}00\rangle + |\hspace{.06cm} 11\rangle \hspace{.05cm}|\hspace{.06cm}00\rangle \hspace{.05cm} \big{)} $$

Then, we apply our $g$ matrix, which will result in some final state based on $f$:

$$ \frac{1}{2} \big{(}\hspace{.16cm} |\hspace{.06cm} 00\rangle \hspace{.06cm} | \hspace{.12cm} f\hspace{.04cm}(\hspace{.05cm}00\hspace{.05cm})\hspace{.05cm}\rangle \hspace{.15cm}+\hspace{.15cm} |\hspace{.06cm} 01\rangle \hspace{.06cm} | \hspace{.12cm} f\hspace{.04cm}(\hspace{.05cm}01\hspace{.05cm})\hspace{.05cm}\rangle \hspace{.15cm}+\hspace{.15cm} |\hspace{.06cm} 10\rangle \hspace{.06cm} | \hspace{.12cm} f\hspace{.04cm}(\hspace{.05cm}10\hspace{.05cm})\hspace{.05cm}\rangle \hspace{.15cm}+\hspace{.15cm} |\hspace{.06cm} 11\rangle \hspace{.06cm} | \hspace{.12cm} f\hspace{.04cm}(\hspace{.05cm}11\hspace{.05cm})\hspace{.05cm}\rangle \hspace{.11cm} \big{)} $$

In addition, the hidden key bit-string $s$ is also printed, just so we can verify that the operation is working as intended.
\\

Just like the algorithms in 5.2, we once again need additional ancilla qubits for this $g$ operation, stemming from the fact that there are higher order control-gates within $\textbf{Blackbox\_g\_S}$. For our goal of understanding Simon's Algorithm, we can ignore these extra ancilla qubits.
\\

I encourage you to run the cell of code above a couple times, to generate different $f$ functions. In the final wavefunction, you should see a result that is similar to our examples earlier, where all four of the initial states in qubits $0$ and $1$ are still present, and there are always two pairs of states for qubits $2$ and $3$ (unless you happen upon the case of $s = \{ 0,0 \}$). The exact final states on qubits $2$ and $3$ will be different each time, but you should always see two pairs.

Now that we have our unitary operator $g$, we can write out the full Simon's Algorithm:

$$ \textmf{prepare} \hspace{.15cm} | \hspace{.06cm} 0\rangle^N \otimes |\hspace{.06cm} 0\rangle^N \hspace{.2cm} \longrightarrow \hspace{.2cm} H^N \hspace{.08cm} | \hspace{.06cm} 0\rangle^N \otimes |\hspace{.06cm} 0\rangle^N \hspace{.2cm } \longrightarrow \hspace{.2cm} g \hspace{.2cm} \longrightarrow \hspace{.2cm} H^N \hspace{.08cm} | \hspace{.06cm} 0\rangle^N \otimes |\hspace{.1cm} f(x) \hspace{.08cm} \rangle^N \hspace{.2cm} \longrightarrow \hspace{.2cm} \textmf{measure}$$

1) Prepare both systems in the state of all 0's: $|00...0\rangle$
\\

2) Apply H$^N$ on system 1
\\

3) Apply the unitary operator $g$
\\

4) Apply H$^N$ on system 1
\\

5) Measure system 1
\\

Using our example code above, let's add in the second Hadamard Transformation and see what we get:

\pagebreak

\begin{figure}[h]
\centering
\includegraphics[scale=.65]{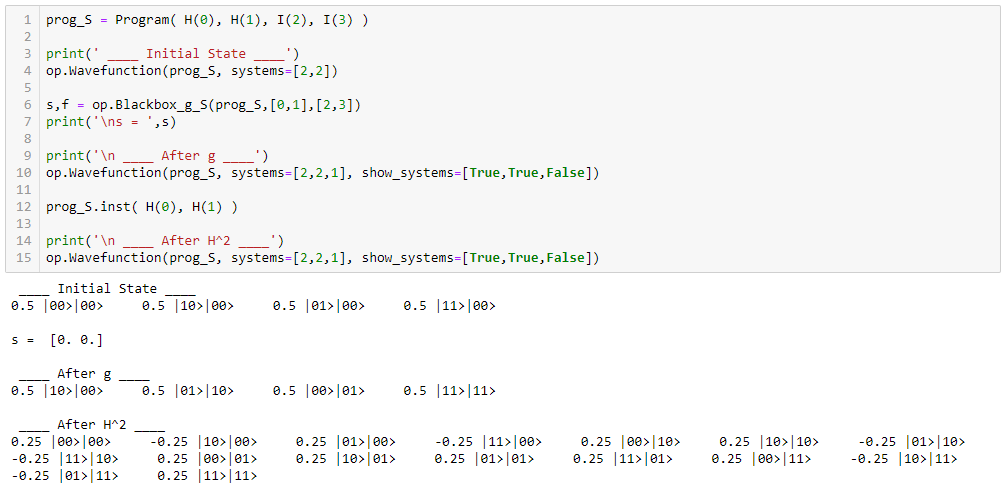}
\end{figure}

That's essentially the full quantum algorithm, so now what? The only thing missing is the final measurement, but seeing the final wavefunction is sufficient here. In the example above, for a non $ \hspace{.08cm} s = \{ 0,0 \} \hspace{.08cm}$ case, you should find that our main system collapses to two possible states, but neither of them necessarily tells us anything about $s$.
\\

Unlike the algorithms in lessons 5.1 and 5.2, Simon's Algorithm will require multiple runs. Essentially, each time we run the quantum algorithm, we will extract a new piece of information. Then, with enough pieces, and some luck, we will arrive at our solution. But before going any further into the solution, we need to revisit a property about Hadamard Transformations first. This is now the 3$^{rd}$ lesson in a row where we've used a Hadamard Transformation as the key ingredient of our algorithm (yeah, we're pros at it now!). In order to appreciate what is happening with Simon's Algorithm here, this time we will go over the effects of the Hadamard Transformation using some new math formalism.
\\

To start off, we know that $H^N$ maps the state $|00...0\rangle$ to an even superposition:

$$ H^2 \hspace{.06cm} | \hspace{.1cm} 00\rangle \hspace{.4cm} = \hspace{.4cm} \frac{1}{2} \big{(}\hspace{.12cm} |\hspace{.1cm} 00\rangle + |\hspace{.1cm} 01\rangle + |\hspace{.1cm} 10\rangle + |\hspace{.1cm} 11\rangle \hspace{.1cm} \big{)} $$

But when applied to any other state, $H^N$ will result in some of the states being negative:

$$ H^2 \hspace{.06cm} | \hspace{.1cm} 01\rangle \hspace{.4cm} = \hspace{.4cm} \frac{1}{2} \big{(}\hspace{.12cm} |\hspace{.1cm} 00\rangle - |\hspace{.1cm} 01\rangle + |\hspace{.1cm} 10\rangle - |\hspace{.1cm} 11\rangle \hspace{.1cm} \big{)} $$

The states which become negative are predictable, following a pattern based on which qubits are in the $|1\rangle$ state. We can determine the final state after an $H^N$ transformation as follows:

$$ H^N \hspace{.06cm} | \hspace{.1cm}x\rangle \hspace{.4cm} = \hspace{.4cm} \frac{1}{\sqrt{2^N}} \sum_{0}^{2^N - 1} (-1)^{x \cdot y} |\hspace{.08cm} y \rangle $$

where the states $|y\rangle$ are just all of the 2$^N$ possible states. For the case of $N$ = 2, the $|y\rangle$ states are just $|00\rangle$, $|01\rangle$, $|10\rangle$ and $|11\rangle$.
\\

The main reason for introducing this notation is because of the $(-1)^{x \cdot y}$ term, which is necessary for our explanation of Simon's Algorithm. Essentially, this term shows exactly how to pick out which states will be negative, based on the dot product $x \cdot y$. Let's show this using the example above:

\pagebreak

$$\hspace{7cm} x \hspace{.4cm} \cdot \hspace{.4cm} y \hspace{3.1cm}$$

$$ \hspace{8.55cm} \{ 0,1 \} \cdot \{ 0,0 \} = 0 \hspace{2.8cm} (-1)^{x \cdot y} = \hspace{.3cm} 1 $$
$$ H \hspace{.1cm} |\hspace{.08cm} 01\rangle \hspace{.2cm} = \hspace{.2cm} \ \frac{1}{2}\big{(} |00\rangle - |01\rangle + |10\rangle - |11\rangle \big{)} :\hspace{2cm} \{ 0,1 \} \cdot \{ 0,1 \} = 1 \hspace{1cm} \longrightarrow \hspace{1cm} (-1)^{x \cdot y} = -1 $$
$$ \hspace{8.55cm} \{ 0,1 \} \cdot \{ 1,0 \} = 0 \hspace{2.8cm} (-1)^{x \cdot y} = \hspace{.3cm} 1 $$

$$ \hspace{8.55cm} \{ 0,1 \} \cdot \{ 1,1 \} = 1 \hspace{2.8cm} (-1)^{x \cdot y} = -1 $$

As we can see in this example, the the dot product ($x \cdot y$) does indeed get all of the correct signs. In general, if this dot product yields an even number, then the state will be positive, and vice versa for an odd number. Note that this result also proves something we pointed out in lesson 5.2, that the effect of the Hadamard Transformation always leaves the state $|00...0\rangle$ positive (a result that led to the interference effects responsible for solving both the Deutsch-Jozsa and Bernstein-Vazirani problems).
\\

Now back to the problem at hand, and the reason why this notation will help us. The key to why we are able to solve Simon's Algorithm faster using a quantum computer comes from the fact that we are guaranteed a key bit-string $s$ in our $f$. We know that there are exactly two inputs that map to the same output, and all of them are correlated by $s$. For example:

$$|\hspace{.06cm} \psi \rangle \hspace{.2cm} = \hspace{.2cm} \frac{1}{2} \big{(}\hspace{.1cm} |\hspace{.06cm} 00\rangle \hspace{.08cm} | \hspace{.1cm}00\rangle \hspace{.15cm} + \hspace{.15cm} \hspace{.1cm} |\hspace{.06cm} 01\rangle \hspace{.08cm} | \hspace{.1cm}00\rangle \hspace{.15cm} + \hspace{.15cm} \hspace{.1cm} |\hspace{.06cm} 10\rangle \hspace{.08cm} | \hspace{.1cm}00\rangle \hspace{.15cm} + \hspace{.15cm} \hspace{.1cm} |\hspace{.06cm} 11\rangle \hspace{.08cm} | \hspace{.1cm}00\rangle \big{)} $$

$$ s \hspace{.2cm} = \hspace{.2cm} \{ 1,1 \} $$

$$ g \hspace{.1cm} | \hspace{.06cm} \psi \rangle \hspace{.2cm} = \hspace{.2cm} \frac{1}{2} \big{(}\hspace{.1cm} |\hspace{.06cm} 00\rangle \hspace{.08cm} | \hspace{.1cm}10\rangle \hspace{.15cm} + \hspace{.15cm} \hspace{.1cm} |\hspace{.06cm} 01\rangle \hspace{.08cm} | \hspace{.1cm}01\rangle \hspace{.15cm} + \hspace{.15cm} \hspace{.1cm} |\hspace{.06cm} 10\rangle \hspace{.08cm} | \hspace{.1cm}01\rangle \hspace{.15cm} + \hspace{.15cm} \hspace{.1cm} |\hspace{.06cm} 11\rangle \hspace{.08cm} | \hspace{.1cm}10\rangle \big{)} $$

Remember that after we apply $g$, we have no more interactions with the ancilla system, so we must understand how the mapping of these ancilla qubits affects our main system. To do this, let's rewrite the state above, grouping terms together that share the same ancilla state:

$$|\hspace{.06cm} \psi \rangle _g = \hspace{.2cm} \frac{1}{2} \Big{(}\hspace{.2cm} \big{(}\hspace{.15cm} |\hspace{.06cm} 00\rangle \hspace{.15cm} + \hspace{.15cm} |\hspace{.06cm} 11\rangle \hspace{.1cm} \big{)} \otimes |\hspace{.06cm} 01\rangle \hspace{.15cm} + \hspace{.15cm} \big{(} \hspace{.15cm} |\hspace{.06cm} 01\rangle \hspace{.15cm} + \hspace{.15cm} |\hspace{.06cm} 10\rangle \hspace{.1cm} \big{)} \otimes |\hspace{.06cm} 10\rangle \hspace{.15cm} \Big{)} $$

Thus, the effect of $g$ can be thought of as a 'regrouping' of our states. Before $g$, all of the states in our main system could interfere with each other. But after $g$, only states that are correlated by $s$ will interfere when we apply the next H$^2$ gate:

$$H^2 \hspace{.1cm}|\hspace{.06cm} \psi \rangle _g = \hspace{.2cm} \frac{1}{2} \Big{(}\hspace{.2cm} \big{(}\hspace{.15cm} H^2 \hspace{.1cm} |\hspace{.06cm} 00\rangle \hspace{.15cm} + \hspace{.15cm} H^2 \hspace{.1cm} |\hspace{.06cm} 11\rangle \hspace{.1cm} \big{)} \otimes |\hspace{.06cm} 01\rangle \hspace{.15cm} + \hspace{.15cm} \big{(} \hspace{.15cm} H^2 \hspace{.1cm} |\hspace{.06cm} 01\rangle \hspace{.15cm} + \hspace{.15cm} H^2 \hspace{.1cm} |\hspace{.06cm} 10\rangle \hspace{.1cm} \big{)} \otimes |\hspace{.06cm} 10\rangle \hspace{.15cm} \Big{)} $$

Now, using our new way of describing the effect of a Hadamard gate, let's see the interference that happens between the states $|00\rangle$ and $|11\rangle$:

$$ H^2 \hspace{.1cm} |\hspace{.06cm} 00\rangle \hspace{.15cm} + \hspace{.15cm} H^2 \hspace{.1cm} |\hspace{.06cm} 11\rangle \hspace{.2cm} = \hspace{.2cm} \sum_y \big{(} (-1)^{ 00 \cdot y } \hspace{.12cm} + \hspace{.12cm} (-1)^{ 11 \cdot y } \big{)} |\hspace{.08cm}y \rangle $$

For each $|y\rangle$, the $|00\rangle$ and $|11$ states are each going to contribute either a 1 or -1, which will lead to some states deconstructively interfering. More specifically, we can show that the states which will deconstructiviley interfere to 0 are related to $s$:

$$ (-1)^{ 00 \cdot y } \hspace{.12cm} + \hspace{.12cm} (-1)^{ 11 \cdot y } $$

$$ = \hspace{.3cm} (-1)^{ 00 \cdot y } \hspace{.12cm} + \hspace{.12cm} (-1)^{ (\hspace{.04cm} 00 \hspace{.05cm} \oplus \hspace{.05cm} s \hspace{.04cm}) \cdot y } $$

$$ \hspace{.8cm} = \hspace{.3cm} (-1)^{ 00 \cdot y } \hspace{.12cm} + \hspace{.12cm} \hspace{.12cm} (-1)^{ 00 \cdot y } \cdot \hspace{.12cm} (-1)^{ s \cdot y } $$

$$ = \hspace{.3cm} (-1)^{ 00 \cdot y } \cdot \big{(} 1 \hspace{.12cm} + \hspace{.12cm} \hspace{.12cm} (-1)^{ s \cdot y } \big{)} $$

which will result in all of the $|y\rangle$ to deconstructively interfere when $ s \cdot y = $ odd. Thus, we have just shown a correlation between the states that will go to 0, and $s$. Regardless of the exact mapping of $f$, the final states of our main system will always be determined by $s$.
\\

For completeness, we skipped the following steps above:

$$ (\hspace{.05cm} x_1 \oplus x_2 \hspace{.05cm}) \cdot y \hspace{.2cm} = \hspace{.2cm} (\hspace{.05cm} x_1 \oplus y \hspace{.05cm}) \cdot (\hspace{.05cm} x_2 \oplus y \hspace{.05cm}) $$

$$ (-1)^{(\hspace{.05cm} x_1 \hspace{.04cm} \oplus \hspace{.04cm} y \hspace{.05cm}) \hspace{.05cm} \cdot \hspace{.05cm} (\hspace{.05cm} x_2 \hspace{.04cm} \oplus \hspace{.04cm} y \hspace{.05cm})} \hspace{.2cm} = \hspace{.2cm} (-1)^{\hspace{.05cm} x_1 \hspace{.04cm} \oplus \hspace{.04cm} y \hspace{.05cm}} \cdot (-1)^{\hspace{.05cm} x_2 \hspace{.04cm} \oplus \hspace{.04cm} y \hspace{.05cm}} $$

which aren't too difficult to verify.
\\

But back to the main point, we now know that the final states of our main system will be entirely determined by $s$, the string of bits which correlates inputs. Which means, there is a direct link between our final measured state, and our unknown blackbox $f$.

\section*{\large{ Classical Solving }}

Now comes part 2 to Simon's Algorithm, the classical component. If you followed along all of the quantum steps up to this point, the hardest part is over. Once you understand $\textit{how}$ and $\textit{why}$ the quantum component of Simon's Algorithm works, the classical steps are much more straightforward.
\\

The final result of our analysis above is that the final state of our main system consists of only $\textit{half}$ of all possible states. Specifically, the states that survive the second Hadamard Transformation correspond to states where $s \cdot y =$ even (we showed that states where $s \cdot y =$ odd all deconstructively go away).
\\

SO THEN, knowing this, we can use our measurement results to try and figure out $s$. Specifically, each new measurement result we get gives us another piece of information about $s$. Once we get enough unique measurement results, we can combine them together in a set of linear equations. For example, let's use our 2-qubit example, and suppose we got the following measurement results:

$$ 1)\hspace{.12cm} | \hspace{.08cm} 00\rangle \hspace{1.4cm} 2)\hspace{.12cm} | \hspace{.08cm} 00\rangle \hspace{1.4cm} 3) \hspace{.12cm} | \hspace{.08cm} 11\rangle $$

which we can then combine into the set of equations:

$$ \hspace{.3cm} \{ 0,0 \} \cdot s = 0 \hspace{7cm} (\textmf{no information})$$
$$ \{ 1,1 \} \cdot s = 0 \hspace{1cm}(\textmf{modulo 2}) \hspace{2cm} \longleftrightarrow \hspace{2cm} s_0 \oplus s_1 = 0$$

On our first two trials we measure the state $|00\rangle$, which gives us no information about $s$ (repeat measurement results is a unavoidable problem of our quantum algorithm). But on the third trial, we find the state $|11\rangle$, which does give us some information. There are two possible solutions to the second equation above: $\hspace{.2cm}$ s=$\{$0,0$\}$ $\hspace{.1cm}$ or $\hspace{.1cm}$ s=$\{$1,1$\}$. The first of these two solutions is $\textit{always}$ a solution, and represents a special case (which we will cover next). The second solution is a valid candidate, and since we've already measured all possible states (because we showed that exactly half of all states survive after the second H$^2$), there's nothing more we can do with our quantum system.
\\

So then, to conclude if our candidate $s$' = $\{$1,1$\}$ is really our $s$, we test our $f$ classically:

$$\hspace{.15cm} f\hspace{.05cm}( \hspace{.1cm}00\hspace{.1cm} ) \hspace{.15cm} = \hspace{.15cm} f\hspace{.05cm}( \hspace{.1cm}s'\hspace{.1cm} ) \hspace{3cm} \therefore \hspace{.18cm} s = s' \hspace{1cm} $$
$$ f\hspace{.05cm}( \hspace{.1cm}00\hspace{.1cm} ) \hspace{.15cm} \neq \hspace{.15cm} f\hspace{.05cm}( \hspace{.1cm}s'\hspace{.1cm} ) \hspace{3cm} \therefore \hspace{.18cm} s = \{ \hspace{.08cm}0,\hspace{.08cm}0\hspace{.08cm} \} $$

By testing our classical $f$ for the cases $s'$ and $0^N$, we will arrive at one of two conclusions about $s$. Following the logic from our set of linear equations, we can narrow down $s$ to only two possibilities: $s'$ and $0^N$. By prompting $f$( $s$ ) and $f$( $0^N$ ), finding that they both give the same output will conclude that $s'$ is indeed our hidden key bit-string. Conversely, if they yield different outputs, then the only possibility for $s$ is the string of all $0$'s.
\\

To fully understand this conclusion requires that you understood all of the preceding quantum steps, so it may take a few reads before it full sinks in. Essentially, because of the way we arrive at the two candidates above, we are $\textit{guaranteed}$ that they are the $\textit{only}$ possibilities.
\\

Since we are ultimately turning our quantum results over to a set of linear equations for solving, it is worth noting how many equations we need to solve for $s$. Our final quantum system will be an even distribution of $2^{N-1}$ states, which gives us $2^{N-1}$ equations. However, $\hspace{.08cm}\{ 0, 0, 0, ...\} \hspace{.08cm}$ may emerge from one of our measurements, and several others may not be linearly independent. Since $f$ is a function of $N$ bits, we need at most $N$ linearly independent equations for a solution (but as we shall see, we can often get away with fewer).
\\

Thus, there's no guarantee on how long it takes our algorithm to arrive at a set of linearly independent equations in order to solve for $s$. While this isn't ideal, it's important to understand that not all quantum algorithms are deterministic. The algorithms in 5.1 and 5.2 solve their respective problems with 100\% success rates, but they are the exception to the rule. Almost all quantum algorithms that we will study from this point on will come with some inherent probabilities of failure, where 'failures' typically mean we have to run the quantum algorithm again.
\\

For Simon's Algorithm, we are ultimately reliant on the final measurements. First, we are probabilistically halted until we measure up to $N$ independent states, and then hope that they are enough to solve for $s$. As $N$ gets bigger however, our probability of getting $N$ out of $2^{N-1}$ possible states goes up. But once we have enough unique measurements, the solving of the linear equations is easy, since we can just let a classical algorithm handle that (much later we shall see that there are even quantum algorithms to handle this portion too!).
\\

Ideally, we would want to perform measurements simultaneously while trying to solve our set of linear equations. This way we perform the fewest number of quantum runs. We will do this later, but for learning purposes now, we will 'overshoot' and run the quantum component of our algorithm more times than needed in order to make sure we get enough unique measurements. After that, we let our custom function $\textbf{Simons\_Solver}$ classically work through all the possible $s'$ candidates, and return back an answer:

\pagebreak

\begin{figure}[h]
\centering
\includegraphics[scale=.65]{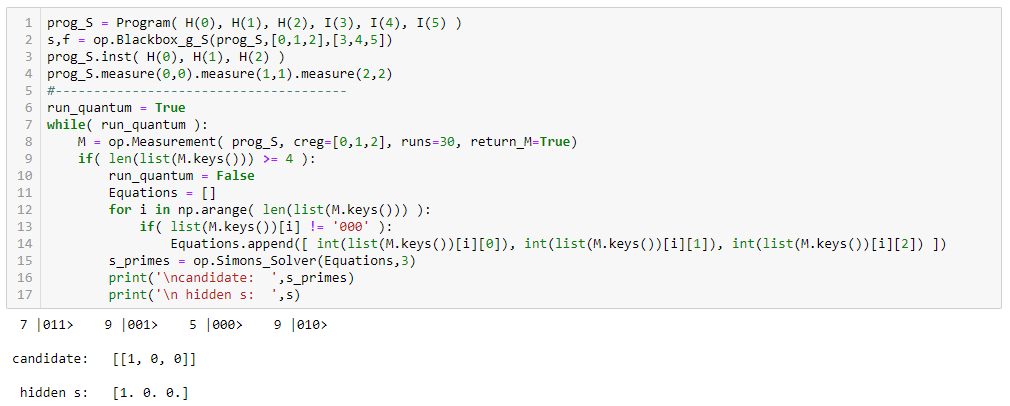}
\end{figure}

Run this example a couple of times and see that our algorithm works! Using our measurement results, the $\textmf{Simons\_Solver}$ function takes all of the linear equations obtained through measurements and returns to us a list of possible candidates for $s$. If we don't provide enough equations, it will return a list of multiple candidates. If our system of equations is sufficient, the function will return a list with a single candidate ($s$'). If we happen to get an $s$ of all $0$'s, it will return a lost of either one or no entries (which is why the final step is to always check $f\hspace{.05cm}( \hspace{.1cm}0^N\hspace{.1cm} ) \hspace{.15cm} = \hspace{.15cm} f\hspace{.05cm}( \hspace{.1cm}s'\hspace{.1cm} )$ ) .
\\

As we can see in the 'Measurement Results' line above, we run our quantum system far more times than needed. If we want to optimize the process, we can run the $\textmf{Simons\_Solver}$ after every unique measurement result, until it returns only a single value:

\begin{figure}[h]
\centering
\includegraphics[scale=.65]{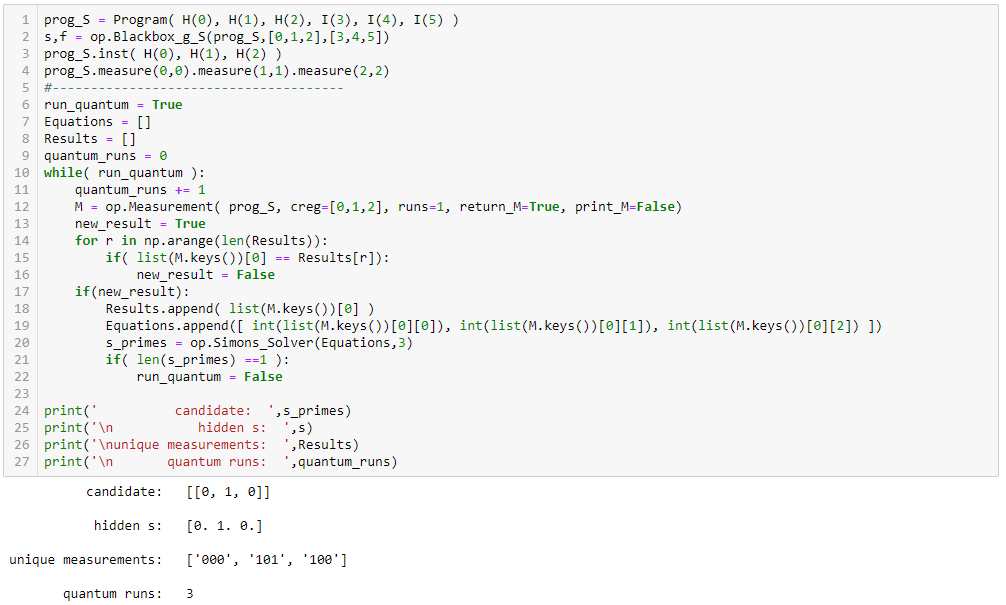}
\end{figure}

Compare the 'unique measurements' line in this example, to the 'Measurement Results' from the previous one. For a 3-qubit system, often times only two unique measurements is sufficient to solve for $s$. By trying to solve for $s$ after each time we get a unique measurement, we ensure that we don't run our quantum system more times than needed. BUT, keep in mind that every time $\textmf{Simons\_Solver}$ doesn't return a single value, we've essentially 'wasted' some computational time (although an argument can be made that these quantum and classical computers work in parallel, and do not bottleneck each other). Thus, for our code example, there's a balance between the number of unique measurements we should acquire before we start solving the linear equations.
\\

This is just one example of how quantum algorithms are not straightforward speedups. Because probability is always involved, we find situations where the exact number of 'steps' isn't constant. For Simon's Algorithm, the 'speed' at which we arrive at our answer is largely determined by how lucky we get at finding unique solutions. At lower problem sizes, this probability actually causes our quantum algorithm to be slower (a very common feature as we shall see). Thus, we need to implement Simon's Algorithm on larger problems if we really want to see a speedup.
\\

To conclude this tutorial, the cell of code below incorporates all of the steps outlined thus far into the function $\textbf{Simons}$. Change $Q$ to be the number of qubits you would like as the main system:

\begin{figure}[h]
\centering
\includegraphics[scale=.65]{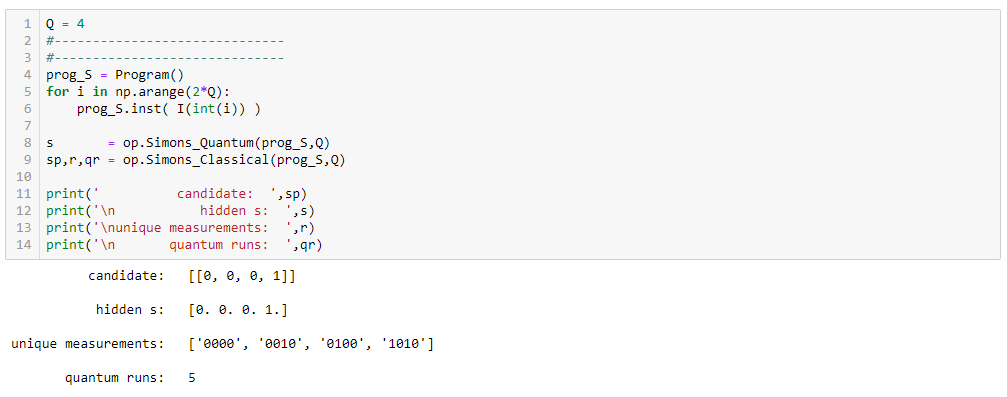}
\end{figure}

--------------------------------------------------------------------------------------------------------------------------------------------------------
\\

This concludes lesson 5.3! The algorithm studied in this lesson is our first taste of what is commonly referred to as a 'hybrid' algorithm, whereby the final solution is reached via a mix of quantum / classical computing. Simon's Algorithm is an excellent introduction to these kinds of algorithms because the final result is still relatively deterministic. That is to say, given enough measurements on the system, we will eventually have enough linear equations to solve for $s$.
\\

--------------------------------------------------------------------------------------------------------------------------------------------------------


\pagebreak

\section*{\Large{ Lesson 5.4 - The Grover Search }}
--------------------------------------------------------------------------------------------------------------------------------------------------------
\\

In this final lesson 5 tutorial, we will cover the Grover Algorithm. Like the previous algorithms we've studied, at the heart of the Grover Algorithm is a Hadamard Transformation. However, this algorithm does not solve a 'blackbox' problem, making it different from our previous three lessons. Instead, we will be solving a searching problem, whereby we would like to locate one particular state with a measurement, out of $2^N$.
\\

Original publication of the algorithm: \cite{G}
\\

--------------------------------------------------------------------------------------------------------------------------------------------------------
\\
In order to make sure that all cells of code run properly throughout this lesson, please run the following cell of code below:

\begin{figure}[h]
\centering
\includegraphics[scale=.65]{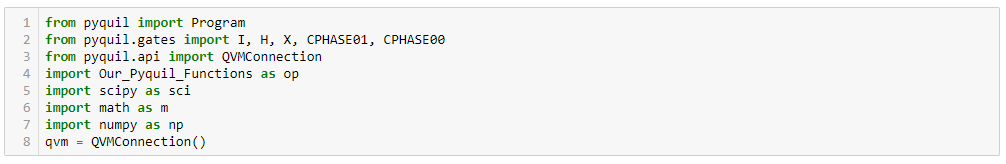}
\end{figure}

\section*{\large{ The Grover Algorithm }}
\centerline{---------------------------------------------------------------------------------------------------------------------------------}

The Grover Algorithm, also referred to as a Grover Search, is a quantum algorithm that can be thought of as searching through an unordered list. Imagine you want to look someone up in a directory, which is alphabetically ordered by last name, but you only have their first name. In this scenario, you are stuck going through each entry one at a time, until you eventually happen upon the person you are looking for.
\\

Exhaustively searching through the database represents the classical approach, which requires on average $\frac{N}{2}$ evaluations, which is of the order $\textit{O}$($N$). By instead using the Grover Algorithm, we can complete this search (with a high success probability) using only $\textit{O}$($\sqrt{N}$) evaluations.

\section*{\large{ Setting Up the Problem }}

Our goal is to create a quantum algorithm that will allow us to pick any state we want (within the $2^{N}$ space), and then attempt to find that state with a single measurement. As we shall see, we will measure our desired state, which we shall refer to as our 'marked state', with a high success probability. In addition, larger systems will result in higher success probabilities, a nice feature that is unique to the quantum approach!
\\

Like the classical search, our quantum algorithm needs to first reflect the problem of having no $\textit{a priori}$ knowledge of where the marked entry is located. For our quantum algorithm, we can represent this by starting our system in an equal superposition of all states. Thus, the starting point for our code will be to specify the size of our problem, and then create an equal superposition:

\begin{figure}[h]
\centering
\includegraphics[scale=.65]{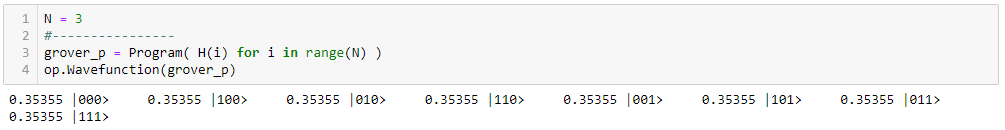}
\end{figure}

In the code above, we specify the size of our problem with the parameter $N$, creating a quantum system of the size 2$^N$. $N$ is the number of qubits we will be using, which means we can create significantly large systems with only a minimal amount of qubits. We prepare our system in an equal superposition of all 2$^N$ states by applying a Hadamard gate to each qubit, creating the following initial state:

$$H^{\otimes N} | \hspace{.05cm} 000...0 \rangle \hspace{.1cm} = \hspace{.1cm} \frac{1}{\sqrt{2^{N}}} \sum_{k=0}^{2^{N}-1} |k \rangle \hspace{.1cm} \equiv \hspace{.1cm} | \hspace{.06cm} s \hspace{.04cm} \rangle $$

Now, let's do some simulated measurements on this state. These measurements represent the classical approach of picking blindly until we happen on our desired state:

\begin{figure}[h]
\centering
\includegraphics[scale=.65]{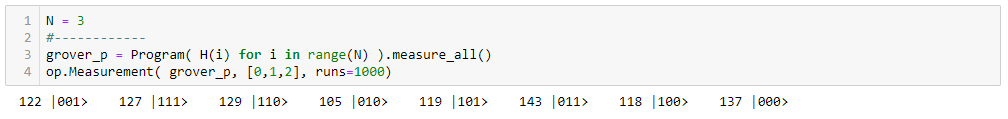}
\end{figure}

Take a look at the measurement counts for each state and verify that all states in the system are equally probable (although it's rare to get a perfectly even distribution). Using a quantum system like this to find a specific state is quite slow, and in fact it's even worse than the classical analog! Consider what the typical method would be if we were to pick states at random classically: suppose we are looking for the state $|000\rangle$, but instead got $|110\rangle$. It would be crazy to put $|110\rangle$ back into the mix and try again. Thus, we would naturally remove is from the problem, thereby improving our odds of finding $|000\rangle$ on the next try.
\\

The main advantage to a classical search is the ability to 'remember' past measurements, and remove them from the problem. By doing so, the classical approach will slowly narrow down the pool of possible entries, until eventually finding the desired one. When using a quantum approach, we can't do this. If we measure the state $|110\rangle$, that's it. Our wavefunction collapses to that state, and we've failed our search. And, when we go to prepare the system the next time, we have no way of removing the state $|110\rangle$ from the system, which means we could get it again!
\\

The difference between the classical and quantum approaches to a search problem are very noteworthy. Since our quantum system has no memory of past measurements, we can only hope to find our desired state with a single attempt. Thus, the goal of the Grover Algorithm will be to boost our chance of measuring the desired state.

\section*{\large{ Implementing an Oracle }}
\centerline{---------------------------------------------------------------------------------------------------------------------------------}

Now that we have our equal superposition of $2^{N}$ states, we can begin to construct our Grover Algorithm.
\\

To do this, the first thing we need is an operator $U_w$, known as an 'oracle.' Simply put, this is an operator that picks out a single state in the system, say $|0101\rangle$, and applies an operation. Specifically, this oracle operator $U_w$ isolates a single state such that it is the $\textit{only}$ state in the system that will then receive the desired operation.
\\

We've worked with similar operators in the past, such as the control gates from lesson 3, and the higher order $\textmf{n\_NOT}$ gates in lesson 4. In essence, that's exactly what we're going to do here as well. By default, control gates only pick out states where all of the control qubits are in the state $|1\rangle$, for example:

$$CCNOT \hspace{.2cm} |\hspace{.06cm}100\rangle \hspace{.16cm}\rightarrow\hspace{.16cm} |\hspace{.06cm}100\rangle $$
$$CCNOT \hspace{.2cm} |\hspace{.06cm}110\rangle \hspace{.16cm}\rightarrow\hspace{.16cm} |\hspace{.06cm}111\rangle $$

For our Grover algorithm, we need our oracle to be able to pick out $\textit{any}$ state, including states with $0$'s on any qubit. Luckily, we've already seen how to pull off this trick before in our past blackbox functions.
\\

In order to make sure that only our marked state is the control-state, we will perform a series of X gates to $\textit{transform}$ our marked state to the state of all $1$'s:$\hspace{.14cm}$ $|111...1\rangle$. Simultaneously, this transformation will also guarantee that our marked state is the $\textit{only}$ state in the system of all $1$'s. Thus, when we apply our N-qubit control gate operation, its effect will $\textit{only}$ get applied to our marked state. Then, we will transform all of the states back to the original basis, using the same X gates:

\pagebreak

\begin{figure}[h]
\centering
\includegraphics[scale=.65]{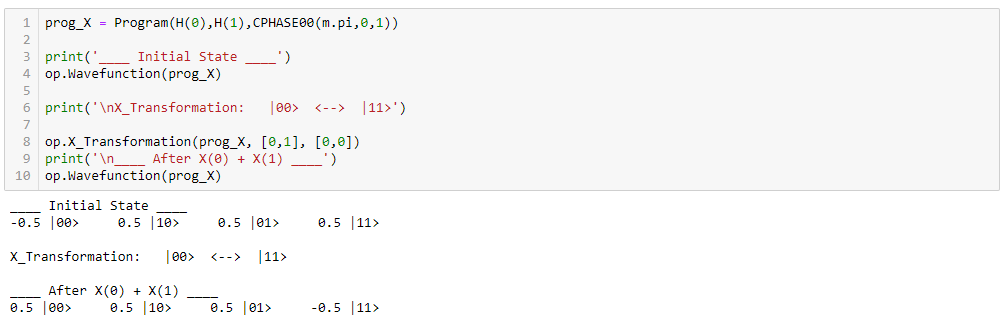}
\end{figure}

In the example above, we transform the state $|00\rangle \rightarrow$ $|11\rangle$ by applying X gates on qubits $0$ and $1$. We mark the $|11\rangle$ state with a negative phase just for clarity here, so we can track which state it gets transformed to (the state that ends up with the negative sign is the original state that maps to $|11\rangle$).
\\

In this example, we use our custom function $\textbf{X\_Transformation}$ to perform the correct X gates, specifying the desired state we want to map to the state of all $1$'s. In general, choosing which X gates to perform is very straightforward, as all we need to do it look at where the $0$'s are for our marked state. In the example above, our marked state would be $|00\rangle$, which has $0$'s in the qubit locations $0$ and $1$, $\textit{therefore}$ we applied the gates $X(0)$ and $X(1)$.
\\

Equally as important as the transformation of the marked state, is the effect of this transformation on the rest of the system. For example, consider the effect of the X Transformation when $|001\rangle$ is the marked state:

$$ \textmf{Applying X(0) + X(1)} $$

$$ |\hspace{.06cm}000\rangle \hspace{.2cm} \rightarrow \hspace{.2cm} |\hspace{.06cm}110\rangle \hspace{1.5cm} |\hspace{.06cm}100\rangle \rightarrow |\hspace{.06cm}010\rangle$$
$$ |\hspace{.06cm}001\rangle \hspace{.2cm} \rightarrow \hspace{.2cm} |\hspace{.06cm}111\rangle \star \hspace{1.15cm} |\hspace{.06cm}101\rangle \rightarrow |\hspace{.06cm}011\rangle$$
$$ |\hspace{.06cm}010\rangle \hspace{.2cm} \rightarrow \hspace{.2cm} |\hspace{.06cm}100\rangle \hspace{1.5cm} |\hspace{.06cm}110\rangle \rightarrow |\hspace{.06cm}000\rangle$$
$$ |\hspace{.06cm}011\rangle \hspace{.2cm} \rightarrow \hspace{.2cm} |\hspace{.06cm}101\rangle \hspace{1.5cm} |\hspace{.06cm}111\rangle \rightarrow |\hspace{.06cm}001\rangle$$

No other state in the system gets mapped to $|111\rangle$, exactly the result we need. If we consider that our marked state is 'unique', in that no other state in the system has the same $0$'s and $1$'s, it makes sense that the transformation to $|111\rangle$ is unique as well, mapping all other states elsewhere.
\\

The last step is the transformation back to our original basis. For our Grover Algorithm, transforming our marked state to $|11...1\rangle$ will allow us to apply a higher order control operation, but afterwards, we must transform back in order to search for the marked state in its original form. Lucky for us, the transformation back to our original basis is just as easy. All we need to do is apply the exact same X gates again:

\pagebreak

\begin{figure}[h]
\centering
\includegraphics[scale=.65]{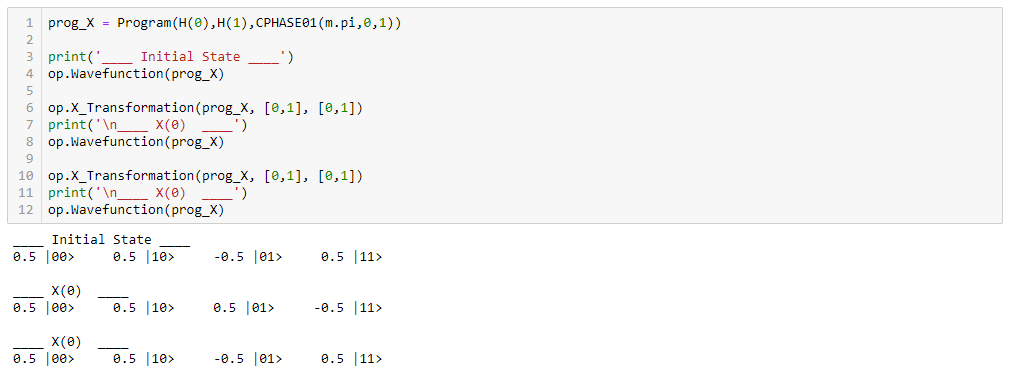}
\end{figure}

In the example above, we successfully transform back and forth between our marked state and $|11...1\rangle$. Next, we are going to use this transformation to effectively apply a higher order control-Z gate to our marked state.

\section*{\large{ Sign Flip on $|$111...1$\rangle$ (The Oracle Function) }}

The first component to our Grover Algorithm, the oracle $U_w$, will achieve the effect of an N-control-Z gate, applied to our marked state. That is to say, it will achieve the effect: $\hspace{.1cm}$ $|11...1\rangle \hspace{.1cm} \rightarrow \hspace{.1cm} -$$|11...1\rangle$. In matrix form, this operator looks like:

$$
\begin{bmatrix}
1 & 0 & 0 & . & . & . & & \\
0 & 1 & 0 & & & & & \\
0 & 0 & 1 & & & & & \\
. & & & . & & & & \\
. & & & & . & & & \\
. & & & & & . & & \\
& & & & & & & & \\
& & & & & & & 1 & 0 \\
& & & & & & & 0 & -1 \\
\end{bmatrix}
$$

There are a couple ways to achieve this operation, but we are going to use the most common method, which involves an ancilla qubit in the state $|-x\rangle$. In fact, we've already seen this trick in lessons 5.1 and 5.2.
\\

Essentially, we will be taking advantage of the effect of an X gate on the state $|-x\rangle$:

$$ X \hspace{.08cm} |-x\rangle \hspace{.12cm}=\hspace{.12cm} -|-x\rangle $$

Since every state in the system will be coupled to this $|-x\rangle$ state, we must be sure that $\textit{only}$ our marked state receives the X gate operation on the ancilla qubit. For example, suppose $|01\rangle$ was our marked state:

$$ |\hspace{.06cm}00\rangle \hspace{.05cm} |-x\rangle \hspace{3cm} |\hspace{.06cm}00\rangle \hspace{.05cm} |-x\rangle \hspace{2cm} |\hspace{.06cm}00\rangle \hspace{.05cm} |-x\rangle $$
$$ |\hspace{.06cm}01\rangle \hspace{.05cm} |-x\rangle \hspace{3cm} |\hspace{.06cm}01\rangle \hspace{.08cm}X\hspace{.05cm} |-x\rangle \hspace{1.1cm} -|\hspace{.06cm}01\rangle \hspace{.05cm} |-x\rangle $$
$$ |\hspace{.06cm}10\rangle \hspace{.05cm} |-x\rangle \hspace{1.1cm} \longrightarrow \hspace{1.1cm} |\hspace{.06cm}10\rangle \hspace{.05cm} |-x\rangle \hspace{.76cm} = \hspace{.76cm} |\hspace{.06cm}10\rangle \hspace{.05cm} |-x\rangle $$
$$ |\hspace{.06cm}11\rangle \hspace{.05cm} |-x\rangle \hspace{3cm} |\hspace{.06cm}11\rangle \hspace{.05cm} |-x\rangle \hspace{2cm} |\hspace{.06cm}11\rangle \hspace{.05cm} |-x\rangle $$

This example above shows the desired effect of our Oracle, essentially causing our marked state to pick up a negative phase. Then, after the negative sign has been applied, we work with our main system only, completing ignoring the ancilla.
\\

To achieve the effect shown above, we will need the combination of our $\textmf{X\_Transformation}$ function with $\textbf{n\_NOT}$, also one of our custom functions. In short, the $\textmf{n\_NOT}$ function is equivalent to any higher order CNOT gate of our choosing. Thus, we will use it to perform an $N^{th}$ order CNOT operation ($N$ being the number of qubits in our system), with the target qubit being the ancilla.
\\

For a refresher on exactly how our $\textmf{n\_NOT}$ operation achieves a higher order CNOT gate, please refer to lesson 4.
\\

In total, the flow of our Oracle function will be as follows:

$$ |\hspace{.06cm} \Psi \hspace{.02cm} \rangle_i \otimes |-x\rangle \hspace{.25cm} \rightarrow \hspace{.25cm} \textmf{X\_Transformation} \hspace{.25cm} \rightarrow \hspace{.25cm} \textmf{n\_NOT} \hspace{.25cm} \rightarrow \hspace{.25cm} \textmf{X\_Transformation} \hspace{.25cm} \rightarrow \hspace{.25cm} |\hspace{.06cm} \Psi \hspace{.02cm} \rangle_f \otimes |-x\rangle $$

Let's see it in code:

\begin{figure}[h]
\centering
\includegraphics[scale=.65]{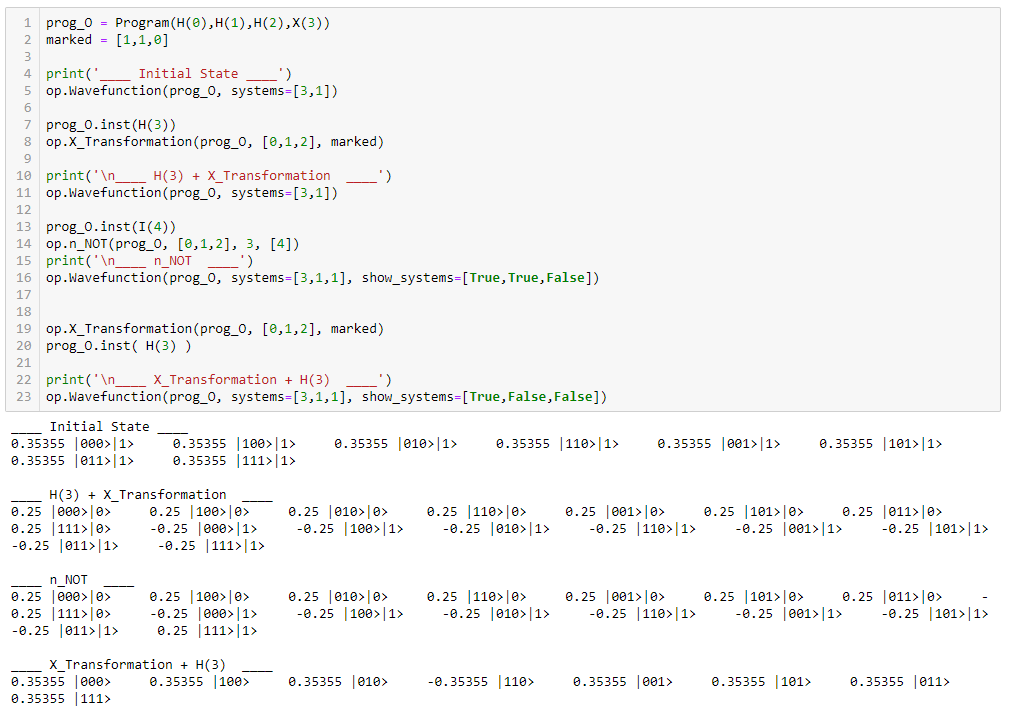}
\end{figure}

The example above follows all of the steps for our Oracle operator. Note that in these steps, the Hadamard gates on the ancilla qubit are separated out to better show the negative sign being applied to the marked state. Feel free to change the array 'marked' in this example, and see that it will always pick out the correct state. Also note that calling upon the $\textmf{n\_NOT}$ function requires the use $N-2$ extra qubits, which we've chosen not to display in our last two wavefunctions.
\\

To avoid clutter, we combine all of the operation steps above into a function called $\textbf{Grover\_Oracle}$:

\pagebreak

\begin{figure}[h]
\centering
\includegraphics[scale=.65]{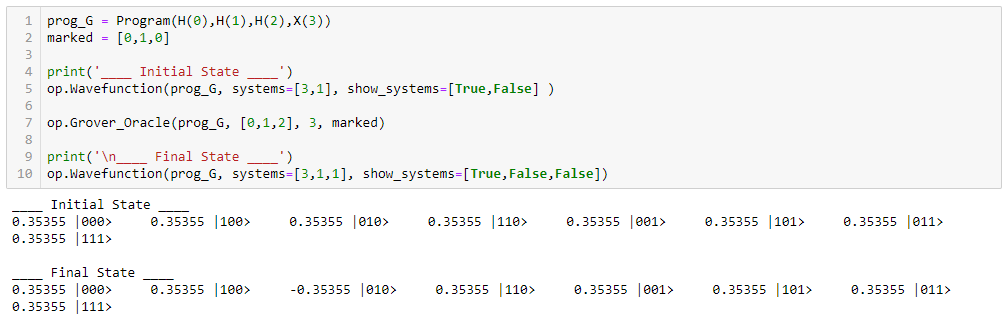}
\end{figure}

In this example we can see that $\textmf{Grover\_Oracle}$ takes care of all the instructions for us, so long as we prepare the system in the correct initial state:

$$|\hspace{.06cm}00...0\hspace{.03cm}\rangle\hspace{.05cm}|\hspace{.06cm}1\rangle$$

With the Oracle operator $U_w$ now in hand, we're ready to move on to the second part of Grover's Algorithm, which will require us to revisit the Hadamard transformation one final time.

\section*{\large{ Reflection About the Average }}

Like Simon's Algorithm from lesson 5.3, Grover's Algorithm will require multiple runs of our quantum system. The difference here, is that we will not be making measurements after each run. Instead, we will perform multiple 'Grover Iterations', followed by a single measurement at the very end.
\\

In one sentence, we can say that mathematically: "One Grover Iteration is equivalent to a reflection about the average amplitude." (Don't worry, we will make sense of this.) Let's start by talking about a reflection. Geometrically, a reflection involves two components: the object who is being reflected, and the point, line, plane, etc. with which we reflect about. For example, consider the diagram below, which illustrates a reflection of state $|\psi \rangle$ about the state $|0\rangle$:

\begin{figure}[h]
\centering
\includegraphics[scale=.8]{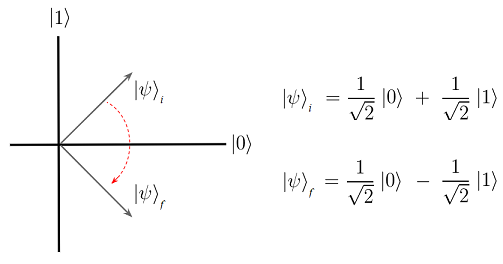}
\end{figure}

In this example, the object being reflected is the state $|\hspace{.04cm}\psi \hspace{.02cm} \rangle$, and the point of reflection is the state $|0\rangle$. We can see that a 'reflection about $|0\rangle$' is equivalent to a sign flip on the $|1\rangle$ state. And in general, a reflection about a single state $|\phi \rangle$ leaves a quantum state's $|\phi \rangle$ component unchanged, while flipping the sign on all other components:

$$ |\hspace{.06cm}\psi \hspace{.02cm} \rangle_i = \frac{1}{2}\big{(} \hspace{.08cm} |\hspace{.06cm}00\rangle + |\hspace{.06cm}01\rangle + |\hspace{.06cm}10\rangle + |\hspace{.06cm}11\rangle \hspace{.08cm} \big{)} \hspace{.8cm} \longrightarrow \hspace{.8cm} \frac{1}{2}\big{(} \hspace{.08cm} |\hspace{.06cm}00\rangle - |\hspace{.06cm}01\rangle - |\hspace{.06cm}10\rangle - |\hspace{.06cm}11\rangle \hspace{.08cm} \big{)}$$

However, Grover's Algorithm will require use to perform reflections about a state for arbitrarily large systems, which translates to implementing many of these sign flips. Needless to say, sign flipping every single state besides just one is a bit tedious, and quite costly in terms of gates. Luckily for us, we can achieve the same net effect by taking the reverse route: only flipping the sign on the single state. Consider our first example again, only this time we will flip the sign on the |0$\rangle$ component:

\begin{figure}[h]
\centering
\includegraphics[scale=.8]{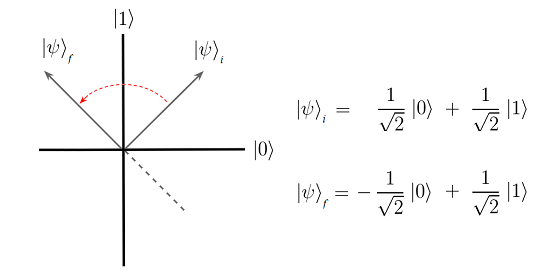}
\end{figure}

In the diagram above, notice how both final states 'align', shown by the dashed line marking where $|\hspace{.06cm}\psi \hspace{.02cm}\rangle _f$ from the first example was. Denoting the final states from the two examples as $|\hspace{.06cm}\psi \hspace{.02cm} \rangle _{1f}$ and $|\hspace{.06cm}\psi \hspace{.02cm} \rangle _{2f}$, we have that $|\hspace{.06cm}\psi \hspace{.02cm} \rangle _{1f} = -$ $|\hspace{.06cm}\psi \hspace{.02cm}\rangle _{2f}$. Or more specifically, the two states are parallel, with opposite phase.
\\

The nice thing about this for us, is that a measurement on the system can't tell the difference between $|\psi \rangle _{1f}$ and $|\psi \rangle _{2f}$. Thus, so long as the opposite phase isn't an issue anywhere else in our algorithm, we are free to use either reflection method as we see fit. And for our Grover Algorithm, we are definitely going to use the second method in order to minimize steps.
\\

Now, let's discuss what it means to 'reflect about the average amplitude'. Perhaps the easiest way to understand this initially, is with a diagram:

\begin{figure}[h]
\centering
\includegraphics[scale=.6]{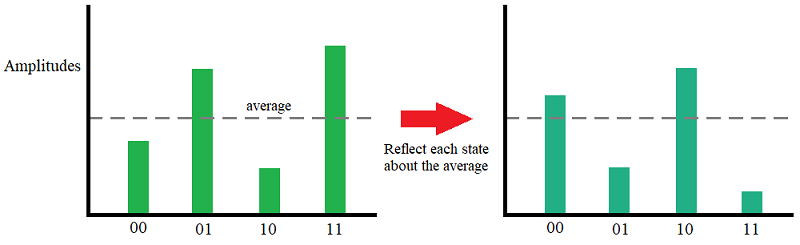}
\end{figure}

This diagram illustrates the effect we are going for: we take the average of all the states' amplitudes, and reflect each state's individual amplitudes about that average. We can see that states with amplitudes above the average get reflected below it, and vice versa. In total, the average amplitude for the system is unchanged, even though all of the states have. Mathematically, this is then a unitary operation:

$$ \alpha_i \equiv \textmf{amplitude of each state} $$

$$ \alpha_{avg} = \frac{\sum_i^N \alpha_i }{N} $$

$$ \sum_i^N \alpha_i^2 = 1 \hspace{.4cm} and \hspace{.4cm} \sum_i^N (\alpha_{avg} + (\alpha_{avg} - \alpha_i))^2 = 1 $$

We won't go through this proof here, but rather provide a simple arithmetic example (which is by no means a proof):

$$ 2^2 + 3^2 + 5^2 + 9^2 = 119$$

$$ \textmf{reflect around the average: 4.75} $$

$$ 7.5^2 + 6.5^2 + 4.5^2 + 0.5^2 = 119 $$

\section*{\large{ U$_s$ - Grover Diffusion Operator }}
\centerline{---------------------------------------------------------------------------------------------------------------------------------}

The operator that is going to achieve this reflection about the average will be $U_s$, often referred to as the Grover Diffusion Operator. We will start by writing out the effect of the operation we want:

$$U \hspace{.06cm}| \hspace{.06cm}\psi \hspace{.02cm} \rangle \hspace{.2cm}=\hspace{.2cm} |\hspace{.06cm} \psi \hspace{.02cm} \rangle - 2\hspace{.03cm}\big{(}\hspace{.08cm}| \hspace{.06cm}\psi \hspace{.02cm} \rangle - |\hspace{.06cm}r \hspace{.02cm} \rangle \hspace{.06cm} \big{)}. $$

where $|\psi \rangle$ is the state of our system, and $|r \rangle$ represents an equal superposition of all states, where each state has an amplitude of $\alpha_{avg}$:

$$ |\hspace{.06cm}r \hspace{.02cm} \rangle \hspace{.2cm} = \hspace{.2cm} \alpha_{avg} \sum_i^N |\hspace{.05cm}i\hspace{.02cm}\rangle $$

This state $|r\rangle$ is most definitely $\textit{not}$ normalized, meaning we can't physically create it, but represents what we want to happen as a result from our operation. Specifically, the operation $|\hspace{.06cm} \psi \hspace{.02cm} \rangle - 2\hspace{.03cm}\big{(}\hspace{.08cm}| \hspace{.06cm}\psi \hspace{.02cm} \rangle - |\hspace{.06cm}r \hspace{.02cm} \rangle \hspace{.06cm} \big{)}$ is written this way in order to understand its two components:
\\

1) take the difference in amplitudes between each state and the average: $\hspace{.15cm}\big{(}\hspace{.08cm}| \hspace{.06cm}\psi \hspace{.02cm} \rangle - |\hspace{.06cm}r \hspace{.02cm} \rangle \hspace{.06cm} \big{)} \hspace{.15cm}$
\\

2) double each of these differences, and subtract them from the initial amplitudes: $|\hspace{.04cm} \psi \hspace{.02cm} \rangle - 2 \big{(} \cdot \cdot \cdot \big{)}$.
\\

For example, suppose we have a system where the amplitude for the state $|01\rangle$ is $\alpha_{01}$ = 0.7, and the average amplitude for the whole system is $\alpha_{avg}$ = 0.45. We want the effect of our operation to do the following:

$$ U_s \hspace{.05cm} | \hspace{.06cm} 01\rangle \hspace{.15cm} \rightarrow \hspace{.15cm} \big{(} 0.7 - 2 \hspace{.04cm} ( \hspace{.05cm}.07 - .45 \hspace{.05cm} ) \hspace{.04cm}\big{)} \hspace{.05cm} | \hspace{.06cm} 01\rangle \hspace{.15cm} = \hspace{.15cm} 0.2 \hspace{.04cm} | \hspace{.06cm} 01\rangle $$

Hopefully this example illustrates what we are going for. We want an operation that uses the difference between each state and the average ( $\alpha_i - \alpha_{avg}$ ), and subtracts double this amount from the initial amplitude. If $\hspace{.06cm} \alpha_i - \alpha_{avg} \hspace{.06cm}$ is positive, then the final amplitude will be smaller (like state $11$ in the diagram above), possibly even negative. Conversely, if $\hspace{.06cm} \alpha_i - \alpha_{avg} \hspace{.06cm}$ is negative, then the final amplitude will be larger (like state $00$ in the diagram), which we shall see happens to our marked state.
\\

Now then, let's see how we can construct this $U_s$ operator. First off, let's do a little rewriting:

$$ \hspace{.04cm} U_s \hspace{.05cm} | \hspace{.06cm}\psi \hspace{.02cm} \rangle \hspace{.3cm} = \hspace{.3cm} |\hspace{.06cm} \psi \hspace{.02cm} \rangle - 2\hspace{.03cm}\big{(}\hspace{.08cm}| \hspace{.06cm}\psi \hspace{.02cm} \rangle - |\hspace{.06cm}r \hspace{.02cm} \rangle \hspace{.06cm} \big{)} $$

$$ = \hspace{.3cm} 2\hspace{.05cm}|\hspace{.06cm}r\hspace{.02cm}\rangle - |\hspace{.06cm}\psi \hspace{.02cm}\rangle $$

The second part of this operation should stand out to you, it's just the Identity operator $I$. Thus, our unitary operator will have the following form:

$$ U \hspace{.2cm} \equiv \hspace{.2cm} \textmf{something} \hspace{.08cm} - \hspace{.08cm} I \hspace{.2cm}$$

This $\textit{something}$, is a operation that when applied to a state $|\hspace{.04cm} \psi \hspace{.02cm} \rangle$, results in the state 2$|\hspace{.04cm}r\hspace{.02cm} \rangle$. As mentioned before, $|\hspace{.04cm}r\hspace{.02cm} \rangle$ is a state that is not guaranteed to be normalized, thus we cannot physically create it. However, the combination of $\hspace{.1cm} 2\hspace{.05cm}|\hspace{.06cm}r\hspace{.02cm}\rangle - |\hspace{.06cm}\psi \hspace{.02cm}\rangle \hspace{.1cm}$ $\textit{will}$ be normalized.
\\

The matrix operation that creates the state $|r \rangle$ is as follows:

$$ |\hspace{.06cm}s \hspace{.02cm} \rangle \hspace{.3cm} \equiv \hspace{.3cm} \frac{1}{\sqrt{N}} \sum_i^N |\hspace{.06cm} i \hspace{.02cm} \rangle \hspace{.35cm} \textmf{(equal superposition of all states)} $$

$$ |\hspace{.06cm}r \hspace{.02cm} \rangle \hspace{.3cm} = \hspace{.3cm} | \hspace{.06cm} s \hspace{.02cm} \rangle \langle \hspace{.02cm} s \hspace{.06cm} | \hspace{.06cm} \psi \hspace{.02cm} \rangle $$

Thus, we can create $|\hspace{.04cm} r \hspace{.02cm}\rangle$ by using the state $|\hspace{.04cm} s \hspace{.02cm}\rangle$, which is definitely a physically realizable state (Hadamard gates on every qubit). However, $|\hspace{.04cm} s \hspace{.02cm}\rangle$ $\langle \hspace{.02cm} s \hspace{.04cm}|$ is not a unitary operator (if it were, it would mean that we could physically create $|\hspace{.04cm} r \hspace{.02cm}\rangle$ ). Let's quickly show how these two quantities are equal:
\\

1) The inner product $\langle s | \psi \rangle $ results in the following sum of all the amplitudes: $\frac{1}{\sqrt{N}}$ $\sum_i^N \alpha_i$
\\

2) We borrow the remaining $\frac{1}{\sqrt{N}}$ term from the other $|s \rangle$ state, giving us our average amplitude: $\frac{1}{N} \sum_i^N \alpha_i = \alpha_{avg}$.
\\

3) This average amplitude $\alpha_{avg}$ is left multiplying all of the states leftover from $|s \rangle$, leaving us with:

$$ | \hspace{.06cm} s \hspace{.02cm} \rangle \langle \hspace{.02cm} s \hspace{.06cm} | \hspace{.06cm} \psi \hspace{.02cm} \rangle \hspace{.25cm} = \hspace{.25cm} \alpha_{avg} \hspace{.05cm}|\hspace{.06cm}000\rangle \hspace{.1cm} + \hspace{.1cm} \alpha_{avg} \hspace{.05cm}|\hspace{.06cm}001\rangle \hspace{.1cm} + \hspace{.1cm} ... \hspace{.1cm} \hspace{.15cm} = \hspace{.15cm} |\hspace{.06cm}r \hspace{.02cm} \rangle $$

Thus, we now have a full mathematical description for $U_s$:

$$ U_s = 2\hspace{.05cm} | \hspace{.06cm} s \hspace{.02cm} \rangle \langle \hspace{.02cm} s \hspace{.06cm} | \hspace{.08cm} - \hspace{.08cm} I $$

\section*{\large{ Implementing U$_s$ via H$^N$ }}

Although we just derived a nice compact form for our Grover Diffusion Operator, implementing it into our quantum algorithm is a tad bit more challenging. As we pointed out, the operator as a whole is unitary, but the individual contributions are physically unrealizable. But fear not, there is an impressively simple way of realizing $U_s$, using a Hadamard Transformation (our favorite).
\\

To start, we must take a slight detour from our algorithm in order to talk about a very important property of the Hadamard Transformation, particularly how it transforms the state of all $0$'s:

$$ H^2 \hspace{.05cm} |\hspace{.06cm}00\rangle = \frac{1}{2}\big{(}\hspace{.05cm}|\hspace{.06cm}00\rangle + |\hspace{.06cm}01\rangle + |\hspace{.06cm}10\rangle + |\hspace{.06cm}11\rangle \hspace{.05cm}\big{)} = |\hspace{.06cm}s\rangle $$

Nothing new here, but we want to take special note of how the Hadamard Transformation is a map between the state of all $0$'s, and the equal superposition state $|s\rangle$:

$$ H^2 \hspace{.05cm} |\hspace{.06cm}s\hspace{.02cm}\rangle \hspace{.25cm} = \hspace{.25cm} |\hspace{.06cm}00\hspace{.02cm}\rangle \hspace{3cm} H^2 \hspace{.05cm} |\hspace{.06cm}00\hspace{.02cm}\rangle \hspace{.25cm} = \hspace{.25cm} |\hspace{.06cm}s\hspace{.02cm}\rangle $$

We just saw in our previous discussion that the state $|\hspace{.06cm}s\hspace{.02cm} \rangle$ was exactly what we needed to create $U_s$, and it's no coincidence that we are seeing it again here via the Hadamard Transformation. This $H^N$ mapping is what is going to allow us to implement the Grover Diffusion Operator.
\\

Although we've seen the Hadamard Transformation at the core of all our previous algorithms, this implementation is a bit different. Previously, we used $H^N$ as a way of simultaneously sampling all possible entries for our blackbox problems. Here, we are using $H^N$ in order to transform our system to a basis where the Grover Diffusion Operator is achievable in one simple operation, and then transforming back. This use of $H^N$ is identical to our use of $\textmf{X\_Transformation}$, where we transform our system to a different basis in order to use control gates.
\\

Consider this somewhat silly example: Imagine you need to lift a 1 ton brick onto a shelf under Earth's gravity, so you $\textit{transform}$ your problem to the moon where gravity is weaker, do the lift, and then transform back to Earth. That's the spirit of what we're going to achieve with this Hadamard Transformation here in the Grover Algorithm.
\\

Before any further explanation, it's more powerful to see it in action first:

\begin{figure}[h]
\centering
\includegraphics[scale=.65]{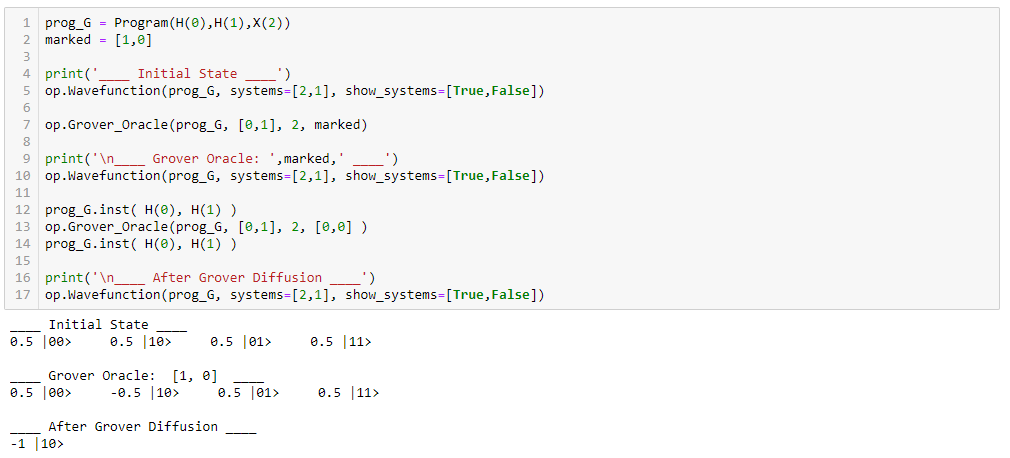}
\end{figure}

And viola! Like magic, we've increased the probability of our marked state, while suppressing all other states. And $N=2$ (four total states) is a special case, where all non-marked states get suppressed to amplitudes of 0! Feel free to change the marked state in the example above, and see that the Grover Algorithm always makes our marked state dominant.
\\

To see why this happened, let's again draw the amplitudes before and after reflecting about the average:

\begin{figure}[h]
\centering
\includegraphics[scale=.8]{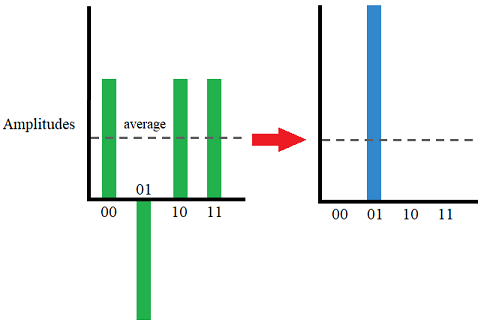}
\end{figure}

Because we flipped the sign on our marked state before $U_s$ (via the Oracle Operator), we effectively changed the average amplitude. Then, because the average amplitude is positive and our marked state is negative, the reflection about this new average results in a huge increase in amplitude. Simultaneously, all of our non-marked states have larger amplitudes than the average, so the reflection causes their amplitudes to decrease.
\\

Now, let's run the code above once more, this time observing the state of our system at each point during the Grover Diffusion Operator:

\begin{figure}[h]
\centering
\includegraphics[scale=.65]{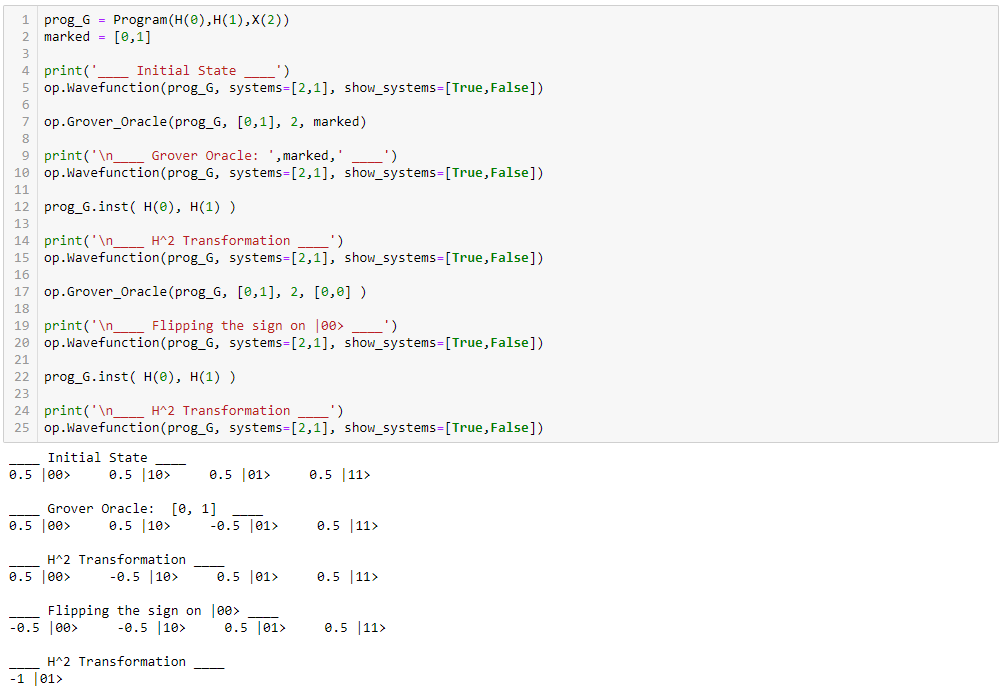}
\end{figure}

And there it is, the full Grover Algorithm. The key here is that once in the transformed basis (after the first $H^N$), all we did was $\textit{flip the sign on the state}$ $|00\rangle$, again via our Oracle function. Then, when we transformed back to our original basis (after the second $H^N$), our main system is in the state $-$$|01\rangle$. Which means... the sign flip on the state $|00\rangle$ $\textit{was}$ the Grover Diffusion Operator. Well, technically no. It is more accurate to say that as a whole:

$$H^N \hspace{.1cm}+\hspace{.1cm} \textmf{Oracle}( \hspace{.08cm} |\hspace{.06cm}00\rangle \hspace{.08cm} ) \hspace{.1cm}+\hspace{.1cm} H^N \hspace{.08cm} \hspace{.3cm} \equiv \hspace{.3cm} \textmf{Grover Diffusion Operator} $$

Remember earlier that we took take special note of the transformation $\hspace{.15cm}|\hspace{.06cm}00...0\rangle \hspace{.12cm}\longleftrightarrow \hspace{.12cm} H^N \hspace{.12cm}\longleftrightarrow \hspace{.12cm} |\hspace{.06cm}s\hspace{.02cm}\rangle$. One way of thinking about unitary transformations, is that operations performed in the two bases can look very different, but turn out to be equivalent. Here, we avoid doing some complicated series of operations in our original basis by using a Hadamard Transformation to achieve the same result with ease, and then transform back.
\\

Let's take a look at one more example:

\pagebreak

\begin{figure}[h]
\centering
\includegraphics[scale=1]{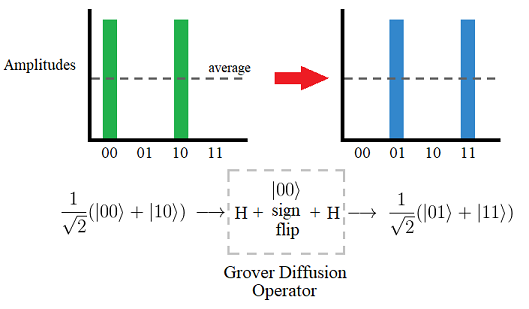}
\end{figure}

\begin{figure}[h]
\centering
\includegraphics[scale=.65]{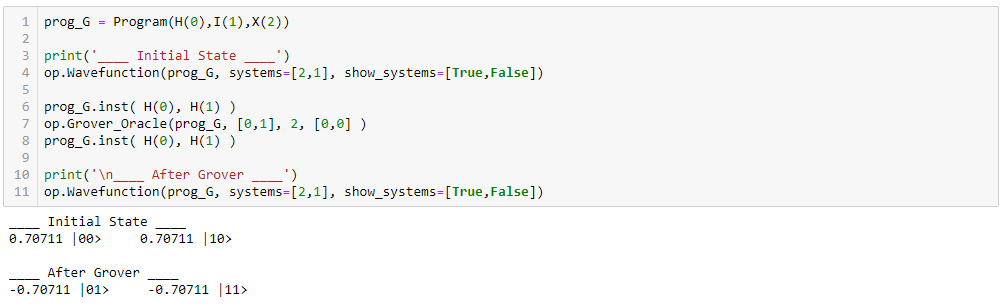}
\end{figure}

In the example above, we start in the state $\hspace{.12cm} \frac{1}{\sqrt{2}} \big{(} \hspace{.08cm} |\hspace{.06cm}00\rangle + |\hspace{.06cm}10\rangle \hspace{.06cm} \big{)} \hspace{.03cm}$, which has an average amplitude of $ \hspace{.1cm} \frac{1}{2\sqrt{2}}$. A reflection of each state about this average results in the states $|00\rangle$ and $|10\rangle$ going to zero, and the states $|01\rangle$ and $|11\rangle$ going to $ \hspace{.1cm} \frac{1}{\sqrt{2}}$.
\\

The code we've written above achieves exactly this, except for one thing. We get the correct final states and amplitudes, but our final states all have negative phases. In fact, you may have already picked up on this in all of our previous examples as well. All of our results are in agreement with the corresponding diagrams, except for their final phases. To understand these results, we must return to the first two diagrams in the "Reflection About an Average" section.
\\

Remember we showed that a reflection about a single state is equivalent to flipping the sign on all other states in the system. For example, a reflection about the state $|10\rangle$:

$$ \frac{1}{2}\big{(}\hspace{.12cm}|\hspace{.06cm}00\rangle + |\hspace{.06cm}01\rangle + |\hspace{.06cm}10\rangle + |\hspace{.06cm}11\rangle \hspace{.08cm}\big{)} \hspace{.4cm} \longrightarrow \hspace{.4cm} \frac{1}{2}\big{(}\hspace{.08cm} -|\hspace{.06cm}00\rangle -|\hspace{.06cm}01\rangle + |\hspace{.06cm}10\rangle -|\hspace{.06cm}11\rangle \hspace{.08cm} \big{)} $$

But we also showed that we can achieve a parallel state by only flipping the sign on the one state:

$$ \frac{1}{2}\big{(}\hspace{.12cm}|\hspace{.06cm}00\rangle + |\hspace{.06cm}01\rangle + |\hspace{.06cm}10\rangle + |\hspace{.06cm}11\rangle \hspace{.08cm}\big{)} \hspace{.4cm} \longrightarrow \hspace{.4cm} \frac{1}{2}\big{(}\hspace{.08cm} |\hspace{.06cm}00\rangle + |\hspace{.06cm}01\rangle - |\hspace{.06cm}10\rangle +|\hspace{.06cm}11\rangle \hspace{.08cm} \big{)} $$

Now, we must apply this concept to the way we flip the $|00...0\rangle$ state in the transformed basis. In particular, we pointed out that the Hadamard transformation is a map between: $\hspace{.15cm}|\hspace{.06cm}00...0\rangle \hspace{.12cm}\longleftarrow \hspace{.12cm} H^N \hspace{.12cm}\longrightarrow \hspace{.12cm} |\hspace{.06cm}s\hspace{.02cm}\rangle$. One way to understand this mapping is to say that these states are equivalent, via the $H^N$ transformation. Thus, performing our reflection about $|00...0\rangle$ in the transformed basis is equivalent to a reflection about $|\hspace{.04cm}s\rangle$ in our original basis. And, since flipping the sign on $|00...0\rangle$ achieves a state parallel to the reflection, transforming back via $H^N$ will also result in the parallel state to the reflection about the average, explaining where our minus signs are coming from.
\\

Understanding how the mapping of $\hspace{.15cm}|\hspace{.06cm}00...0\rangle \hspace{.12cm}\longleftarrow \hspace{.12cm} H^N \hspace{.12cm}\longrightarrow \hspace{.12cm} |\hspace{.06cm}s\hspace{.02cm}\rangle$ produces our reflection about the average is really the most important topic in this lesson, and will likely take a little time to fully sink in. As another example, let's suppose we wanted to perform the proper reflection about the average, without picking up the final phase difference on our state. As we showed earlier, to do this we need to to flip the sign on all other states in the system:

\begin{figure}[h]
\centering
\includegraphics[scale=.65]{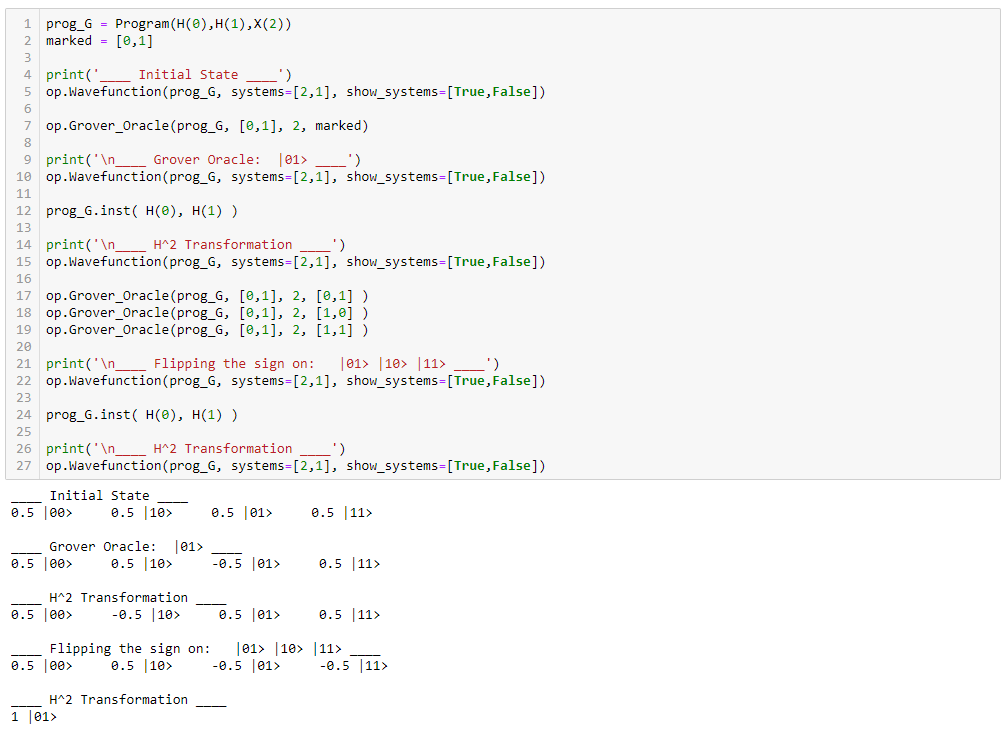}
\end{figure}

In the example above, we have achieved a true reflection about the average, which results in the same final state as predicted by our amplitude diagrams. We can see that the true reflection about $|00\rangle$ comes from flipping the sign on all other states in the system. And when we transform back to our original basis via $H^N$, sure enough we get the expected result. However, going through all the trouble of flipping extra states just for a final phase isn't really worth the extra cost in quantum steps. Thus, when we go to implement our Grover Algorithm later, we will opt for the more efficient method of only flipping $|00...0\rangle$ in the transformed basis.
\\

A reflection about a single state is much easier to understand at first than a reflection about the average, but we can express them in a similar way. Earlier we showed that a reflection about some state $|X_i\rangle$ will leave $|X_i\rangle$ unchanged, while flipping the sign on all other states. Let's show that this property holds true for our reflection about the average as well, using our 2-qubit example. To do this, we will need to manually separate out the average state from our system (which is $|\hspace{.04cm}r\hspace{.02cm}\rangle$ from earlier), and flip the sign on everything else:

\pagebreak

$$ |\hspace{.06cm}\psi \hspace{.04cm}\rangle_i \hspace{.1cm} = \hspace{.1cm} \frac{1}{2} \big{(} \hspace{.12cm} |\hspace{.06cm}00\rangle \hspace{.08cm}-\hspace{.08cm} |\hspace{.06cm}01\rangle \hspace{.08cm}+\hspace{.08cm} |\hspace{.06cm}10\rangle \hspace{.08cm}+\hspace{.08cm} |\hspace{.06cm}11\rangle \hspace{.08cm} \big{)} $$

$$ |\hspace{.06cm}r \hspace{.04cm}\rangle \hspace{.1cm} = \hspace{.1cm} \frac{1}{4} \big{(} \hspace{.12cm} |\hspace{.06cm}00\rangle \hspace{.08cm}+\hspace{.08cm} |\hspace{.06cm}01\rangle \hspace{.08cm}+\hspace{.08cm} |\hspace{.06cm}10\rangle \hspace{.08cm}+\hspace{.08cm} |\hspace{.06cm}11\rangle \hspace{.08cm} \big{)} $$

$$ \hspace{1.5cm} |\hspace{.06cm}\psi \hspace{.04cm}\rangle_i \hspace{.1cm} = \hspace{.1cm} |\hspace{.06cm}r \hspace{.04cm}\rangle \hspace{.15cm} + \hspace{.15cm} \frac{1}{4} \big{(} \hspace{.12cm} |\hspace{.06cm}00\rangle \hspace{.08cm}-3\hspace{.08cm} |\hspace{.06cm}01\rangle \hspace{.08cm}+\hspace{.08cm} |\hspace{.06cm}10\rangle \hspace{.08cm}+\hspace{.08cm} |\hspace{.06cm}11\rangle \hspace{.08cm} \big{)} $$

$$ \ast \hspace{.1cm} reflection \hspace{.2cm} about \hspace{.2cm} |\hspace{.06cm}r \hspace{.04cm}\rangle \hspace{.1cm} \ast $$

$$ \hspace{1.5cm} |\hspace{.06cm}\psi \hspace{.04cm}\rangle_f \hspace{.1cm} = \hspace{.1cm} |\hspace{.06cm}r \hspace{.04cm}\rangle \hspace{.15cm} - \hspace{.15cm} \frac{1}{4} \big{(} \hspace{.12cm} |\hspace{.06cm}00\rangle \hspace{.08cm}-3\hspace{.08cm} |\hspace{.06cm}01\rangle \hspace{.08cm}+\hspace{.08cm} |\hspace{.06cm}10\rangle \hspace{.08cm}+\hspace{.08cm} |\hspace{.06cm}11\rangle \hspace{.08cm} \big{)} $$

$$ |\hspace{.06cm}\psi \hspace{.04cm}\rangle_f \hspace{.1cm} = \hspace{.1cm} \hspace{.12cm} \big{(}\hspace{.06cm} \frac{1}{4} - \frac{1}{4} \hspace{.06cm}\big{)} \hspace{.08cm} |\hspace{.06cm}00\rangle \hspace{.3cm}+\hspace{.3cm} \big{(}\hspace{.06cm} \frac{1}{4} + \frac{3}{4} \hspace{.06cm}\big{)} \hspace{.08cm} |\hspace{.06cm}01\rangle \hspace{.3cm}+\hspace{.3cm} \big{(}\hspace{.06cm} \frac{1}{4} - \frac{1}{4} \hspace{.06cm}\big{)} \hspace{.08cm} |\hspace{.06cm}10\rangle \hspace{.3cm}+\hspace{.3cm} \big{(}\hspace{.06cm} \frac{1}{4} - \frac{1}{4} \hspace{.06cm}\big{)} \hspace{.08cm} |\hspace{.06cm}11\rangle \hspace{.08cm} $$

$$ |\hspace{.06cm}\psi \hspace{.04cm}\rangle_f \hspace{.1cm} = \hspace{.1cm} | \hspace{.06cm} 01\rangle $$

Here we can see that the average state $|\hspace{.04cm} r \hspace{.02cm} \rangle$ is unchanged through this reflection, just like our earlier example with the state $|0\rangle$. Although $|\hspace{.04cm} r \hspace{.02cm} \rangle$ is an unphysical state, hopefully this example helps illuminate what it means to reflect about the average.

\section*{\large{ The Full Grover Search }}
\centerline{---------------------------------------------------------------------------------------------------------------------------------}

In the coding examples above, we were able to fully pick out our marked state with only one application of our Grover Diffusion Operator. Two qubits is a special case, and in general we will need many more applications in order to make our marked state significantly probable. Specifically, we will need a certain number of Grover Iterations, based on the size of the problem. To remind ourselves, a single Grover Iteration is defined as: $\hspace{.1cm}$ 1) Flipping the sign on our marked state via the Oracle Operator $\hspace{.15cm}$ 2) Applying the Grover Diffusion Operator.
\\

The reason we will need many Grover Iterations as our problem size gets larger, is because each individual iteration will only boost the probability of our marked state by so much. Consider the diagram below, which shows that a single Grover Iteration is not enough to give our marked state a significant probability:

\begin{figure}[h]
\centering
\includegraphics[scale=.8]{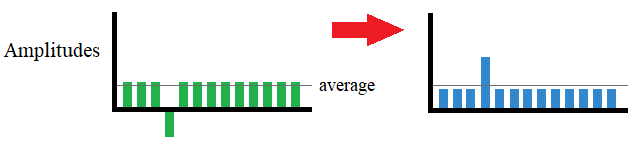}
\end{figure}

While one step does indeed make our marked state $\textit{more}$ probable, it isn't significant enough to where it is worth making a measurement. And as the size of our problem increases, this first step will be less and less impactful. However, we can simply repeat the process as many times as we need to, until we reach a desirable probability distribution. For example, let's apply one more Grover Iteration to our diagram example:

\pagebreak

\begin{figure}[h]
\centering
\includegraphics[scale=.8]{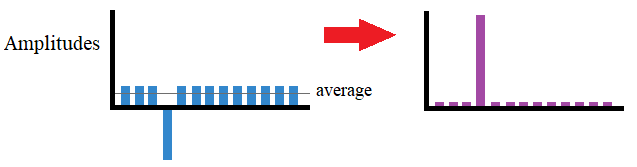}
\end{figure}

By applying a second Grover Iteration, we are essentially starting with a state where the amplitude of our marked state is already larger than all the rest. This in turn causes the average amplitude to be smaller, which further decreases all the non-marked states. Thus, after two Grover Iterations, we reach a state where a measurement on the system will find our marked state with a high probability of success.
\\

But, we must point out something very important here. The Grover Iteration is not a magical operation that $\textit{always}$ boosts the amplitude of our marked state. The trick relies on the average amplitude, and at a certain point, the Grover Iteration actually works against us. Let's continue our diagram example with one more iteration to show this negative effect:

\begin{figure}[h]
\centering
\includegraphics[scale=.8]{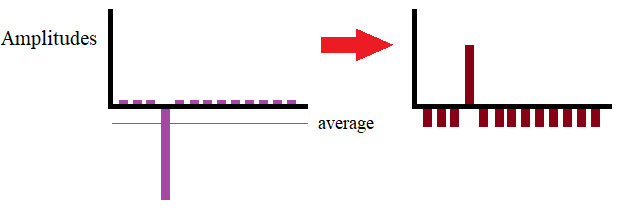}
\end{figure}

Take note of where the average amplitude is located in this third step. Because our marked state's amplitude is so large, it actually weighs the average down below 0 after we flip its sign. It is at this point where the Grover Iteration is working against us. This negative average amplitude causes all of our non-marked states to $\textit{increase}$ in amplitude, which comes at the cost of our marked state.
\\

Even worse yet, try and visualize where the next average amplitude would be after we flip the marked state. Because all of the non-marked states now have negative amplitudes, a fourth Grover Iteration will result in an even lower probability on our marked state, eventually leading to a point where the marked state is the $\textit{least}$ probable state in the system.
\\

Thus, this example has highlighted the final piece to the Grover Algorithm: when to stop. Too many Grover Iterations will make things worse, so we need to never go over the optimal amount. Luckily for us, there is a well known trend that tells us when to stop, for a system of $N$ states:

$$ \textmf{optimal steps:} \hspace{.3cm} \approx \hspace{.3cm} \frac{\pi}{4} \sqrt{N} $$

There is an 'exact optimal' number of steps for any given $N$, which may not be exactly $\frac{\pi}{4} \sqrt{N}$. But once $N$ is large enough, applying $\frac{\pi}{4} \sqrt{N}$ Grover Iterations will always be nearly optimal. The more problematic cases are for smaller $N$'s, but these aren't really too concerning since using a quantum algorithm for a search on a list of say 4 or 8 entries, is a bit of an overkill. The real merit of this algorithm is for searching on very large lists, where the $\sqrt{N}$ factor is a significant speedup.
\\

Now that we've seen the effect of too many Grover Iternations, let's see it in a coding example. To do this, we will import $\textbf{Grover\_Diffusion}$ from $\textmf{Our\_Pyquil\_Functions}$:

\pagebreak

\begin{figure}[h]
\centering
\includegraphics[scale=.65]{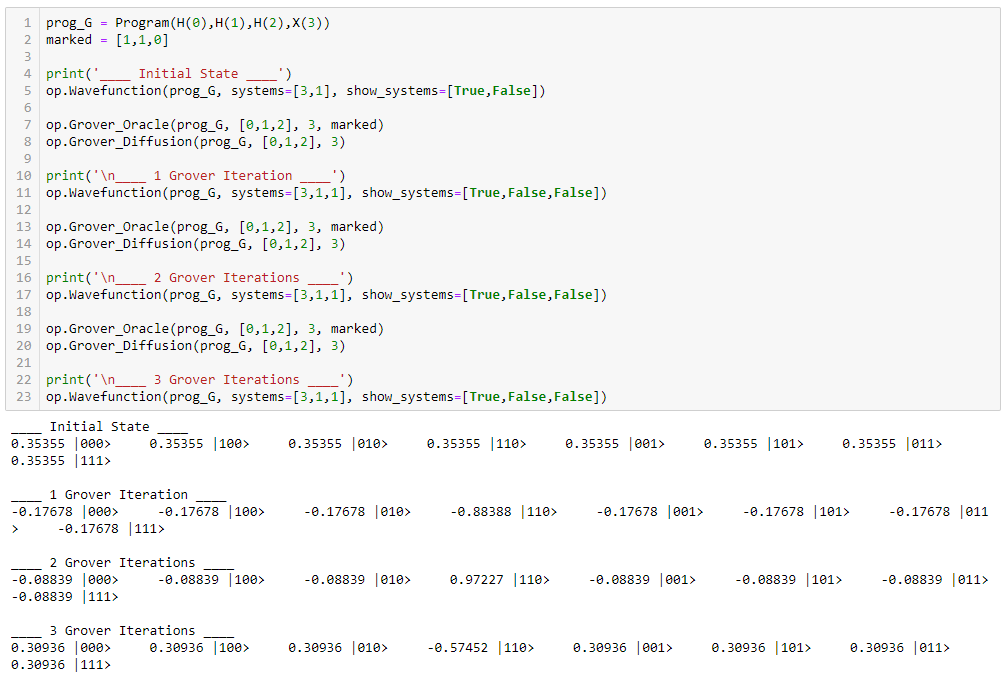}
\end{figure}

Take a look at the amplitudes displayed above. After 1 Grover Iteration, we have a 78\% percent chance of measuring our marked state. After the second Grover Iteration, this probability jumps to over 94\%! But, if we apply a third iteration, our probability of measuring the marked state plummets to a measly 33\% (but is still the highest single state). If we had carried out a fourth iteration, we would find our marked state with a 1\% probability, the complete opposite of what we set out to do!
\\

The Grover Algorithm is cyclic in the way it increases / decreases the probability of our marked state. $\frac{\pi}{4} \sqrt{N}$ represents the first peak, which corresponds to half of the cycle. If we perform $\frac{\pi}{2} \sqrt{N} $ iterations, we will find the point where our marked state is $\textit{least}$ probable. But from there, the probabilities will begin to increase again, peaking at $\frac{3\pi}{4} \sqrt{N}$, and so on. But for the purpose of our searching algorithm, we will only ever aim for the first $\frac{\pi}{4} \sqrt{N}$ peak.
\\

The cell of code below is our complete Grover Algorithm, combining all of the steps we've covered thus far into the function $\textbf{Grover}$. Change $Q$ to be any number of qubits you like, corresponding to a system of the size $2^Q$, and pick a corresponding marked state of length $Q$:

\pagebreak

\begin{figure}[h]
\centering
\includegraphics[scale=.65]{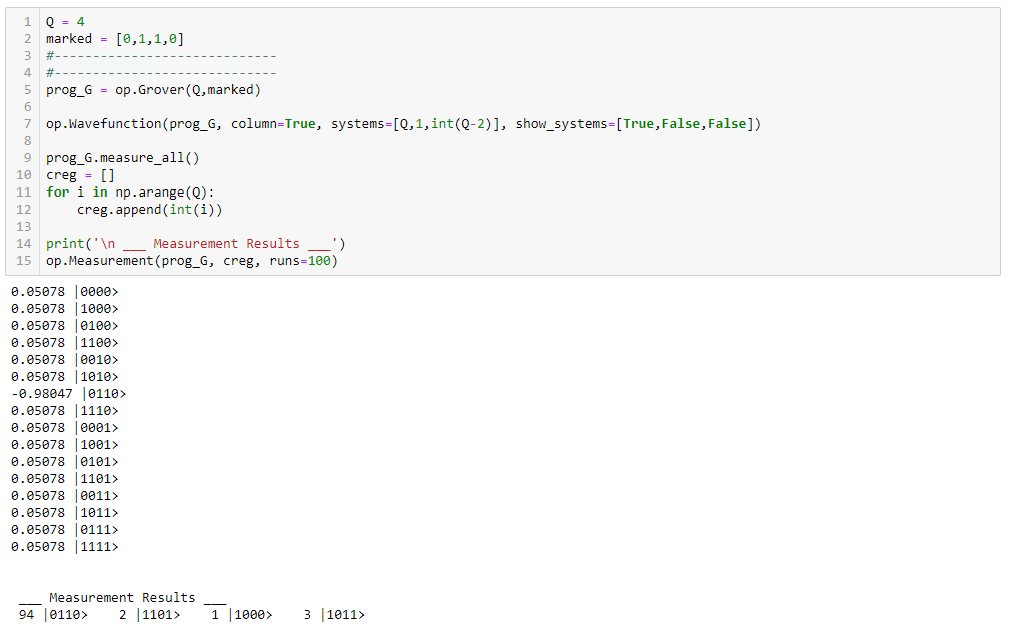}
\end{figure}

--------------------------------------------------------------------------------------------------------------------------------------------------------
\\

This concludes lesson 5.4, and our series of introductory quantum algorithms! We have now seen four lessons worth of Hadamard Transformations, and the various problems it can solve. We saved this algorithm for last because of the way in which we used $H^N$, which is very analogous to the next lesson to come. In general, the use of unitary transformations are at the core of a lot of the most successful quantum algorithms to date.
\\

--------------------------------------------------------------------------------------------------------------------------------------------------------


\pagebreak

\section*{\Large{ Lesson 6 - Quantum Fourier Transformation }}
--------------------------------------------------------------------------------------------------------------------------------------------------------
\\

In this final tutorial, we will cover an important transformation used at the heart of many successful quantum algorithms: the Quantum Fourier Transformation ($\textmf{QFT}$). Much like how the Hadamard Transformation was the basis for all of the algorithms studied in lessons 5.1 - 5.4, the $QFT$ plays a major role in algorithms like Shor's, Quantum Phase Estimation, Variational Quantum Eigensolver, and many more. At their core, the two transformations share a lot of similarities, both in their effect and usage in quantum algorithms.
\\

Original publication of the algorithm: \cite{QFT}
\\

--------------------------------------------------------------------------------------------------------------------------------------------------------
\\
In order to make sure that all cells of code run properly throughout this lesson, please run the following cell of code below:

\begin{figure}[h]
\centering
\includegraphics[scale=.65]{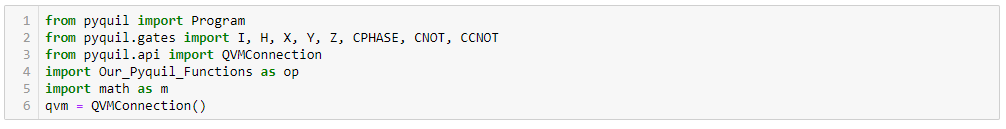}
\end{figure}

\section*{\large{ Importance of Unitary Transformations }}
\centerline{---------------------------------------------------------------------------------------------------------------------------------}

If we think back to lessons 5.1 - 5.4, we should ask ourselves: what was it about the Hadamard Transformation that allowed all of those algorithms to be successful. For the blackbox problems, we would say that it allowed us to work with all possible states at once, thus out performing classical algorithms that are forced to check only one input at a time. And for the Grover Algorithm, the vital role of the Hadamard Transformation was that it allowed us to perform a 'reflection about the average' by transforming to a different basis.
\\

The success of any transformation can always be traced to $\textit{the way}$ it maps states. In particular, by studying the way a certain transformation maps individual states, as well as how it maps combinations of states, we can learn about what types of advantages it can achieved. Or in other words, a transformation provide us with two 'domains' in which to work, where we can use the advantages of each to solve complex problems. Visually, moving to a transformed basis in order to achieve some desired effect looks like:

\begin{figure}[h]
\centering
\includegraphics[scale=.65]{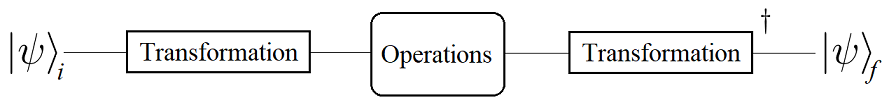}
\end{figure}

The operations we perform 'inside' the transformation are dependent on the algorithm, and what type of problem we are trying to solve. Sometimes, we need to perform transformations $\textit{within}$ transformations in order to get a certain effect. For example, the Grover Diffusion Operator from lesson 5.4 is essentially and X Transformation inside of a H Transformation, in order to flip the sign on the state $|00...0\rangle$.
\\

Another important property of transformations are the operators that map back and forth between the bases. For the Hadamard Transformation, the same operator is used for both transformations, but in general this is not always the case. In the figure above, this is represented by the $\textmf{Transformation}$ and $\textmf{Transformation}^{\dagger}$ operations. As an example, consider the role of orthogonality when using a Hadamard Transformation:

\pagebreak

$$ \langle 01 \hspace{.08cm} | \hspace{.08cm} 10 \rangle = 0 $$

$$ H^2 \hspace{.06cm} | \hspace{.08cm} 01 \rangle \hspace{.3cm} = \hspace{.3cm} \frac{1}{2} \big{(}\hspace{.1cm} | \hspace{.08cm} 00 \rangle - | \hspace{.08cm} 01 \rangle + | \hspace{.08cm} 10 \rangle - | \hspace{.08cm} 11 \rangle \hspace{.08cm} \big{)} \hspace{1.5cm} H^2 \hspace{.06cm} | \hspace{.08cm} 10 \rangle \hspace{.3cm} = \hspace{.3cm} \frac{1}{2} \big{(}\hspace{.1cm} | \hspace{.08cm} 00 \rangle + | \hspace{.08cm} 01 \rangle - | \hspace{.08cm} 10 \rangle - | \hspace{.08cm} 11 \rangle \hspace{.08cm} \big{)}$$

$$ \frac{1}{4} \big{(}\hspace{.1cm} \langle \hspace{.08cm} 00 \hspace{.08cm} | - \langle 01 \hspace{.08cm} | + \langle 10 | \hspace{.08cm} - \langle 11 \hspace{.08cm} | \hspace{.08cm} \big{)} \hspace{.08cm} \big{(}\hspace{.1cm} | \hspace{.08cm} 00 \rangle + | \hspace{.08cm} 01 \rangle - | \hspace{.08cm} 10 \rangle - | \hspace{.08cm} 11 \rangle \hspace{.08cm} \big{)} \hspace{.6cm} = \hspace{.6cm} \frac{1}{4} \big{(} 1 - 1 - 1 + 1 \big{)} \hspace{.6cm} = \hspace{.6cm}0$$

Or written in a more elegant way:

$$ \langle 01 \hspace{.08cm} | \hspace{.1cm} H^{\dagger 2} \hspace{.1cm} H^2 \hspace{.1cm} | \hspace{.08cm} 10 \rangle \hspace{.4cm} = \hspace{.4cm} \langle 01 \hspace{.08cm} | \hspace{.16cm} (H^{\dagger}H)\otimes(H^{\dagger}H) \hspace{.16cm} | \hspace{.08cm} 10 \rangle $$

$$ \hspace{3.3cm} = \hspace{.4cm} \langle 01 \hspace{.08cm} | \hspace{.16cm} (H^{\dagger}H)\otimes(H^{\dagger}H) \hspace{.16cm} | \hspace{.08cm} 10 \rangle $$

$$ \hspace{1.6cm} = \hspace{.4cm} \langle 01 \hspace{.08cm} | \hspace{.16cm} I\otimes I \hspace{.16cm} | \hspace{.08cm} 10 \rangle $$

$$ \hspace{1.6cm} = \hspace{.4cm} \langle 01 \hspace{.08cm} | \hspace{.08cm} 10 \rangle \hspace{.3cm}=\hspace{.3cm} 0 $$

What's important to note in the second example is the property $H^{\dagger}H = 1$. This is true of all unitary operators, $\textit{but}$, not all unitary operators are their own complex conjugate like $H^N$. That is to say, the Hadamard transformation is special in that $H = H^{\dagger}$, a property known as being Hermitian, which means that we can apply the same operation to transform back and forth between bases. And since an $H^N$ transformation is essentially $N$ individual 1-qubit Hadamard Transformations in parallel: $\hspace{.2cm}$ $H \otimes H \otimes H \hspace{.01cm} \cdot \cdot \hspace{.1cm} \cdot$, the net result is that ${H^N}^{\dagger} = H^N$.
\\

If we have an operation that acts on $N$ qubits, and can be decomposed into $N$ individual Hermitian operators: $O_0 \otimes O_1 \otimes O_2 \cdot \cdot \hspace{.1cm} \cdot$, then the total operator is Hermitian as well. For example:

\begin{figure}[h]
\centering
\includegraphics[scale=.65]{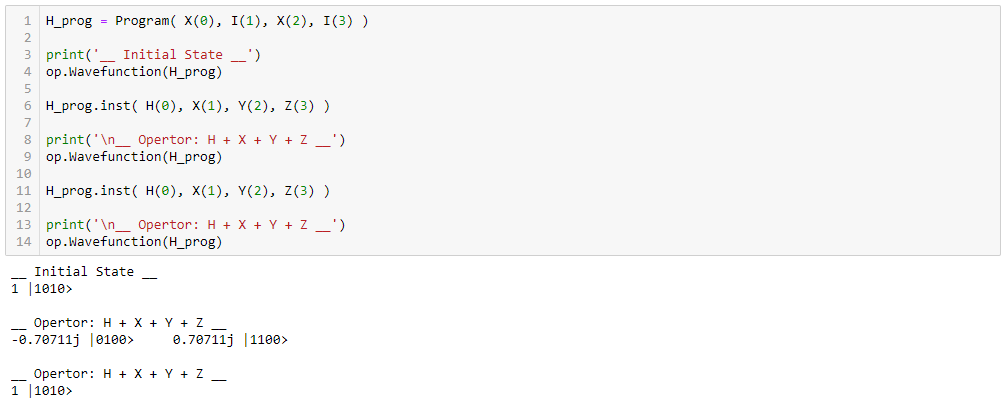}
\end{figure}

In this example, we make up a 4-qubit operator, which can be decomposed as: $\hspace{.2cm} H_0 \otimes X_1 \otimes Y_2 \otimes Z_3$. Each of the individual components is Hermitian, therefore the total operator is Hermitian as well. This is demonstrated by the fact that two applications of this operator return us back to our original state.
\\

However, as we pointed out earlier, not all multi-qubit operations are their own complex conjugate. For example, we've already seen such an operator in lesson 4 when we showed how to construct an $N$-NOT gate. This is because the $N$-NOT operation uses a specific ordering of gates. And in linear algebra, the order of operators is not always interchangeable. For example, consider a single qubit operator that can be decomposed as: $\hspace{.2cm} X_0 \otimes Z_0$

\pagebreak

\begin{figure}[h]
\centering
\includegraphics[scale=.65]{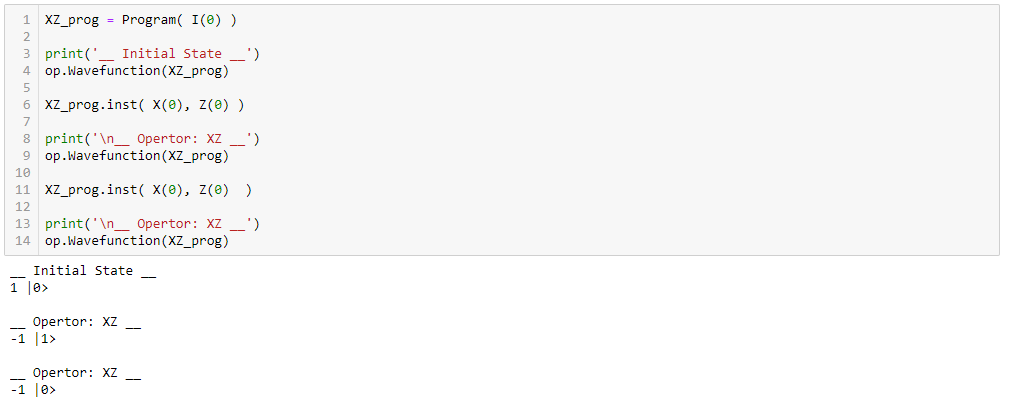}
\end{figure}

As we can see, applying this operator twice does not return us to our original state. Thus, $\hspace{.1cm} X_0 \otimes Z_0$ is not a Hermitian operation, $\textit{even though}$ it is made up of Hermitian components. If we define an operation that contains several gates that must act on the same qubit in a specific order, then chances are it won't be Hermitian. So then, if our algorithm requires us to use such an operator as a transformation, then we will need to find a $\textit{different}$ operator if we want to transform back. Specifically, we will need the complex conjugate of the operator.
\\

Luckily, if we know how to decompose an operation like the one in our example above, then finding the complex conjugate is as simple as reversing the order (with one caveat that we will see later):

\begin{figure}[h]
\centering
\includegraphics[scale=.65]{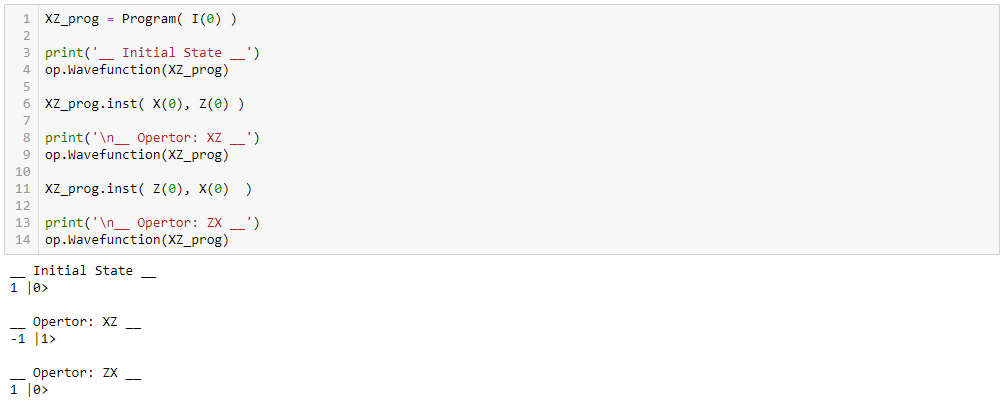}
\end{figure}

As you may have guess, the reason we've gone out of our way to discuss non-Hermitian operators is because the transformation we will be studying in this lesson is exactly that. The Quantum Fourier Transformation (QFT), which we will be using as the core of the next couple lessons, is an example where $QFT$ and $QFT^{\dagger}$ are different operations. As we shall see, the relation between these two transformations is very straightforward, and is analogous to the way we constructed our $\textmf{n\_NOT}$ gate in lesson 4.

\section*{\large{ Discrete Fourier Transformation }}
\centerline{---------------------------------------------------------------------------------------------------------------------------------}

The $QFT$ is essentially the Discrete Fourier Transformation (DFT), but applied to the states of our quantum system. Thus, we will begin with a quick review the DFT. Formally written, the Discrete Fourier Transformation looks like this:

$$ X = \{ \hspace{.05cm} x_0, \hspace{.1cm} ... \hspace{.05cm}, \hspace{.05cm} x_k, \hspace{.1cm} ...\hspace{.05cm}, \hspace{.05cm} x_{N-1} \hspace{.05cm} \} $$
$$ \tilde{X} = \{ \hspace{.05cm} \tilde{x}_0, \hspace{.1cm} ... \hspace{.05cm}, \hspace{.05cm} \tilde{x}_k, \hspace{.1cm} ...\hspace{.05cm}, \hspace{.05cm} \tilde{x}_{N-1} \hspace{.05cm} \} $$

$$ \tilde{x}_k \hspace{.2cm} = \hspace{.2cm} \sum_{j=0}^{N-1} x_j \cdot e^{ 2 \pi i \frac{k \cdot j}{N}}$$

Where the DFT maps all of the numbers in $X$ to $\tilde{X}$, and $\hspace{.16cm} e^{\pm i\theta} \hspace{.08cm} = \hspace{.08cm} cos(\theta) \hspace{.05cm} \pm \hspace{.05cm} i \hspace{.06cm} sin(\theta)$.

$$ X \hspace{.6cm} - DFT \rightarrow \hspace{.6cm} \tilde{X} $$

The DFT is defined by the sum above, which shows that each output value $\tilde{x}_k$, receives a contribution from each input value $x_k$. Specifically, each input value is multiplied by a complex number of the form $e^{i \theta}$, which are then all summed together. The value of each $\theta$ is determined by the multiplication of $k \cdot j$. Let's see a quick example:

$$ X = \big{[} 1 \hspace{.4cm} -1 \hspace{.4cm} -1 \hspace{.6cm} 1 \big{]} $$

$$\hspace{.18cm} \tilde{x}_1 \hspace{.2cm} = \hspace{.2cm} \sum_{j=0}^{3} x_j \cdot e^{ 2 i \pi \frac{1 \cdot j}{4}} $$

$$\hspace{3.7cm}= \hspace{.2cm} 1 \cdot e^{ 0 } -1 \cdot e^{ \frac{ i\pi}{2} } -1 \cdot e^{ i \pi } + 1 \cdot e^{ \frac{3 i\pi}{2} }$$

$$\hspace{.9cm}= \hspace{.2cm} 1 -i +1 -i \hspace{.3cm} $$

$$ = \hspace{.1cm} 2 - 2i \hspace{.3cm}$$

and the full transformation:

$$ X = \big{[} 1 \hspace{.4cm} -1 \hspace{.4cm} -1 \hspace{.6cm} 1 \big{]} \hspace{1cm} \longrightarrow \hspace{1cm} \tilde{X} = \big{[} \hspace{.08cm} 0 \hspace{.4cm} 2 - 2i \hspace{.4cm} 0 \hspace{.6cm} 2 + 2i \hspace{.08cm} \big{]} $$

These $e^{i\theta}$ terms are derived from the concept of taking the roots of -1, which we will not cover here. I encourage you to work through all of the example above, as you will want to really develop a good feel for these transformations if you plan to continue onto the lesson 7 algorithms. For our goal of understanding the $QFT$, we will only be taking from the DFT what we need.
\\

In particular, let's see what this DFT looks like in a matrix representation:

$$ \begin{bmatrix} 1 & 1 & 1 & 1 \\ 1 & i & -1 & -i \\ 1 & -1 & 1 & -1 \\ 1 & -i & -1 & i \end{bmatrix} \begin{bmatrix} 1 \\ -1 \\ -1 \\ 1 \end{bmatrix} \hspace{.7cm} = \hspace{.7cm} \begin{bmatrix} 0 \\ 2-2i \\ 0 \\ 2+2i \end{bmatrix} $$

where the values in the matrix above can all be expressed in terms of:

$$ \omega \equiv e^{ \frac{2i \pi}{N} } \hspace{2cm} \textmf{DFT matrix: } \hspace{.2cm}F_4 = \begin{bmatrix} \omega^0 & \omega^0 & \omega^0 & \omega^0 \\ \omega^0 & \omega^1 & \omega^2 & \omega^3 \\ \omega^0 & \omega^2 & \omega^4 & \omega^6 \\ \omega^0 & \omega^3 & \omega^6 & \omega^9 \end{bmatrix} $$

The powers on all of the $\omega$'s come from the the products of $k \cdot j$, and $N$ refers to the total number of values being transformed ($N=4$ for our example):

$$ k \cdot j : \hspace{1cm} 0 \cdot 1 \hspace{1cm} 1 \cdot 2 \hspace{1cm} 3 \cdot 1$$

$$ \omega^{k \cdot j}: \hspace{1.15cm} \omega^{0} \hspace{1.3cm} \omega^{2} \hspace{1.3cm} \omega^{3} \hspace{.1cm} $$

$$ \hspace{.6cm} = \hspace{1.2cm} e^{0} \hspace{1.35cm} e^{i \pi} \hspace{1.35cm} e^{\frac{3i \pi}{2}} $$

$$ \hspace{.75cm} = \hspace{1.3cm} 1 \hspace{1.1cm} -1 \hspace{1.1cm} -i \hspace{.6cm} $$

We could go on and on about the things one can do with DFT, but we will end our discussion here. I encourage you to read other references about the Discrete Fourier Transformation, and the various things it can be used for. Doing so will help you get a deeper understanding for why the $\textmf{QFT}$ is so powerful.

\section*{\large{ Quantum Fourier Transformation }}
\centerline{---------------------------------------------------------------------------------------------------------------------------------}

We now have a formal definition for the Discrete Fourier Transformation, so how do we make it quantum? Well, we've already shown how to represent the DFT as a matrix, so our task is to implement it as an operator. Since we are dealing with quantum systems, we will naturally gravitate towards transformations of the size $2^N$.
\\

Let's use a 2-qubit example so illustrate how the DFT will look on a quantum system:

$$ \hspace{.8cm} | \hspace{.08cm} \psi \rangle \hspace{.2cm} = \hspace{.2cm} \frac{1}{2} \big{(} \hspace{.15cm} |\hspace{.08cm} 00\rangle - |\hspace{.08cm} 01\rangle - |\hspace{.08cm} 10\rangle + |\hspace{.08cm} 11\rangle \hspace{.1cm} \big{)} $$

$$ \hspace{.6cm} F_4 \hspace{.1cm} | \hspace{.08cm} \psi \rangle \hspace{.2cm} = \hspace{.2cm} \frac{1}{2} \big{(} \hspace{.2cm} (1-i) \hspace{.1cm} |\hspace{.08cm} 10\rangle \hspace{.15cm}+\hspace{.15cm} (1+i) \hspace{.1cm} |\hspace{.08cm} 11\rangle \hspace{.1cm} \big{)} $$

This example is the quantum version of our $\hspace{.08cm} X \hspace{.1cm} \rightarrow \hspace{.1cm} \tilde{X} \hspace{.08cm}$ transformation from earlier. Our initial state corresponds to $X$, and our final state is $\tilde{X}$. Verifying that is operation is indeed unitary is simple enough, which means that $F_4$ is a legitimate quantum operation. And in general, any DFT matrix is guaranteed to be unitary.
\\

For clarity, the vector representing the state of our system is in the following order:

$$ \begin{bmatrix} \hspace{.1cm} |\hspace{.08cm} 00\rangle \hspace{.1cm} \\ |\hspace{.08cm} 10\rangle \\ |\hspace{.08cm} 01\rangle \\ |\hspace{.08cm} 11\rangle \end{bmatrix} $$

\section*{\large{ Implementing a QFT }}

At this point, we have the structure for generating our $QFT$ matrices, and the corresponding vector representations of our states. From the mathematical perspective, we have the full picture for the $QFT$. However, as we've already seen with past algorithms, simply writing it down doesn't do it justice. And, if we want to actually run a $QFT$ in our quantum algorithms, we need a way of translating the mathematical picture into gates.
\\

The way in which we are going to achieve our $QFT$'s is quite elegant, and by no means obvious at first. As it turns out, the only gates we need in order to construct a $2^N$ $QFT$ are $H$ and $R_{\phi}$: our trusty Hadamard gate along with some control-phase gates. Even better yet, we will not require any additional ancilla qubits.
\\

Below is the general template for how to construct a $QFT$ circuit on $N$ qubits, acting on a $2^N$ space of states:

\begin{figure}[h]
\centering
\includegraphics[scale=.55]{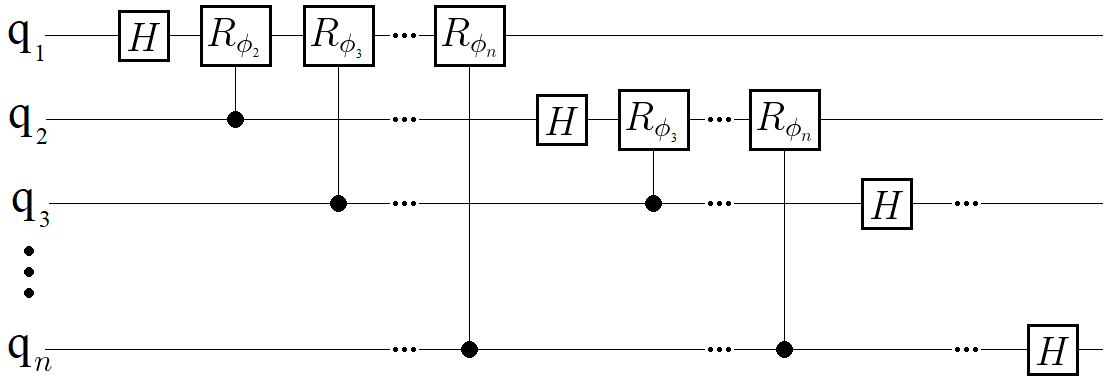}
\end{figure}

where

\begin{figure}[h]
\centering
\includegraphics[scale=.8]{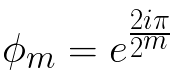}
\end{figure}

At first glance, this circuit may look a bit complex, but it is actually quite straightforward. Each qubit in the system undergoes the same process: a Hadamard gate followed by a series of control-phase gates. The number of $R_{\phi}$ gates that a qubit experiences is determined by its index. The first qubit in the system receives $N - 1$, while the last qubit doesn't receive any. In addition, the phase of each $R_{\phi}$ is determined by which qubit acts as the control, as shown by the equation above (note that we typically start our first qubit as 0, but here we are starting with 1).
\\

Now, it isn't immediately obvious why the circuit above works, but we're going to first test it out with a coding example:

\begin{figure}[h]
\centering
\includegraphics[scale=.65]{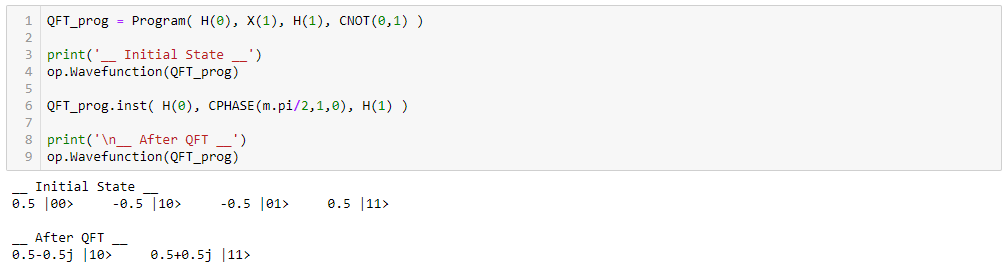}
\end{figure}

Try and match the pattern in the template above, with the steps we've implemented in this code example.:

$$ 1)\hspace{.4cm} H_0 \hspace{1.4cm} 2) \hspace{.4cm} R_{\frac{\pi}{2} \hspace{.03cm} 10} \hspace{1.4cm} 2) H_1 $$

where the $10$ subscript on the control-phase gate represents qubit $1$ is the control, and qubit $0$ is the target. Confirm for yourself that these are indeed the steps written into our coding example, and that they match the $QFT$ template.
\\

Next, we will do one more example, this time with 3 qubits:

$ \hspace{3cm}QFT \hspace{.08cm}| \hspace{.08cm} 001 \hspace{.02cm} \rangle \hspace{.1cm} = \hspace{.1cm} \frac{1}{4} \Big{(}\hspace{.1cm} \sqrt{2}\hspace{.12cm} |\hspace{.08cm}000\hspace{.02cm}\rangle \hspace{.06cm}-\hspace{.06cm} \sqrt{2}\hspace{.12cm} |\hspace{.08cm}001\hspace{.02cm}\rangle \hspace{.06cm}+\hspace{.06cm} (\hspace{.05cm} 1+i \hspace{.05cm})\hspace{.08cm} |\hspace{.08cm}010\hspace{.02cm}\rangle \hspace{.06cm}-\hspace{.06cm} (\hspace{.05cm}1+i \hspace{.05cm})\hspace{.08cm} |\hspace{.08cm}011\hspace{.02cm}\rangle$
\\

$ \hspace{5cm}+\hspace{.06cm} (\hspace{.05cm} 1+i \hspace{.05cm})\hspace{.08cm} |\hspace{.08cm}100\hspace{.02cm}\rangle \hspace{.06cm}-\hspace{.06cm} (\hspace{.05cm}1+i \hspace{.05cm})\hspace{.08cm} |\hspace{.08cm}101\hspace{.02cm}\rangle \hspace{.06cm}+\hspace{.06cm} \sqrt{2}i\hspace{.12cm} |\hspace{.08cm}110\hspace{.02cm}\rangle \hspace{.06cm}-\hspace{.06cm} \sqrt{2}i\hspace{.12cm} |\hspace{.08cm}111\hspace{.02cm}\rangle \hspace{.1cm} \Big{)} $

$\ast$ For an extra exercise, try deriving this result by writing out the full 8$\times$8 matrix for a 3-qubit transformation via our definitions earlier.
\\

Now to implement this transformation in code:

\begin{figure}[h]
\centering
\includegraphics[scale=.65]{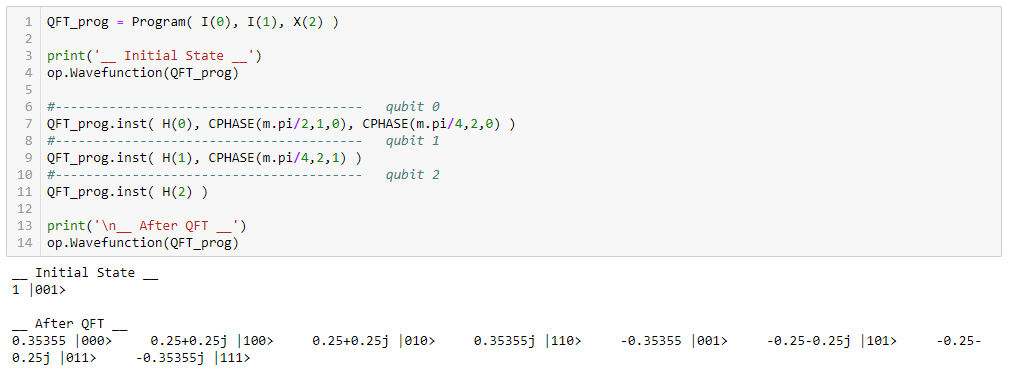}
\end{figure}

In this example, we've broken up the $QFT$ instructions into three sections, where each section incorporates all of the operations being applied to one of the three qubits. Just like in the $QFT$ template shown above, the number of operations decreases by 1 per qubit, where the last qubit only receives a single $H$.
\\

Ultimately, writing out all the steps for a QFT is a tedious task, so just like the $\textmf{n\_NOT}$ function, we will use the function $\textbf{QFT}$ from $\textmf{Our\_Pyquil\_Functions}$ instead:

\begin{figure}[h]
\centering
\includegraphics[scale=.65]{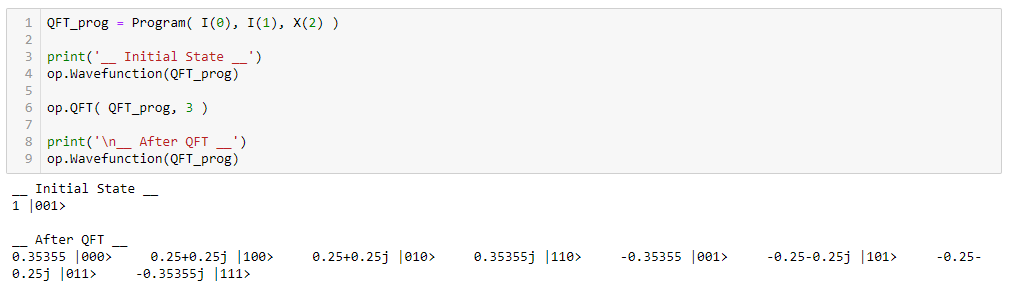}
\end{figure}

\section*{\large{ Why the QFT Circuit Works }}

Now that we have shown that we $\textit{can}$ implement a $QFT$, let's talk about why it works. If you followed along the derivation of the DFT matrix at the beginning of this lesson, then the way we are achieving these operations may seem surprisingly simple. For example, take a look at all of the complexity happening in the 2-qubit $QFT$ matrix from earlier, and then note that we achieve all of this with only 2 $H$'s and one $R_{\phi}$.
\\

To make sense of how these gates are achieving all the desired phases, we will work through a 3-qubit example:

\begin{figure}[h]
\centering
\includegraphics[scale=.65]{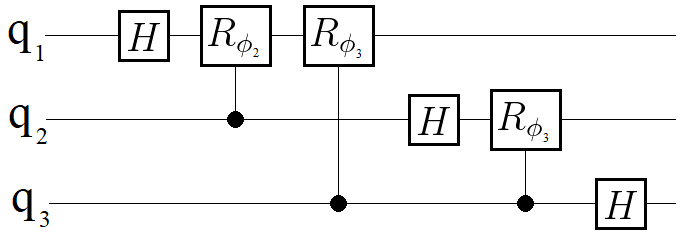}
\end{figure}

In particular, let's start with $q_1$, and see what its final state will look like at the end. We want to be general here, so we will say that our qubit starts off in the state $|q_1\rangle$, where $q_1$ is either a $0$ or $1$. Following along with all of the operations that $q_1$ receives:

$$\hspace{.3cm} H: \hspace{.15cm}\frac{1}{\sqrt{2}}\hspace{.08cm}\big{(}\hspace{.12cm} |\hspace{.06cm}0\hspace{.02cm}\rangle \hspace{.1cm}+\hspace{.1cm} e^{q_1 \cdot i \pi} \hspace{.08cm} |\hspace{.06cm}1\hspace{.02cm}\rangle \hspace{.1cm} \big{)} \hspace{2.3cm}$$

$$\hspace{.2cm} R_{\phi_{2}}: \hspace{.15cm}\frac{1}{\sqrt{2}}\hspace{.08cm}\big{(}\hspace{.12cm} |\hspace{.06cm}0\hspace{.02cm}\rangle \hspace{.1cm}+\hspace{.1cm} e^{q_1 \cdot i \pi} \cdot e^{q_2 \cdot \frac{i \pi}{2}} \hspace{.08cm} |\hspace{.06cm}1\hspace{.02cm}\rangle \hspace{.1cm} \big{)} \hspace{1.3cm}$$

$$R_{\phi_{3}}: \hspace{.15cm}\frac{1}{\sqrt{2}}\hspace{.08cm}\big{(}\hspace{.12cm} |\hspace{.06cm}0\hspace{.02cm}\rangle \hspace{.1cm}+\hspace{.1cm} e^{q_1 \cdot i \pi} \cdot e^{q_2 \cdot \frac{i \pi}{2}} \cdot e^{q_3 \cdot \frac{i \pi}{4}} \hspace{.08cm} |\hspace{.06cm}1\hspace{.02cm}\rangle \hspace{.1cm} \big{)} $$

First, take a look at how we've chosen to write the effect of our Hadamard gate on $q_1$: $\hspace{.08cm}\frac{1}{\sqrt{2}}\hspace{.08cm}\big{(}\hspace{.12cm} |\hspace{.06cm}0\hspace{.02cm}\rangle \hspace{.1cm}+\hspace{.1cm} e^{q_1 \cdot i \pi} \hspace{.08cm} |\hspace{.06cm}1\hspace{.02cm}\rangle \hspace{.1cm} \big{)} $. Typically we would write this with something like $\hspace{.05cm}(-1\hspace{.04cm})^{q_1}$, where the state of $q_1$ determines whether or not the the Hadamard gate results in a positive or negative $|1\rangle$ state. Here however, we've chosen to express $-1$ as $e^{i \pi}$, in order to be consistent with the other gate effects.
\\

Next are the control-phase gates, which produce a similar effect to that of the Hadamard gate at first glance, but have an important difference. Remember that control-phase gates only apply an effect when both the target and control qubits are in the $|1\rangle$ state. This is why a $H$ gate is necessary before any of the $R_{\phi}$'s, to ensure that $q_1$ is in a superposition state of both $|0\rangle$ and $|1\rangle$. Then, effect of the $R_{\phi}$ gate applies an additional phase to the $|1\rangle$ component of $q_1$.
\\

However, because this is a control gate, and we must take into account that $q_2$ and $q_3$ may not be in the $|1\rangle$ state, so there is an extra term multiplying each of the added phases, for example: $e^{q_2 \cdot \frac{i \pi}{2}}$. We can understand this extra term as our condition that $q_2$ is in the $|1\rangle$ state. If it is, then $1$ times the rest of the power will leave it unchanged. But if $q_2$ is in the $|0\rangle$ state, then we will get $e^{0}$, which is just a multiplication of $q_1$ by $1$, meaning that no phase is applied to $q_1$'s $|1\rangle$ component.
\\

This pattern continues for each qubit, all the way down to the last. Each qubit receives a number of phases added to their $|1\rangle$ component, which will then all be multiplied together in the final state:

$$ |\hspace{.08cm}\psi\hspace{.04cm}\rangle_f \hspace{.3cm} = \hspace{.3cm} \big{(} \hspace{.08cm} q_{1f} \hspace{.08cm} \big{)} \otimes \big{(} \hspace{.08cm} q_{2f} \hspace{.08cm} \big{)} \otimes \big{(} \hspace{.08cm} q_{3f} \hspace{.08cm} \big{)} \hspace{11.3cm}$$

$$ \hspace{.3cm} = \hspace{.3cm}\frac{1}{2\sqrt{2}} \hspace{.06cm} \big{(}\hspace{.12cm} |\hspace{.06cm}0\hspace{.02cm}\rangle \hspace{.1cm}+\hspace{.1cm} e^{q_1 \cdot i \pi} \hspace{.08cm} |\hspace{.06cm}1\hspace{.02cm}\rangle \hspace{.1cm} \big{)} \otimes \big{(}\hspace{.12cm} |\hspace{.06cm}0\hspace{.02cm}\rangle \hspace{.1cm}+\hspace{.1cm} e^{q_1 \cdot i \pi} \cdot e^{q_2 \cdot \frac{i \pi}{2}} \hspace{.08cm} |\hspace{.06cm}1\hspace{.02cm}\rangle \hspace{.1cm} \big{)} \otimes \big{(}\hspace{.12cm} |\hspace{.06cm}0\hspace{.02cm}\rangle \hspace{.1cm}+\hspace{.1cm} e^{q_1 \cdot i \pi} \cdot e^{q_2 \cdot \frac{i \pi}{2}} \cdot e^{q_3 \cdot \frac{i \pi}{4}} \hspace{.08cm} |\hspace{.06cm}1\hspace{.02cm}\rangle \hspace{.1cm} \big{)}$$

This is how we are able to achieve all of the various phases shown in the $QFT$ matrices from earlier. Multiplying the states and phases of each qubit together results in our normal $2^N$ states, where each state will be a unique combination of phases, contributed by the $|0\rangle$'s and $|1\rangle$'s that make up the state. The math is still a little cumbersome, even for just three qubits, but hopefully this illustrates the idea behind why we are able to achieve a $QFT$ with this quantum circuit.
\\

As a final optional exercise, I would encourage you to prove for yourself that mathematically our circuit representation is equal to our matrix representation:

$$ \textmf{show that} \hspace{.6cm} H_1 \hspace{.05cm} R_{\phi} \hspace{.05cm} H_0 \hspace{.1cm}|\hspace{.08cm} \psi \rangle \hspace{.4cm} = \hspace{.4cm} \begin{bmatrix} 1 & 1 & 1 & 1 \\ 1 & -1 & i & -i \\ 1 & 1 & -1 & -1 \\ 1 & -1 & -i & i \end{bmatrix} \hspace{.15cm} \begin{bmatrix} |\hspace{.08cm} 00\rangle \\ |\hspace{.08cm} 01\rangle \\ |\hspace{.08cm} 10\rangle \\ |\hspace{.08cm} 11\rangle \end{bmatrix} \hspace{1cm} \phi = \frac{\pi}{2}$$

$$ \textmf{hint:} \hspace{.2cm} \textmf{don't forget to represent }H_0 \textmf{ and } H_1 \textmf{ as 4x4 matrices!} \hspace{.5cm} \longrightarrow \hspace{.5cm} H_0 \equiv H_0 \otimes I_1$$

\section*{\large{ Inverse QFT }}

Now that we have a way of transforming our system via a $QFT$, and hopefully a better intuition as to why it works, next we need to be able to transform back. As we mentioned earlier, the power of using transformations in quantum algorithms relies on being able to transform back and forth between bases. And as we've also mentioned already, our $QFT$ transformation is not Hermitian, so the same construction of gates will not transform us back.
\\

Just to verify this, let's try to use our $\textmf{QFT}$ function twice:

\begin{figure}[h]
\centering
\includegraphics[scale=.65]{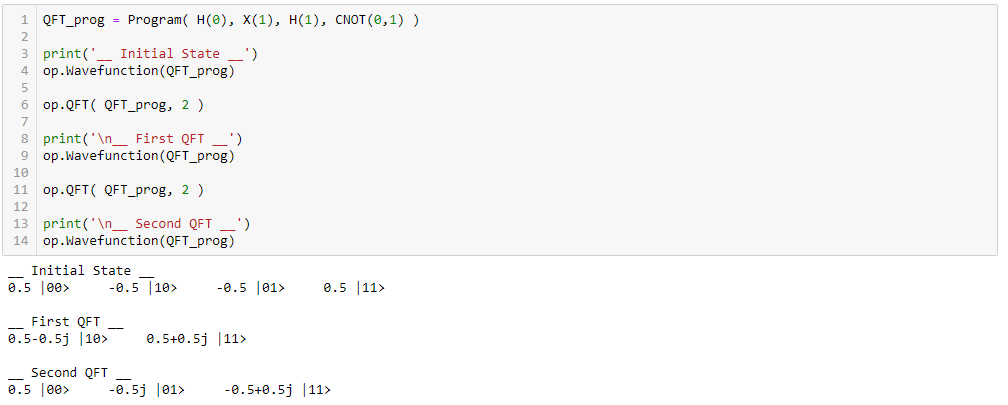}
\end{figure}

Sure enough, we do not return to our original state. From our quantum computing perspective, we can understand why the $QFT$ doesn't transform us back to our original state if we look at two $QFT$s in a row:

\begin{figure}[h]
\centering
\includegraphics[scale=.65]{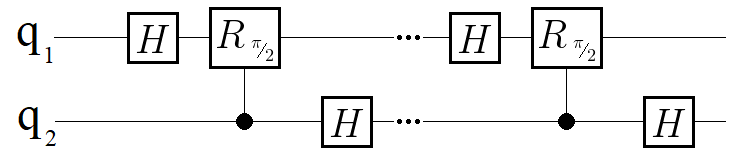}
\end{figure}

What should jump out at you is the apparent lack of symmetry here. Recall our example earlier of the gate $XZ$, and that the correct inverse transformation was to change the order: $ZX$. Here, if we want to implement the inverse of our $QFT$, we will need to invoke the same strategy of reversing the order of all the gates. In essence, imagine placing a mirror after our $QFT$, and the reflection will be our inverse $QFT$, with one slight change:

\begin{figure}[h]
\centering
\includegraphics[scale=.65]{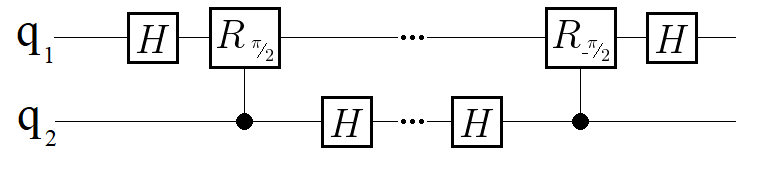}
\end{figure}

The slight change here is that our second $R_{\phi}$ has the opposite sign of our first. Conceptually, this should make sense: if our original transformation applies a phase $\theta$, then our inverse should apply the opposite phase, -$\theta$. As we pointed out earlier, the inverse of a transformation needs to be the complex conjugate of the original, which is why we need negative phases on all of the $\theta$'s. All together, our inverse $QFT$ must be the $\textit{exact}$ reverse ordering our $QFT$, with all opposite phases on the $R_{\phi}$ gates:

\begin{figure}[h]
\centering
\includegraphics[scale=.65]{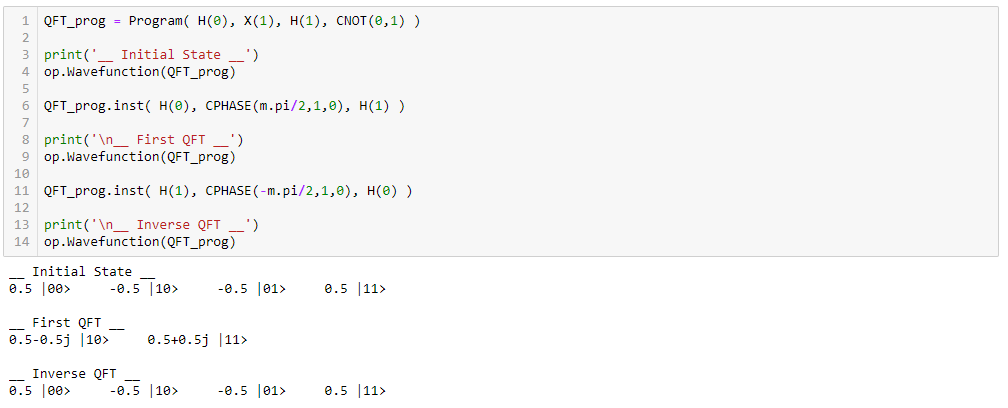}
\end{figure}

Sure enough, we recover our original state, which means that we performed the correct inverse transformation. And like our $\textmf{QFT}$ function, we can use $\textbf{QFT\_dgr}$ from $\textmf{Our\_Pyquil\_Functions}$ to implement our inverse $QFT$:

\begin{figure}[h]
\centering
\includegraphics[scale=.65]{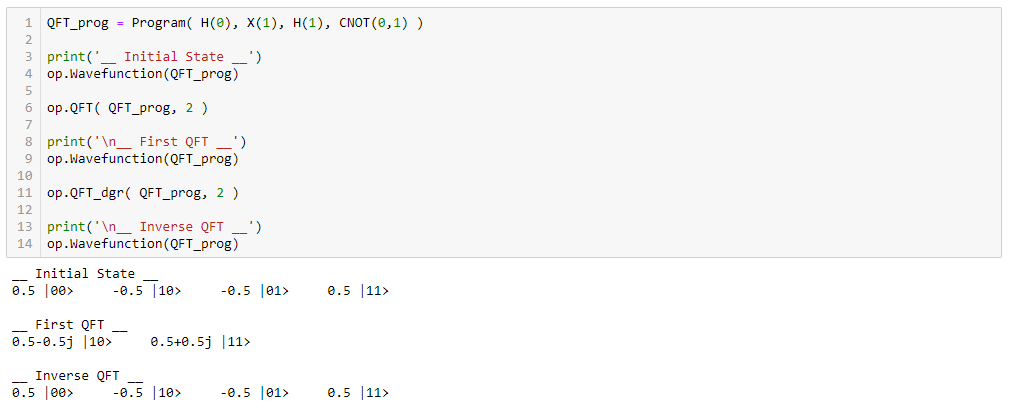}
\end{figure}

Now that we have $\textmf{QFT}$ and $\textmf{QFT\_dgr}$, we are finished covering the basics of the Quantum Fourier Transformation. In the next couple lessons, we will be using these $QFT$'s as the basis for some very important algorithms. If you would like to proceed to those lessons now, this is a sufficient concluding spot in the tutorial. The next and final section is an aside about the $QFT$, comparing some of its properties to the Hadamard transformation.

\section*{\large{ Aside: Comparing QFT and H Transformations }}
\centerline{---------------------------------------------------------------------------------------------------------------------------------}

Now that we have built up an understanding of how to use a $QFT$, let's discuss its similarities with the Hadamard Transformation. First off, if you remove all of the $R_{\phi}$ gates from the $QFT$ template, you're left with just a Hadamard Transformation. And in fact, our last qubit in the system only receives a single $H$. What this means, is that we can think of the $QFT$ as a 'more complex' version of a Hadamard Transformation in some sense, where the extra bit of complexity is the additional phases. To see this, let's compare the 4$\times$4 unitary matrices for the $QFT$ and Hadamard Transformation on two qubits:

$$ QFT \hspace{5.2cm} H \hspace{.5cm} $$

$$ \begin{bmatrix} 1 & 1 & 1 & 1 \\ 1 & i & -1 & -i \\ 1 & -1 & 1 & -1 \\ 1 & -i & -1 & i \end{bmatrix} \hspace{3cm} \begin{bmatrix} 1 & 1 & 1 & 1 \\ 1 & 1 & -1 & -1 \\ 1 & -1 & 1 & -1 \\ 1 & -1 & -1 & 1 \end{bmatrix} $$

The two transformations are nearly identical, except for the extra presences of a couple $i$'s in the $QFT$. These $i$'s represent the extra complexity of the $QFT$ for the 2-qubit case. And when we look at larger transformations, we will see more and more unique amplitudes accompanying states in the system.
\\

However, regardless of size, one property that both the Hadamard Transformation and $QFT$ share is the way they map the state of all $0$'s:

$$ |\hspace{.08cm} 00...0 \rangle \hspace{1cm} \longleftrightarrow \hspace{1cm} \frac{1}{\sqrt{2^N}} \big{(} \hspace{.14cm} |\hspace{.08cm} 00...0 \rangle \hspace{.1cm} + \hspace{.1cm} \cdot \cdot \cdot \hspace{.1cm} + \hspace{.1cm} |\hspace{.08cm} 11...1 \rangle \hspace{.1cm} \big{)} $$

Both transformations map the state of all $0$'s to an equal superposition, where all the states have the same positive phase. For $H^N$, we've shown that this result comes from the fact that $H\hspace{.04cm}|0\rangle$ and $H\hspace{.04cm}|1\rangle$ both produce a state where the $|0\rangle$ component is positive. Similarly, if we return to our earlier example where we broke down all of the gate operations for the 3-qubit QFT, we get the exact same result. Because each qubit initially receives a $H$ gate followed by all control-gates, the $|0\rangle$ component for every qubit will always be positive. Simultaneously, since we are dealing with the state $|\hspace{.05cm}00...0\rangle$, none of the $R_{\phi}$ are applying any phases.
\\

This mapping was the core ingredient for the the Grover Algorithm. Specifically, we used this mapping of $|\hspace{.05cm}00...0\rangle$ as our way of achieving a reflection about the average. Thus, since our $QFT$ also has this mapping property, we should be able to perform the Grover Algorithm using a $QFT$ in place of the $H^N$ transformations:

\pagebreak

\begin{figure}[h]
\centering
\includegraphics[scale=.65]{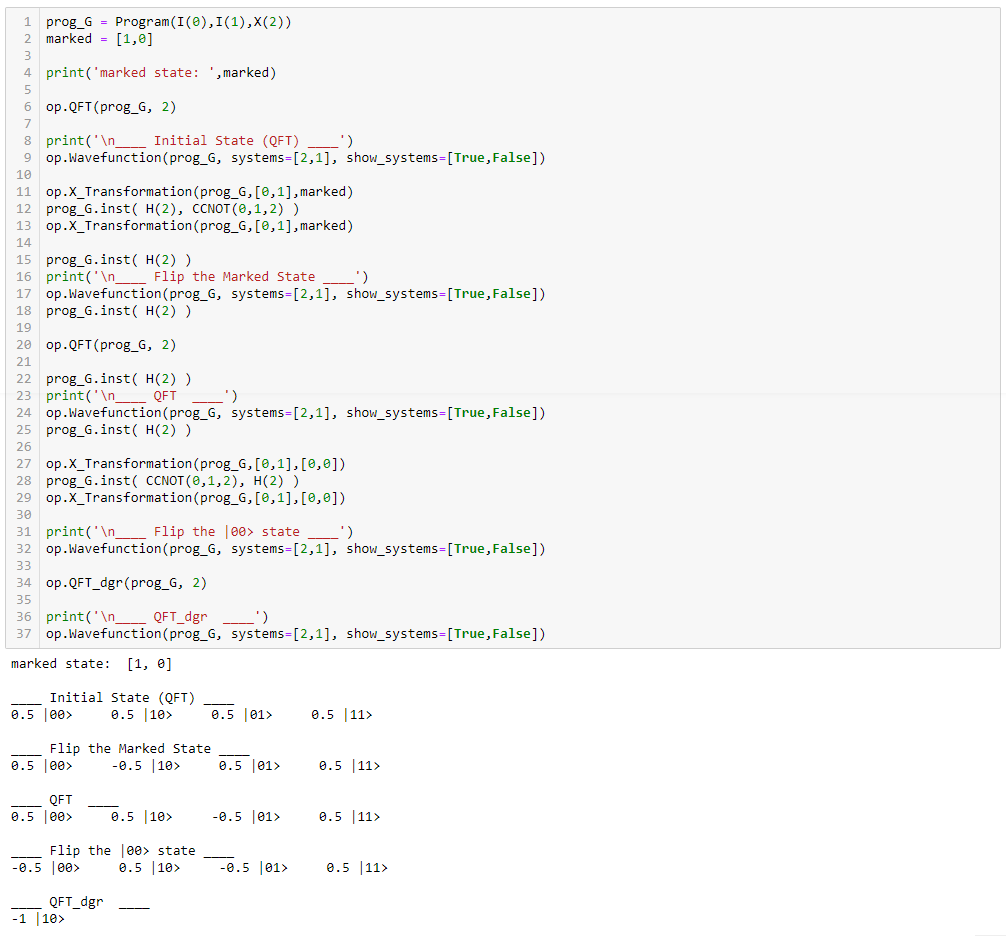}
\end{figure}

Success! By using the $QFT$ and inverse $QFT$, we are able to perform a Grover Search for a marked state. For an explanation on the Grover Algorithm, please refer to lesson 5.4. Note that in this coding example there are a lot of added steps, in order to display all of the individual steps nicely.
\\

Hopefully this example gives you an idea of just how similar the $QFT$ and Hadamard transformation are at their core. But, the reason we will be able to use the $QFT$ to accomplish some more complex algorithms, comes from the fact that the states it maps to contain more phase differences. Or another way of saying that is, the $QFT$ allows us to create 'more orthogonal' states (not literally), where the extra phases will prove very useful.
\\

--------------------------------------------------------------------------------------------------------------------------------------------------------
\\

This concludes lesson 6 and all of the tutorials in this series! Understanding the $QFT$ is a bit tricky at first, so don't worry if everything in this tutorial doesn't feel second nature yet. Just like all of the practice we got with the Hadamard Transformation in lessons 5.1 - 5.4, you will need to see the $QFT$ in action a few times to truly understand and appreciate its role in quantum algorithms
\\

--------------------------------------------------------------------------------------------------------------------------------------------------------
\\

This concludes all of the lessons in this tutorial series, but there is still much to learn about quantum algorithms! If you are looking to continue your learning endeavors, I encourage you to take a look at the following more advanced algorithms:
\\

-Quantum Phase Estimation
\\

-Shor's Algorithms
\\

-Variational Quantum Eigensolver (VQE)
\\

-Quantum Approximate Optimization Algorithm (QAOA)
\\

In addition to these, there are many many more quantum algorithms available at The Quantum Algorithm Zoo:
\\

http://quantumalgorithmzoo.org/

\pagebreak

\pagebreak

\section*{\Large{ Appendix }}
\centerline{---------------------------------------------------------------------------------------------------------------------------------}

Below are all of the functions contained within the accompanying python file $\textmf{Our\_Pyquil\_Functions.py}$. These functions were written to be compatible with Pyquil 2.0

\begin{figure}[h]
\centering
\includegraphics[scale=.65]{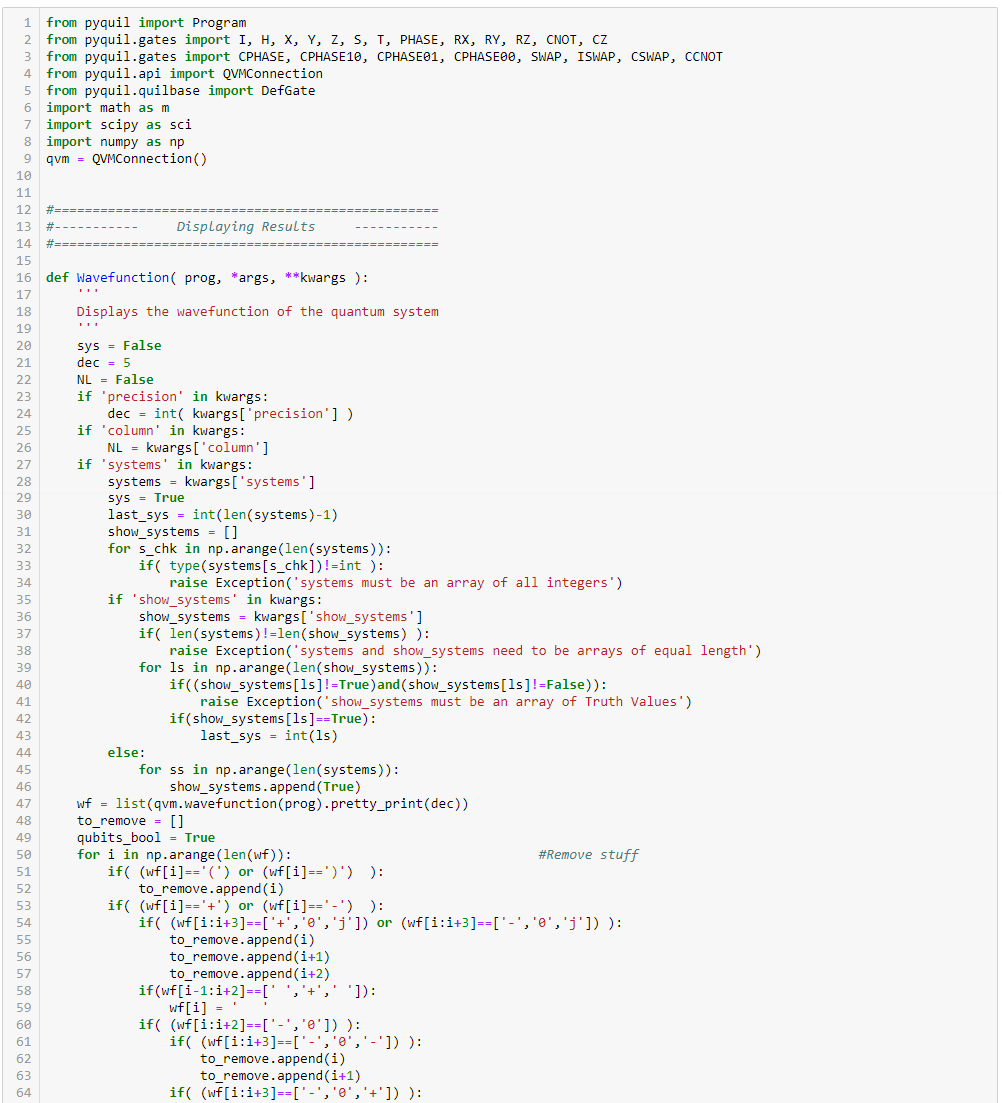}
\end{figure}

\begin{figure}[h]
\centering
\includegraphics[scale=.65]{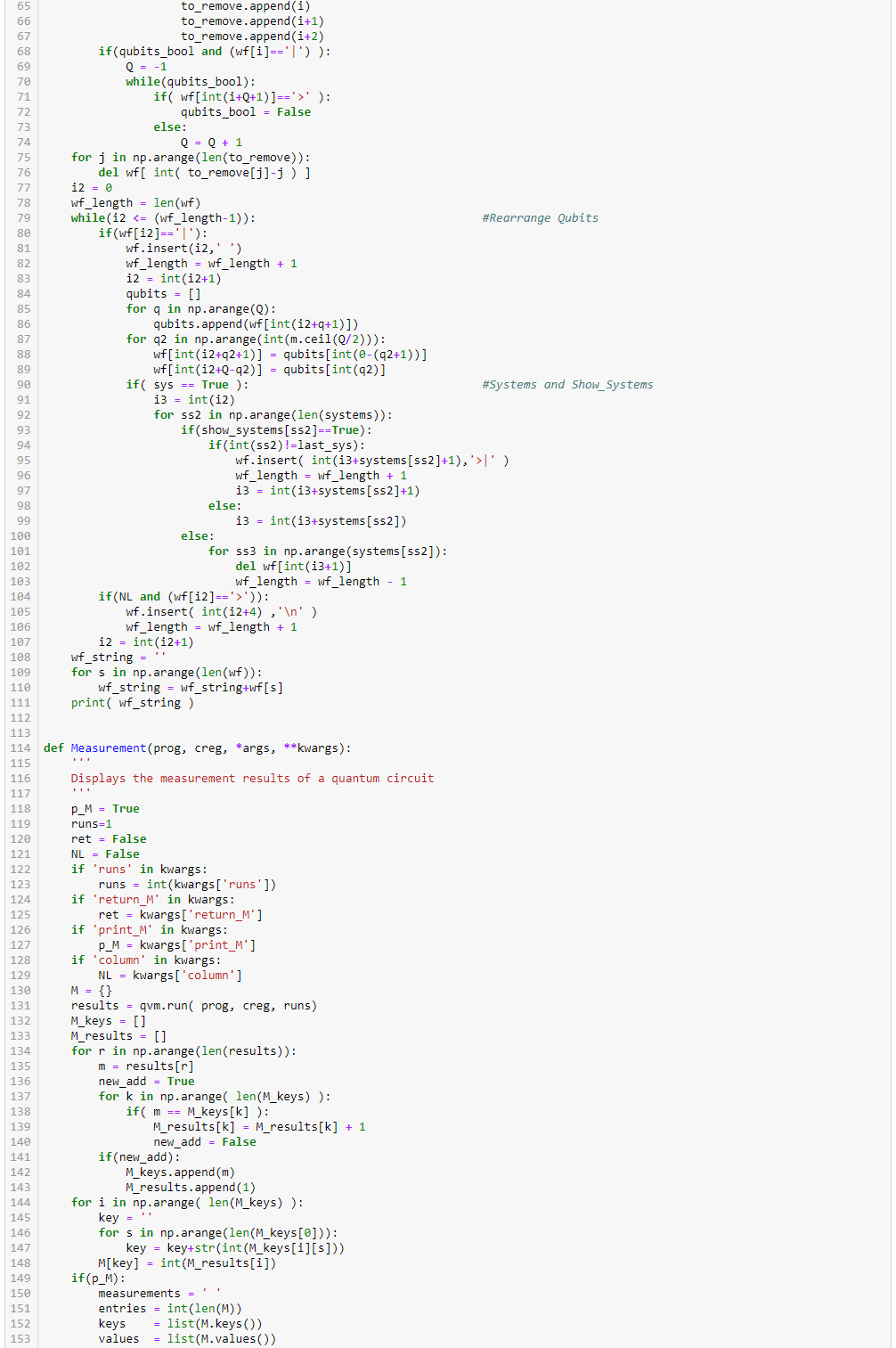}
\end{figure}

\begin{figure}[h]
\centering
\includegraphics[scale=.65]{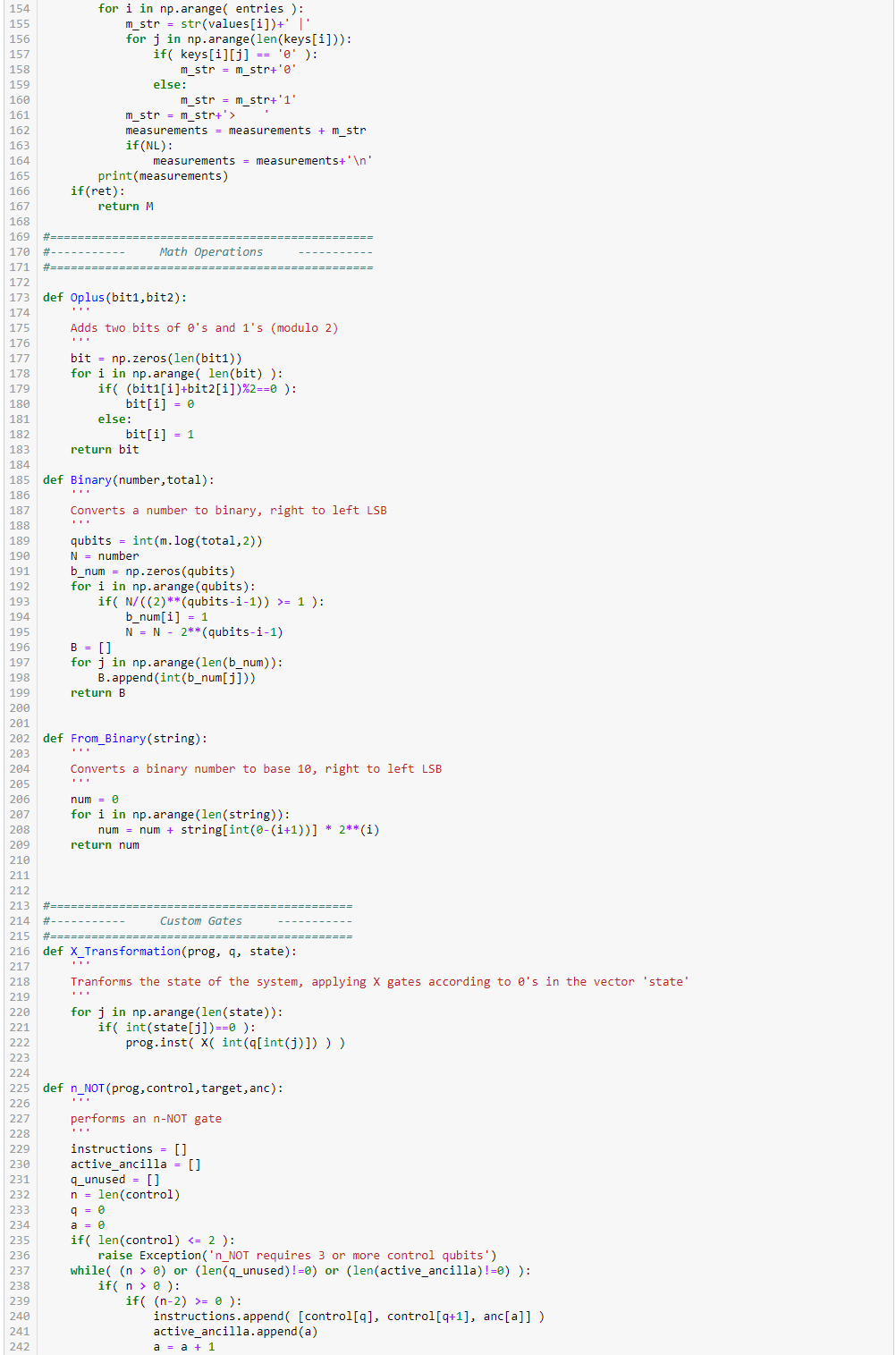}
\end{figure}

\begin{figure}[h]
\centering
\includegraphics[scale=.65]{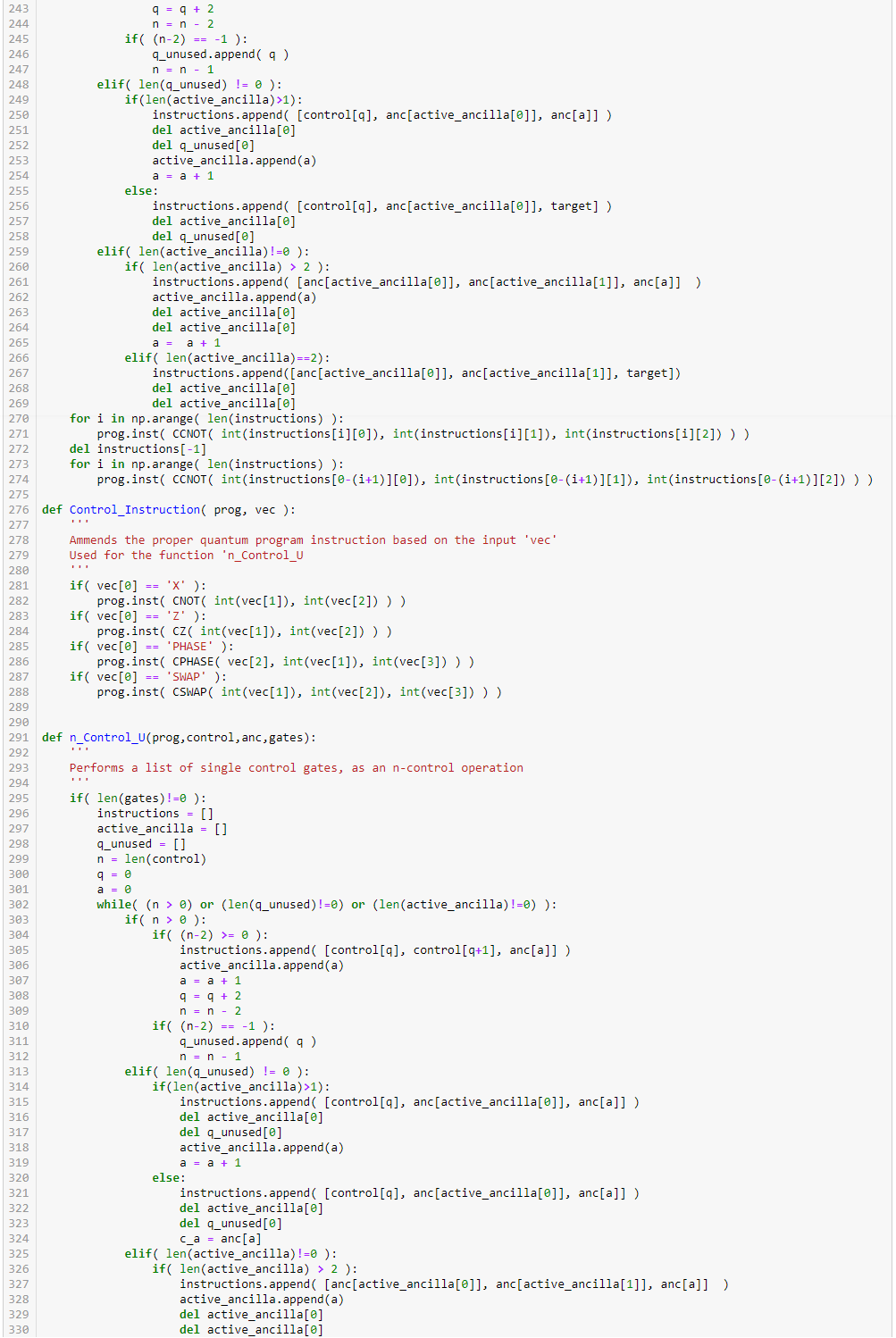}
\end{figure}

\begin{figure}[h]
\centering
\includegraphics[scale=.65]{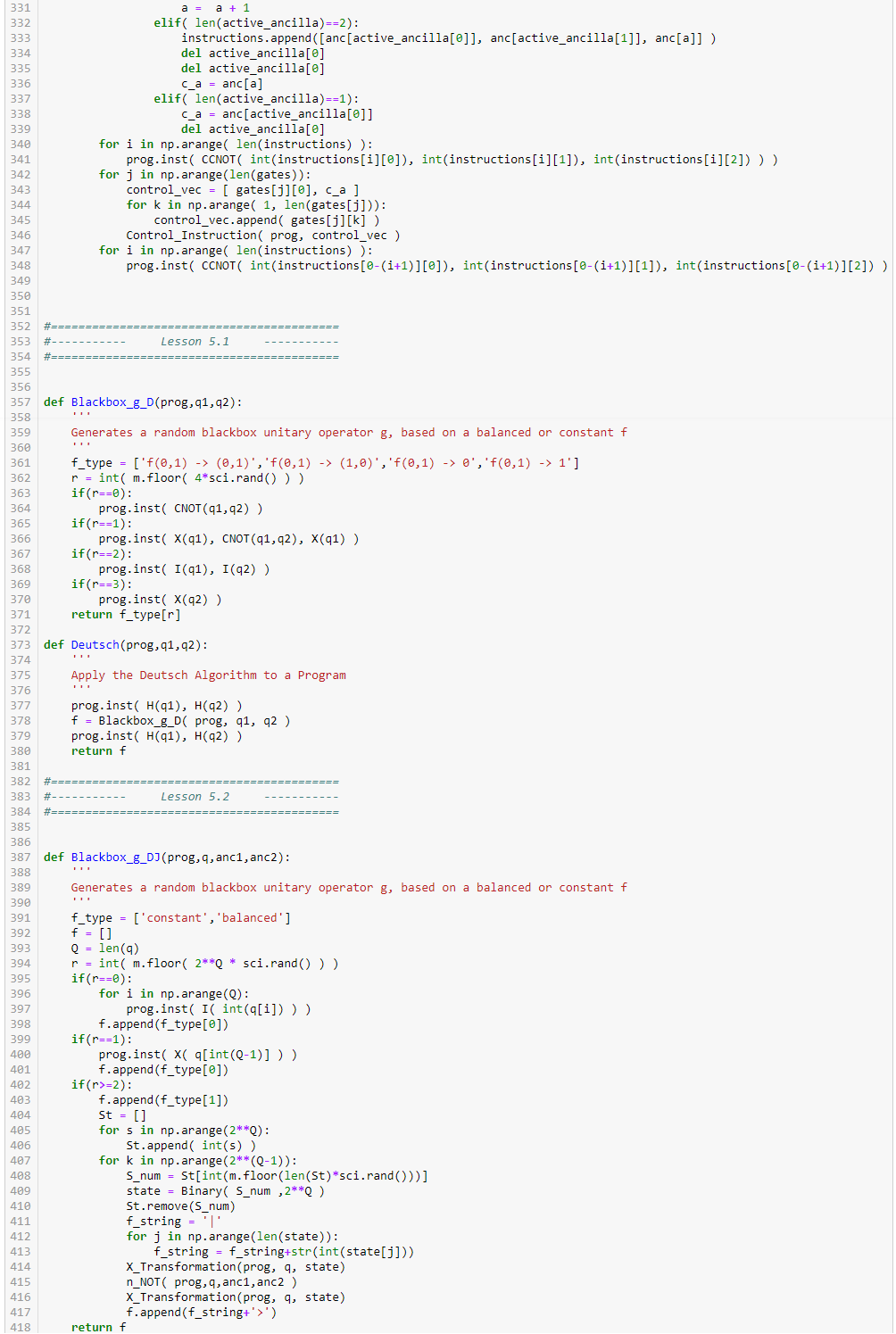}
\end{figure}

\begin{figure}[h]
\centering
\includegraphics[scale=.65]{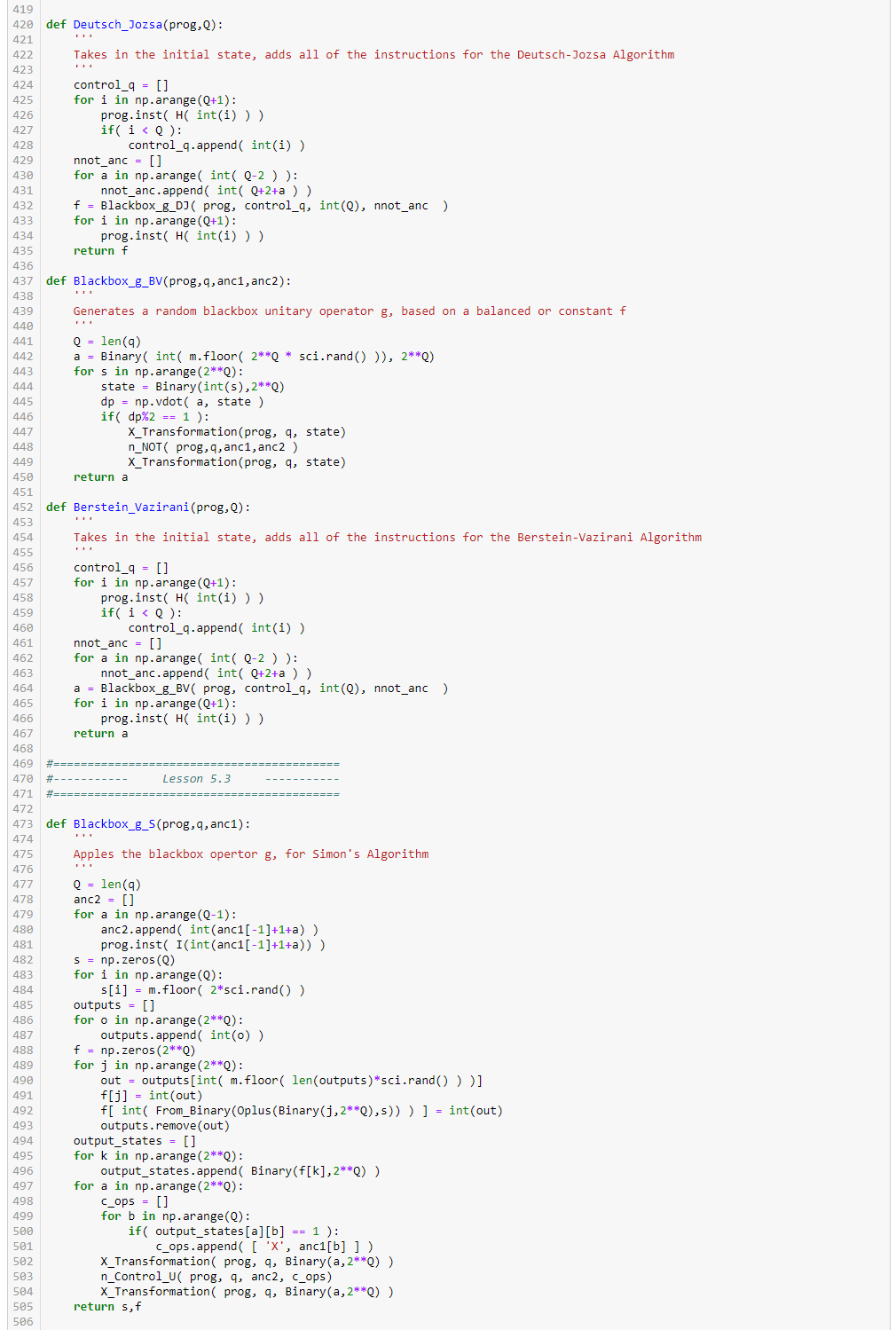}
\end{figure}

\begin{figure}[h]
\centering
\includegraphics[scale=.65]{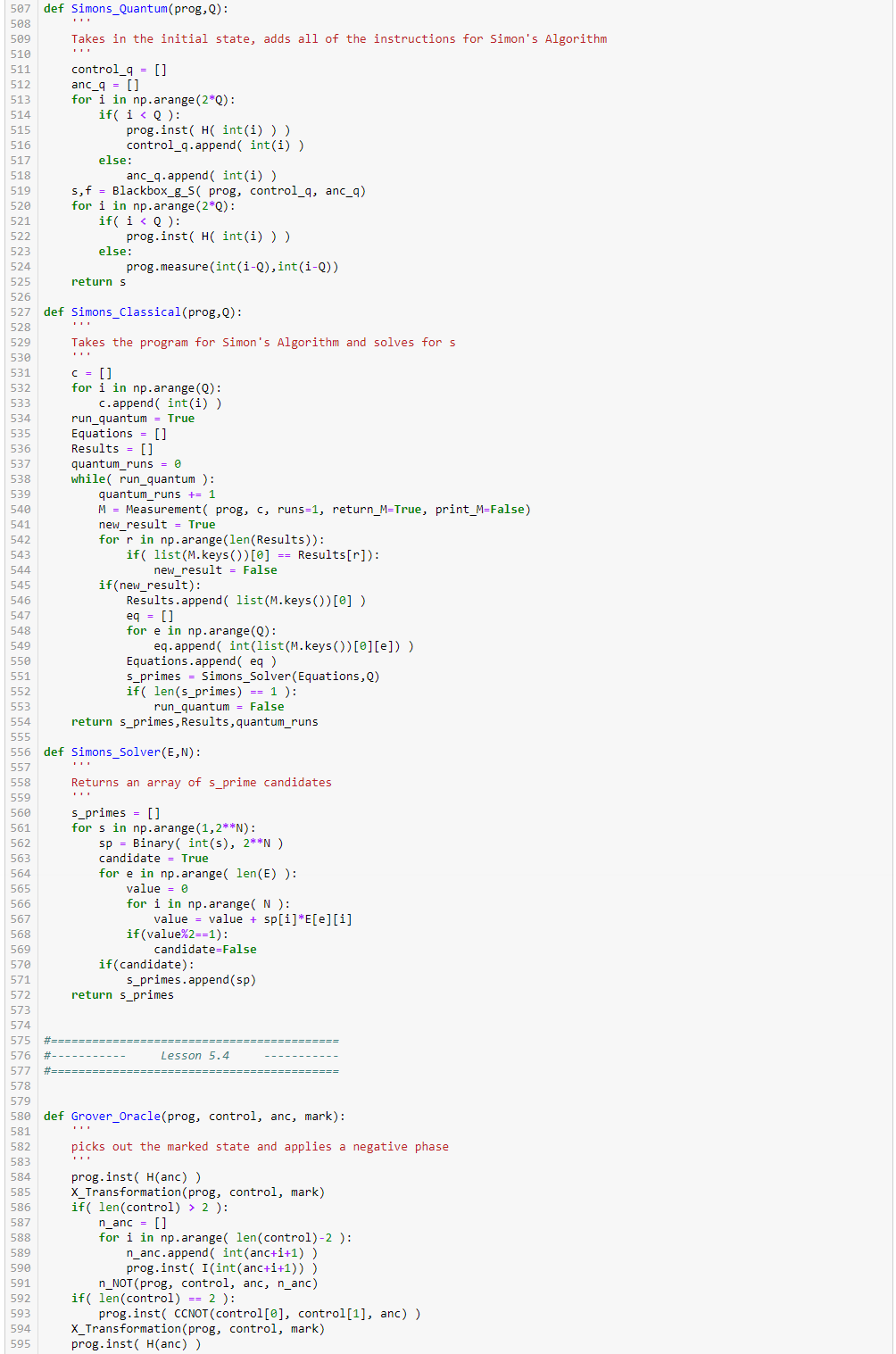}
\end{figure}

\begin{figure}[h]
\centering
\includegraphics[scale=.65]{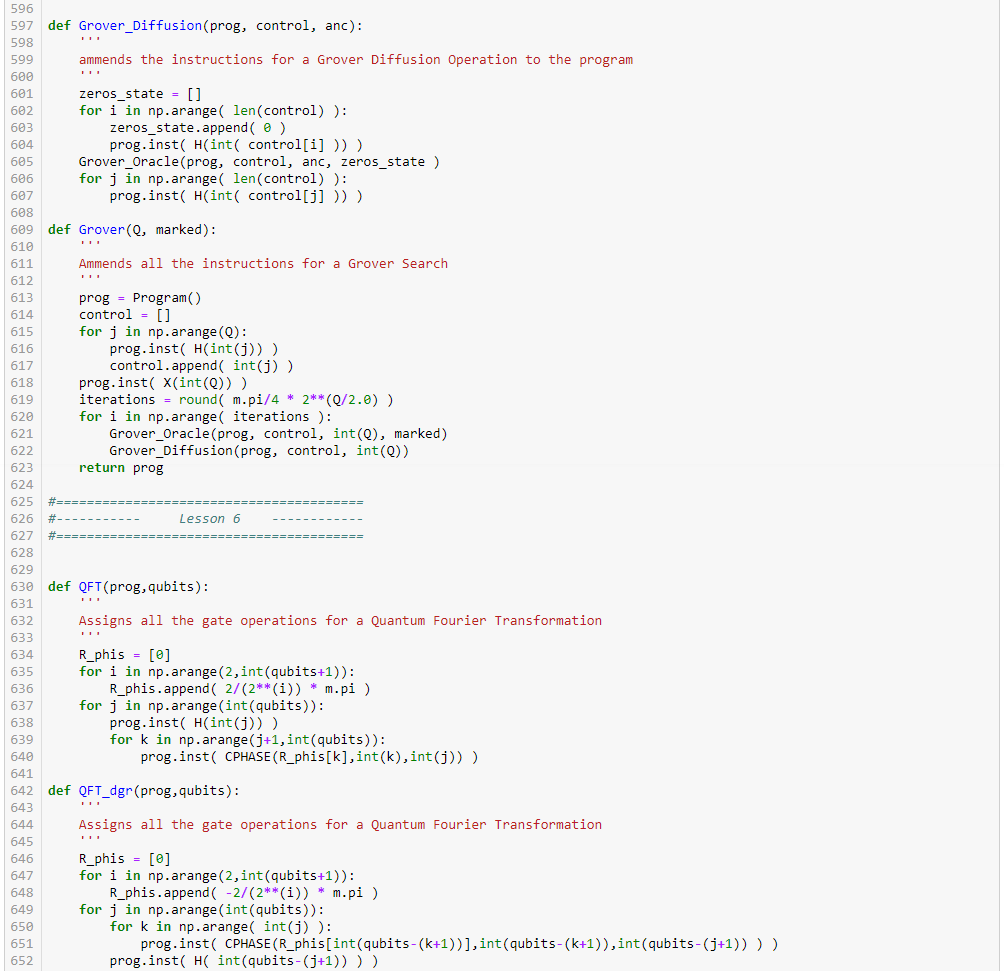}
\end{figure}

\end{document}